\documentclass[12pt,twoside,a4paper]{article}
\pdfoutput=1

\usepackage{amsmath, amsthm, amsfonts, amssymb}
\usepackage{graphicx}
\usepackage{multirow}
\usepackage{float}
\usepackage{color}
\usepackage{cite}
\usepackage[width=\textwidth,font=small,labelfont=bf,
labelsep=endash]{caption}
\usepackage{titlesec}

\titleformat{\section}
  {\normalfont\fontsize{13}{16}\bfseries}{\thesection}{1em}{}
\titleformat{\subsection}
  {\normalfont\fontsize{11}{14}\bfseries}{\thesubsection}{1em}{}

\normalsize
\newcommand\Lame {Lam\'e\ }
\newcommand\Backlund {B\"{a}cklund }

\newcommand\tr {\mathrm{Tr}}
\newcommand\diag {\mathrm{diag}}
\newcommand\sign {\mathrm{sgn}}

\newcommand\lcm {\mathrm{lcm}}

\renewcommand{\theequation}{\arabic{section}.\arabic{equation}}

\begin{document}

\title{\textbf{Salient Features of Dressed Elliptic String Solutions on $\mathbb{R}\times$S$^2$}}
\author{Dimitrios Katsinis$^{1,2}$, Ioannis Mitsoulas$^3$ and Georgios Pastras$^2$}
\date{\small $^1$Department of Physics, National and Kapodistrian University of Athens,\\University Campus, Zografou, Athens 15784, Greece\\
$^2$NCSR ``Demokritos'', Institute of Nuclear and Particle Physics,\\Aghia Paraskevi 15310, Attiki, Greece\\
$^3$Department of Physics, School of Applied Mathematics and Physical Sciences,\\National Technical University, Athens 15780, Greece\linebreak \vspace{8pt}
\texttt{dkatsinis@phys.uoa.gr, mitsoula@central.ntua.gr, pastras@inp.demokritos.gr}}

\vskip .5cm

\maketitle

\begin{abstract}
We analyse several physical aspects of the dressed elliptic strings propagating on $\mathbb{R} \times \mathrm{S}^2$ and of their counterparts in the Pohlmeyer reduced theory, i.e. the sine-Gordon equation. The solutions are divided into two wide classes; kinks which propagate on top of elliptic backgrounds and those which are non-localised periodic disturbances of the latter. The former class of solutions obey a specific equation of state that is in principle experimentally verifiable in systems which realize the sine-Gordon equation. Among both of these classes, there appears to be a particular class of interest the closed dressed strings. They in turn form four distinct subclasses of solutions. Unlike the closed elliptic strings, these four subclasses, exhibit interactions among their spikes. These interactions preserve a carefully defined turning number, which can be associated to the topological charge of the sine-Gordon counterpart. One particular class of those closed dressed strings realizes instabilities of the seed elliptic solutions. The existence of such solutions depends on whether a superluminal kink with a specific velocity can propagate on the corresponding elliptic sine-Gordon solution. Finally, the dispersion relations of the dressed strings are studied. A qualitative difference between the two wide classes of dressed strings is discovered. This would be an interesting subject for investigation in the dual field theory.
\end{abstract}

\newpage

\tableofcontents

\newpage

\setcounter{equation}{0}
\section{Introduction}
\label{sec:introduction}

Classical string solutions \cite{GKP_string,BMN,Giant_Magnons,Dyonic_Giant_Magnons,single_spike_rs2,dual_spikes,spiky_string,helical,multi,Kruczenski:2006pk,tseytlin_review} have enlightened several interesting features of the holographic duality \cite{ads_malda,ads_GKP,ads_witten} and have provided a framework for non-trivial checks of its validity. Classical string solutions in symmetric spaces, relevant to holography, such as the sphere or the AdS spacetime, as well as their tensor product, have been the subject of extensive study in the literature. Furthermore, the study of such solutions facilitates the qualitative understanding of the classical dynamics of the system whose quantum version is the only known consistent quantum theory of gravity.

The sigma models that describe string propagation on symmetric spaces have the additional interesting feature of integrability, which provides several non-trivial tools for the construction of string solutions. Taking advantage of integrability methods, one can find general expressions for string solutions on specific symmetric spaces, in terms of hyperelliptic functions \cite{Kazakov:2004qf,Beisert:2005bm}. These suggest a natural classification of the solution in terms of the genus of the relevant algebraic curve. Although this kind of treatment has the advantage of being very generic, the understanding of the physical properties of the solutions in this language is rather limited. This is due to fact that the behaviour of the hyperelliptic functions is not as well documented as that of simpler elliptic or trigonometric/hyperbolic functions. As an exception the genus one solutions, i.e. elliptic solutions, can be expressed in terms of elliptic functions and their properties have been extensively studied.

Another signature of the system's integrability is the fact that the sigma models, which describing the propagation of strings in symmetric spaces, are reducible to integrable systems, which belong to the family of the sine-Gordon equation, the so called symmetric space sine-Gordon models (SSSG) \cite{Barbashov:1980kz,DeVega:1992xc,Larsen:1996gn,Grigoriev:2007bu}. This property is formally called \emph{the Pohlmeyer reduction} \cite{Pohlmeyer:1975nb,Zakharov:1973pp}. It can, in principle, facilitate the study of string propagation, as the SSSGs have been quite extensively studied in the literature. Many non-trivial solutions of those are known. However, the Pohlmeyer reduction is a non-local, many-to-one mapping that is difficult to invert.

The Pohlmeyer reduction sheds new light on another interesting property of the aforementioned sigma models. It is possible to construct new string solutions given an initial seed one, by means of solving a simpler auxiliary system instead of the original equations of motion. This procedure is the so called ``dressing method'' \cite{Zakharov:1980ty,Harnad:1983we,Pohl_avatars} and it is the analog of the \Backlund transformations on the side of the reduced integrable theory. The dressing transformation, as well as the \Backlund transformations add one extra genus on the initial solution. This is a degenerate one, as one of the related periods is divergent.

Although the Pohlmeyer reduction is difficult to invert, a systematic approach for its inversion in the case of elliptic solutions has been developed recently \cite{bakas_pastras}, in the case of strings propagating on AdS$_3$ or dS$_3$. We have extended the method for the case of strings propagating on the sphere $\mathrm{S}^2$ \cite{part1}. It was shown that the method leads to a unified and simple description of all elliptic solutions in terms of the Weierstrass elliptic function. In a subsequent work \cite{Katsinis:2018ewd}, we took advantange of this simple decription to construct dressed elliptic solutions through the dressing method, i.e. degenerate genus two solutions. An advantage of this construction is the expression of the solutions in terms of simple elliptic and trigonometric functions, which makes many of their physical properties accessible to study. In the present work we focus on salient aspects of the above solutions, such as spike interactions, implications to the stability of the seed solutions and their dispersion relations.

An interesting feature of the elliptic string solutions is the fact that they have several singular points, which are spikes. These can be kinematically understood, as points of the string that propagate at the speed of light \cite{GKP_string} due to the initial conditions. As they cannot change velocity, no matter what the forces are which are exerted on them, they continue to exist indefinitely, as long as they do not interact with each other. In the already studied spiky string solutions \cite{dual_spikes,spiky_string,multi,helical,Kruczenski:2006pk,part1}, the spikes rotate around the sphere with the same angular velocity, and thus, never interact. Interacting spikes emerge in higher genus solutions. The simplest possible such solutions are those which are constructed via the dressing of elliptic strings \cite{Katsinis:2018ewd}.

Furthermore, the stability of the elliptic strings is closely related to the stability of their Pohlmeyer counterparts, which are trains of kinks or trains of kinks-antikinks. Although the later is known \cite{SGstability}, it is not easy to construct an explicit non-perturbative solution exposing the instability of the elliptic strings. Naively, such a solution has to be a degenerate genus two solution. In this case, one of the two periods must coincide to the periodicity of the original elliptic solution under study. On the other hand, the degenerate one will describe the infinite evolution which either asymptotically leads to or away from the elliptic solution. Therefore, the dressed elliptic strings are conducive to the determination and study of the instabilities of the elliptic ones.

The structure of the present paper is as follows: In section \ref{sec:review}, we review some elements of the construction of the dressed elliptic strings on $\mathrm{S}^2$ that are necessary for the study of their physical properties. In section \ref{sec:SG_properties}, we elucidate the properties of the sine-Gordon counterparts of the dressed elliptic string solutions, in order to both facilitate the study of the latter and furthermore establish a mapping between the properties of the string solutions and their counterparts. In section \ref{sec:string_asymptotics}, we study the constraints which have to be imposed on the dressed string solutions, so that they are closed. In effect they emerge to belong to four distinct classes. In section \ref{sec:spike_interactions} we study the time evolution of the string solutions focusing on the interaction of spikes. In section \ref{sec:instabilities} we study a specific class of dressed string solutions that reveals instabilities of a subset of the elliptic string solutions. In section \ref{sec:dispersion}, we calculate the energy and angular momentum of the dressed elliptic strings, which have great interest in the context of the holographic dualities. In section \ref{sec:discussion} we discuss our results. Finally, there is an appendix containing some technical details on the asymptotic behaviour of the dressed strings and the calculation of the conserved quantities.

\setcounter{equation}{0}
\section{Review of Dressed Elliptic String Solutions}
\label{sec:review}

String propagation in symmetric spaces can be described by a sigma model action, whose target space $\Sigma$ is the respective symmetric space, supplemented by the Virasoro constrains. In this work, we focus on the case $\Sigma=\mathbb{R}^t \times$S$^2$, namely strings propagating on a two-dimensional sphere.  The target space can be embedded in a higher dimensional flat space, namely, $\mathbb{R}^{(1,3)}$. Then, the sigma model action reads
\begin{equation}
S = \int d\xi^+ d\xi^- \left( \left( \partial_+ X \right)\cdot \left( \partial_- X \right) + \lambda \left( \vec X \cdot \vec X - 1 \right) \right) ,
\label{eq:Pohlmeyer_action}
\end{equation}
where $\xi_\pm \equiv \left( \xi^1 \pm \xi^0 \right) / 2$ are the usual light-cone worldsheet coordinates. Four-vectors are denoted by $X$, whereas the notation $\vec X$ is used for the three-vector composed by the spatial components of $X$. The inner product of two four-vectors $A$ and $B$ with respect to the Minkowski metric $g = \diag \{ -1 ,1 ,1 ,1 \}$ is denoted as $A\cdot B$.

A well-known method for constructing solutions of the equations of motion of \eqref{eq:Pohlmeyer_action} is the so-called dressing method \cite{Zakharov:1980ty,Harnad:1983we,Pohl_avatars}, which connects solutions of the sigma model equations in pairs. Given a solution of the latter, henceforth called the seed solution, one can apply the above method in order to generate a new non-trivial one \cite{Combes:1993rw,Spradlin:2006wk,Kalousios:2006xy,Jevicki:2007pk,Jevicki:2007aa}. Recently, an application of the dressing method appeared \cite{Katsinis:2018ewd}, where elliptic string solutions \cite{part1} were used as seeds. The study of the physical properties of the above dressed solutions is the subject of this work.

The previously mentioned elliptic string solutions on $\mathbb{R}^t \times$S$^2$ were obtained in \cite{part1} through the inversion of Pohlmeyer reduction. For this purpose, it is convenient to adopt a more general gauge selection than the usual static gauge, the linear gauge, i.e. $X^0 = m_+ \xi^+ + m_- \xi^-$, with $m_\pm$ constants. In this gauge, the equations of motion of the sigma model \eqref{eq:Pohlmeyer_action} can be mapped to the sine-Gordon equation. This mapping is known as Pohlmeyer reduction and in general it is non-invertible. However, a systematic way of inverting this mapping was developed in \cite{part1}, for the case of elliptic solutions of the sine-Gordon equation. The latter can be written in a form depending on only one of the worldsheet coordinates $\xi^0$ or $\xi^1$ and will be called translationally invariant and static respectively. The translationally invariant solutions read,
\begin{equation}
\varphi \left(\xi^0; E \right) = \begin{cases}
{\left( { - 1} \right)^{\left\lfloor {\frac{\xi^0}{{2{\omega _1}}}} \right\rfloor }}\arccos \left( - {\frac{{2\wp \left( {\xi^0 + {\omega _2}} \right) + \frac{E}{3}}}{{{\mu ^2}}}} \right), & E < \mu^2 , \\
{\left( { - 1} \right)^{\left\lfloor {\frac{\xi^0}{{{\omega _1}}}} \right\rfloor }}\arccos \left( - {\frac{{2\wp \left( {\xi^0 + {\omega _2}} \right) + \frac{E}{3}}}{{{\mu ^2}}}} \right) + 2\pi \left\lfloor {\frac{{\xi^0} + {\omega _1}}{{2{\omega _1}}}} \right\rfloor , & E > \mu^2,
\end{cases}
\label{eq:elliptic_solution_phi_0}
\end{equation}
where $\mu^2=-m_+m_-$. Interchanging $\xi^0$ with $\xi^1$ in \eqref{eq:elliptic_solution_phi_0}, accompanied by an overall shift of the solution by $\pi$, yields the static solutions. The parameter $E$ is an integration constant that can take any value larger than $-\mu^2$. The solutions with $E<\mu^2$ are called oscillatory, whereas the ones with $E>\mu^2$ are called rotating, in an analogy to the simple pendulum that was established in \cite{part1}. Clearly there are four classes of solutions characterized as translationally invariant-oscillating or -rotating and static-oscillating or -rotating.

The resulting string solutions are expressed in terms of the Weierstrass functions $\wp$, $\sigma$ and $\zeta$. The associated cubic polynomial has always three real roots, and, thus, the one of the periods of the elliptic function $\wp$ is real, whereas the other one is purely imaginary. In the following, they will be denoted as $2\omega_1$ and $2\omega_2$, respectively. The string solutions read,
\begin{align}
t_{0/1} &= \sqrt {{x_2} - \wp \left( a \right)} {\xi ^0} + \sqrt {{x_3} - \wp \left( a \right)} {\xi ^1} , \label{eq:elliptic_solutions_t} \\
\vec X_{0/1} &= \left( {\begin{array}{*{20}{c}}
{{F_1}\left( {{\xi^{0/1}}} \right)\cos \left(\ell {\xi^{1/0}} - \Phi \left( \xi^{0/1} ; a \right)\right)}\\
{{F_1}\left( {{\xi^{0/1}}} \right)\sin \left(\ell {\xi^{1/0}} - \Phi \left( \xi^{0/1} ; a \right)\right)}\\
{{F_2}\left( {{\xi^{0/1}}} \right)}
\end{array}} \right) ,
\label{eq:elliptic_solutions_review}
\end{align}
where
\begin{equation}
{F_1}\left( {{\xi}} \right) = \sqrt {\frac{{\wp \left( {{\xi} + {\omega _2}} \right) - \wp \left( a \right)}}{{{x_1} - \wp \left( a \right)}}} , \quad {F_2}\left( {{\xi}} \right) = \sqrt {\frac{{{x_1} - \wp \left( {{\xi} + {\omega _2}} \right)}}{{{x_1} - \wp \left( a \right)}}} , \label{eq:dressed_strings_F1_F2} 
\end{equation}
and
\begin{equation}
{\Phi }\left( {{\xi };a} \right) :=  - \frac{i}{2}\ln \frac{{\sigma \left( {{\xi } + {\omega _2} + a} \right)\sigma \left( {{\omega _2} - a} \right)}}{{\sigma \left( {{\xi } + {\omega _2} - a} \right)\sigma \left( {{\omega _2} + a} \right)}} + i\zeta \left( a \right){\xi } .
\label{eq:elliptic_solutions_lame_phase_def}
\end{equation}
The indices $0$ and $1$ account for the Pohlmeyer counterpart of the solution being translationally invariant or static respectively. The function $\Phi$ has the quasi-periodicity property
\begin{equation}
\Phi \left( {{\xi ^0} + 2{\omega _1}; a} \right) = \Phi \left( {{\xi ^0}; a} \right) + 2 i \left( {\zeta \left( { a} \right){\omega _1} - \zeta \left( {{\omega _1}} \right) a} \right) ,
\label{eq:elliptic_solutions_lame_phase_quasiper}
\end{equation}
The moduli of the Weierstrass elliptic function, which are usually denoted in the literature by $g_2$ and $g_3$, take the values
\begin{equation}
g_2 = \frac{E^2}{3} + \mu^4 , \quad g_3 = \frac{E}{3} \left( \left( \frac{E}{3} \right)^2 - \mu^4 \right) ,
\end{equation}
whereas the parameters $\ell$ and $\wp \left( a \right)$ appearing in \eqref{eq:elliptic_solutions_review} read
\begin{equation}
\ell^2 = {{x_1} - \wp \left( a \right)} = \frac{{m_ + ^2 + m_ - ^2}}{{4{R^2}}} + \frac{{E}}{2},
\end{equation}
where $x_1=E/3$ is one of the three roots of the cubic polynomial associated with the Weierstrass elliptic function. The other two roots read $x_{2/3} = -E/6 \pm \mu^2 /2$. The parameter $a$ takes values on the imaginary axis and it is a free parameter of the solution, which reflects the fact that Pohlmeyer reduction is a many to one mapping. The value of $a$ is specified by demanding that the string \eqref{eq:elliptic_solutions_review} obeys the correct periodicity conditions, so that the solution is a finite closed string. Furthermore, $a$ is connected to the parameter $\ell$ through
\begin{equation}
- i \frac{\wp '\left( a \right)}{\ell} = \frac{{m_ + ^2 - m_ - ^2}}{2} ,
\label{eq:elliptic_solutions_sign_of_a}
\end{equation}
which determines the relevant sign of $a$ and $\ell$.

Taking appropriate limits of the parameters involved in \eqref{eq:elliptic_solutions_review} results in some well known string solutions. In the case of a static Pohlmeyer counterpart, the giant magnons can be recovered in the limit $E=\mu^2$ and the GKP strings in the limit $a=\omega_2$. Had one considered the solutions with translationally invariant counterpart, one would obtain the BMN particle for $E=-\mu^2$ and the single spike solution for $E=\mu^2$. The reader is referred to \cite{part1} for further details on the elliptic string solutions.

The two-dimensional sphere S$^2$ is trivially isomorphic to the coset $\mathrm{SO}(3)/\mathrm{SO}(2)$, and, thus, instead of the action  \eqref{eq:Pohlmeyer_action} one could consider the non linear sigma model action
\begin{equation}
S=\frac{1}{8}\int\,d\xi_+d\xi_-\, \tr\left(\partial_+ f^{-1}\partial_- f\right) ,
\label{eq:dressing_review_nlsm_action}
\end{equation}
where the group valued field $f$ is appropriately constrained in the aforementioned coset space. The mapping
\begin{equation}
f = \left( I - 2{X_0}X_0^T \right) \left( {I - 2X{X^T}} \right) ,
\label{eq:dressing_mapping_coset}
\end{equation}
where $X_0^T=(0\;0\;1)$, provides the corresponding isomorphism. The equations of motion following from the action \eqref{eq:dressing_review_nlsm_action} are
\begin{equation}
\left[\partial_+-\frac{\partial_+ ff^{-1}}{1+\lambda}  , \partial_--\frac{\partial_- ff^{-1}}{1-\lambda}  \right] = 0 ,
\label{eq:dressing_review_zero_curvature}
\end{equation}
with $\lambda$ being the spectral parameter. These equations can be considered as the compatibility condition of the auxiliary system of first order differential equations
\begin{equation}
{\partial _ \pm }\Psi \left( \lambda  \right) = \frac{\partial_\pm ff^{-1}}{{1 \pm \lambda }} \Psi \left( \lambda  \right) .
\label{eq:dressing_review_auxiliary}
\end{equation}
The existence of the system \eqref{eq:dressing_review_auxiliary} can be attributed to the integrability properties of the non linear sigma model and, furthermore, it has the property that if $\Psi \left( \lambda  \right)$ is a solution, $\Psi \left( 0 \right)$ satisfies the equation of motion of the non linear sigma model. 

The dressing method involves finding the auxiliary field $\Psi \left( \lambda  \right)$ corresponding to a known seed solution $f$ of the non linear sigma model, by solving the system \eqref{eq:dressing_review_auxiliary}, supplemented  by the condition $\Psi \left( 0  \right)=f$. Then, one constructs a new solution of the auxiliary system of the form $\Psi' \left( \lambda  \right)=\chi(\lambda)\Psi \left( \lambda  \right)$, by utilizing an appropriate \textit{dressing factor} $\chi(\lambda)$. Finally, taking $\lambda$ to zero, one obtains a new solution of the sigma  model equations of motion, namely, $f'=\chi(0)\Psi \left( 0  \right)$. The latter is called the dressed solution. The dressing factor, which is in general a meromorphic function  of the complex parameter $\lambda$, must obey certain conditions, that ensure that the dressed solution is still an element of the coset $\mathrm{SO}(3)/\mathrm{SO}(2)$. The simplest possible dressing factor, which fulfils these requirements, has only two poles, complex conjugate to each other, that lie on the unit circle. It reads
\begin{equation}
\chi \left( \lambda  \right) = I + \frac{Q_1}{{\lambda  - {\lambda _1}}} + \frac{\bar Q_1}{{\lambda  - {{\bar \lambda }_1}}} ,
\label{eq:dress_review_dressing_factor_two_conjugate_poles}
\end{equation}
where
\begin{equation}
Q_1=\left(\lambda_1-\bar\lambda_1\right)P,\quad\text{and}\quad P = \frac{{\Psi \left( {{{\bar \lambda }_1}} \right)p{p^\dag }{\Psi ^{ - 1}}\left( {{\lambda _1}} \right)}}{{{p^\dag }{\Psi ^{ - 1}}\left( {{\lambda _1}} \right)\Psi \left( {{{\bar \lambda }_1}} \right)p}}
\label{eq:dress_review_dressing_factor_two_conjugate_poles_projector}
\end{equation}
and the vector $p$ is any constant complex vector obeying the consitions ${p^T}p = 0$ and $\bar p = \left( I - 2{X_0}X_0^T \right) p$.  The dressing transformation also induces a change in the (left) sigma model charge\footnote{The right charge is not independent in the case of a symmetric space sigma model.} of the seed solution
\begin{equation}
\Delta \mathcal{Q}_L = \sum_j \int d \xi^1 \partial_1 Q_j.
\label{eq:dressing_review_change_of_charge}
\end{equation}
Since the sigma model charge is proportional to the angular momentum of the string, the above formula connects the angular momenta of the seed and dressed solutions.

The change of variables
\begin{equation}
X := U X_0 ,\quad f = \theta U\theta \hat f{U^T} \quad\text{and}\quad \Psi \left( \lambda  \right): = \theta U\theta \hat \Psi \left( \lambda  \right) ,
\label{eq:review_hat_defs}
\end{equation}
where
\begin{equation}
\theta=\left(1-2X_0X_0^T\right),\quad \hat f: = \theta \left( {I - 2\hat X{{\hat X}^T}} \right),
\end{equation}
has been proven useful for the solution of the equations \eqref{eq:dressing_review_auxiliary}, in the case the seed solution is the elliptic string solution \eqref{eq:elliptic_solutions_review} \cite{Katsinis:2018ewd}. In the case of a seed solution with static Pohlmeyer counterpart, the matrix $U$ reads
\begin{equation}
{U} = \left( {\begin{array}{*{20}{c}}
{\cos \varphi }&{ - \sin \varphi }&0\\
{\sin \varphi }&{\cos \varphi }&0\\
0&0&1
\end{array}} \right) \left( {\begin{array}{*{20}{c}}
{{F_2}}&0&{{F_1}}\\
0&1&0\\
{{-F_1}}&0&{{F_2}}
\end{array}} \right),
\label{eq:review_U}
\end{equation}
where $\varphi \left( {{\xi^0},{\xi^1}} \right) = \sqrt {{x_1} - \wp \left( a \right)} {\xi^0} - \Phi \left( \xi^1 ; a \right)$. 

The $\{i,j\}$-element of the solution of the auxiliary system $\Psi \left( \lambda \right)$ reads
\begin{equation}
\Psi_{ij} \left( \lambda  \right) =  - E_j^i,
\label{eq:dressed_string_auxiliary_solution}
\end{equation}
where
\begin{align}
{E_1} &:= \cos \left( {\sqrt{\Delta} {\xi^0} - \Phi \left( {{\xi^1}; {\tilde{a}}} \right)} \right){e_1} + \sin \left( {\sqrt{\Delta} {\xi^0} - \Phi \left( {{\xi^1}; {\tilde{a}}} \right)} \right){e_2} , \label{eq:review_E1}\\
{E_2} &:= - \cos \left( {\sqrt{\Delta} {\xi^0} - \Phi \left( {{\xi^1}; {\tilde{a}}} \right)} \right){e_2} + \sin \left( {\sqrt{\Delta} {\xi^0} - \Phi \left( {{\xi^1}; {\tilde{a}}} \right)} \right){e_1} , \label{eq:review_E2}\\
{E_3} &:= {e_3} . \label{eq:review_E3}
\end{align}
The vectors $E_j$ are in turn expressed in terms of the vectors $e_i$
\begin{equation}
{e_i} = \left\{ {\frac{{{X_0} \times {\kappa _0}}}{{\sqrt {{{\left( {{X_0} \times {\kappa _0}} \right)}^T}\left( {{X_0} \times {\kappa _0}} \right)} }} \times \frac{{{\kappa _0}}}{{\sqrt {\kappa _0^T{\kappa _0}} }},\frac{{{X_0} \times {\kappa _0}}}{{\sqrt {{{\left( {{X_0} \times {\kappa _0}} \right)}^T}\left( {{X_0} \times {\kappa _0}} \right)} }},\frac{{{\kappa _0}}}{{\sqrt {\kappa _0^T{\kappa _0}} }}} \right\} 
\label{eq:review_es}
\end{equation}
and the parameters
\begin{equation}
\Delta =\kappa^T_0 \kappa_0= \frac{E}{2} + \frac{m_ + ^2}{4}{\left( {\frac{{1 - \lambda }}{{1 + \lambda }}} \right)^2} + \frac{m_ - ^2}{4}{\left( {\frac{{1 + \lambda }}{{1 - \lambda }}} \right)^2}
\end{equation}
and $\tilde{a}$, which play a role analogous to the parameters $E$ and $a$ of the elliptic solution \eqref{eq:elliptic_solutions_review}. The parameter $\tilde{a}$ is defined through the equations
\begin{align}
\wp \left( \tilde{a} \right) &=  - \frac{E}{6} - \frac{{m_ + ^2}}{4}{\left( {\frac{{1 - \lambda }}{{1 + \lambda }}} \right)^2} - \frac{{m_ - ^2}}{4}{\left( {\frac{{1 + \lambda }}{{1 - \lambda }}} \right)^2}, \label{eq:review_p_atilde}\\
\wp ' \left( \tilde{a} \right) &= i {\sqrt{\Delta} \left( {\frac{{m_ + ^2}}{2}{{\left( {\frac{{1 - \lambda }}{{1 + \lambda }}} \right)}^2} - \frac{{m_ - ^2}}{2}{{\left( {\frac{{1 + \lambda }}{{1 - \lambda }}} \right)}^2}} \right)} \label{eq:review_pprime_atilde}
\end{align}
Finally, the vector $\kappa_0$ is defined through the equations
\begin{align}
\kappa _{0/1}^3 &= - k_{0/1}^3 ,\label{eq:defs_kappa_1}\\
\kappa _{0/1}^{1/2} &=  - \frac{{1 + {\lambda ^2}}}{{1 - {\lambda ^2}}}k_{0/1}^{1/2} + \frac{{2\lambda }}{{1 - {\lambda ^2}}}k_{1/0}^{1/2} .\label{eq:defs_kappa_2}
\end{align}
The vectors $k_{0/1}$ are determined by the matrix $U$, via the equations ${U^T}\left( {{\partial _i}U} \right) = k_i^j{T_j}$, where $T_i$ are the generators of the lie group $\mathrm{SO}(3)$ defined as usual.

Application of the dressing factor \eqref{eq:dress_review_dressing_factor_two_conjugate_poles} yields the dressed solution 
\begin{equation}
\hat f' = I - \frac{{{\lambda _1} - 1/{\lambda _1}}}{{{\lambda _1}}}\frac{{{X_-}X_+^T}}{{X_+^T{X_-}}} - \frac{{1/{\lambda _1} - {\lambda _1}}}{{1/{\lambda _1}}}\frac{{{X_+}X_-^T}}{{X_+^T{X_-}}} ,
\end{equation}
where
\begin{equation}
{X_+} = \hat \Psi \left( {{\lambda _1}} \right)\theta p ,\quad {X_-} = \theta \hat \Psi \left( {{\lambda _1}} \right)\theta p .
\label{eq:dressed_string_Xpm_def}
\end{equation}
In terms of the variable $\hat X'$ the solution reads
\begin{equation}
\hat X ' = \sqrt {\frac{1}{{2X_ + ^T{X_ - }}}} \sin \theta_1 \left( {{X_ + } + {X_ - }} \right) + \cos \theta_1 {X_0} := \sin \theta_1 {X_1} + \cos \theta_1 {X_0},
\label{eq:dressed_solution_x_hat_prime}
\end{equation}
where $\lambda_1=\exp{i\theta_1}$, and, finally,  expressing the dressed solution in terms of the non-hatted variables $X$, yields $X'=U\hat X'$. The time component $t'$ of the dressed solution is the same as the one of the seed solution \eqref{eq:elliptic_solutions_review}. This is expected, since the dressing transformation acts only on the S$^2$-part, i.e. the spatial part of the seed solution. In order to get the corresponding solution, in the case of a seed with translationally invariant Pohlmeyer counterpart, one should simply exchange $\xi^0$ with $\xi^1$.

The Pohlmeyer counterparts of the dressed solutions were also obtained in \cite{Katsinis:2018ewd}. It turns out that they correspond to the solutions of the sine-Gordon equation obtained after the application of a single \Backlund transformation to the Pohlmeyer counterpart of the seed solution \eqref{eq:elliptic_solution_phi_0}, revealing the deep connection between the dressing method and the \Backlund transformations
\begin{equation}
{\partial _ + }\frac{{\varphi  + \tilde \varphi }}{2} = a\mu \sin \frac{{\varphi  - \tilde \varphi }}{2} ,\quad
{\partial _ - }\frac{{\varphi  - \tilde \varphi }}{2} = \frac{1}{a}\mu \sin \frac{{\varphi  + \tilde \varphi }}{2} 
\label{eq:SG_dress_Backlund}
\end{equation}
of the sine-Gordon equation. The corresponding \Backlund parameter $a$ is related to the position of the poles of the dressing factor through
\begin{equation}
a = \sqrt { - \frac{{{m_ + }}}{{{m_ - }}}} \tan \frac{\theta_1 }{2} .
\label{eq:dress_vs_Backlund}
\end{equation}
The Pohlmeyer counterpart of the dressed solution in the case of a translationally invariant seed reads 
\begin{equation}
\tilde \varphi  = \begin{cases}
{\hat \varphi} + 4\arctan \left[ \frac{{A + B}}{D}\tanh \frac{{D{\xi ^1} + i \Phi \left( {{\xi ^0};\tilde a} \right)}}{2} \right] , & D^2 > 0 , \\
{\hat \varphi} + 4\arctan \left[ \frac{{ 1 - s_c}}{2} B\left( {{\xi ^1} + i \Phi \left( {{\xi ^0};\tilde a} \right)} \right) \right] , & D^2 = 0 , \\
{\hat \varphi} + 4\arctan \left[ \frac{{A + B}}{i D}\tan \frac{{i D{\xi ^1} - \Phi \left( {{\xi ^0};\tilde a} \right)}}{2} \right] , & D^2 < 0 ,
\end{cases}
\label{eq:SG_final_expression_cases}
\end{equation}
where 
\begin{align}
D^2&= \frac{1}{4} \left[ {{\mu ^2}{{\left( {a^2 + {a^{ - 2}}} \right)}} - 2 E } \right] =-\Delta\left(e^{i\theta_1}\right), \label{eq:SG_D2_def} \\
\wp \left( {\tilde a} \right) &=  - \frac{E}{6} + \frac{{{\mu ^2}}}{4}\left( {{a^2} + {a^{ - 2}}} \right) = \wp \left( {\tilde{a}_{NLSM}(\exp{i\theta_1})} \right). \label{eq:SG_patilde}
\end{align}
The functions $\hat \varphi$, $A$ and $B$ are completely determined by the seed solution of the sine-Gordon equation as
\begin{align}
\hat \varphi &= 2\arctan \left( {\frac{{a - {a^{ - 1}}}}{{a + {a^{ - 1}}}}\tan \frac{\varphi }{2}} \right) + \left( {2k - 1} \right)\pi  + {\mathop{\rm sgn}} \left( {{a^2} - 1} \right)2\pi \left\lfloor {\frac{\varphi }{{2\pi }} + \frac{1}{2}} \right\rfloor , \label{eq:ell_tan_phi0} \\
A &= s_c \frac{\mu}{2} \sqrt {{a^2} + {a^{ - 2}} + 2\cos \varphi } , \quad B\left( {{\xi ^0}} \right) =  - {\partial _0}\frac{\varphi }{2},\quad s_c : = {{\left( { - 1} \right)}^k}{\mathop{\rm sgn}} a\label{eq:ell_A_and_B_and_s_c}
\end{align}
and $k\in\mathbb Z$.

\setcounter{equation}{0}
\section{Properties of the Sine-Gordon Counterparts of the Dressed Elliptic Strings}
\label{sec:SG_properties}

It has been shown that many physical properties of the elliptic strings solutions are directly connected to properties of their sine-Gordon counterparts \cite{part1}. The establishment of this mapping enhances the intuitive understanding of the dynamics of string propagation on the sphere via the dynamics of the sine-Gordon equation, which is a much simpler system. For this purpose, in this section, we will study some basic properties of the sine-Gordon counterparts of the dressed elliptic string solutions reviewed in section \ref{sec:review}.

The dressed strings, as well as their sine-Gordon counterparts can be classified into two large categories depending on the sign of the constant $D^2$. When $D^2 > 0$ (or equivalently when $\tilde{a}$ lies on the real axis), equation \eqref{eq:SG_final_expression_cases} describes a localized kink travelling on top of an elliptic background. The position of the kink can be identified with the position where the argument of the tanh in equation \eqref{eq:SG_final_expression_cases} vanishes, namely ${{\xi ^1} = - i \Phi \left( {{\xi ^0};\tilde a} \right)} / D$, where it holds that $\varphi = \hat{\varphi}$. Far away from this region, the solution assumes a form that is determined solely by the seed solution. As we have commented in section \ref{sec:review}, a \Backlund transformation increases the genus of the solution by one, adding a degenerate hole to the relevant torus, which corresponds to a diverging period. This is evident in this case, where the two periods appearing in the solution are the one of the seed solution and the infinite time/space required to accommodate the kink.

The minimum value of the parameter $D^2$ is  $D_{\text{min}}^2=(\mu^2-E)/2$. Thus, when a rotating seed is considered, it is possible that $D^2 < 0$ (or equivalently $\tilde{a}$ lies on the imaginary axis shifted by the real half period $\omega_1$). In such a case, the hyperbolic tangent function appearing in the dressed solution becomes trigonometric tangent. As a result, the effect of the dressing on the solution is not localized in the position where the argument of this function vanishes, but it is rather spread everywhere in a periodic fashion. It follows that these solutions do not describe a kink propagating on an elliptic background. They should be rather understood as a periodic structure of oscillating deformations on top of a rotating elliptic background. Such solutions contain two periods; one of the seed solution and one imposed by the aforementioned trigonometric tangent. However, it is the imaginary period of the trigonometric tangent that is divergent, and, thus, these solutions are still degenerate genus two solutions, in this manner similar to the solutions of the $D^2>0$ class.

It follows that a bifurcation of the qualitative characteristics of the dressed solution occurs at $D^2 = 0$.

\subsection{$D^2>0$: Kink-Background Interaction}
\label{subsec:SG_asymptotics}

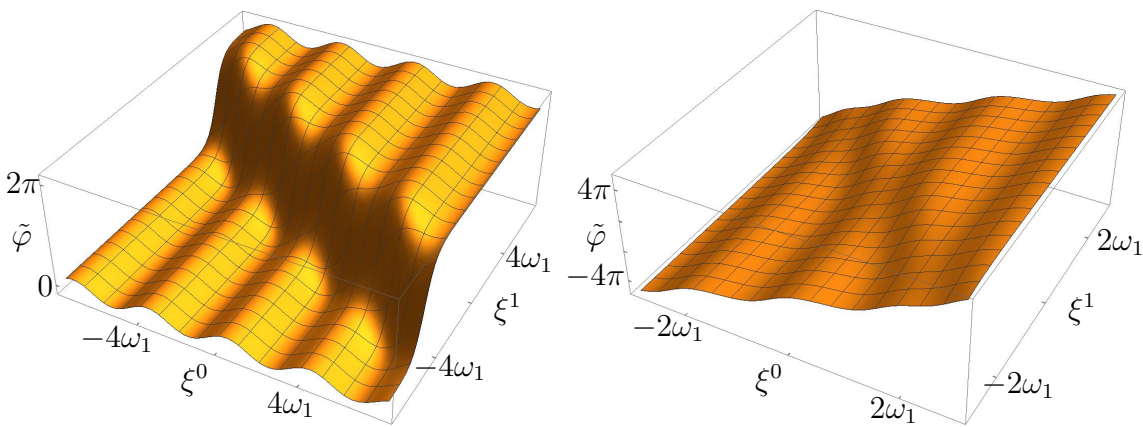
\begin{figure}[ht]
\vspace{10pt}
\begin{center}
\begin{picture}(100,40)
\put(2.5,0){\includegraphics[width = 0.45\textwidth]{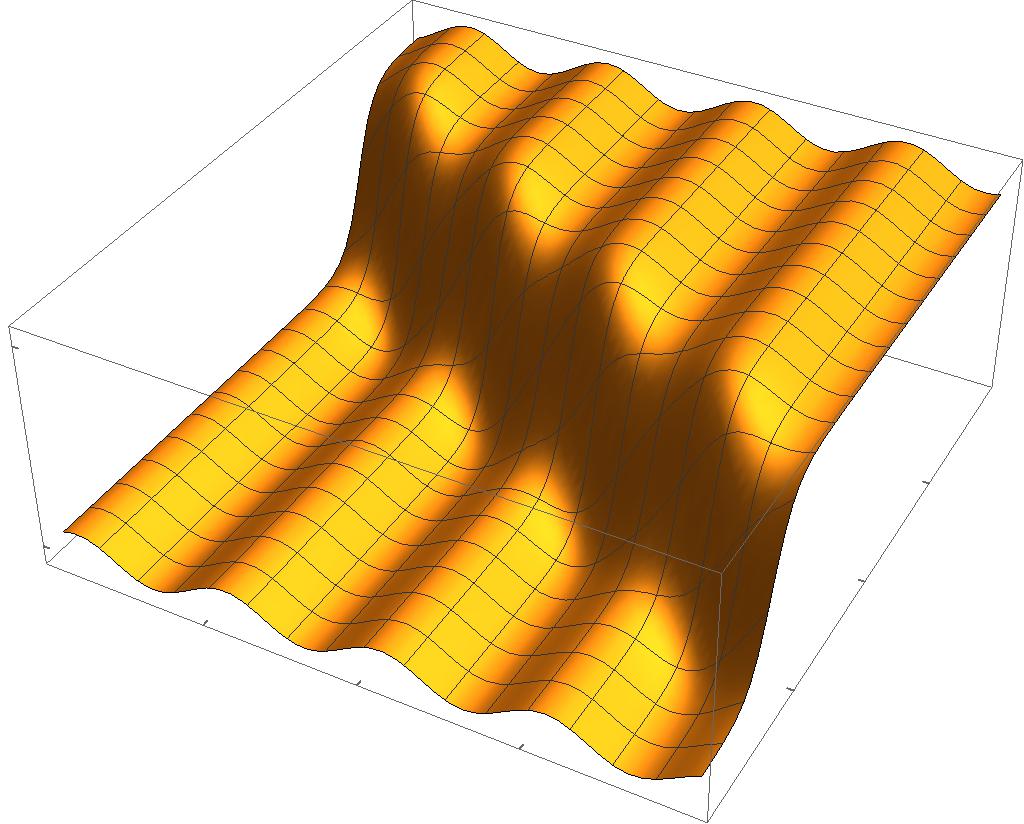}}
\put(52.5,0){\includegraphics[width = 0.45\textwidth]{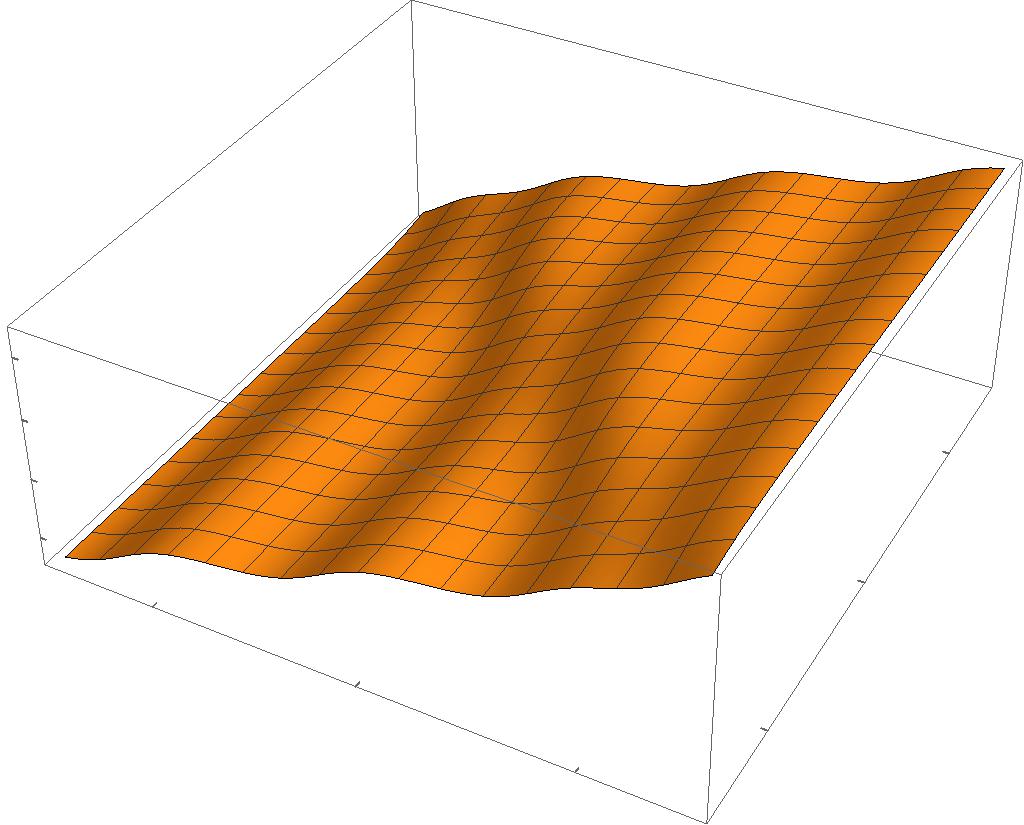}}
\put(0.5,15.75){$\tilde{\varphi}$}
\put(42.5,9.25){$\xi^1$}
\put(15.25,3){$\xi^0$}
\put(43.5,13.75){$4\omega_1$}
\put(35.5,4.25){$-4\omega_1$}
\put(22.75,1){$4\omega_1$}
\put(6.5,6.75){$-4\omega_1$}
\put(2.75,11.25){$0$}
\put(0,20.25){$2\pi$}
\put(50.5,15.75){$\tilde{\varphi}$}
\put(92.5,9.25){$\xi^1$}
\put(65.25,3){$\xi^0$}
\put(95,15.25){$2\omega_1$}
\put(85,2.75){$-2\omega_1$}
\put(75.25,0.5){$2\omega_1$}
\put(54.5,7.75){$-2\omega_1$}
\put(48.75,12.25){$-4\pi$}
\put(49.75,19.75){$4\pi$}
\end{picture}
\end{center}
\vspace{-10pt}
\caption{The dressed sine-Gordon solution for a translationally invariant oscillating seed with $E = - 9 \mu^2 /10$ and a translationally invariant rotating seed with $E = 11 \mu^2 /10$. In both cases, the \Backlund parameter equals $a = 2$.}
\vspace{5pt}
\label{fig:ti_pos_1}
\end{figure}

Let us start our analysis considering solutions whose seeds are translationally invariant. Figure \ref{fig:ti_pos_1} depicts two such dressed solutions of the sine-Gordon equation, one with an oscillatory seed and one with a rotating seed.

It is evident from the form of the solution \eqref{eq:SG_final_expression_cases}, as well as figure \ref{fig:ti_pos_1}, that the solutions with $D^2>0$ have the form of a localized kink at ${\xi ^1} =  - i\Phi \left( {{\xi ^0};\tilde a} \right)/D$ propagating on top of an elliptic background. Let us determine, whether the kink is left- or right-moving. This is determined by the monotonicity of the function $- i \Phi \left( {{\xi ^0};\tilde a} \right) / D$. It turns out that
\begin{equation}
\frac{d}{{d{\xi ^0}}}\left( { - \frac{{i\Phi \left( {{\xi ^0};\tilde a} \right)}}{D}} \right) = \frac{{{\mu ^2}}}{4}\frac{{{a^2} - {a^{ - 2}}}}{{\wp \left( {{\xi ^0} + {\omega _2}} \right) - \wp \left( {\tilde a} \right)}} ,
\end{equation}
implying that the direction of the motion of the kink is determined by the sign of ${a^2} - {a^{ - 2}}$, or equivalently by $s_d : = \sign \left( \left| a \right| - 1 \right)$. Since ${\wp \left( {{\xi ^0} + {\omega _2}} \right) < \wp \left( {\tilde a} \right)}$, as the former takes values between the two smaller roots and the former is larger than the largest root, it turns out that the regime $\left|a\right| > 1 $ corresponds to the left-moving kinks and the regime $\left|a\right| < 1 $ corresponds to the right-moving ones, similarly to the usual analysis for kinks built on top of the sine-Gordon vacuum.

Moreover, equation \eqref{eq:SG_final_expression_cases} implies that far away from the kink location, the solution depends solely on $\xi^0$. This is also visible in figure \ref{fig:ti_pos_1}. As this is the defining property of the elliptic solutions of the sine-Gordon equation \cite{part1}, we expect that asymptotically the solution assumes the form of an elliptic solution. One can easily check, either directly or calculating the energy density far away from the kink location (see section \ref{subsec:SG_kink_energy_momentum}), that this is not an arbitrary elliptic solution, but the seed one up to a time shift (and a possible reflection). This time shift may be different before and after the passage of the kink. It is a matter of algebra to show that
\begin{equation}
\begin{split}
\mathop {\lim }\limits_{D{\xi ^1} + i \Phi \left( {{\xi ^0};\tilde a} \right) \to  \pm \infty } s_d \tilde \varphi &= s_d \left( {\hat \varphi  \pm 4\arctan \frac{{A + B}}{D}} \right)\\
&= \varphi \left( {{\xi ^0} \pm {\tilde a}} \right) + s_d \left( { \left( {2k - 1} \right) \pm s_c} \right)\pi  .
\end{split}
\label{eq:kink_asymptotics}
\end{equation}
Thus, indeed the asymptotic form of the solution is a shifted version of the seed solution, being reflected depending on the sign $s_d$. In the following, taking advantage of the reflection symmetry $\varphi \to - \varphi$ of the sine-Gordon equation, we will avoid this reflection, considering the properties of the solution $s_d \tilde{\varphi}$. The above asymptotic expression \eqref{eq:kink_asymptotics} determines $\hat \varphi$ and $4\arctan ({A + B})/{D}$ in terms of the seed solution, allowing the re-expression of the dressed solution \eqref{eq:SG_final_expression_cases} in terms of the latter as
\begin{multline}
s_d \tilde \varphi = \frac{1}{2}\left( {\varphi \left( {{\xi ^0} +  {\tilde a} } \right) + \varphi \left( {{\xi ^0} -  {\tilde a} } \right)} \right) + s_d \left( {2k - 1} \right)\pi \\
 + 4 s_d \arctan \left[ {\tan \left( { \frac{1}{8}\left( {\varphi \left( {{\xi ^0} +  {\tilde a} } \right) - \varphi \left( {{\xi ^0} -  {\tilde a} } \right)} \right) + s_c \frac{\pi }{4}} \right)\tanh \frac{{D{\xi ^1} + i \Phi \left( {{\xi ^0};\tilde a} \right)}}{2}} \right] .
\label{eq:kink_in_terms_of_asymptotics}
\end{multline}

Equation \eqref{eq:kink_asymptotics} obviously implies $\mathop {\lim }\limits_{{\xi ^0} \to  \pm \infty } s_d \tilde \varphi = \varphi \left( {{\xi ^0} \mp \left| {\tilde a} \right|} \right) + 2{n_ \pm }\pi $, where $ n_\pm \in \mathbb{Z}$. Therefore, the passage of the kink effectively causes a delay to the translationally invariant motion of the system equal to
\begin{equation}
\Delta \xi^0 = 2 \left| \tilde{a} \right| .
\label{eq:kink_ti_time_delay}
\end{equation}
This observation provides a nice physical meaning to the parameter $\tilde{a}$. This time delay quantifies the effect of the interaction of the kink imposed by the \Backlund transformation with the elliptic background.
\begin{figure}[ht]
\vspace{10pt}
\begin{center}
\begin{picture}(60,40)
\put(2.5,2.5){\includegraphics[width = 0.55\textwidth]{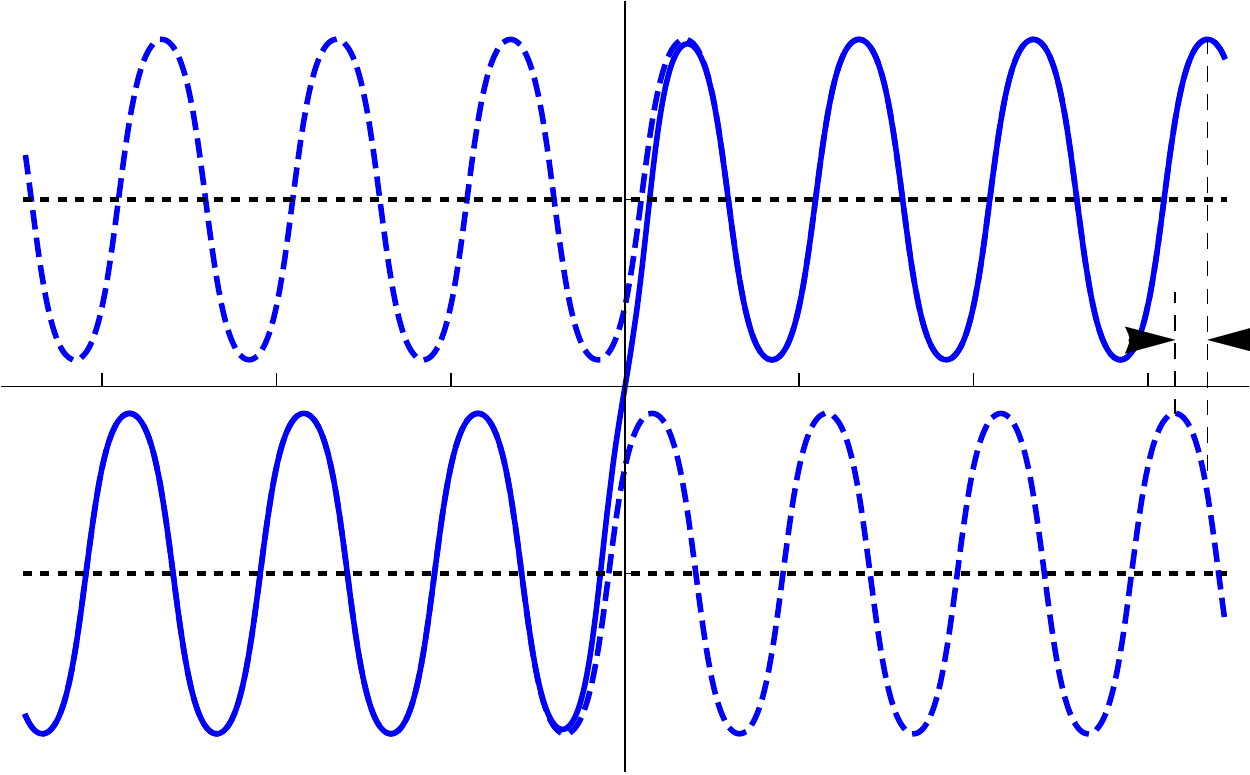}}
\put(1.75,10.5){$0$}
\put(0.25,27.25){$2\pi$}
\put(29,37.25){$\tilde{\varphi}_+$}
\put(58,18.5){$\xi^0$}
\put(56,22.5){$2 \left| \tilde{a} \right|$}
\put(37.75,20.25){$4\omega_1$}
\put(20,20.25){$-4\omega_1$}
\end{picture}
\end{center}
\vspace{-10pt}
\caption{The dressed solution for an oscillating seed with $E = 9 \mu^2 /10$ and \Backlund parameter $a = 2$ at $\xi^1 = 0$. The dashed lines indicate the asymptotic behaviour $\varphi \left( \xi^0 \pm \tilde{a} \right)$.}
\vspace{5pt}
\label{fig:time_delay}
\end{figure}
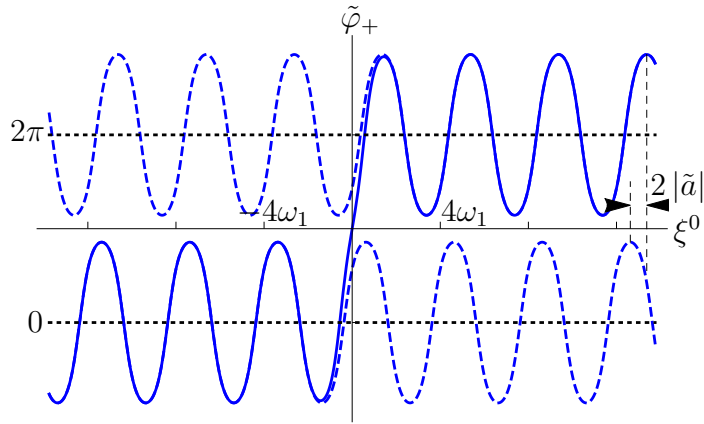

Finally, studying the average value of $\tilde{\varphi}$ in a full period of the seed solution at spatial infinity, we find that
\begin{equation}
\begin{split}
\left\langle {\mathop {\lim }\limits_{{\xi ^1} \to  + \infty } s_d \tilde \varphi  - \mathop {\lim }\limits_{{\xi ^1} \to  - \infty } s_d \tilde \varphi } \right\rangle  &= \left\langle {\varphi \left( {{\xi ^0} - \left| {\tilde a} \right|} \right) - \varphi \left( {{\xi ^0} + \left| {\tilde a} \right|} \right)} \right\rangle  + 2\pi s_c \\
 &= \begin{cases}
2\pi s_c , & E < \mu^2 , \\
2\pi s_c - 2\pi \frac{{\left| {\tilde a} \right|}}{{{\omega _1}}} , & E > \mu^2 ,
 \end{cases}
\end{split}
\end{equation}
implying that the solution is a kink or antikink depending on the sign $s_c$. Notice that in the case of a rotating background, the jump in the rotation induced by the kink is not an integer multiple of $2\pi$, but it ranges in $\left[ { - 4\pi , - 2\pi } \right] \cup \left[ {0,2\pi } \right]$; it is actually $\pm 2 \pi$ minus a quantity induced by the delay to the background rotation. The asymmetry is due to the fact that we have considered the rotating elliptic seed solutions to be always increasing functions of time.
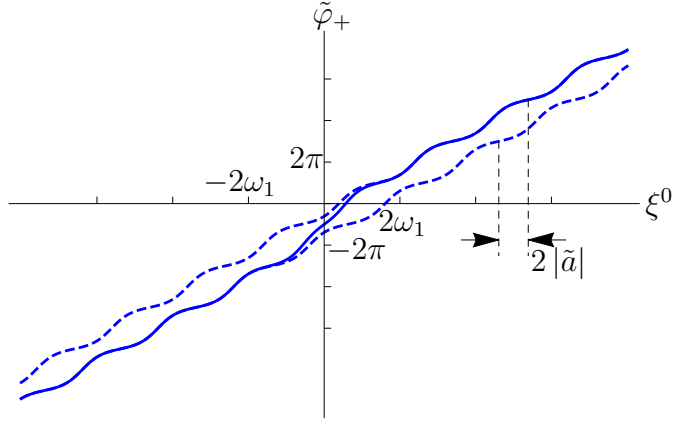
\begin{figure}[ht]
\vspace{10pt}
\begin{center}
\begin{picture}(60,40)
\put(2.5,2.5){\includegraphics[width = 0.55\textwidth]{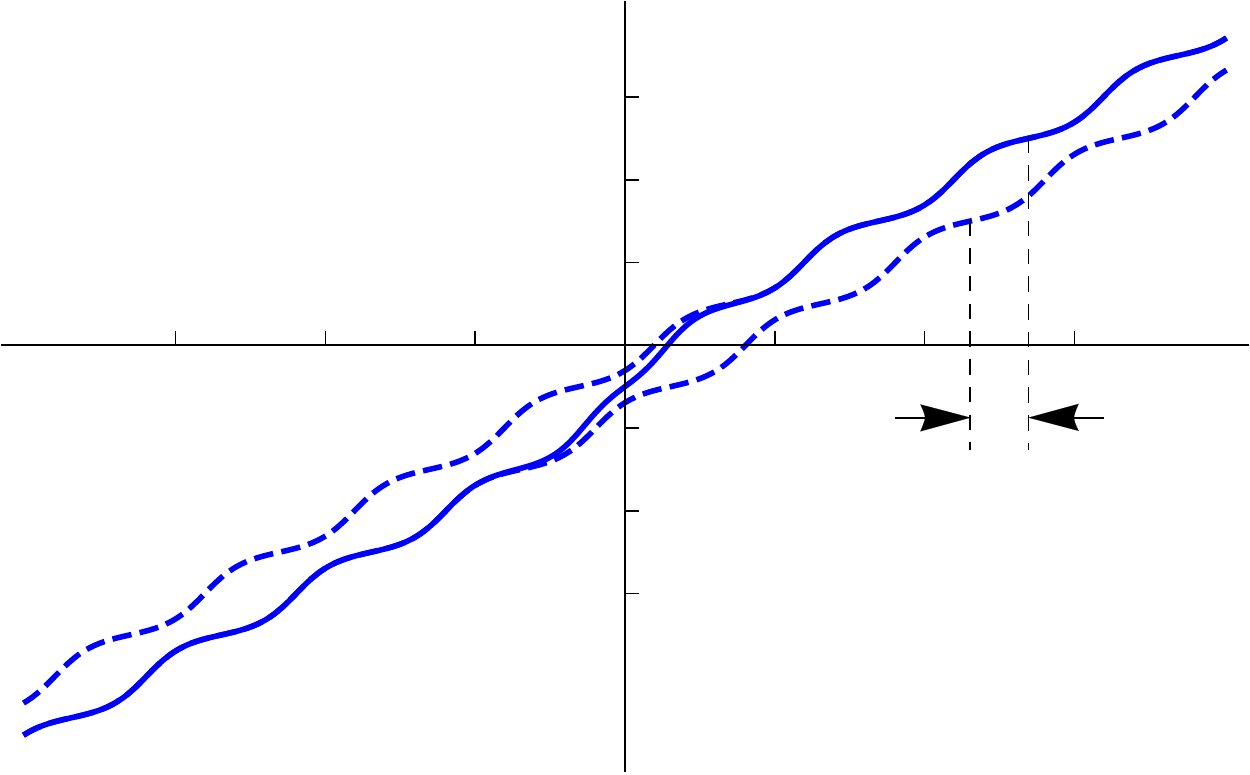}}
\put(30.25,16.5){$-2\pi$}
\put(27,24.5){$2\pi$}
\put(29,37.25){$\tilde{\varphi}_+$}
\put(58,20.5){$\xi^0$}
\put(48,15.5){$2 \left| \tilde{a} \right|$}
\put(34.75,19){$2\omega_1$}
\put(19.5,22.5){$-2\omega_1$}
\end{picture}
\end{center}
\vspace{-10pt}
\caption{The dressed solution for a rotating seed with $E = 11 \mu^2 /10$ and \Backlund parameter $a = 2$ at $\xi^1 = 0$. The dashed lines indicate the asymptotic behaviour $\varphi \left( \xi^0 \pm \tilde{a} \right)$. The jump due to the kink is positive, but smaller than $2\pi$, as a result of the delay in the background motion.}
\vspace{5pt}
\label{fig:phi_jump}
\end{figure}
All cases are summarized in table \ref{tb:kink_cases_ti}.
\begin{table}[ht]
\centering
\begin{tabular}{|c|c|c|c|c|}
\hline
parity of $k$ & $a \in \left( - \infty , -1 \right)$ & $a \in \left( - 1 , 0 \right)$ & $a \in \left( 0 , 1 \right)$ & $a \in \left( 1 , \infty \right)$\\
\hline\hline
$k$ even & left moving & right moving & right moving & left moving \\
 & antikink & antikink & kink & kink \\
\hline
$k$ odd & left moving & right moving & right moving & left moving \\
 & kink & kink & antikink & antikink \\
\hline
\end{tabular}
\caption{The translationally invariant background kink solutions for all $a$ and $k$.}
\label{tb:kink_cases_ti}
\end{table}
These four classes of solutions are the physical depiction of the fact that the same value of $D^2$ can be acquired for four distinct \Backlund parameters $a$. The definition of the sign of $A$ \eqref{eq:ell_A_and_B_and_s_c} has been made so that all four classes of solutions can be accessed with the same formula, simply varying the parameter $a$, in a similar manner to the usual analysis of kinks built using the vacuum as the seed solution. The special case $a = \pm 1$ corresponds to static kinks/antikinks leading to only two physical distinct cases.

The situation is similar in the case of static seed solutions. In this case, we find that $\mathop {\lim }\limits_{{\xi ^0} \to  \pm \infty } s_d \tilde \varphi  = \varphi \left( {{\xi ^1} \pm {\tilde a}} \right) + 2{n_ \pm }\pi $, where $ n_\pm \in \mathbb{Z}$. Thus, the effect of the passage of the kink is a displacement of the background static configuration by
\begin{equation}
\Delta \xi_\pm^1 = \mp 2 \tilde{a} .
\label{eq:kink_static_spacial_displacement}
\end{equation}

Furthermore, considering the average value of $\tilde{\varphi}$ in a full spatial period of the background solution at spatial infinity, we find that
\begin{equation}
\left\langle {\mathop {\lim }\limits_{{\xi ^1} \to  + \infty } \left( s_d \tilde \varphi - \varphi \right) } \right\rangle  - \left\langle {\mathop {\lim }\limits_{{\xi ^1} \to  - \infty } \left( s_d \tilde \varphi -\varphi \right) } \right\rangle = \begin{cases}
 2{s_c}\pi  , & E < \mu^2 , \\
 2\pi \frac{{ {\tilde a} }}{{{\omega _1}}} + 2{s_c}\pi  , & E > \mu^2 .
 \end{cases}
\end{equation}
This implies that again the solution is a kink or an antikink depending on the sign $s_c$. All cases are summarized in table \ref{tb:kink_cases_static}.
\begin{table}[ht]
\centering
\begin{tabular}{|c|c|c|c|c|}
\hline
parity of $k$ & $a \in \left( - \infty , -1 \right)$ & $a \in \left( - 1 , 0 \right)$ & $a \in \left( 0 , 1 \right)$ & $a \in \left( 1 , \infty \right)$\\
\hline\hline
$k$ even & right moving & left moving & left moving & right moving \\
 & antikink & kink & antikink & kink \\
\hline
$k$ odd & right moving & left moving & left moving & right moving \\
 & kink & antikink & kink & antikink \\
\hline
\end{tabular}
\caption{The static background kink solutions for all $a$ and $k$.}
\label{tb:kink_cases_static}
\end{table}

\subsection{$D^2>0$: Kink Velocity}
\label{subsec:SG_kink_velocity}
Let us consider the class of kinks propagating on a translationally invariant elliptic background. A naive way to define the kink velocity is
\begin{equation}
v_0 = {\left. {\frac{{d{\xi ^1}}}{{d{\xi ^0}}}} \right|_{D{\xi ^1} + i \Phi \left( {{\xi ^0};\tilde a} \right) = c }} = \frac{1}{{2D}}\frac{{\wp '\left( {\tilde a} \right)}}{{\wp \left( {{\xi ^0} + {\omega _2}} \right) - \wp \left( {\tilde a} \right)}} .
\label{eq:kinks_instant_velocity}
\end{equation}
The above velocity is not constant but rather it is a periodic function of time. Its range is
\begin{equation}
\sqrt {\frac{{\wp \left( {\tilde a} \right) - {x_2}}}{{\wp \left( {\tilde a} \right) - {x_3}}}} = \left| {\frac{{a - {a^{ - 1}}}}{{a + {a^{ - 1}}}}} \right| < \left| v_0 \right| < \left| {\frac{{{a^2} - {a^{ - 2}}}}{{ {a^2 + {a^{ - 2}}}  -  {E/{\mu ^2} } }}} \right| = \frac{{\sqrt {\left( {\wp \left( {\tilde a} \right) - {x_2}} \right)\left( {\wp \left( {\tilde a} \right) - {x_3}} \right)} }}{{\wp \left( {\tilde a} \right) - {x_1}}}
\end{equation}
for oscillating backgrounds and
\begin{equation}
\sqrt {\frac{{\wp \left( {\tilde a} \right) - {x_2}}}{{\wp \left( {\tilde a} \right) - {x_3}}}} = \left| {\frac{{a - {a^{ - 1}}}}{{a + {a^{ - 1}}}}} \right| < \left| v_0 \right| < \left| {\frac{{a + {a^{ - 1}}}}{{a - {a^{ - 1}}}}} \right| = \sqrt {\frac{{\wp \left( {\tilde a} \right) - {x_3}}}{{\wp \left( {\tilde a} \right) - {x_2}}}} 
\end{equation}
for the rotating ones. The minimum value is always smaller than the speed of light, whereas the maximum value is always larger than the speed of light in the case of rotating backgrounds. In the case of oscillating backgrounds, the maximum instant velocity is always smaller than that of light when $E<0$, whereas when $E>0$, it is so only when the \Backlund parameter satisfies
\begin{equation}
\frac{{{\mu ^2}}}{E} > {a^2} > \frac{E}{{{\mu ^2}}}  \Leftrightarrow {D^2} < \frac{{{\mu ^4} - {E^2}}}{{2E}}.
\end{equation}

The velocity defined above is a notion of instant velocity. Within a period of the elliptic background, the propagation of the kink is quite complicated, since the shape of the kink is also fluctuating periodically. A more physical definition of the kink velocity is the mean velocity in a period, $\bar v$, defined as
\begin{equation}
{\bar v} _0 = \frac{\Phi \left( {{\xi ^0} + 2{\omega _1};\tilde a} \right) - \Phi \left( {{\xi ^0};\tilde a} \right)}{2 i \omega_1 D}.
\end{equation}
The function ${\Phi \left( {{\xi ^0};\tilde a} \right)}$ is a quasi-periodic function. Its property \eqref{eq:elliptic_solutions_lame_phase_quasiper} implies that the mean velocity of the kink equals
\begin{equation}
{\bar v} _0 = \frac{{\zeta \left( {\tilde a} \right){\omega _1} - \zeta \left( {{\omega _1}} \right)\tilde a}}{{{\omega _1}D}} .
\label{eq:kinks_mean_velocity}
\end{equation}
This velocity should not be necessarily conceived as the kink velocity. Any of these solutions can be boosted to an arbitrary frame, altering the kink velocity. It should rather be understood as a parameter of the family of dressed elliptic solutions of the sine-Gordon equation, being equal to the velocity of the kink at the specific frame, where the background is translationally invariant.
\begin{figure}[ht]
\vspace{10pt}
\begin{center}
\begin{picture}(100,43)
\put(2.5,0){\includegraphics[width = 0.45\textwidth]{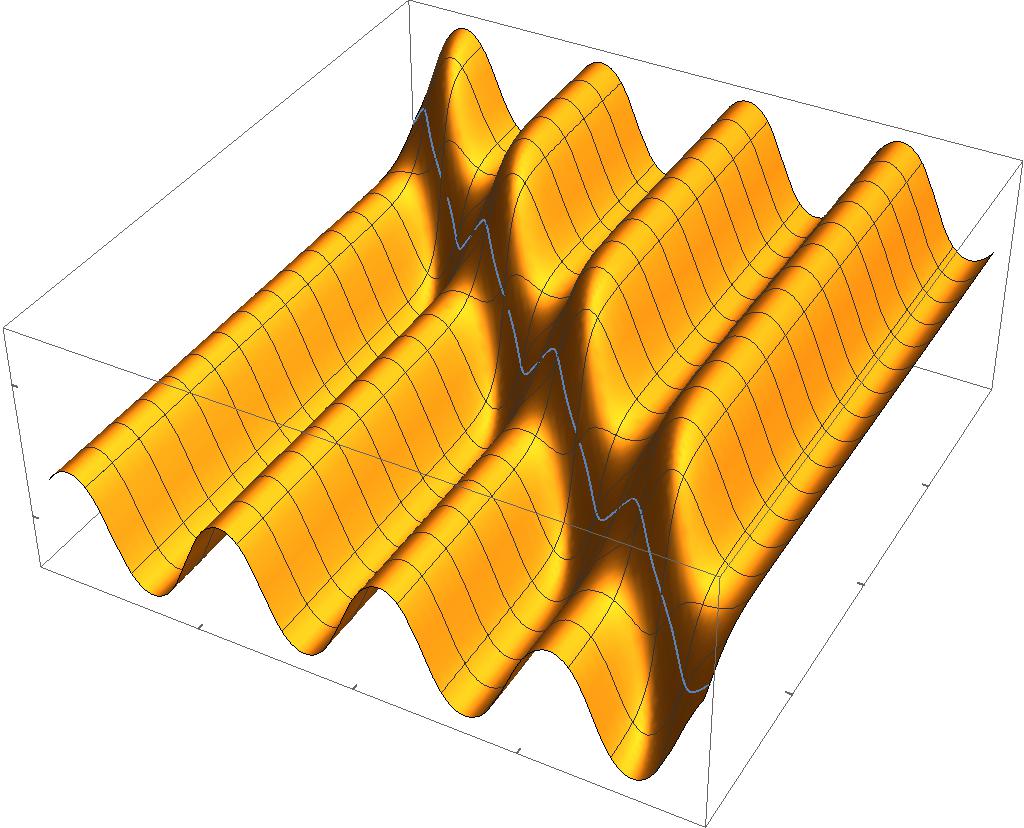}}
\put(55,0){\includegraphics[width = 0.4\textwidth]{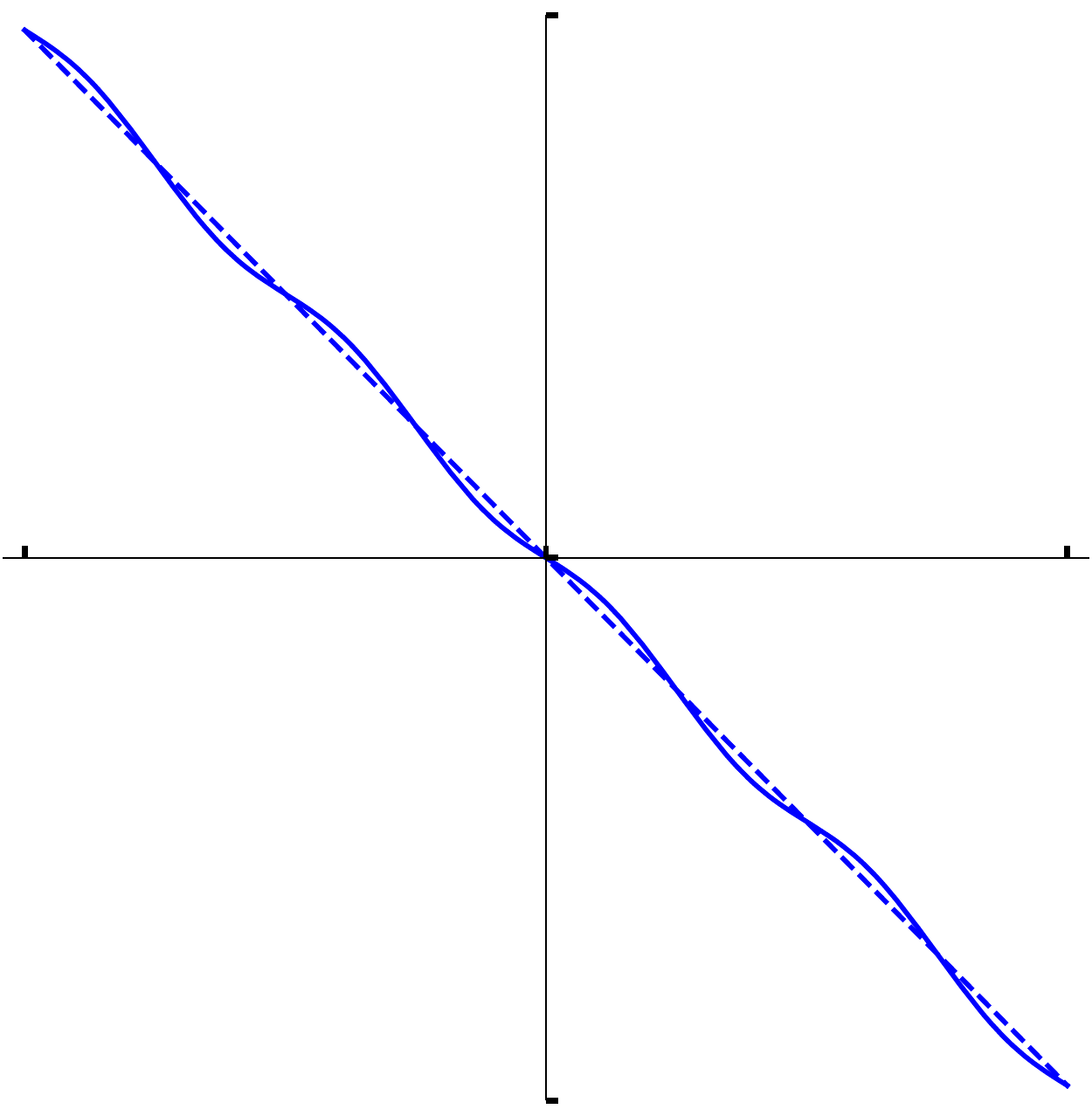}}
\put(0.5,15.75){$\tilde{\varphi}$}
\put(42.5,9.25){$\xi^1$}
\put(15.25,3){$\xi^0$}
\put(43.5,13.75){$4\omega_1$}
\put(35.5,4.25){$-4\omega_1$}
\put(22.75,1){$4\omega_1$}
\put(6.5,6.75){$-4\omega_1$}
\put(2.75,11.25){$0$}
\put(0,20.25){$2\pi$}
\put(74,41.25){$\xi^1$}
\put(96,19.5){$\xi^0$}
\put(92,21.25){$4\omega_1$}
\put(51.75,21.25){$-4\omega_1$}
\put(75.25,0.5){$-4\omega_1$}
\put(75.5,39.5){$4\omega_1$}
\end{picture}
\end{center}
\vspace{-10pt}
\caption{The solution for an oscillating background with $E = 7 \mu^2 /10$ and \Backlund parameter $a = 2$. The blue line indicates the position of the kink. The dashed line is the average position. Its inclination is the mean velocity of the kink.}
\vspace{5pt}
\label{fig:kink_velocity3d}
\end{figure}
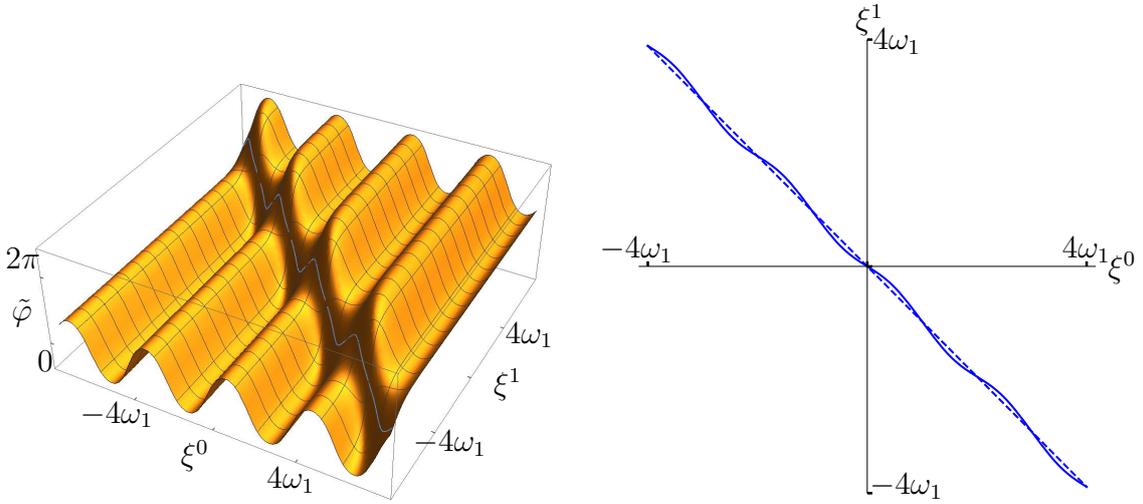

For the solutions with $D^2 > 0$, the parameter $\tilde a$ takes values on the real axis between $-\omega_1$ and $\omega_1$. The mean velocity is a decreasing function of $\tilde a$ for energies smaller than a critical value $E_c \simeq 0.65223 \mu^2$ defined through the equation
\begin{equation}
6\zeta \left( {{\omega _1}\left( {{E_c}} \right);{g_2}\left( {{E_c}} \right),{g_3}\left( {{E_c}} \right)} \right) = {E_c}{\omega _1}\left( {{E_c}} \right)
\label{eq:kink_E_criterion}
\end{equation}
and an increasing function for $E > \mu^2$. In the intermediate range of constants $E$ there is a global maximum. Bearing in mind the pendulum picture for the translationally invariant elliptic solution of the sine-Gordon equation, the criterion \eqref{eq:kink_E_criterion} is equivalent to demanding that the mean potential energy of the pendulum vanishes.

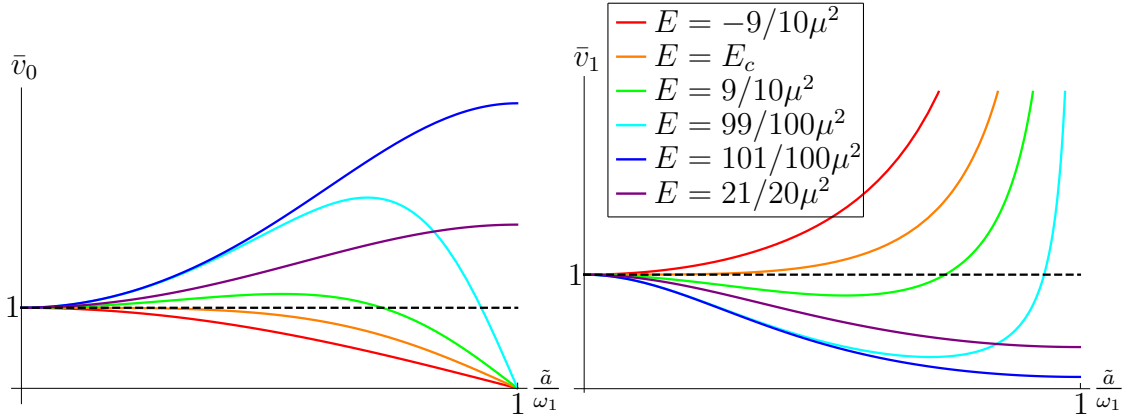
\begin{figure}[ht]
\vspace{10pt}
\begin{center}
\begin{picture}(100,37)
\put(2,1){\includegraphics[width = 0.45\textwidth]{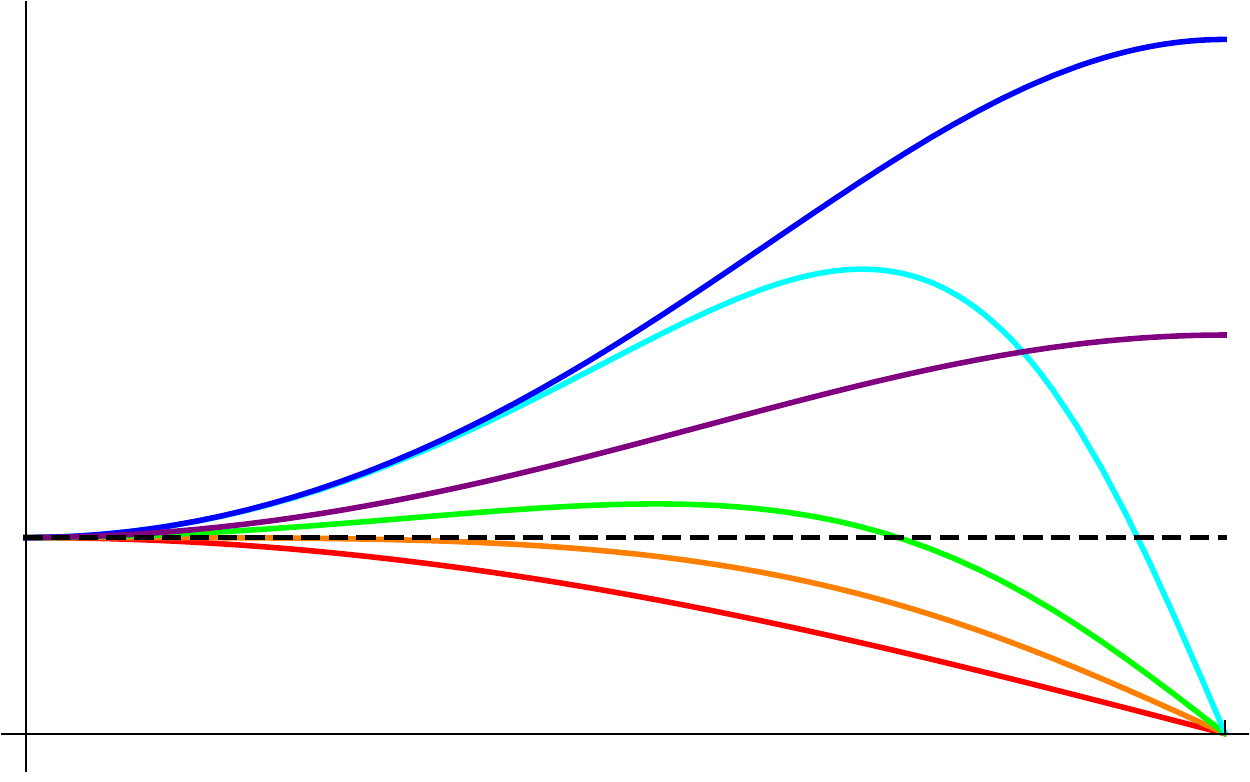}}
\put(45.5,0.5){$1$}
\put(1.5,9){1}
\put(2,30.25){${\bar v}_0$}
\put(47.25,2){$\frac{\tilde{a}}{\omega_1}$}
\put(51,1){\includegraphics[width = 0.45\textwidth]{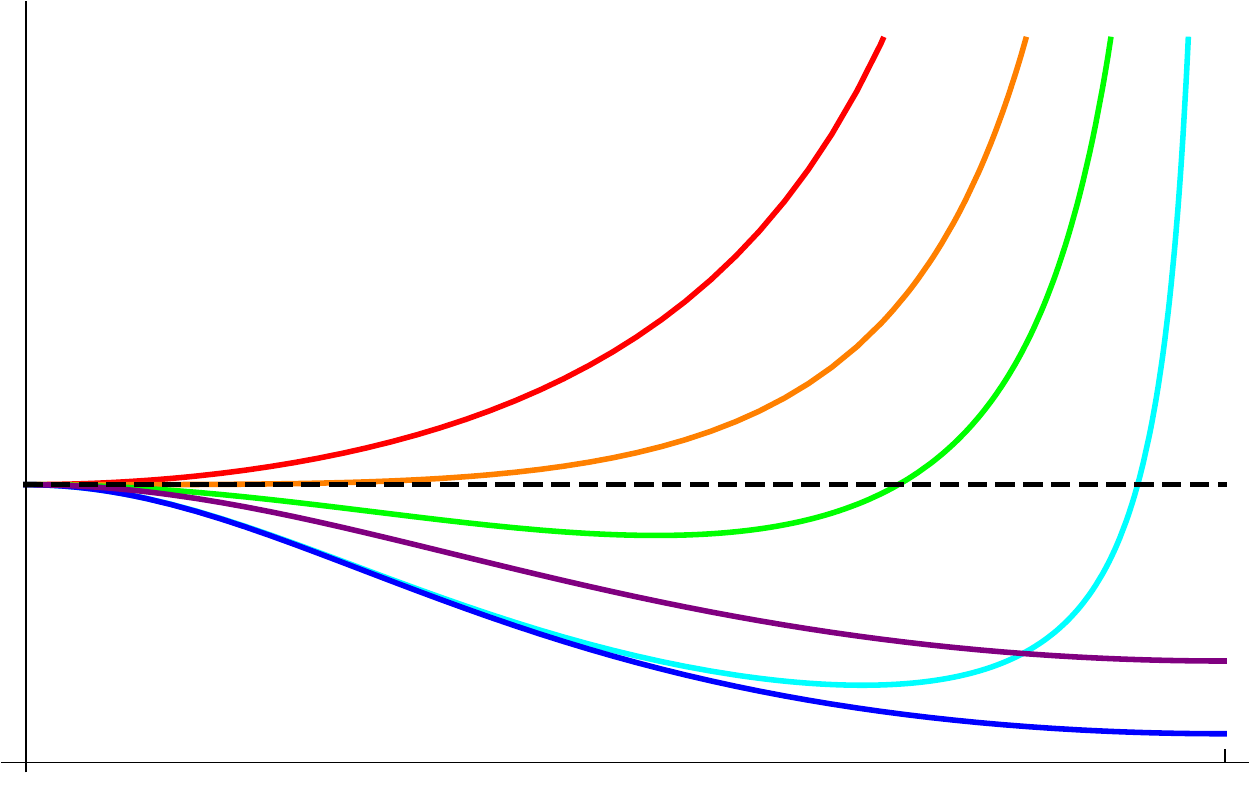}}
\put(94.5,0.5){$1$}
\put(50.5,11.75){1}
\put(51,31){${\bar v}_1$}
\put(96.25,2){$\frac{\tilde{a}}{\omega_1}$}
\put(54,17.75){\includegraphics[height = 0.1875\textwidth]{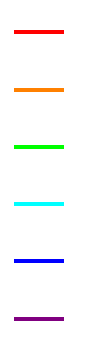}}
\put(54,17.75){\line(0,1){18.75}}
\put(54,17.75){\line(1,0){22.25}}
\put(76.25,17.75){\line(0,1){18.75}}
\put(54,36.5){\line(1,0){22.25}}
\put(58,19){$E=21/20\mu^2$}
\put(58,22){$E=101/100\mu^2$}
\put(58,25){$E=99/100\mu^2$}
\put(58,28){$E=9/10\mu^2$}
\put(58,31){$E=E_c$}
\put(58,34){$E=-9/10\mu^2$}
\end{picture}
\end{center}
\vspace{-10pt}
\caption{The mean velocity as function of $\tilde{a}$ for translationally invariant seeds (left) and static seeds (right) for various values of the energy constant $E$}
\vspace{5pt}
\label{fig:mean_velocity}
\end{figure}

Furthermore, 
\begin{equation}
\mathop {\lim }\limits_{\tilde a \to 0} {\bar v}_0 = 1 .
\end{equation}
In the case of an oscillating background, it is also trivial that
\begin{equation}
\mathop {\lim }\limits_{\tilde a \to \omega_1} {\bar v}_0 = 0 .
\end{equation}
Thus, all possible velocities between $0$ and $1$ relative to the translationally invariant background are allowed. In the case of rotating backgrounds though, the expression for the velocity \eqref{eq:kinks_mean_velocity} is undetermined at the limit $\tilde a \to \omega_1$ and it turns out that
\begin{equation}
\mathop {\lim }\limits_{\tilde a \to {\omega _1}} {\bar v}_0 = \frac{{{{\zeta \left( {{\omega _1}} \right)}} / {{{\omega _1}}} + {x_1}}}{{\sqrt {\left( {{x_1} - {x_2}} \right)\left( {{x_1} - {x_3}} \right)} }} \equiv \bar v_{\max} > 1 ,
\label{eq:elliptic_kink_vmax}
\end{equation}
implying that all kinks on a rotating background are moving with speeds larger than the speed of light and up to the value given by \eqref{eq:elliptic_kink_vmax}. The left panel of figure \ref{fig:mean_velocity} depicts the dependence of the mean velocity on the modulus $\tilde{a}$ for various values of the other modulus $E$. To sum up, only when $E<E_c$, all kinks moving on the elliptic background are subluminal. When $E>E_c$, there is always a range of $\tilde{a}$ corresponding to superluminal kinks.

When kinks propagating on a static elliptic background solution are considered, both the instant and the mean velocity are simply the inverse of the ones calculated for the translationally invariant backgrounds as given by equations \eqref{eq:kinks_instant_velocity} and \eqref{eq:kinks_mean_velocity}, i.e.
\begin{equation}
{\bar v} _1 = \frac{{{\omega _1}D}}{{\zeta \left( {\tilde a} \right){\omega _1} - \zeta \left( {{\omega _1}} \right)\tilde a}}.
\label{eq:kinks_mean_velocity_static}
\end{equation}
Therefore, kinks propagating on an oscillating static background are always superluminal, when $E<E_c$, but there are kinks moving with velocities under the speed of light when $E>E_c$, whereas kinks propagating on a rotating static background move with velocities smaller than the speed of light. However, they cannot move with an arbitrarily small velocity. The minimum velocity is the inverse of $\bar v_{\max}$ as given by \eqref{eq:elliptic_kink_vmax}. The right panel of figure \ref{fig:mean_velocity} depicts the dependence of the mean velocity on the modulus $\tilde{a}$.
In the case of the static seed, only when $E>\mu^2$ all kinks propagating on the elliptic background are subluminal. When $E<\mu^2$, there is always a range of $\tilde{a}$ giving rise to superluminal kinks.

\subsection{$D^2>0$: Periodic Properties}
\label{subsec:SG_kink_periodicity}

The elliptic solutions of the sine-Gordon equation have specific periodic properties. These are critical in the determination of the appropriate periodicity conditions for the construction of the corresponding elliptic strings solutions. The translationally invariant elliptic solutions obey
\begin{equation}
\varphi \left( {{\xi ^0} + 4{\omega _1},{\xi ^1} + \delta {\xi ^1}} \right) = \varphi \left( {{\xi ^0},{\xi ^1}} \right) ,
\label{eq:elliptic_periodic_osc}
\end{equation}
when they are oscillatory, and
\begin{equation}
\varphi \left( {{\xi ^0} + 2{\omega _1},{\xi ^1} + \delta {\xi ^1}} \right) = \varphi \left( {{\xi ^0},{\xi ^1}} \right) + 2\pi ,
\label{eq:elliptic_periodic_rot}
\end{equation}
when they are rotating. The above properties hold for any value of $\delta \xi^1$, which is a result of the fact that $\varphi$ does not depend on $\xi^1$. The static solutions have similar periodic properties that are given by the relations above after the interchange $\xi^0 \leftrightarrow \xi^1$.

The periodic properties of the dressed elliptic solutions have been disturbed due to the presence of the kink, which needs infinite time to complete. However, the new solution has also some interesting periodic properties.

Firstly, in the region far away from the location of the kink $\left| {D{\xi ^1} + i\Phi \left( {{\xi ^0};\tilde a} \right)} \right| \gg 1$, the solution tends to a shifted version of the elliptic seed solution. Therefore, at this region, the periodic properties \eqref{eq:elliptic_periodic_osc} and \eqref{eq:elliptic_periodic_rot} are approximately recovered.

Secondly, as the shape of the kink also alters periodically in time, an observer that follows the kink thinks that the sine-Gordon field alters periodically in all positions. This is evident in equation \eqref{eq:kink_in_terms_of_asymptotics}, which implies
\begin{equation}
\tilde \varphi \left( {{\xi ^0} + 4{\omega _1},{\xi ^1} + 4{{\bar v}_0}{\omega _1}} \right) = \tilde \varphi \left( {{\xi ^0},{\xi ^1}} \right)
\label{eq:dressed_elliptic_periodic_Dpos_osc}
\end{equation}
for solutions with oscillatory seeds and
\begin{equation}
\tilde \varphi \left( {{\xi ^0} + 2{\omega _1},{\xi ^1} + 2{{\bar v}_0}{\omega _1}} \right) = \tilde \varphi \left( {{\xi ^0},{\xi ^1}} \right) + 2\pi
\label{eq:dressed_elliptic_periodic_Dpos_rot}
\end{equation}
for solutions with rotating seeds.

Trivially, one can acquire the corresponding periodic properties of the dressed elliptic solutions with static kinks, after the interchange $\xi^0 \leftrightarrow \xi^1$.

\subsection{$D^2>0$: Energy and Momentum}
\label{subsec:SG_kink_energy_momentum}

The energy-momentum tensor of the sine-Gordon theory is given by
\begin{align}
{T^{00}} &= \frac{1}{2}{\left( {{\partial _0}\varphi } \right)^2} + \frac{1}{2}{\left( {{\partial _1}\varphi } \right)^2} - {\mu ^2}\cos \varphi \equiv \mathcal{H} ,\\
{T^{01}} &=  - \left( {{\partial _0}\varphi } \right)\left( {{\partial _1}\varphi } \right) \equiv \mathcal{P} \equiv J_{\mathcal{H}},\\
{T^{11}} &= \frac{1}{2}{\left( {{\partial _0}\varphi } \right)^2} + \frac{1}{2}{\left( {{\partial _1}\varphi } \right)^2} + {\mu ^2}\cos \varphi \equiv J_{\mathcal{P}}.
\end{align}
Since the static solutions can be derived from the translationally invariant ones via the interchange of the variables $\xi^0$ and $\xi^1$ and a shift of $\varphi$ by $\pi$, it holds that if $T_{{\rm{st}}}^{00} = {f^0}\left( {{\xi ^0},{\xi ^1}} \right)$, then $T_{{\rm{ti}}}^{11} = {f^0}\left( {{\xi ^1},{\xi ^0}} \right)$ and similarly if $T_{{\rm{st}}}^{01} = {f^1}\left( {{\xi ^0},{\xi ^1}} \right)$, then $T_{{\rm{ti}}}^{01} = {f^1}\left( {{\xi ^1},{\xi ^0}} \right)$.

The elliptic solutions of the sine-Gordon equation lead to simple expressions for most of the elements of the energy-momentum tensor (see e.g. \cite{part1}). More specifically, $T_{{\rm{ti}}}^{00} = T_{{\rm{st}}}^{11} = E$ and $T_{{\rm{ti/st}}}^{01} = 0$. However, the elements $T_{{\rm{ti}}}^{11}$ and $T_{{\rm{st}}}^{00}$ are non-trivial functions of $\xi^0$ and $\xi^1$ respectively.

Let us now study the energy and momentum of the dressed solutions of the sine-Gordon equation. We initiate our analysis considering the kinks propagating on a translationally invariant elliptic background. It is a matter of algebra to calculate the energy density to find,
\begin{equation}
\mathcal{H} = 2DA\frac{{\sin \left[ {4\arctan \left( {\frac{{A + B}}{D}\tanh \frac{{D{\xi ^1} + i \Phi \left( {{\xi ^0};\tilde a} \right)}}{2}} \right)} \right]}}{{\sinh \left( {D{\xi ^1} + i \Phi \left( {{\xi ^0};\tilde a} \right)} \right)}} + E.
\end{equation}
Therefore, the energy density, far away from the kink position assumes the same constant value that matches the energy density of the seed solution. This is not surprising, since we have seen that the asymptotics of the dressed solution far away from the kink is the seed solution shifted by an appropriate time/position. Actually, we could also have deduced the above fact by the form of the energy density.

Defining the kink energy density as the difference of the energy densities of the dressed solution and the background solution, we can calculate the energy of the kink and find it equal to
\begin{equation}
E_{\textrm{kink}} = \int {d{\xi ^1} \left( \mathcal{H} - E \right) = 8D} .
\label{eq:kink_ti_energy}
\end{equation}
The above formula reveals the physical meaning of the constant $D$. It is now clear why the quantity $D^2$ is a decreasing function of the energy constant $E$, since the larger the background energy, the smaller the necessary energy for a kink to jump from the region of one vacuum to the region of the neighbouring one. Furthermore, it is also physically expected that the kink energy is a decreasing function of the background time delay $2 \tilde {a}$. As the latter gets larger approaching $\omega_1$, the jump is facilitated and less energy is required for this purpose (see figure \ref{fig:time_delay}).

\begin{figure}[ht]
\vspace{10pt}
\begin{center}
\begin{picture}(100,40)
\put(2.5,0){\includegraphics[width = 0.45\textwidth]{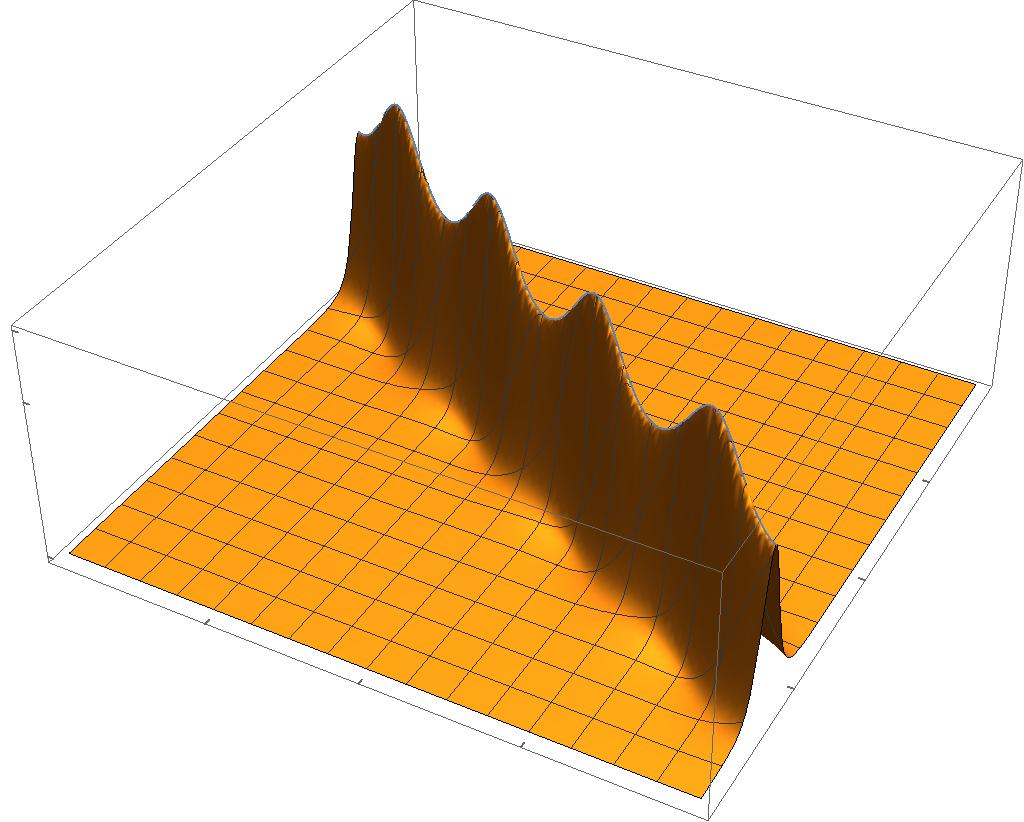}}
\put(52.5,0){\includegraphics[width = 0.45\textwidth]{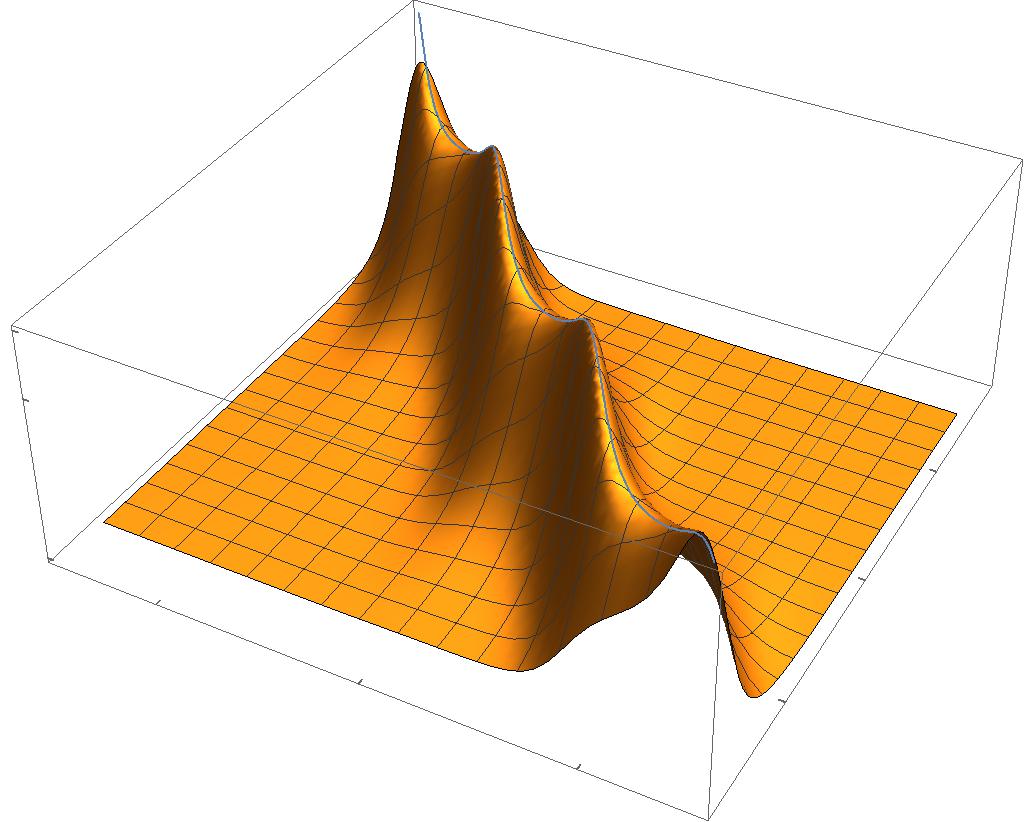}}
\put(0.5,15.75){$\mathcal{H}$}
\put(42.5,9.25){$\xi^1$}
\put(15.25,3){$\xi^0$}
\put(43.5,13.75){$4\omega_1$}
\put(35.5,4.25){$-4\omega_1$}
\put(22.75,1){$4\omega_1$}
\put(6.5,6.75){$-4\omega_1$}
\put(2.75,11.25){$E$}
\put(50.5,15.75){$\mathcal{H}$}
\put(92.5,9.25){$\xi^1$}
\put(65.25,3){$\xi^0$}
\put(93.5,13.75){$4\omega_1$}
\put(85.5,4.25){$-4\omega_1$}
\put(72.75,1){$4\omega_1$}
\put(56.5,6.75){$-4\omega_1$}
\put(52.75,11.25){$E$}
\end{picture}
\end{center}
\vspace{-10pt}
\caption{The energy densities of the dressed elliptic solutions with translationally invariant seeds depicted in figure \ref{fig:ti_pos_1}}
\vspace{5pt}
\label{fig:osc_energy}
\end{figure}
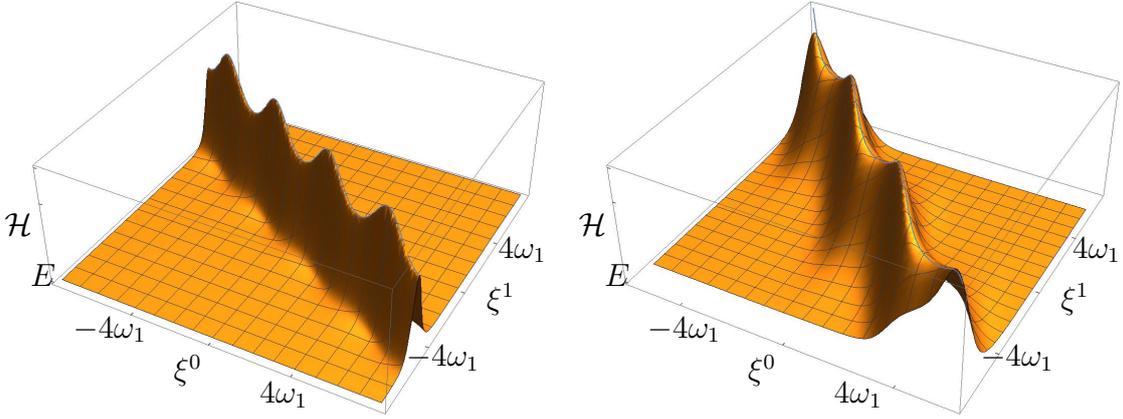

As the kink propagates, it periodically changes shape, due to the interaction with the non-trivial elliptic background. This is also depicted in the profile of the energy density. One measure that quantifies this phenomenon is the peak energy density at the location of the kink. The latter equals
\begin{equation}
\mathcal{H}_{\textrm{peak}} - E = 4 A \left({A + B}\right),
\end{equation}
which obviously is a periodic function of time.
In the limit $E \to - \mu^2$, the energy density of the peak becomes constant as expected from the physics of the kinks propagating on the vacuum. Figure \ref{fig:osc_energy} depicts the energy density for the two solutions depicted in figure \ref{fig:ti_pos_1}.

In a similar manner, we may calculate the momentum density of the kink solution
\begin{multline}
\mathcal{P} = - A\frac{{\sin \left[ {4\arctan \left( {\frac{{A + B}}{D}\tanh \frac{{D{\xi ^1} - \Phi \left( {{\xi ^0};\tilde a} \right)}}{2}} \right)} \right]}}{{\sinh \left( {D{\xi ^1} - \Phi \left( {{\xi ^0};\tilde a} \right)} \right)}}\frac{{\wp '\left( {\tilde a} \right)}}{{\wp \left( {{\xi ^0 + \omega_2}} \right) - \wp \left( {\tilde a} \right)}}\\
 - \frac{{2{\mu ^2}D}}{A}\frac{{{{\sin }^2}\left[ {2\arctan \left( {\frac{{A + B}}{D}\tanh \frac{{D{\xi ^1} - \Phi \left( {{\xi ^0};\tilde a} \right)}}{2}} \right)} \right]}}{{\sinh \left( {D{\xi ^1} - \Phi \left( {{\xi ^0};\tilde a} \right)} \right)}}\sin \varphi .
\end{multline}
The momentum density vanishes far away from the location of the kink, a fact which is expected since the momentum density of the elliptic background solution vanishes. We define the kink momentum as the integral of the momentum density over all space to find
\begin{equation}
P_{\textrm{kink}} = 4\frac{{\wp '\left( {\tilde a} \right)}}{{{A^2}}} =  - 4\frac{{\wp '\left( {\tilde a} \right)}}{{\wp \left( {{\xi ^0} + {\omega _2}} \right) - \wp \left( {\tilde a} \right)}} = 8 D v_0 = E_{\textrm{kink}} v_0,
\label{eq:kink_ti_momentum}
\end{equation}
as one would expect for a particle. Like the instant velocity, the kink momentum is not constant in time. One could define the mean kink momentum as
\begin{equation}
\bar P_{\textrm{kink}} = 8 D {\bar v}_0 = E_{\textrm{kink}} {\bar v}_0 .
\end{equation}

It may appear surprising that the momentum of the kink is not conserved, although the theory possesses translational symmetry. This is due to the asymptotic behaviour of $T^{11}$ in the case of translationally invariant seeds. The conservation law ${\partial _0}{T^{01}} + {\partial _1}{T^{11}} = 0$ implies that
\begin{equation}
{\partial _0}P = {T^{11}}\left( {{\xi ^1} \to  + \infty } \right) - {T^{11}}\left( {{\xi ^1} \to  - \infty } \right) .
\end{equation}
Asymptotically, the solution acquires the form of the translationally invariant seed solution, with a time shift, which is different at plus and minus infinities. As the element $T^{11}$ is a non-trivial periodic function of time in this case, it follows that the kink momentum cannot be conserved. On the contrary, the energy is conserved, since
\begin{equation}
{\partial _0}E = \mathcal{P}\left( {{\xi ^1} \to  + \infty } \right) - \mathcal{P}\left( {{\xi ^1} \to  - \infty } \right) = 0 ,
\end{equation}
as the momentum density of the seed solution vanishes.

When we consider kinks propagating on static backgrounds, it is not easy to repeat the above calculations, since the dependence of the dressed solution on the space-like coordinate $\xi^1$ is highly non-trivial. However, we may adopt a different approach, calculating the total flow of energy or momentum that passes through a given location. Converting from static to translationally invariant backgrounds, leaves the expression of the momentum density the same, apart from an interchange of $\xi^0$ and $\xi^1$. It follows that the flow of energy $E_{\textrm{st}}^{\textrm{flow}} = \int {d{\xi ^0}{\mathcal{P}_{\textrm{st}}}\left( {{\xi ^0},{\xi ^1}} \right)} $ through a given point can be derived from the total momentum of the kink on a translationally invariant background ${P_{\textrm{ti}}} = \int {d{\xi ^1}{\mathcal{P}_{\textrm{ti}}}\left( {{\xi ^0},{\xi ^1}} \right)} $ after the same interchange, Thus,
\begin{equation}
{E_{\textrm{flow}}} =  - 4\frac{{\wp '\left( {\tilde a} \right)}}{{\wp \left( {{\xi ^1} + {\omega _2}} \right) - \wp \left( {\tilde a} \right)}} .
\end{equation}
Naturally, this is not constant. As we have already commented in section \ref{subsec:SG_asymptotics}, the passage of the kink has translated the static background, and as the latter has a non-trivial energy density profile, it has translated energy. In this case, the effect of the interaction of the kink with the background is not limited to a time delay, but it extends to the energy density. The kink energy can be identified as the mean energy flow per spatial period. Bearing in mind that the kink velocity on a static background is the inverse of that on a translationally invariant background with the same \Backlund parameter $a$, the above imply
\begin{equation}
{{ E}_{\textrm{kink}}} = {{\bar E}_{\textrm{flow}}} = 8\left( {\zeta \left( {{\omega _1}} \right)\frac{{\tilde a}}{{{\omega _1}}} - \zeta \left( {\tilde a} \right)} \right) = \frac{{8D}}{{{\bar v}_1}} .
\label{eq:kink_static_energy}
\end{equation}
In a similar manner the flow of momentum from a given point in the case of a static background $P_{\textrm{st}}^{\textrm{flow}} = \int {d{\xi ^0}T_{\textrm{st}}^{11}\left( {{\xi ^0},{\xi ^1}} \right)} $ can be deduced from the energy in the case of a translationally invariant background ${E_{\textrm{ti}}} = \int {d{\xi ^1}{\mathcal{H}_{\textrm{ti}}}\left( {{\xi ^0},{\xi ^1}} \right)} $, after an interchange of $\xi_0$ and $\xi^1$. Subtracting the momentum flow of the background solution, in order to define the kink momentum, yields
\begin{equation}
P_{\textrm{kink}}  = \int {d{\xi ^0}\left( {{T^{11}}\left( {{\xi ^0},{\xi ^1}} \right) - E} \right)} = 8 D = E_{\textrm{kink}} {\bar v}_1 .
\label{eq:kink_static_momentum}
\end{equation}

Notice that in the case of a static seed, both the energy and momentum of the kink are conserved quantities, since
\begin{align}
{\partial _0}E &= \mathcal{P}\left( {{\xi ^1} \to  + \infty } \right) - \mathcal{P}\left( {{\xi ^1} \to  - \infty } \right) = 0 ,\\
{\partial _0}P &= {T^{11}}\left( {{\xi ^1} \to  + \infty } \right) - {T^{11}}\left( {{\xi ^1} \to  - \infty } \right) = E - E = 0 .
\end{align}

The algebra of the \Backlund transformations results in dressed elliptic solutions that are naturally expressed in terms of the parameters $D$ and $\tilde{a}$ \cite{Katsinis:2018ewd}. Interestingly, both parameters have a simple physical meaning. The parameter $D$ is directly related to the energy of the kink in the case of a translationally invariant seed solution (equation \eqref{eq:kink_ti_energy}) or its momentum in the case of a static one (equation \eqref{eq:kink_static_momentum}). The parameter $\tilde{a}$ directly measures the degree of interaction of the kink with the elliptic background. In the case of a translationally invariant seed, it is directly related to the time delay in the background field oscillation induced by the kink (equation \eqref{eq:kink_ti_time_delay}); in the case of a static seed, it is related to the spatial displacement of the static background (equation \eqref{eq:kink_static_spacial_displacement}). Bearing in mind that there are not two independent parameters in this class of solutions, but only one (the \Backlund parameter $a$), there is a relation connecting the energy/momentum of the kink to the effect that it has on the background. This reads
\begin{equation}
\frac{{E_{{\rm{kink}}}^2}}{{64}} = \wp \left( {\frac{{\Delta {\xi ^0}}}{2};\frac{{{E^2}}}{3} + {\mu ^4},\frac{E}{3}\left( {\frac{{{E^2}}}{9} - {\mu ^4}} \right)} \right) - \frac{E}{3} ,
\end{equation}
for translationally invariant backgrounds and 
\begin{equation}
\frac{{P_{{\rm{kink}}}^2}}{{64}} = \wp \left( {\frac{{\Delta {\xi ^1}}}{2};\frac{{{E^2}}}{3} + {\mu ^4},\frac{E}{3}\left( {\frac{{{E^2}}}{9} - {\mu ^4}} \right)} \right) - \frac{E}{3} ,
\end{equation}
for the static ones. The above relations can in principle be verified experimentally in physical systems realizing the sine-Gordon equation, such as coupled torsion pendula, Josephson junctions, spin waves in magnetics (see e.g. \cite{Kevrekidis_book}).

\subsection{$D^2<0$: Periodicity}
\label{subsec:SG_breathers_periodicity}

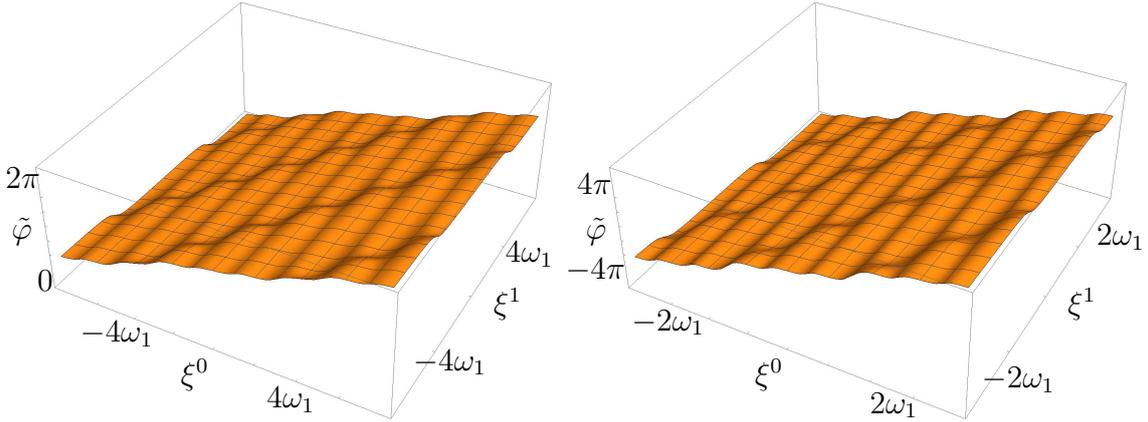
\begin{figure}[ht]
\vspace{10pt}
\begin{center}
\begin{picture}(100,37)
\put(2.5,0){\includegraphics[width = 0.45\textwidth]{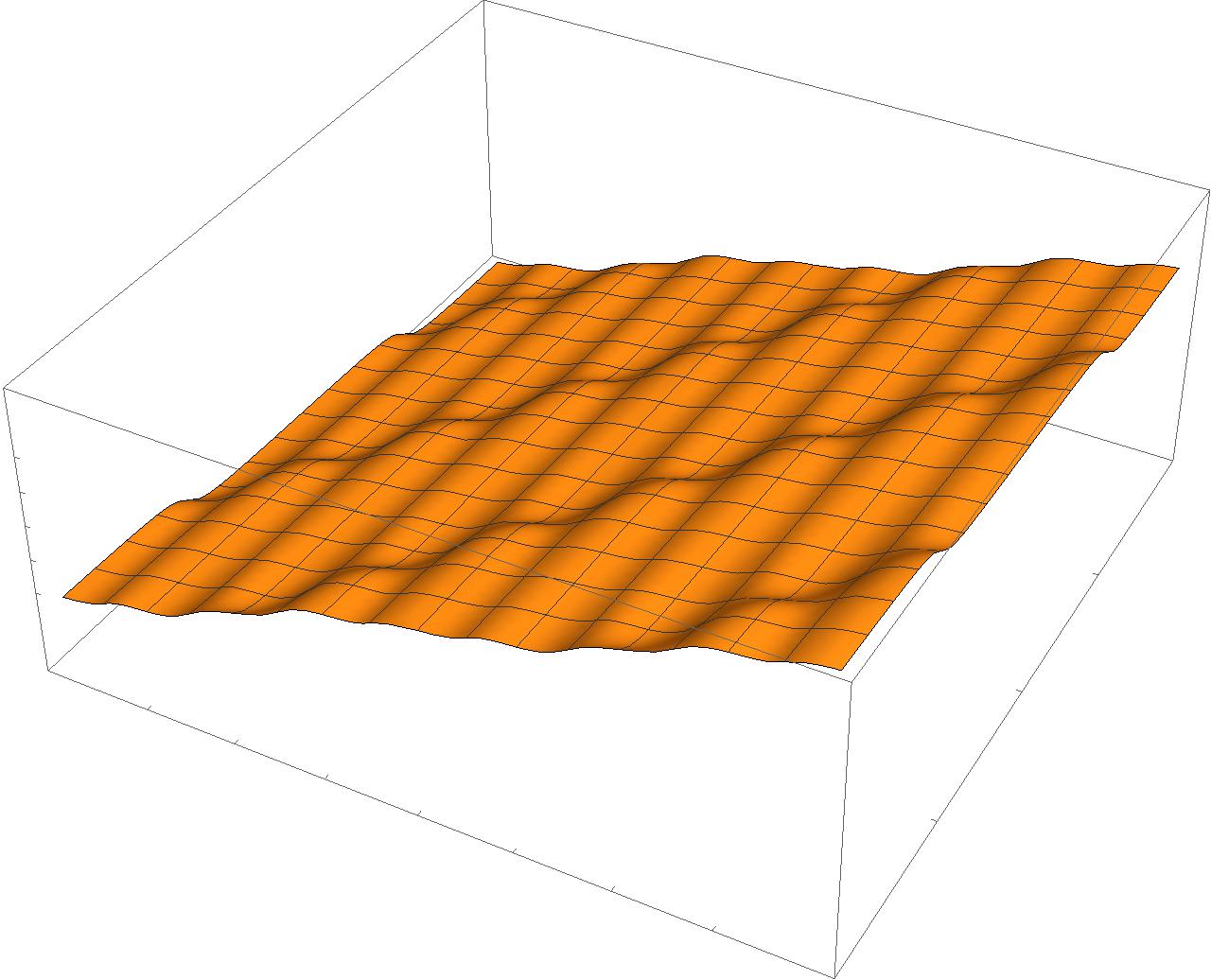}}
\put(52.5,0){\includegraphics[width = 0.45\textwidth]{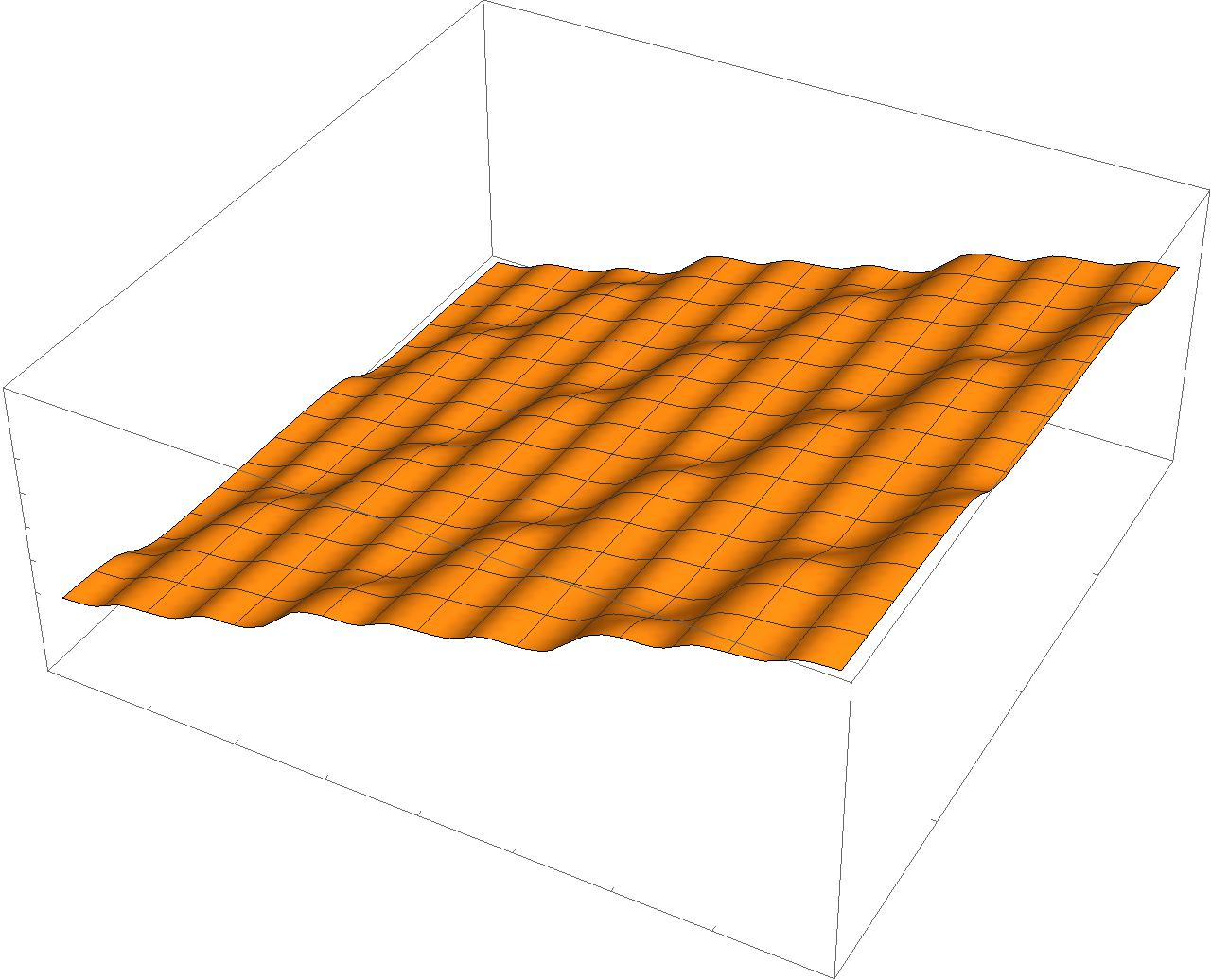}}
\put(0.5,15.75){$\tilde{\varphi}$}
\put(42.5,9.25){$\xi^1$}
\put(15.25,3){$\xi^0$}
\put(43.5,13.75){$4\omega_1$}
\put(35.5,4.25){$-4\omega_1$}
\put(22.75,1){$4\omega_1$}
\put(6.5,6.75){$-4\omega_1$}
\put(2.75,11.25){$0$}
\put(0,20.25){$2\pi$}
\put(50.5,15.75){$\tilde{\varphi}$}
\put(92.5,9.25){$\xi^1$}
\put(65.25,3){$\xi^0$}
\put(95,15.25){$2\omega_1$}
\put(85,2.75){$-2\omega_1$}
\put(75.25,0.5){$2\omega_1$}
\put(54.5,7.75){$-2\omega_1$}
\put(48.75,12.25){$-4\pi$}
\put(49.75,19.75){$4\pi$}
\end{picture}
\end{center}
\vspace{-10pt}
\caption{The solution with $D^2<0$ for two distinct \Backlund parameters. The background solution has energy density $E = 3 \mu^2 /2$ and the \Backlund parameter take the value $a = 1.45482$ on the left and $a = 1.36771$ on the right.}
\vspace{5pt}
\label{fig:ti_non_prop_3d}
\end{figure}

When $D^2 < 0$, the solution acquires the form
\begin{equation}
\tilde \varphi  = 
{\hat \varphi} + 4\arctan \left[ \frac{{A + B}}{i D}\tan \frac{{i D{\xi ^1} - \Phi \left( {{\xi ^0};\tilde a} \right)}}{2} \right] .
\label{eq:D2_negative_solution}
\end{equation}
Figure \ref{fig:ti_non_prop_3d} depicts two example cases of such solutions.
These solutions do not describe a localized kink propagating on top of an elliptic background. They are actually a periodic disturbance propagating on top of a translationally invariant rotating elliptic background. This transition of the qualitative characteristics of the solution is in a sense similar to the well-known behaviour of the solutions that occur after the action of two \Backlund transformations of the vacuum. These solutions form two classes; one class of two-kink scattering solutions and one class of bound states, the so called breathers. Having this picture in mind, we may understand the \Backlund transformed elliptic solutions with $D^2>0$ as the analogue of the scattering solutions, since the kink induced by the \Backlund transformation propagates on top of the train of kinks that forms the elliptic background, interacting with it, causing a delay/translation. On the contrary, the solutions with $D^2<0$ are the analogue of the breathers. Of course instead of a single oscillating breather, these solutions are a whole periodic structure of such oscillating formations, a ''train of breathers''.

The solution \eqref{eq:D2_negative_solution} is obviously periodic in $\xi^1$ since
\begin{equation}
\tilde \varphi \left( \xi^0 , \xi^1 + 2 \pi / \left(i D \right) \right) = \tilde \varphi \left( \xi^0 , \xi^1 \right) .
\label{eq:D2_negative_periodicity_trivial}
\end{equation}
Furthermore, the quasi-periodic properties of the function $\Phi$ imply that
\begin{equation}
\tilde \varphi \left( {{\xi ^0} + 2{\omega _1},{\xi ^1} + 2\frac{{\zeta \left( {\tilde a} \right){\omega _1} - \zeta \left( {{\omega _1}} \right)\tilde a}}{D}} \right) = \tilde \varphi \left( {{\xi ^0},{\xi ^1}} \right) + 2\pi .
\label{eq:D2_negative_periodicity_non_trivial}
\end{equation}

It follows that the solutions with $D^2<0$ are periodic/quasi-periodic under translations in an non-orthogonal two-dimensional lattice. One of the two directions of the lattice coincides with the space-like (in the case of translationally invariant seeds) or time-like (in the case of static seeds) directions. The other is determined by a velocity, which is the average velocity of the periodic disturbances. This velocity equals
\begin{equation}
{v_0^{{\rm{tb}}}} = \frac{{\zeta \left( {\tilde a} \right){\omega _1} - \zeta \left( {{\omega _1}} \right)\tilde a}}{{{\omega _1}D}}
\label{eq:velocity_tb_ti}
\end{equation}
and it is the analytic continuation of the kink mean velocity \eqref{eq:kinks_mean_velocity}.

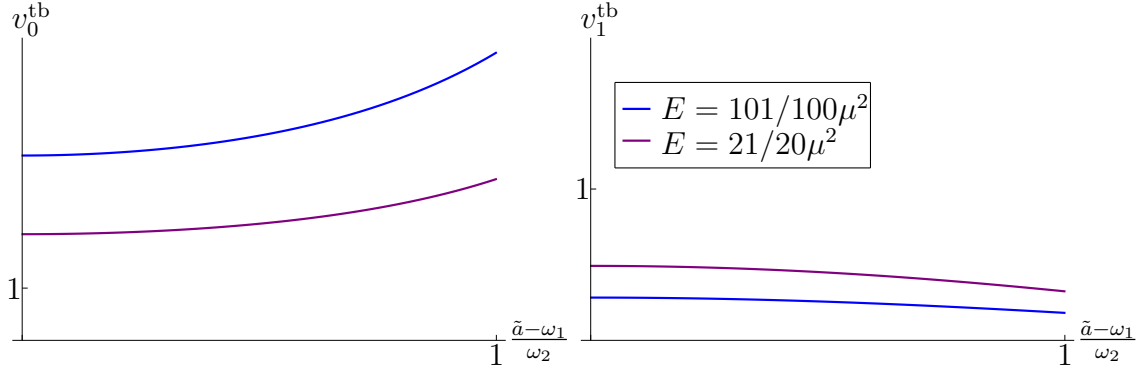
\begin{figure}[ht]
\vspace{10pt}
\begin{center}
\begin{picture}(100,37)
\put(1.5,1){\includegraphics[width = 0.43\textwidth]{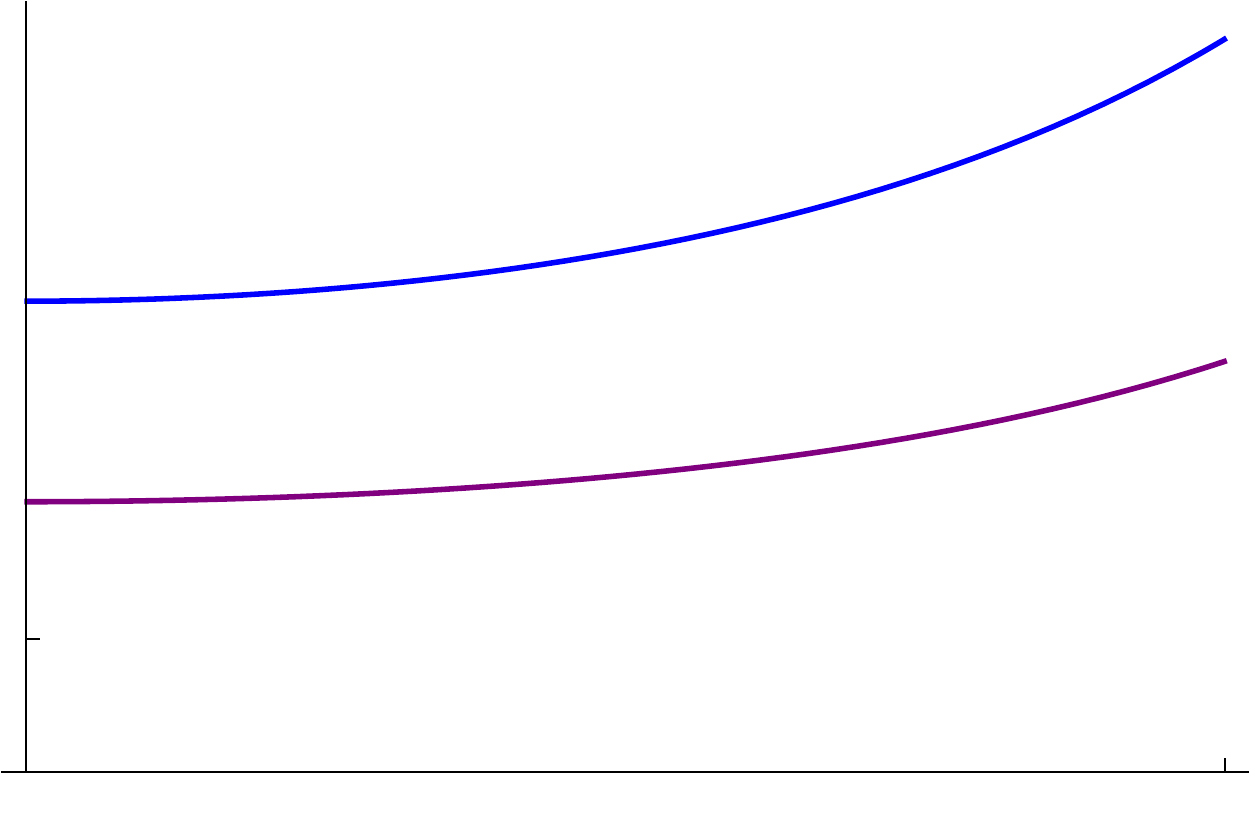}}
\put(43,0.5){$1$}
\put(1,6.25){1}
\put(1.5,30){${v}_0^{{\rm{tb}}}$}
\put(44.75,2){$\frac{\tilde{a}-\omega_1}{\omega_2}$}
\put(51,1){\includegraphics[width = 0.43\textwidth]{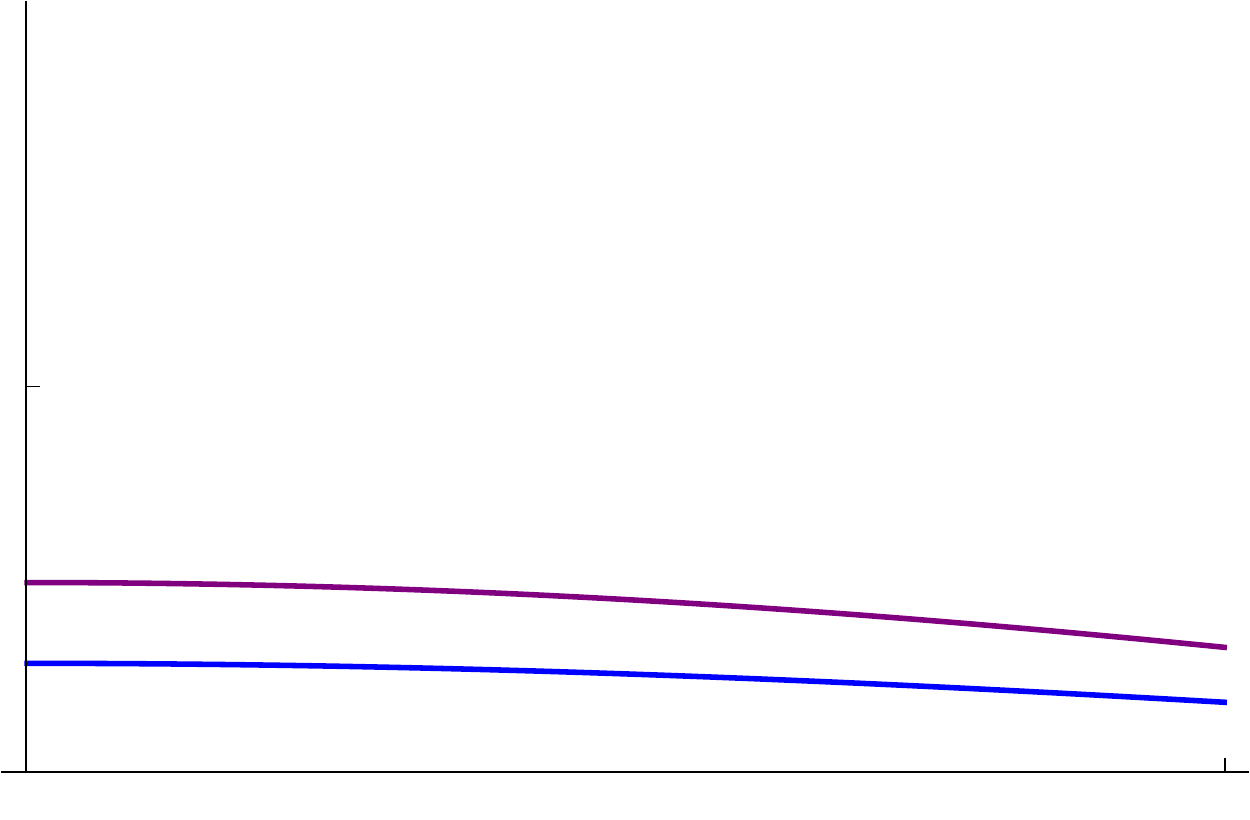}}
\put(92.5,0.5){$1$}
\put(50.5,15){1}
\put(51,30){${v}_1^{{\rm{tb}}}$}
\put(94.25,2){$\frac{\tilde{a}-\omega_1}{\omega_2}$}
\put(54,18.25){\includegraphics[height = 0.0625\textwidth]{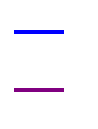}}
\put(54,17.75){\line(0,1){7.5}}
\put(54,17.75){\line(1,0){22.25}}
\put(76.25,17.75){\line(0,1){7.5}}
\put(54,25.25){\line(1,0){22.25}}
\put(58,19){$E=21/20\mu^2$}
\put(58,22){$E=101/100\mu^2$}
\end{picture}
\end{center}
\vspace{-10pt}
\caption{The velocity of the periodic disturbances as function of $\tilde{a}$ for translationally invariant seeds (left) and static seeds (right) for various values of the energy constant $E$. These curves are a smooth continuation of the corresponding ones of figure \ref{fig:mean_velocity} with the same color.}
\vspace{5pt}
\label{fig:mean_velocity_tb}
\end{figure}

As $\tilde{a}$ moves from $\omega_1$ to $\omega_3$, the velocity of the periodic disturbances ${v_0^{{\rm{tb}}}}$ increases. It also obeys
\begin{equation}
\mathop {\lim }\limits_{\tilde a \to {\omega _3}} {v_0^{{\rm{tb}}}} = \frac{\pi }{{{\omega _1}\sqrt {2\left( {E - {\mu ^2}} \right)} }}
\end{equation}
and
\begin{equation}
\mathop {\lim }\limits_{\tilde a \to {\omega _1}} {v_0^{{\rm{tb}}}} = \mathop {\lim }\limits_{\tilde a \to {\omega _1}} {v_{\rm{0}}} = \frac{{\zeta \left( {{\omega _1}} \right)/{\omega _1} + {x_1}}}{{\sqrt {\left( {{x_1} - {x_2}} \right)\left( {{x_1} - {x_3}} \right)} }} ,
\end{equation}
which implies that ${v_0^{{\rm{tb}}}}$ is always larger than the speed of light. In a similar manner, in the case of a static seed solution, the velocity of the periodic disturbances is given by the inverse of equation \eqref{eq:velocity_tb_ti}
\begin{equation}
{v_1^{{\rm{tb}}}} = \frac{{{\omega _1}D}}{{\zeta \left( {\tilde a} \right){\omega _1} - \zeta \left( {{\omega _1}} \right)\tilde a}}.
\label{eq:velocity_tb_st}
\end{equation}
The velocity ${v_1^{{\rm{tb}}}}$ decreases as $\tilde{a}$ moves from $\omega_1$ to $\omega_3$ and it is always smaller than the speed of light. The above are displayed in figure \ref{fig:mean_velocity_tb}.

\begin{figure}[ht]
\vspace{10pt}
\begin{center}
\begin{picture}(100,31)
\put(3.5,0){\includegraphics[width = 0.44\textwidth]{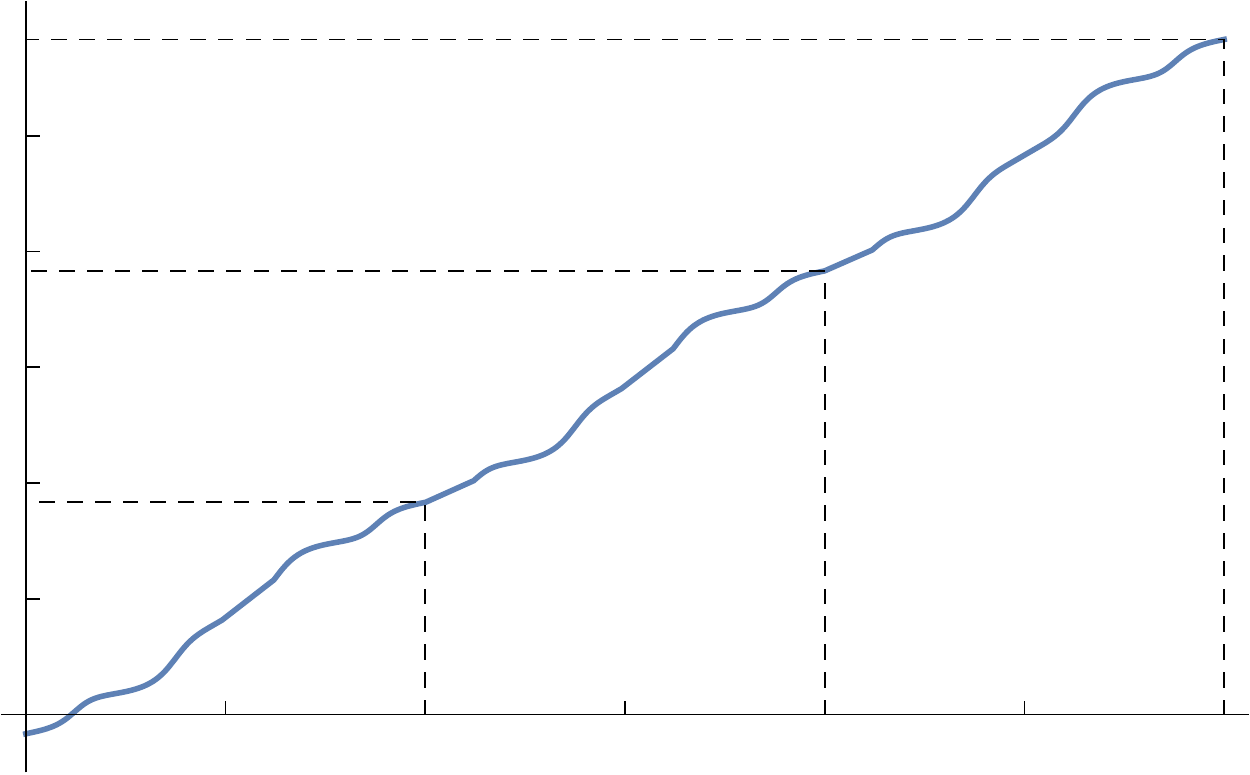}}
\put(52.5,0.5){\includegraphics[width = 0.44\textwidth]{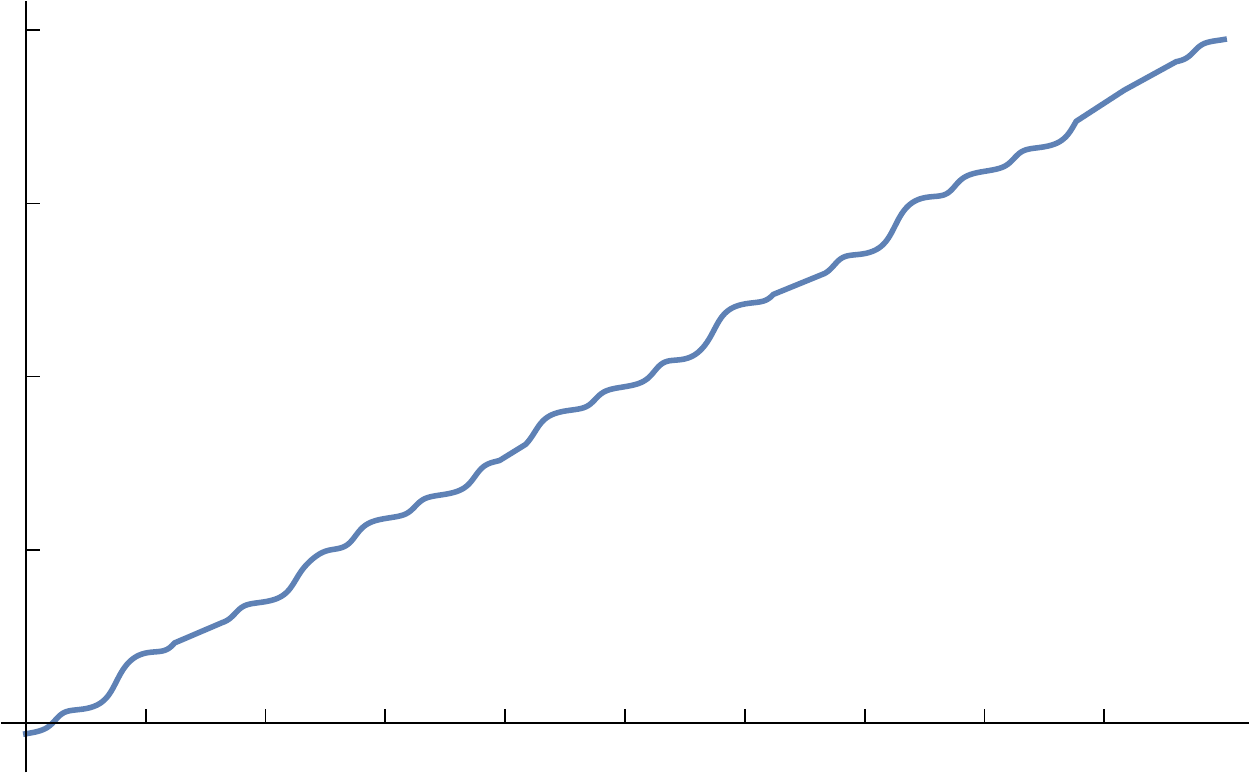}}
\put(3.5,28.25){$\tilde{\varphi}$}
\put(48,1.5){$\xi^1$}
\put(16.5,0){$8\omega_1$}
\put(29.75,0){$16\omega_1$}
\put(0,17.75){$12\pi$}
\put(1.25,9.75){$6\pi$}
\put(52.5,28.25){$\tilde{\varphi}$}
\put(97,1.5){$\xi^1$}
\put(63.5,0){$12\omega_1$}
\put(76,0){$24\omega_1$}
\put(88.5,0){$36\omega_1$}
\put(49,26){$64\pi$}
\put(49,14){$32\pi$}
\end{picture}
\end{center}
\vspace{-10pt}
\caption{The solution with $D^2<0$ at $\xi^1=0$ for two distinct \Backlund parameters. The background solution has energy density $E = 3 \mu^2 /2$ and the \Backlund parameter take the value $a = 1.45482$ on the left, corresponding to a periodic solution with $c=1/2$ and $a = 1.36771$ on the right, corresponding to a non-periodic solution with $c=(\sqrt{5}-1)/2$.}
\vspace{5pt}
\label{fig:ti_non_prop_2d}
\end{figure}
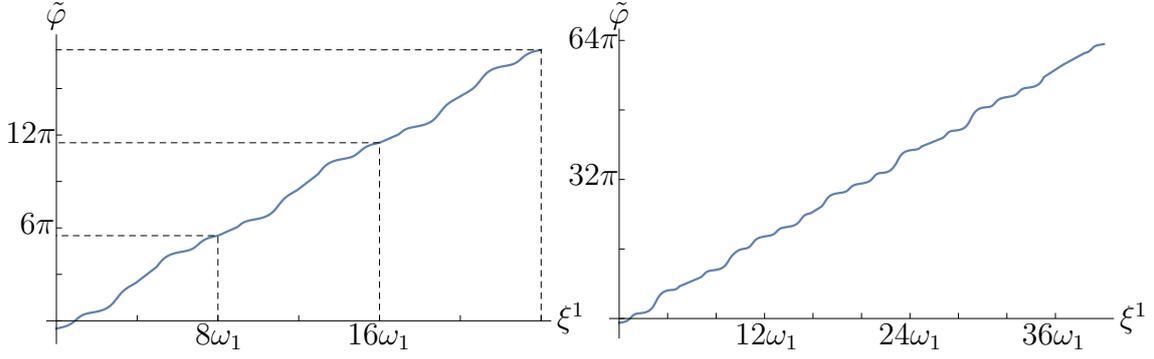

It is not obvious, whether the solution \eqref{eq:D2_negative_solution} is a periodic function of $\xi^0$. In general we have that
\begin{multline}
{{\tilde \varphi } }\left( {{\xi ^0} + 2{\omega _1},{\xi ^1}} \right) = 2 \pi + \hat \varphi \left( {{\xi ^0},{\xi ^1}} \right) \\
+ 4\arctan \left[ {\frac{{A\left( {{\xi ^0}} \right) + B\left( {{\xi ^0}} \right)}}{{iD}}\tan \left( {\frac{{iD{\xi ^1} - \Phi \left( {{\xi ^0};\tilde a} \right)}}{2} - i\left( {\zeta \left( {\tilde a} \right){\omega _1} - \zeta \left( {{\omega _1}} \right)\tilde a} \right)} \right)} \right] .
\end{multline}
The quantity ${\zeta \left( {\tilde a} \right){\omega _1} - \zeta \left( {{\omega _1}} \right)\tilde a} $ is the Bloch phase of the valence band states of the $n=1$ \Lame problem. It is always purely imaginary and its imaginary part decreases monotonically from $0$ to $-i \pi /2$ as $\tilde{a}$ moves from $\omega_1$ to $\omega_3$. Therefore, $ - i{\left( {\zeta \left( {\tilde a} \right){\omega _1} - \zeta \left( {{\omega _1}} \right)\tilde a} \right)} = - c \pi / 2$, where $c$ is a number between 0 and 1. The periodicity of the solution under translations in time is determined by number theoretic properties of the number $c$. If this number is a rational number of the form $\alpha / \beta$, where $\gcd \left( {\alpha ,\beta } \right) = 1$, then the solution ${{\tilde \varphi }}$ will be quasiperiodic in $\xi_0$ with period $4 \beta \omega_1$ and the quasiperiodicity property ${{\tilde \varphi } }\left( {{\xi ^0} + 4\beta{\omega _1},{\xi ^1}} \right) = \hat \varphi \left( {{\xi ^0},{\xi ^1}} \right) + 2\pi \left( \alpha + \beta \right)$. If $c$ is irrational, the solution ${{\tilde \varphi }}$ will not be periodic in $\xi_0$. In figure \ref{fig:ti_non_prop_2d} a periodic and a non-periodic example are shown.

Similarly, if a static background is considered, the solution is always periodic in $\xi^0$, but not always periodic in $\xi^1$, obeying the periodicity properties
\begin{align}
&\tilde \varphi \left( {{\xi ^0} + 2\pi /\left( {iD} \right),{\xi ^1}} \right) = \tilde \varphi \left( {{\xi ^0},{\xi ^1}} \right) , \label{eq:D2_negative_periodicity_trivial_static}\\
&\tilde \varphi \left( {{\xi ^0} + 2\frac{{\zeta \left( {\tilde a} \right){\omega _1} - \zeta \left( {{\omega _1}} \right)\tilde a}}{D} + 2{\omega _1},{\xi ^1} + 2{\omega _1}} \right) = \tilde \varphi \left( {{\xi ^0},{\xi ^1}} \right) + 2\pi . \label{eq:D2_negative_periodicity_non_trivial_static}
\end{align}
In this case the velocity of the periodic disturbances equals
\begin{equation}
{v_{{\rm{tb}}}} = \frac{{D{\omega _1}}}{{\zeta \left( {\tilde a} \right){\omega _1} - \zeta \left( {{\omega _1}} \right)\tilde a}} ,
\end{equation}
which is the analytic continuation of equation \eqref{eq:kinks_mean_velocity_static}.

\subsection{$D^2<0$: Energy and Momentum}
\label{subsec:SG_breathers_energy}

\begin{figure}[ht]
\vspace{10pt}
\begin{center}
\begin{picture}(100,37)
\put(2.5,0){\includegraphics[width = 0.45\textwidth]{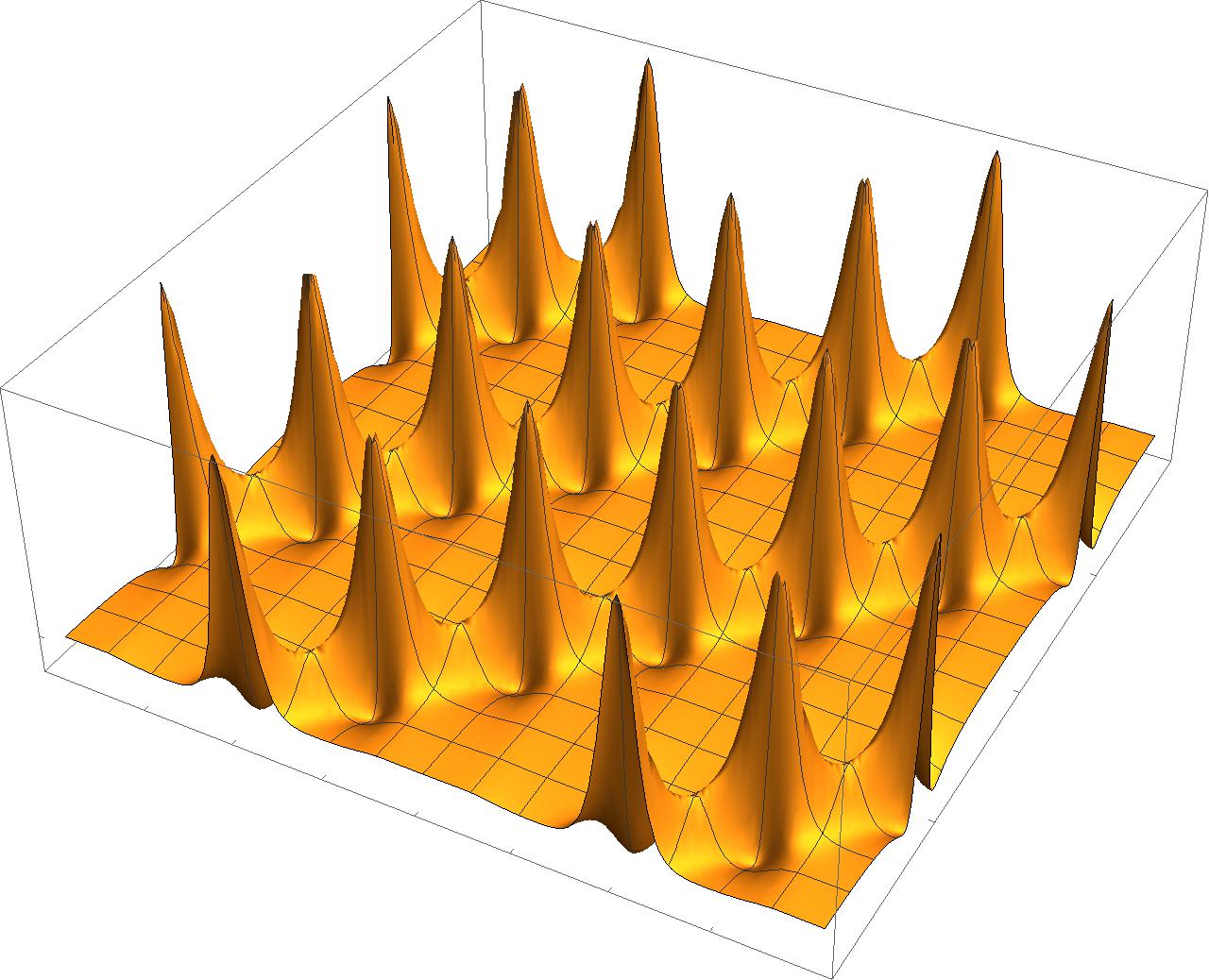}}
\put(52.5,0){\includegraphics[width = 0.45\textwidth]{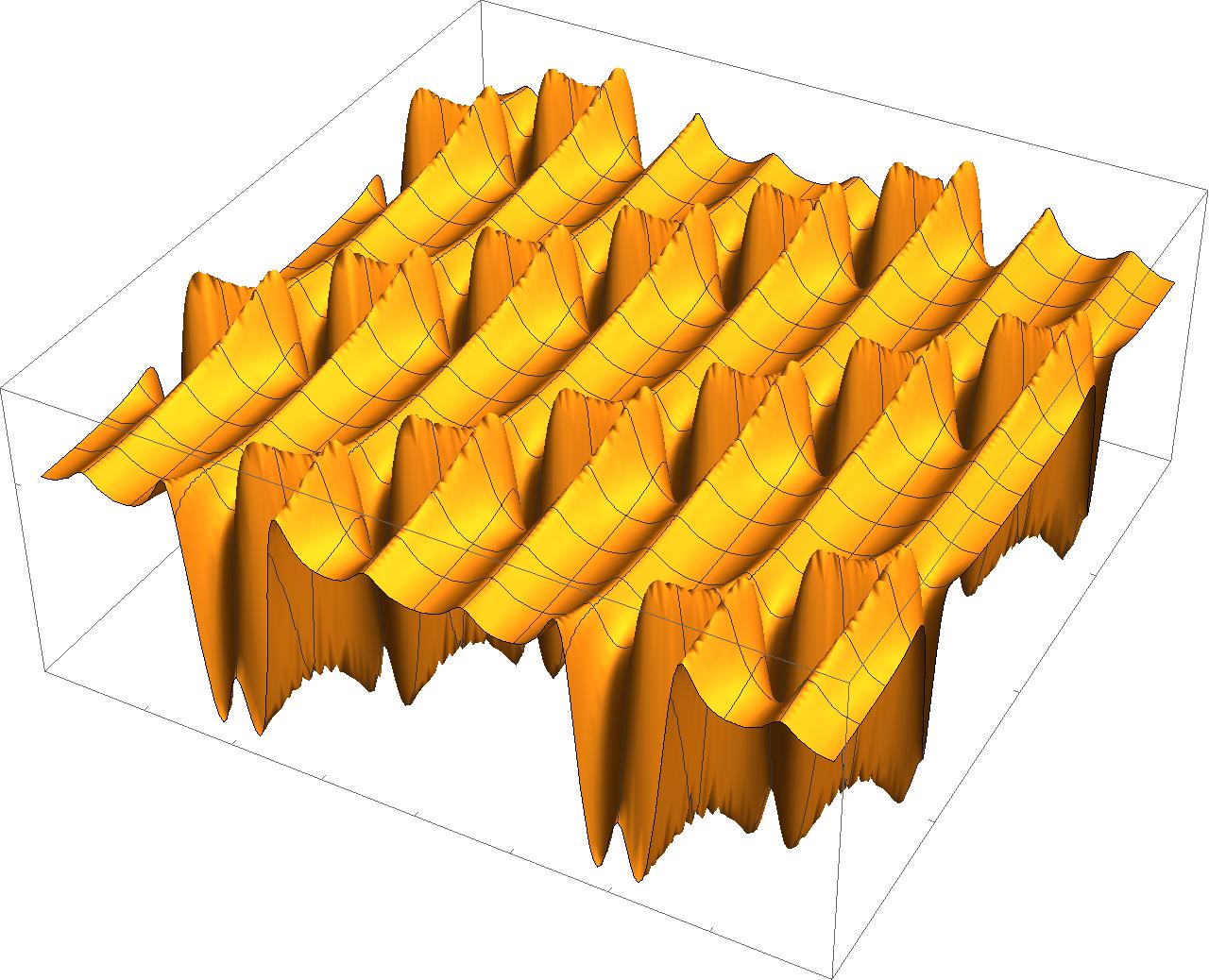}}
\put(0.5,15.75){$\mathcal{H}$}
\put(42.5,9.25){$\xi^1$}
\put(15.25,3){$\xi^0$}
\put(43.5,13.75){$4\omega_1$}
\put(35.5,4.25){$-4\omega_1$}
\put(19.75,1){$\pi / (i D)$}
\put(3.5,6.75){$-\pi / (i D)$}
\put(50.5,15.75){$\mathcal{P}$}
\put(92.5,9.25){$\xi^1$}
\put(65.25,3){$\xi^0$}
\put(93.5,13.75){$4\omega_1$}
\put(85.5,4.25){$-4\omega_1$}
\put(69.75,1){$\pi / (i D)$}
\put(53.5,6.75){$-\pi / (i D)$}
\end{picture}
\end{center}
\vspace{-10pt}
\caption{The energy and momentum density for a solution with $D^2<0$, background energy density $E = 3 \mu^2 /2$ and \Backlund parameter $a = 1.45482$, corresponding to a periodic solution with $c=1/2$.}
\vspace{5pt}
\label{fig:ti_non_prop_en_mom}
\end{figure}
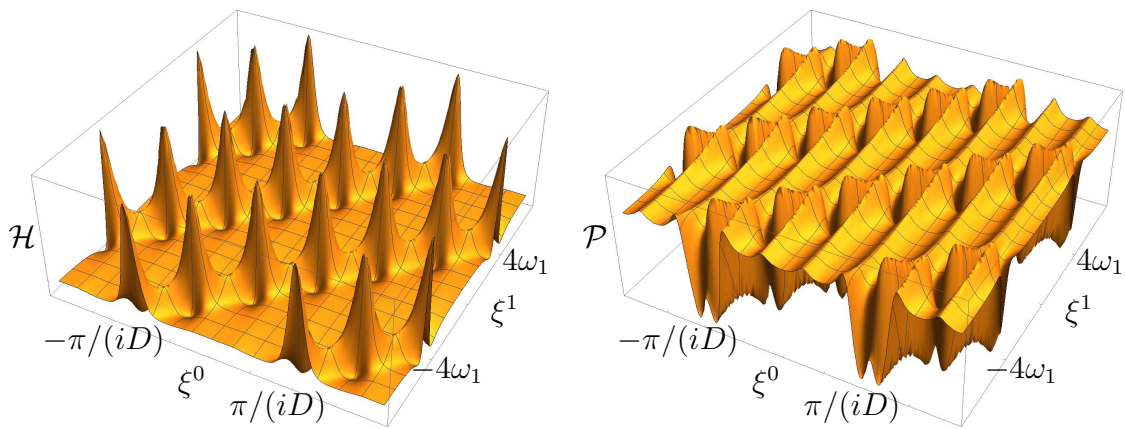

Once again, we first consider a translationally invariant seed solution. As we showed in section \ref{subsec:SG_breathers_periodicity}, these solutions are always periodic in space. Therefore, they cannot have a finite energy difference to the energy of the background solution. However, we may study the average energy density per spatial period of the new solution. It turns out that
\begin{equation}
\left< \mathcal{H} \right> = \frac{{iD}}{2 \pi }\int_{{\xi ^1}}^{{\xi ^1} + \frac{2 \pi }{{iD}}} {d{\xi ^1} \mathcal{H}}  = E .
\end{equation}
Thus, the solution has on average the same energy density as the background solution. In a similar manner the average momentum density vanishes, also similarly to the background solution.
\begin{equation}
\left< \mathcal{P} \right> = \frac{{iD}}{2 \pi }\int_{{\xi ^1}}^{{\xi ^1} + \frac{2 \pi }{{iD}}} {d{\xi ^1} \mathcal{P}}  = 0 .
\end{equation}
Figure \ref{fig:ti_non_prop_en_mom} shows the energy density and the momentum density for a periodic solution.

The relevant solutions whose seed is a static elliptic solution are not manifestly periodic in space. They are periodic in time. One can show that the average current of momentum and energy through a given point is identical to those of the seed solutions, namely,
\begin{align}
\left< \mathcal{P} \right> &= \frac{{iD}}{2 \pi }\int_{{\xi ^0}}^{{\xi ^0} + \frac{2 \pi }{{iD}}} {d{\xi ^0} \mathcal{P}}  = 0 ,\\
\left< \mathcal{T}^{11} \right> &= \frac{{iD}}{2 \pi }\int_{{\xi ^0}}^{{\xi ^0} + \frac{2 \pi }{{iD}}} {d{\xi ^0} \mathcal{T}^{11}}  = E .
\end{align}

\subsection{The $D \to 0$ Limit}
\label{subsec:SG_D0}

In the limit $D \to 0$, the solution degenerates to the form
\begin{equation}
\tilde \varphi = {\hat \varphi} + 4\arctan \left[ \frac{{1 - s_c}}{2} B\left( {{\xi ^1} - i s_d \Phi_0 \left( {{\xi ^0}} \right)} \right) \right] , 
\end{equation}
where
\begin{equation}
{\Phi _0}\left( {{\xi ^0}} \right) = \Phi \left( {{\xi ^0};\omega_1} \right) = \frac{i}{{\sqrt {{E^2} - {\mu ^4}} }}\left( {\zeta \left( {{\xi ^0} + {\omega _3}} \right) - \zeta \left( {{\omega _2}} \right) + {x_1}{\xi ^0}} \right) .
\end{equation}
There are four such solutions, as there are four distinct values of $a$ that set $D$ equal to zero, namely $a =  \pm \sqrt {E \pm \sqrt {{E^2} - {\mu ^4}} } /\mu $. Half of those correspond to a localized solution that generates an overall jump to the background solution equal to $- 4\pi$. For the other half, the solution is equal to $\hat \varphi$, thus a periodic, translationally invariant solution. It turns out that in this specific case, $\hat \varphi$ coincides with an elliptic solution, as the corresponding parameter $\tilde{a}$ is equal to $\pm \omega_1$, namely,
\begin{equation}
\hat \varphi  = \frac{1}{2}{s_d}\left( {\varphi \left( {{\xi ^0} + {\omega _1}} \right) + \varphi \left( {{\xi ^0} - {\omega _1}} \right)} \right) + \left( {2k - 1} \right)\pi  = {s_d}\varphi \left( {{\xi ^0} + {s_d}\left( {2k - 1} \right){\omega _1}} \right) .
\end{equation}

Interestingly enough, the limit $D\to 0$ separating the localized and non-localized solutions comprises of two localized and two non-localized solutions, the latter coinciding with the background solution shifted by an odd number of half-periods.

The total energy and momentum of these solutions exactly match those of the seeds in this limit, not only in the case the dressed solution is a trivial displacement of the seed, but also in the non-trivial cases.

\setcounter{equation}{0}
\section{Asymptotics and Periodicity of the Dressed Elliptic Strings}
\label{sec:string_asymptotics}

In this and the following three sections, we will study some properties of the dressed elliptic string solutions \cite{Katsinis:2018ewd} that we reviewed in section \ref{sec:review} and compare them to the properties of their Pohlmeyer counterpart that we presented in section \ref{sec:SG_properties}. Here, we determine the appropriate values of the moduli that result in closed string solutions.

In order to visualize the string solutions, first we have to select the static gauge, so that the time-like world sheet coordinate $\sigma^0$ is proportional to the physical time $X^0$. This is equivalent to a boost in the worldsheet of the form
\begin{align}
{\xi ^0} &= \gamma \sigma^0  - \gamma \beta \sigma^1, \\
{\xi ^1} &= \gamma \sigma^1  - \gamma \beta \sigma^0, 
\end{align}
where
\begin{align}
\gamma \beta  &= \frac{1}{\mu }\sqrt {{x_3} - \wp \left( a \right)} ,\\
\gamma  &= \frac{1}{\mu }\sqrt {{x_2} - \wp \left( a \right)} .
\end{align}
In this gauge, the time coordinate assumes the form $X^0 = \mu \sigma^0$ and it is easier to study a time snapshot of the string solution in order to determine the periodic properties that the string obeys. It is also easier to visualize the time evolution of the string, which will become handy in sections \ref{sec:spike_interactions} and \ref{sec:instabilities}.

The dressed string solutions, as well as their elliptic seeds, are naturally infinite string solutions. They are parametrized by the spacelike coordinate taking values in the whole real axis. However, the periodic properties of the sine-Gordon counterparts of the elliptic strings \eqref{eq:elliptic_periodic_osc} and \eqref{eq:elliptic_periodic_rot} imply that the string solution obeys appropriate periodicity conditions for specific values of the moduli \cite{part1}, giving rise to finite string solutions.

In the case of the dressed elliptic strings with $D^2>0$, the sine-Gordon counterparts cease to obey periodicity conditions of the form \eqref{eq:elliptic_periodic_osc} and \eqref{eq:elliptic_periodic_rot} due to the existence of the extra kink that propagates on the non-trivial elliptic background. However, the above periodic properties are recovered in the region far away from the position of the kink, as the sine-Gordon solution tends to a shifted version of the elliptic seed. This asymptotic behaviour can be used to construct approximate finite closed dressed elliptic string solutions in the same manner as the closed finite elliptic strings. In order to do so, we first need to study the asymptotics of the dressed elliptic string solutions with $D^2>0$.

Even though the dressed solutions do not have the extended periodicity properties of their elliptic seeds, they still obey the periodic properties \eqref{eq:dressed_elliptic_periodic_Dpos_osc} and \eqref{eq:dressed_elliptic_periodic_Dpos_rot} in the case $D^2>0$, as well as \eqref{eq:D2_negative_periodicity_trivial} and \eqref{eq:D2_negative_periodicity_non_trivial} in the case $D^2<0$. One can take advantage of these periodic properties in order to construct exact finite closed string solutions. It has to be noted that the above equations are expressed in the linear gauge; however, the closed string solution should exhibit appropriate periodicity in their dependence on the spacelike coordinate in the static gauge. In the following, we present all these classes of closed string solutions and derive the appropriate constraints that the moduli should obey for each class.

\subsection{$D^2>0$: The Asymptotics of String Solutions}
\label{subsec:asymptotics}

Having in mind the asymptotic behaviour of the sine-Gordon counterparts of the dressed elliptic strings with $D^2>0$, described in section \ref{subsec:SG_asymptotics}, it is not surprising that in the region far away from the location of the kink, they tend to a rotated version of the seed, elliptic string solution. Assume that the seed solution is written in spherical coordinates, in parametric form as,
\begin{align}
{\theta _{0/1}} &= {\theta _{{\rm{seed}}}}\left( {{\sigma ^0},{\sigma ^1}} \right) ,\\
{\varphi _{0/1}} &= {\varphi _{{\rm{seed}}}}\left( {{\sigma ^0},{\sigma ^1}} \right) .
\end{align}
The functions ${\theta _{{\rm{seed}}}}$ and ${\varphi _{{\rm{seed}}}}$ have the properties \cite{part1}
\begin{align}
{\theta _{{\rm{seed}}}}\left( {{\sigma ^0},{\sigma ^1} + \delta {\sigma _{0/1}}} \right) &=  \pm {\theta _{{\rm{seed}}}}\left( {{\sigma ^0},{\sigma ^1}} \right) , \label{eq:elliptic_periodic_theta}\\
{\varphi _{{\rm{seed}}}}\left( {{\sigma ^0},{\sigma ^1} + \delta {\sigma _{0/1}}} \right) &= {\varphi _{{\rm{seed}}}}\left( {{\sigma ^0},{\sigma ^1}} \right) + \delta \varphi , \label{eq:elliptic_periodic_phi}
\end{align}
where the $\pm$ sign in the first equation applies in the case of rotating/oscillating counterparts, $\delta \varphi$ is the angular opening of the elliptic string, i.e. 
the azimuthal angular distance between two consecutive spikes of the seed solution,
\begin{equation}
\delta {\varphi _{0/1}} =  \mp 2i{\omega _1}\left( {\zeta \left( {{\omega _1}} \right)\frac{a}{{{\omega _1}}} + \zeta \left( {{\omega _{{x_{3/2}}}}} \right) - \zeta \left( {a + {\omega _{{x_{3/2}}}}} \right)} \right) 
\end{equation}
and
\begin{align}
\delta \sigma_0  &= \frac{2{\omega _1}}{\gamma \beta}, \label{eq:periodicity_range_ti}\\
\delta \sigma_1  &= \frac{2{\omega _1}}{\gamma } . \label{eq:periodicity_range_st}
\end{align}
Furthermore, we define the function
\begin{equation}
\tilde \Phi \left( {{\sigma ^0},{\sigma ^1}} \right) := D{\xi ^{1/0}} + i\Phi \left( {{\xi ^{0/1}};\tilde a} \right) .
\end{equation}
The kink which propagates on the elliptic background is located in the region $\tilde{\Phi} \simeq 0$. Several periods away from the kink position, one may use the quasiperiodicity property of the function $\Phi$ to show that
\begin{equation}
\tilde \Phi \left( {{\sigma ^0},{\sigma ^1}} \right) \simeq \begin{cases} D\gamma \left( {1 + \beta {{\bar v}_0}} \right)\left( {{\sigma ^1} - \frac{{\beta  + {{\bar v}_0}}}{{1 + \beta {{\bar v}_0}}}{\sigma ^0}} \right) , &\textrm{for translationally invariant seeds,}\\
- D\gamma \left( {\beta  + \frac{1}{{{{\bar v}_1}}}} \right)\left( {{\sigma ^1} - \frac{{\beta  + {{\bar v}_1}}}{{1 + \beta {{\bar v}_1}}}{\sigma ^0}} \right) , &\textrm{for static seeds.}
\end{cases}
\label{eq:phi_tilde_mean}
\end{equation}
The parameters ${{\bar v}_{0/1}}$ are the mean velocity of the kink relatively to the elliptic background, in the case of a translationally invariant and static seed, respectively, which are given by equations \eqref{eq:kinks_mean_velocity} and \eqref{eq:kinks_mean_velocity_static}. Notice that the above approximations are exact whenever $\sigma^1 = n \delta \sigma_{0/1}$, with $n \in \mathbb{Z}$.

Then, one can show that in the region far away from the kink position, the dressed solution assumes the form
\begin{align}
\mathop {\lim }\limits_{\tilde \Phi  \to  \pm \infty } {\theta _0}\left( {{\sigma ^0},{\sigma ^1}} \right) &= {\theta _{{\rm{seed}}}}\left( {{\sigma ^0},{\sigma ^1} \mp \frac{{\tilde a}}{{2{\omega _1}}}\delta {\sigma _0}} \right) , \label{eq:asymptotic_form_theta_ti}\\
\mathop {\lim }\limits_{\tilde \Phi  \to  \pm \infty } {\varphi _0}\left( {{\sigma ^0},{\sigma ^1}} \right) &= {\varphi _{{\rm{seed}}}}\left( {{\sigma ^0},{\sigma ^1} \mp \frac{{\tilde a}}{{2{\omega _1}}}\delta {\sigma _0}} \right) \pm \Delta {\varphi _0} , \label{eq:asymptotic_form_phi_ti}
\end{align}
for translationally invariant seeds and
\begin{align}
\mathop {\lim }\limits_{\tilde \Phi  \to  \pm \infty } {\theta _1}\left( {{\sigma ^0},{\sigma ^1}} \right) &= {\theta _{{\rm{seed}}}}\left( {{\sigma ^0},{\sigma ^1} \pm \frac{{\tilde a}}{{2{\omega _1}}}\delta {\sigma _1}} \right) , \label{eq:asymptotic_form_theta}\\
\mathop {\lim }\limits_{\tilde \Phi  \to  \pm \infty } {\varphi _1}\left( {{\sigma ^0},{\sigma ^1}} \right) &= {\varphi _{{\rm{seed}}}}\left( {{\sigma ^0},{\sigma ^1} \pm \frac{{\tilde a}}{{2{\omega _1}}}\delta {\sigma _1}} \right) \pm \Delta {\varphi _1} , \label{eq:asymptotic_form_phi}
\end{align}
for static seeds, respectively. An overall reflection $\theta \to \pi - \theta$, $\varphi \to \varphi + \pi$ may be present and we will comment on it later on. The angle $\Delta \varphi_{0/1}$ is equal to
\begin{equation}
\begin{split}
\Delta \varphi_{0/1} &= \arg \left( \ell  + iD \right) + \arg \sigma \left( {\tilde{a} + a} \right) + i\left( {\zeta \left( a + \omega_{x_{3/2}} \right) - \zeta \left( \omega_{x_{3/2}} \right)} \right) \tilde{a} \\
&= \arg \left( {\ell  + iD} \right) + \arg \sigma \left( {\tilde a + a} \right) + \left( {\zeta \left( {{\omega _1}} \right)a \pm \frac{{\delta {\varphi _{0/1}}}}{2}} \right)\frac{{\tilde a}}{{{\omega _1}}} .
\end{split}
\label{eq:asymptotic_form_DPhi}
\end{equation}
The half-period $\omega_{x_{i}}$ is the half-period corresponding to the root $x_i$. More specifically, $\omega_{x_{3}}$ is always the imaginary half-period $\omega_2$, whereas $\omega_{x_{2}}$ is equal to the real half-period $\omega_1$ for oscillating seeds and to $\omega_3 = \omega_1 + \omega_2$ for rotating seeds. The details of the above derivations are included in the appendix \ref{sec:app_asymptotics}.

The above approximation is valid at the region $\left| \tilde{\Phi} \right| \gg 1$ or in other words in the region
\begin{align}
\left| {{\sigma ^1} - \frac{{\beta  + {{\bar v}_0}}}{{1 + \beta {{\bar v}_0}}}{\sigma ^0}} \right| &\gg \frac{1}{D\gamma \left| { 1 + \beta {{\bar v}_0} } \right|} , \label{eq:asymptotics_condition_ti} \\
\left| {{\sigma ^1} - \frac{{\beta  + {{\bar v}_1}}}{{1 + \beta {{\bar v}_1}}}{\sigma ^0}} \right| &\gg \frac{1}{D\gamma \left| { \beta  + \frac{1}{{{{\bar v}_1}}} } \right|} ,\label{eq:asymptotics_condition_st}
\end{align}
for each case respectively. The above inequalities are expressed in terms of the static gauge worldsheet coordinates, and, thus, they describe which region of the dressed elliptic string in any time snapshot is indeed well-approximated by a rotated version of the seed solution. Notice also that one has to be careful in the correspondence between the $\sigma^1$ and $\tilde{\Phi}$ infinite limits. This is determined by whether the kink velocity is larger or smaller than the inverse of the velocity of the boost connecting the linear and static gauges. We define the sign $s_\Phi$ as
\begin{equation}
\mathop {\lim }\limits_{{\sigma ^1} \to  \pm \infty } \tilde \Phi  =  \pm {s_\Phi }\infty .
\end{equation}
The equation \eqref{eq:phi_tilde_mean} implies that
\begin{equation}
{s_\Phi } = \begin{cases}
{\mathop{\rm sgn}} \left( {1 + \beta {{\bar v}_0}} \right) , &\textrm{for translationally invariant seeds,}\\
 - {\mathop{\rm sgn}} \left( {\beta + \frac{1}{{{{\bar v}_1}}}} \right) , &\textrm{for static seeds.}
\end{cases}
\end{equation}
The dependence of the sign $s_\Phi$ on the moduli of the dressed string solutions is exhaustively studied in the appendix \ref{sec:appendix_dispersion}. In the special case where ${1 + \beta {{\bar v}_0}} = 0$ or ${\beta + {1}/{{{{\bar v}_1}}}} = 0$, the string does not exhibit this kind of asymptotic behaviour. This is a interesting special case, which is studied in section \ref{subsec:string_finite_exact_special}.

The equation \eqref{eq:asymptotic_form_DPhi} implies that the angle ${\Delta \varphi}$ obeys
\begin{equation}
\mathop {\lim }\limits_{\tilde{a} \to {\omega _1}} {\Delta \varphi_{0/1} } = \arg \left( {\ell  + iD } \right) \pm \frac{\delta \varphi_{0/1}}{2} ,
\end{equation}
where the angle $\delta \varphi$ is the angular opening of the elliptic string. Further details are provided in appendix \ref{sec:app_asymptotics}.

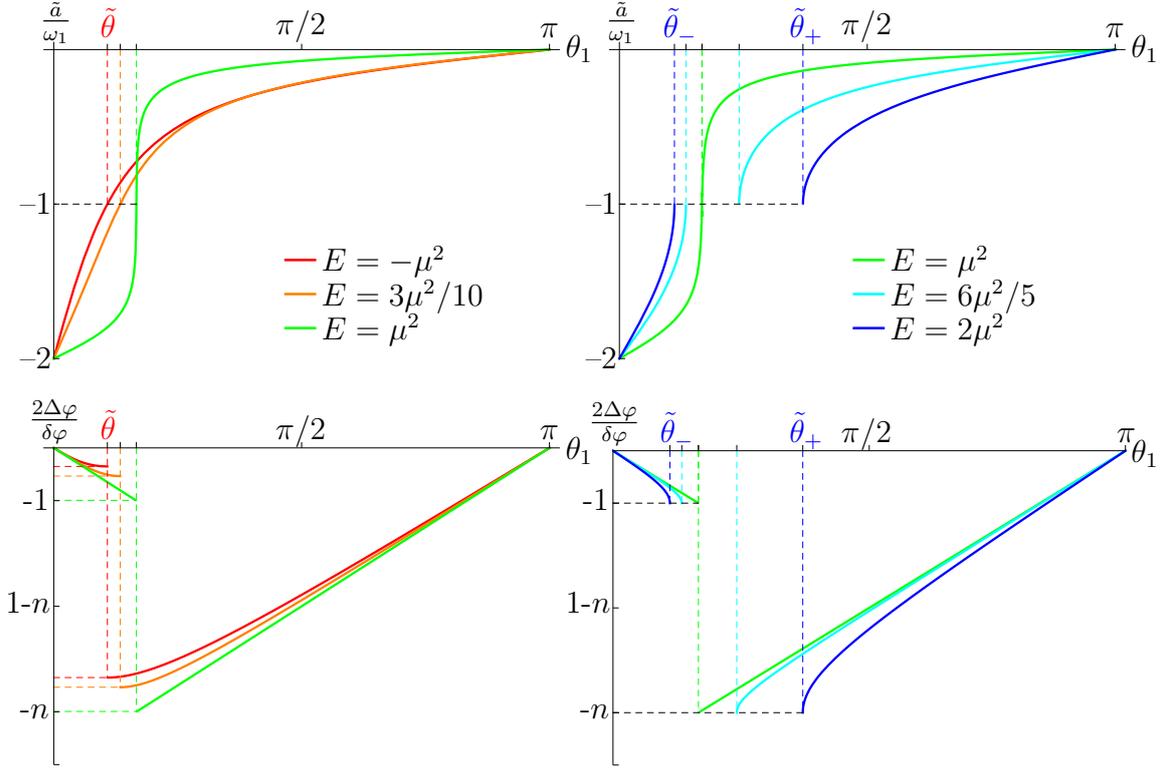
\begin{figure}[ht]
\vspace{10pt}
\begin{center}
\begin{picture}(100,69)
\put(3.25,1){\includegraphics[width = 0.45\textwidth]{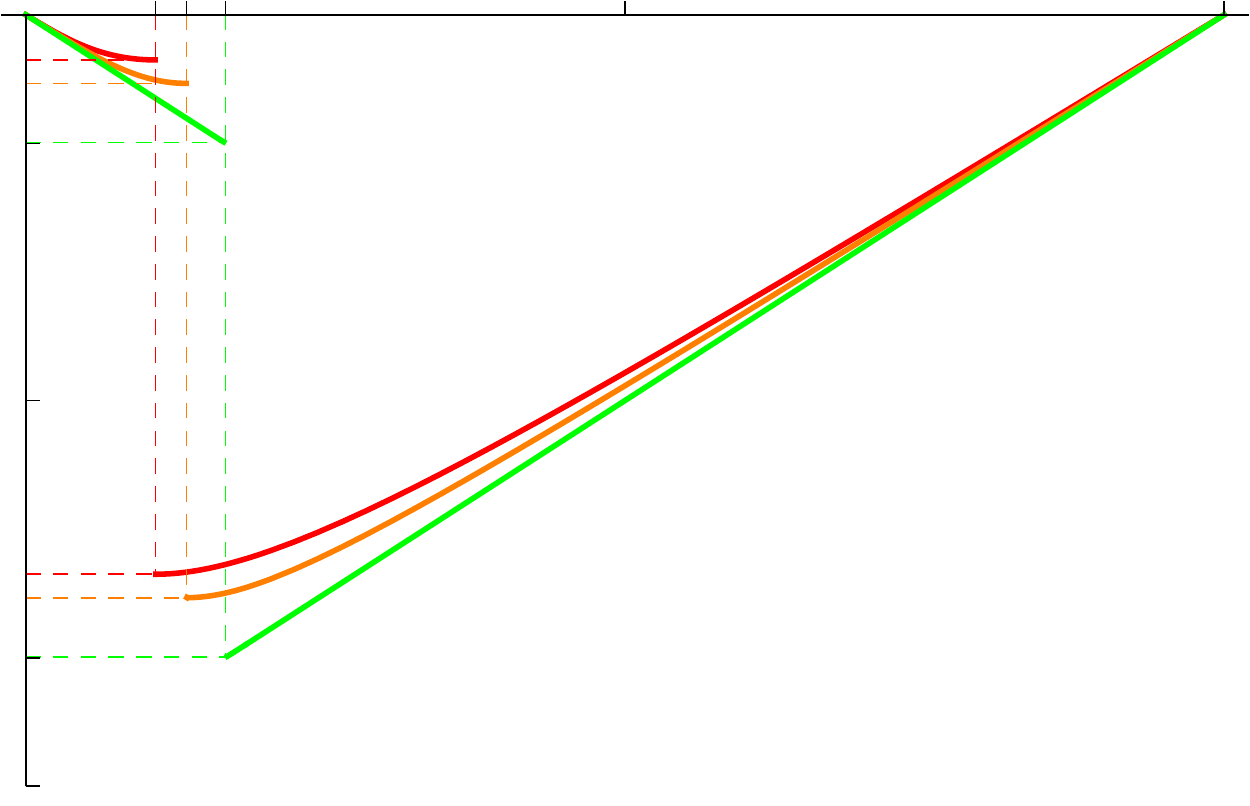}}
\put(2,31.25){$\frac{2\Delta\varphi}{\delta \varphi}$}
\put(1.5,24){-1}
\put(0,15){1-$n$}
\put(1.25,5.75){-$n$}
\put(48.75,28.5){$\theta_1$}
\put(8.25,30.25){\color{red}$\tilde{\theta}$}
\put(23.5,30.25){$\pi/2$}
\put(46.5,30.25){$\pi$}
\put(52.5,1){\includegraphics[width = 0.45\textwidth]{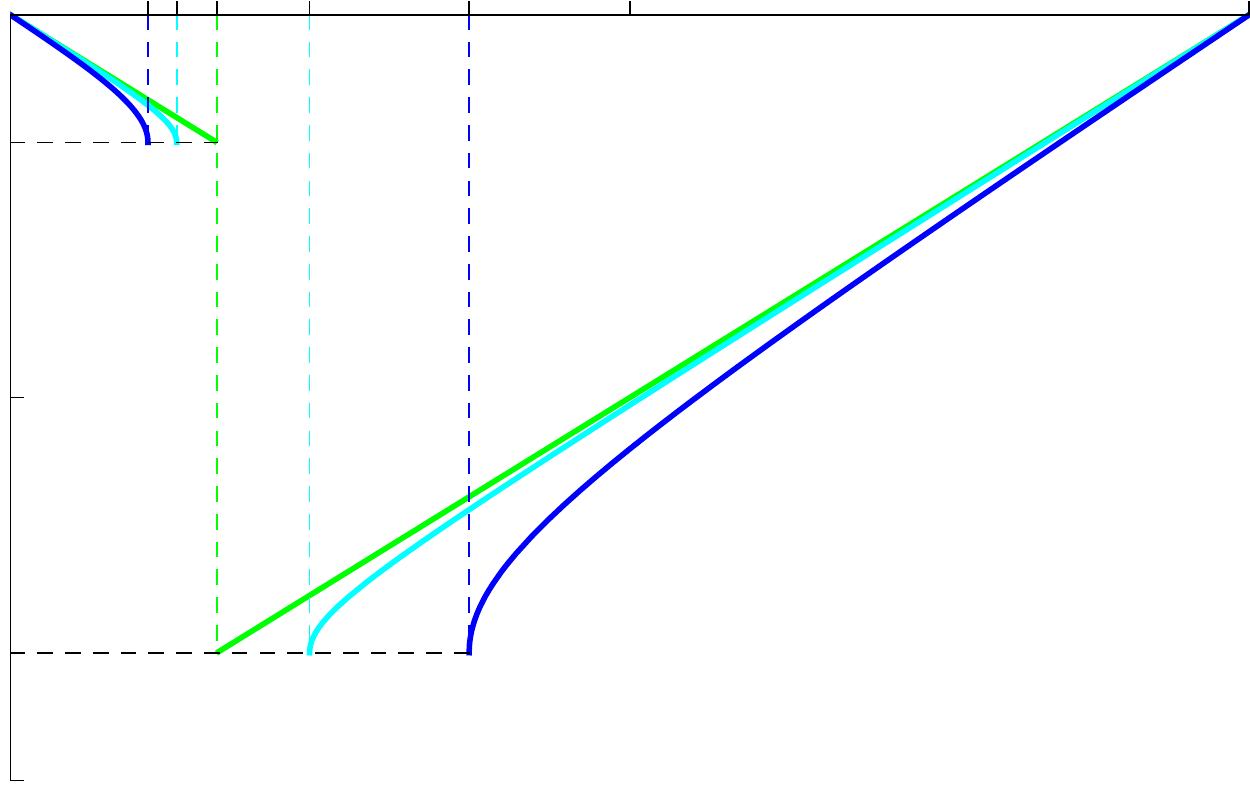}}
\put(50.75,31){$\frac{2\Delta\varphi}{\delta \varphi}$}
\put(50.25,24){-1}
\put(48.75,15){1-$n$}
\put(50,5.75){-$n$}
\put(98,28.25){$\theta_1$}
\put(57,30){\color{blue}$\tilde{\theta}_-$}
\put(68.25,30){\color{blue}$\tilde{\theta}_+$}
\put(72.75,30){$\pi/2$}
\put(96.75,30){$\pi$}
\put(3.25,36.5){\includegraphics[width = 0.45\textwidth]{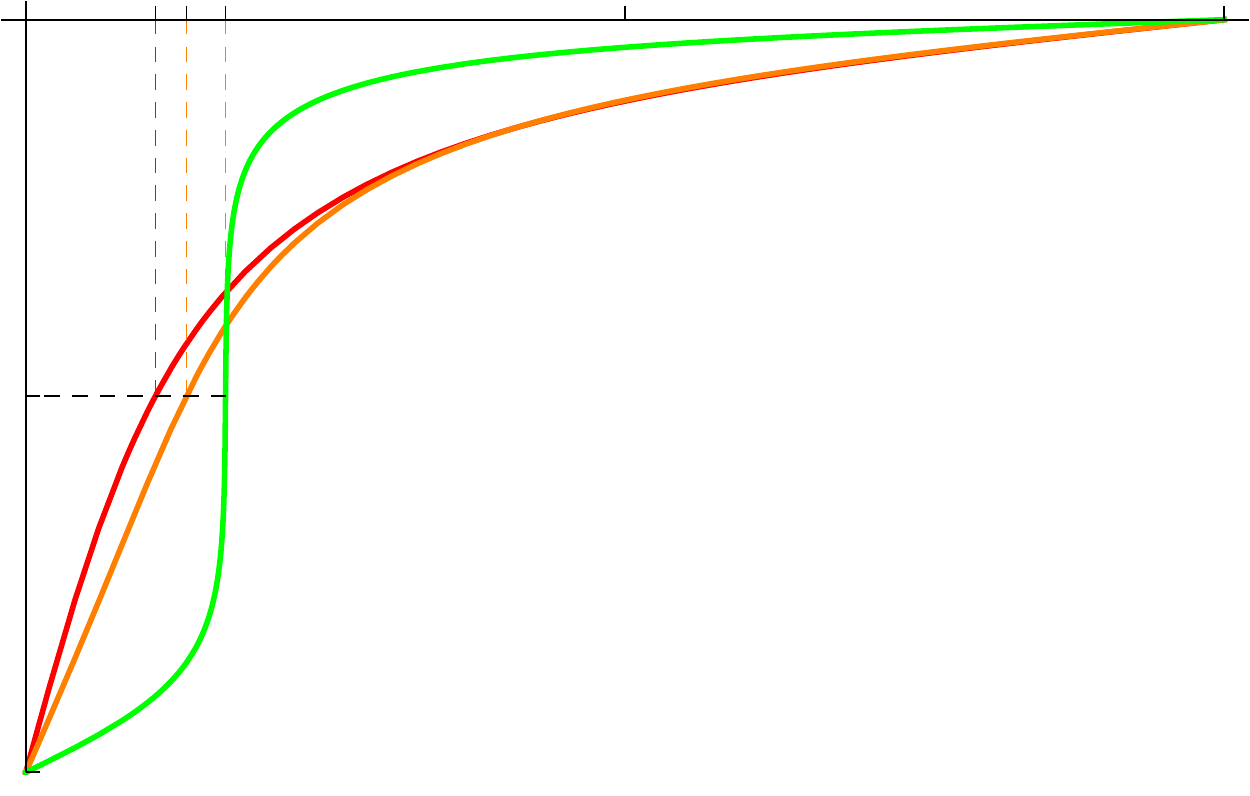}}
\put(3,66.25){$\frac{\tilde{a}}{\omega_1}$}
\put(1.25,36.5){--2}
\put(1.25,50){--1}
\put(48.75,63.75){$\theta_1$}
\put(8.25,65.5){\color{red}$\tilde{\theta}$}
\put(23.5,65.75){$\pi/2$}
\put(46.5,65.5){$\pi$}
\put(52.5,36.5){\includegraphics[width = 0.45\textwidth]{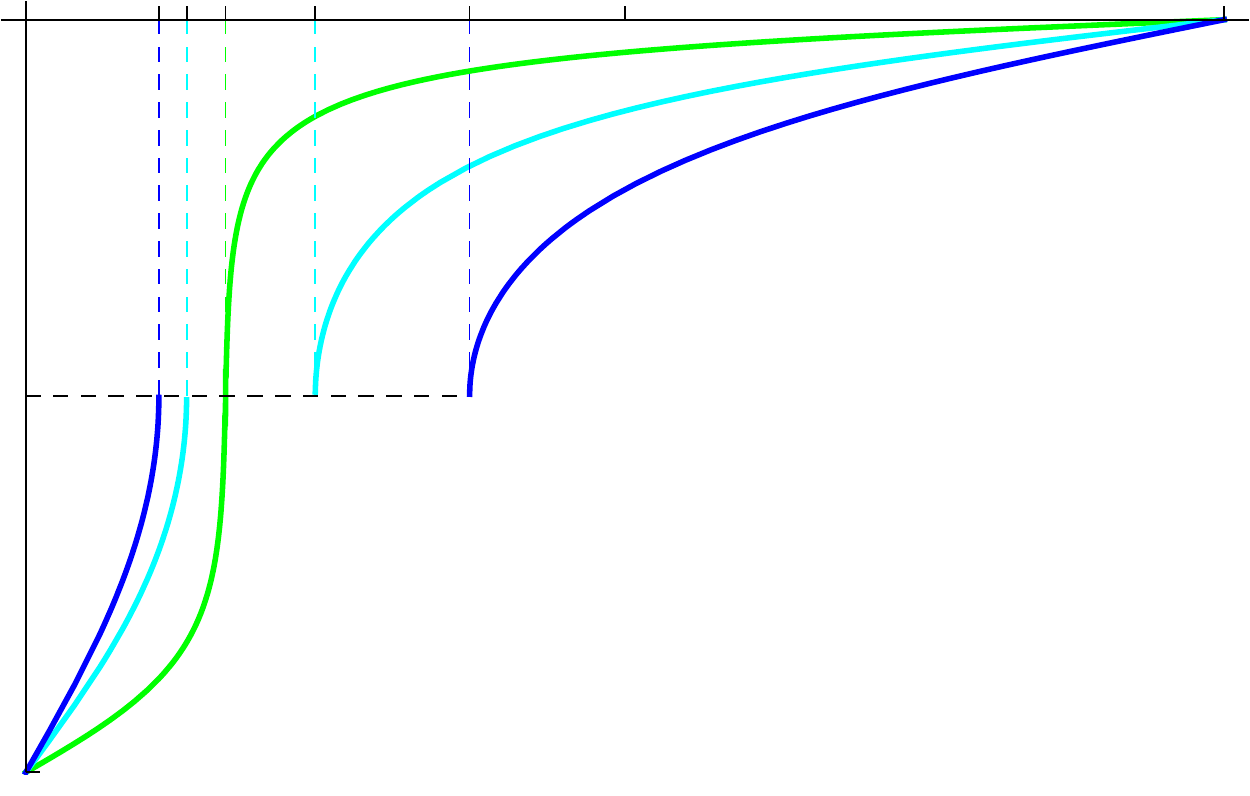}}
\put(52.25,66.25){$\frac{\tilde{a}}{\omega_1}$}
\put(50.5,36.5){--2}
\put(50.5,50){--1}
\put(98,63.75){$\theta_1$}
\put(57.25,65.5){\color{blue}$\tilde{\theta}_-$}
\put(68.25,65.5){\color{blue}$\tilde{\theta}_+$}
\put(72.75,65.75){$\pi/2$}
\put(95.75,65.5){$\pi$}
\put(23.5,38){\includegraphics[width = 0.054\textwidth]{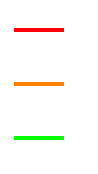}}
\put(27.5,45){$E=-\mu^2$}
\put(27.5,42){$E=3\mu^2/10$}
\put(27.5,39){$E=\mu^2$}
\put(73,38){\includegraphics[width = 0.054\textwidth]{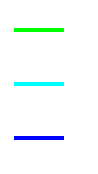}}
\put(77,45){$E=\mu^2$}
\put(77,42){$E=6\mu^2/5$}
\put(77,39){$E=2\mu^2$}
\end{picture}
\end{center}
\vspace{-10pt}
\caption{The parameters $\tilde{a}$ and $\Delta\varphi$ determining the asymptotic behaviour of the dressed solutions as function of the angle $\theta_1$. The parameter $a$ of the seed elliptic solution is selected so that the latter obeys appropriate periodicity conditions with $n=6$.}
\vspace{5pt}
\label{fig:dressed_strings_static_phi}
\end{figure}

The behaviour of $\tilde{a}$ and ${\Delta \varphi}$ as functions of $\theta_1$ is shown in figure \ref{fig:dressed_strings_static_phi}.
For solutions with $D^2 > 0$, usually we select $\tilde{a}$ to lie on the real axis in the segment $\left( - \omega_1 , \omega_1 \right)$. However, in figure \ref{fig:dressed_strings_static_phi} it is selected to lie in the segment $\left( - 2 \omega_1 , 0 \right)$ to show the continuity of its dependence on the position of the poles of the dressing factor. In the case of a seed solution, with an oscillating counterpart, there is a special value of $\theta_1 = \tilde \theta$, equal to
\begin{equation}
\tilde \theta  = 2\arctan \sqrt { - \frac{{{m_ - }}}{{{m_ + }}}} ,
\end{equation}
where $\tilde{a}$ equals the real half period $\omega_1$. At $\theta_1 = \tilde \theta$, $\Delta\varphi$ is stationary and at the same time discontinuous. It perform a jump by $\pi - \delta \varphi$, which is related to the inversion of the asymptotics of the solution. In the case of a seed solution with a rotating counterpart, there are two such special values for $\theta_1$, namely,
\begin{equation}
{{\tilde \theta }_ \pm } = 2\arctan \sqrt {\frac{{E \pm \sqrt {{E^2} - \mu ^4} }}{{m_ + ^2}}} ,
\end{equation}
where $\tilde{a}$ equals the real half period $\omega_1$. When $\theta_1$ acquires these two values, $D^2$ vanishes and $\Delta \varphi$ is maximum and equal to $\delta \varphi / 2$ and $\pi - \delta \varphi / 2$, respectively. For values of $\theta_1$ between these two, it turns out that $D^2<0$ and the solution has a Pohlmeyer counterpart being a periodic disturbance on a rotating background that we will study in section \ref{subsec:string_finite_exact_D_negative}.

It follows that, in the case of rotating backgrounds, the dressed solutions with Pohlmeyer counterparts, which are kinks or antikinks propagating on top of a train of kinks, have been separated into two classes. As expected from the epicycle description of the action of the dressing on the string\footnote{The dressed string solutions with the simplest dressing factor, as those presented here, have an interesting geometric relation to their seeds. Every point of the dressed string is connected to the point of the seed solution with the same worldsheet coordinates, via an arc of a maximum circle equal to $\theta_1$. Therefore, the dressed string can be considered drawn by a point on an epicycle of constant arc radius $\theta_1$ whose center is running on the seed solution.} \cite{Katsinis:2018ewd}, their difference is the following: the class with $\theta_1 < \tilde \theta_-$ asymptotically tends to the seed solution rotated around the $z$-axis by an appropriate angle; the class with $\theta_1 > \tilde \theta_+$ asymptotically tends to the seed solution, first inverted with respect to the origin of the enhanced space and then rotated appropriately around the $z$-axis. Finally, notice that $\Delta \varphi$ tends to $0$ at the limits $\theta_1 \to 0$ and $\theta_1 \to \pi$ as expected, since the epicycle becomes a point.

\subsection{$D^2>0$: Approximate Finite Closed Strings}
\label{subsec:string_finite_approximate}

Strictly speaking, it is not possible to fix the parameters of the solution, so that a dressed string with $D^2 > 0$ satisfies appropriate periodicity conditions (except for very specific cases that we will study in section \ref{subsec:string_finite_exact_special}). In the elliptic strings case, the functions ${\theta _{{\rm{seed}}}}$ and ${\varphi _{{\rm{seed}}}}$ have the periodic properties \eqref{eq:elliptic_periodic_theta} and \eqref{eq:elliptic_periodic_phi}. Therefore, arranging the solution parameters so that $\delta \varphi = 2\pi /n$ where $n \in \mathbb{Z}$, in the case of a rotating counterpart, and $n \in 2\mathbb{Z}$ in the case of an oscillating one, results in a well defined, finite length closed string, parametrized by $\sigma^1 \in \left[ 0 , n \delta \sigma_{0/1} \right)$. However, when dressed strings with a Pohlmeyer counterpart being a kink propagating on an elliptic background are considered, in general these functions are not periodic/quasi-periodic due to the presence of the kink.

Nevertheless, we have shown that the dressed solutions asymptotically approach a rotated version of the seed elliptic ones. This is due to the fact that the effect of the kink is exponentially damped with the distance from its center. Therefore, as long as the characteristic length of the exponential damping of the kink is much smaller that the number of periods appearing in the seed solution, we can claim that we may adjust the periodicity conditions in order to find a string solution that is not exactly a closed finite string, but nevertheless an exponentially good approximation of such a solution. For such a purpose, the parameters of the solution should obey a modified periodicity condition, due to the asymptotic behaviour \eqref{eq:asymptotic_form_phi_ti}, \eqref{eq:asymptotic_form_phi} of the dressed solution, namely,
\begin{equation}
\left( {{n_1}\delta \varphi  + 2 s_\Phi {\Delta \varphi}} \right){n_2} = 2\pi , \quad n_1 , n_2 \in \mathbb{Z} .
\label{eq:finite_approx_condition}
\end{equation}
It has to be noted that in general dressed elliptic string solutions that satisfy the condition \eqref{eq:finite_approx_condition} have elliptic seeds which do not obey the appropriate periodicity conditions, and, thus, they are not finite closed strings. This holds in the simple case that we considered here, were the strings perform only one winding around the $z$-axis and thus, it is possible that they do not contain self-intersection. In general, one could consider a generalization of \eqref{eq:finite_approx_condition} where the left hand side is $2 \pi m$, where $m \in \mathbb{Z}$. In such a case, the seed and the dressed solutions are both closed, as long as the ratio $\Delta \varphi / \delta \varphi$ is rational; however, they correspond to different ranges of the spacelike parameter $\sigma^1$. The simplest case of this kind is the limit $\tilde{a} \to \omega_1$ for rotating seeds, where the angle $\Delta \varphi$ tends to $\delta \varphi / 2$.

\begin{figure}[p]
\vspace{10pt}
\begin{center}
\begin{picture}(80,110)
\put(0,75){\includegraphics[width = 0.35\textwidth]{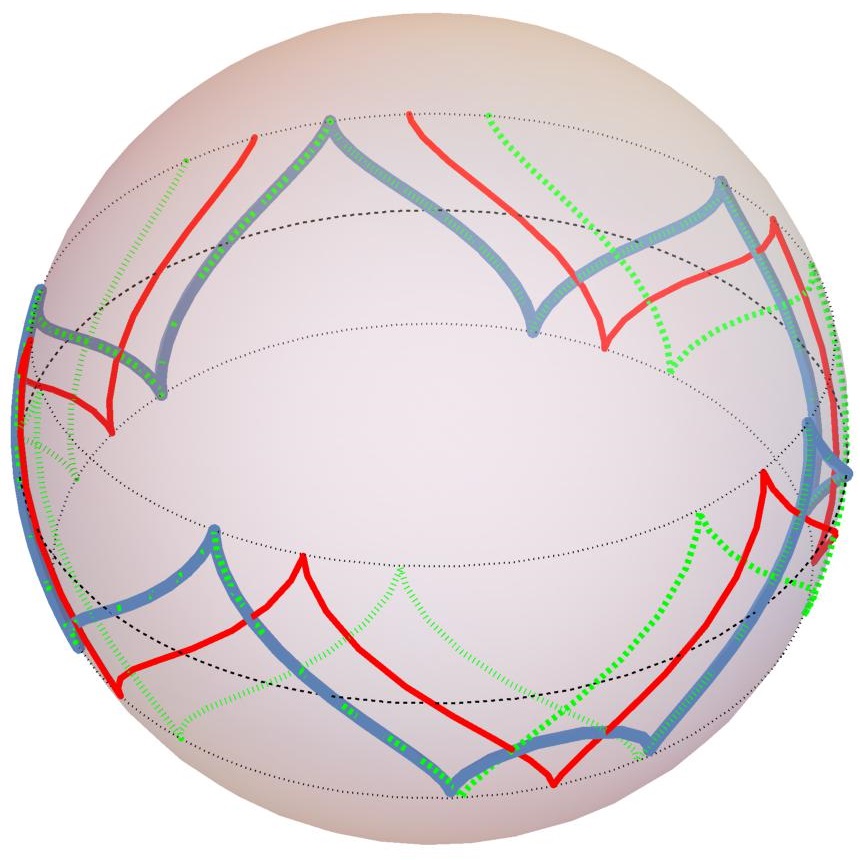}}
\put(45,75){\includegraphics[width = 0.35\textwidth]{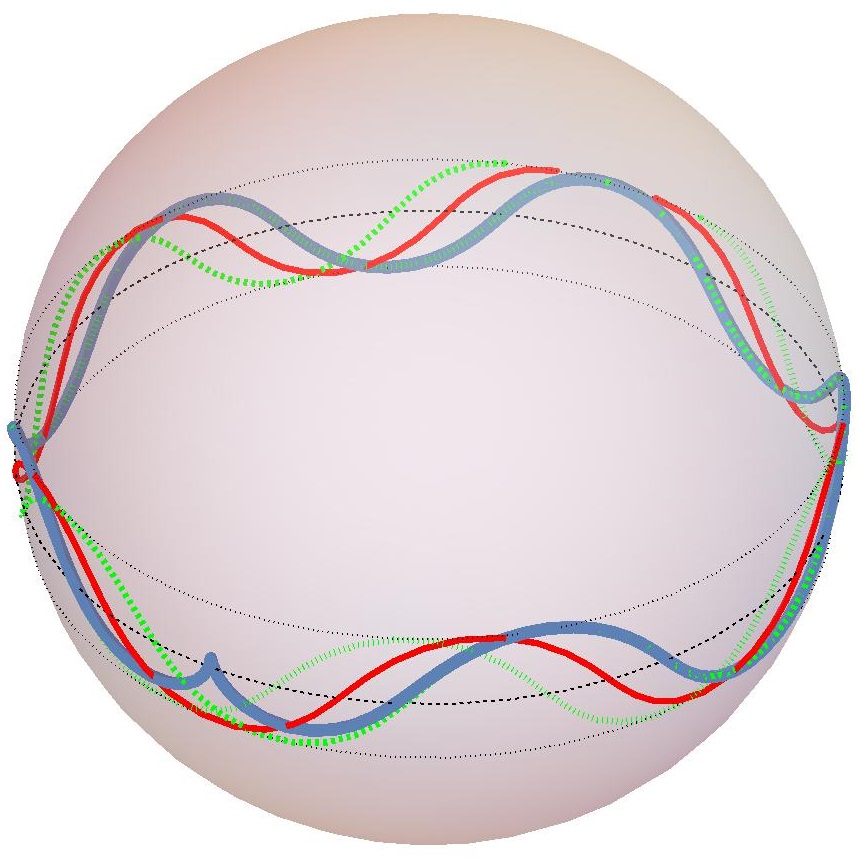}}
\put(0,40){\includegraphics[width = 0.35\textwidth]{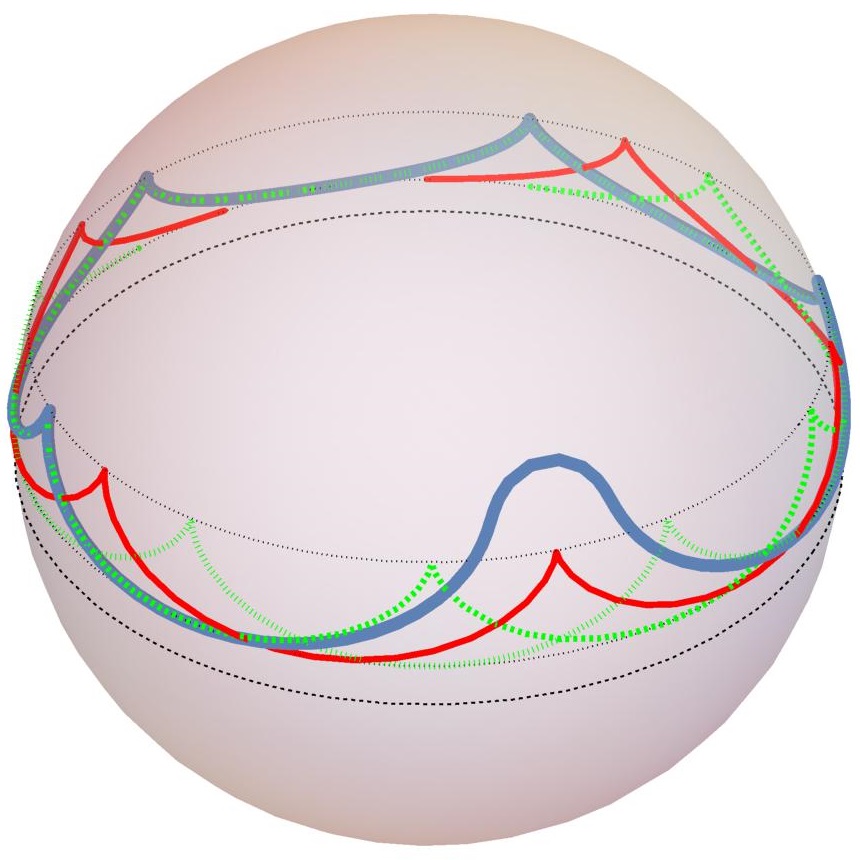}}
\put(45,40){\includegraphics[width = 0.35\textwidth]{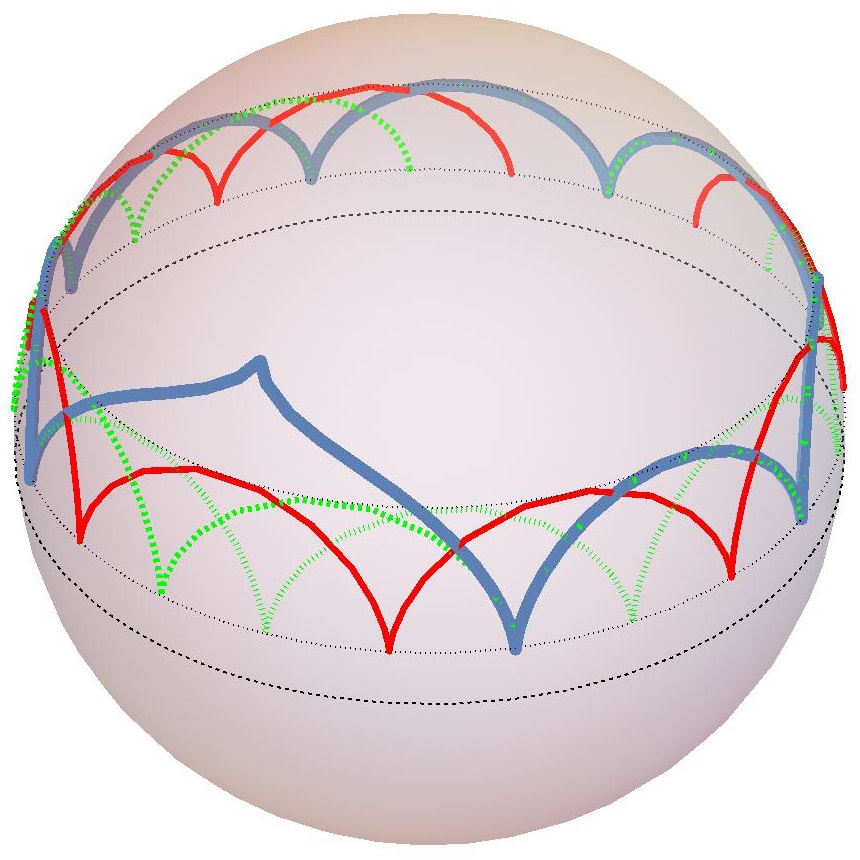}}
\put(0,5){\includegraphics[width = 0.35\textwidth]{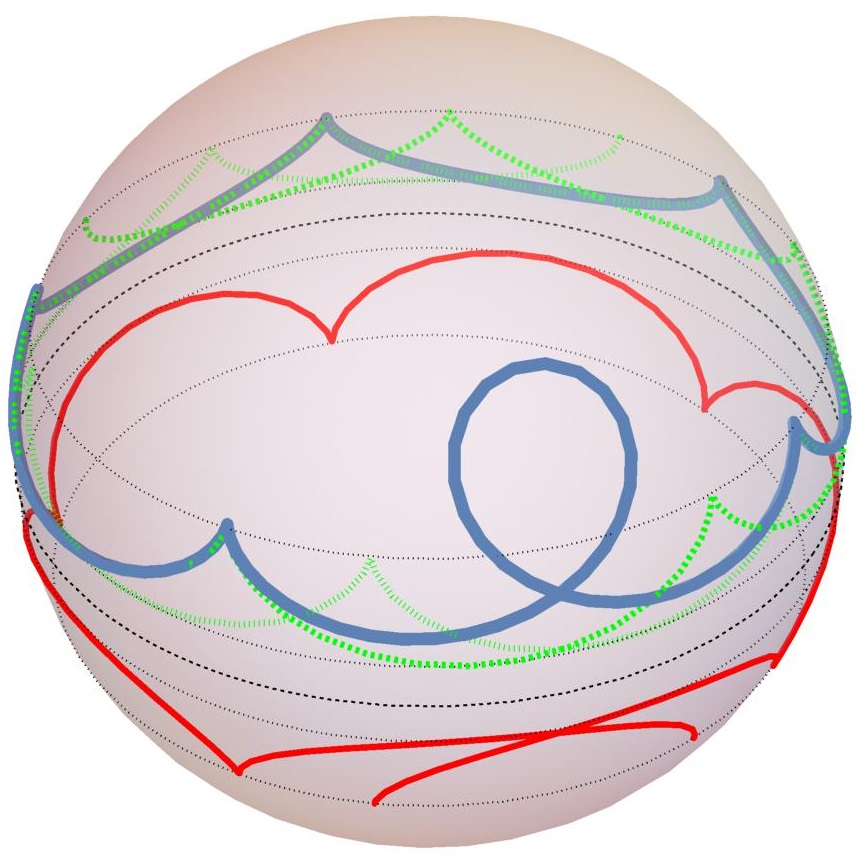}}
\put(45,5){\includegraphics[width = 0.35\textwidth]{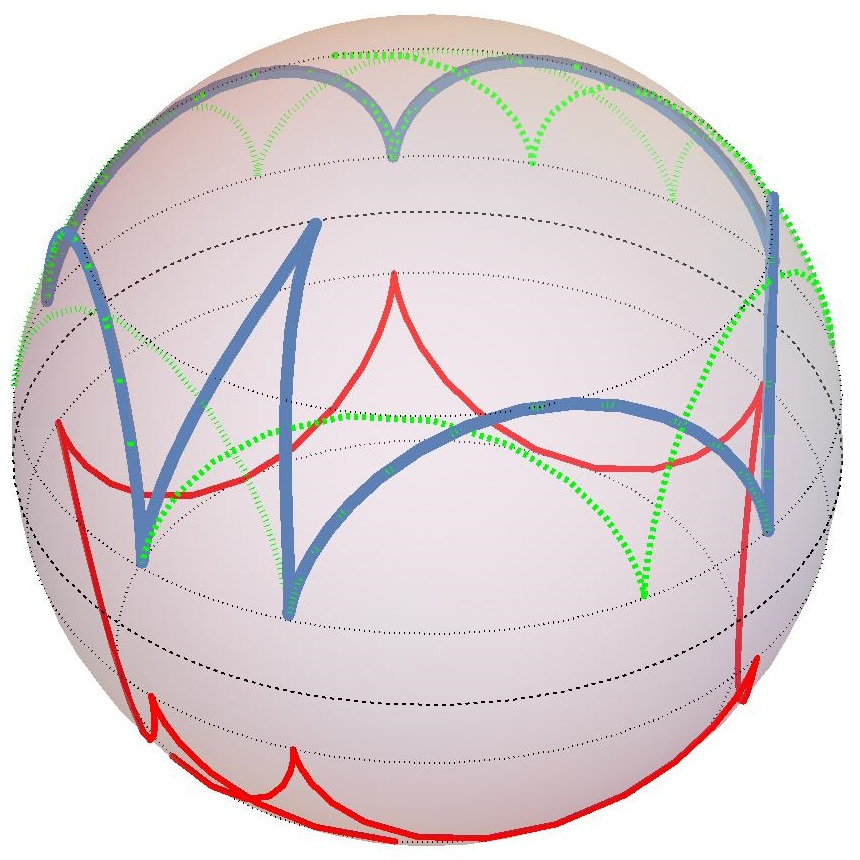}}
\put(27.5,0){\includegraphics[width = 0.05\textwidth]{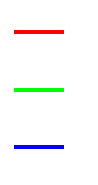}}
\put(32.5,7.25){seed solution}
\put(32.5,4.25){rotated seed solution}
\put(32.5,1.25){dressed solution}
\end{picture}
\end{center}
\vspace{-10pt}
\caption{The finite dressed string solution with approximate periodicity conditions. The left and right column solutions have seeds with translationally invariant and static Pohlmeyer counterparts, respectively. On the first row the seed solution has an oscillating counterpart with $E=\mu^2 / 10$ and $a$ selected so that $n_1=10$ and $n_2=1$. On the second and third rows the seed solution has a rotating counterpart with $E=6\mu^2 /5$ and $a$ selected so that $n_1=7$ and $n_2=1$. On the first and second rows $\theta_1 = {\pi}/{12}$, whereas on the third row $\theta_1 = {7\pi}/{8}$. The solutions of the second and third row belong to the $\theta_1 < \tilde{\theta}_-$ and $\theta_1 > \tilde{\theta}_+$ classes of solutions, respectively.}
\vspace{5pt}
\label{fig:dressed_strings_approx}
\end{figure}
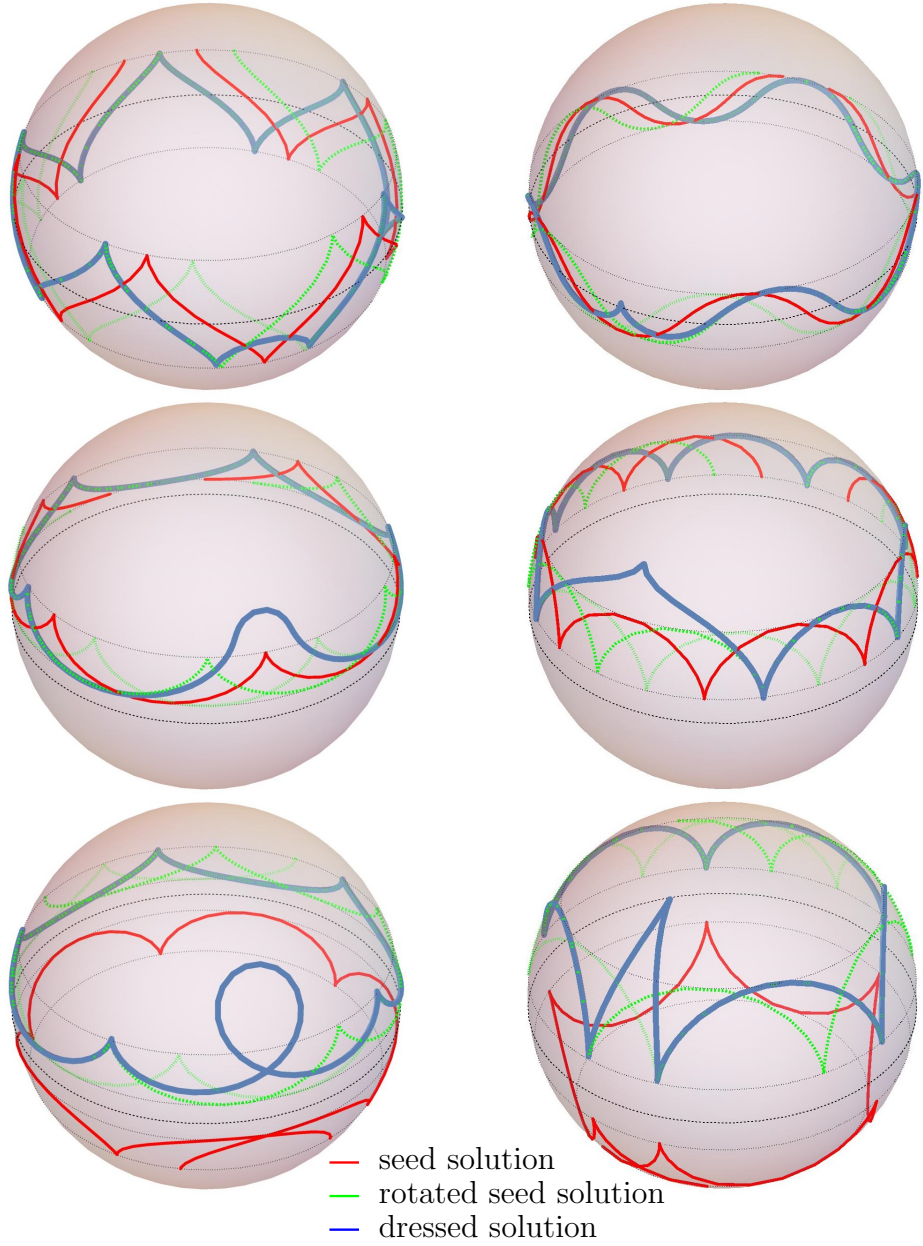
\begin{figure}[ht]
\vspace{10pt}
\begin{center}
\begin{picture}(75,35)
\put(0,0){\includegraphics[width = 0.35\textwidth]{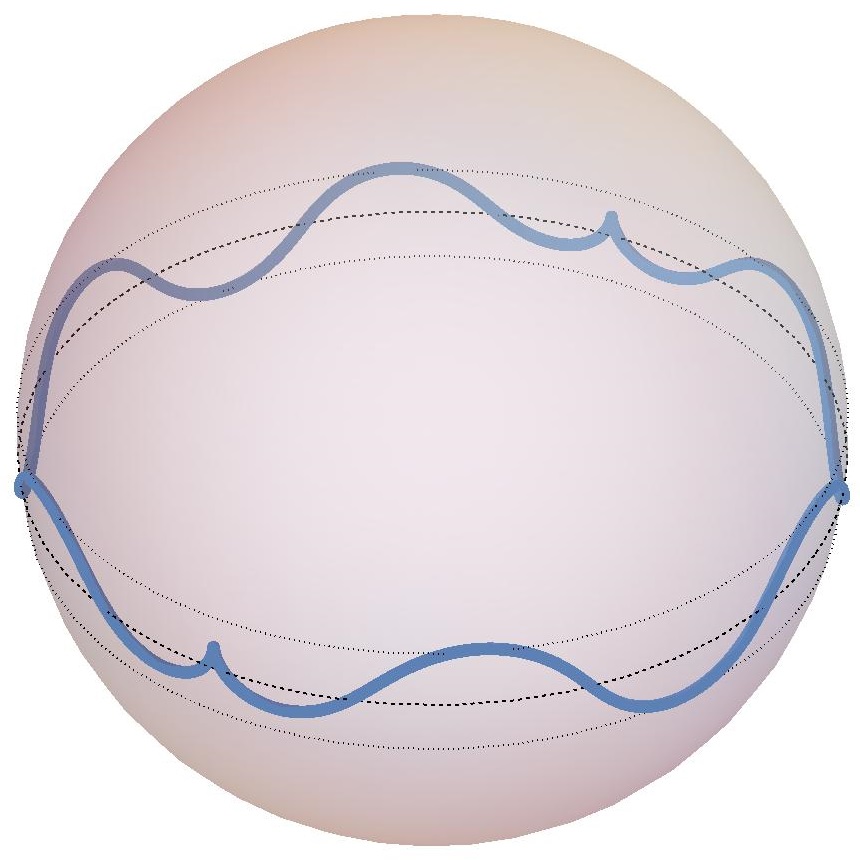}}
\put(40,0){\includegraphics[width = 0.35\textwidth]{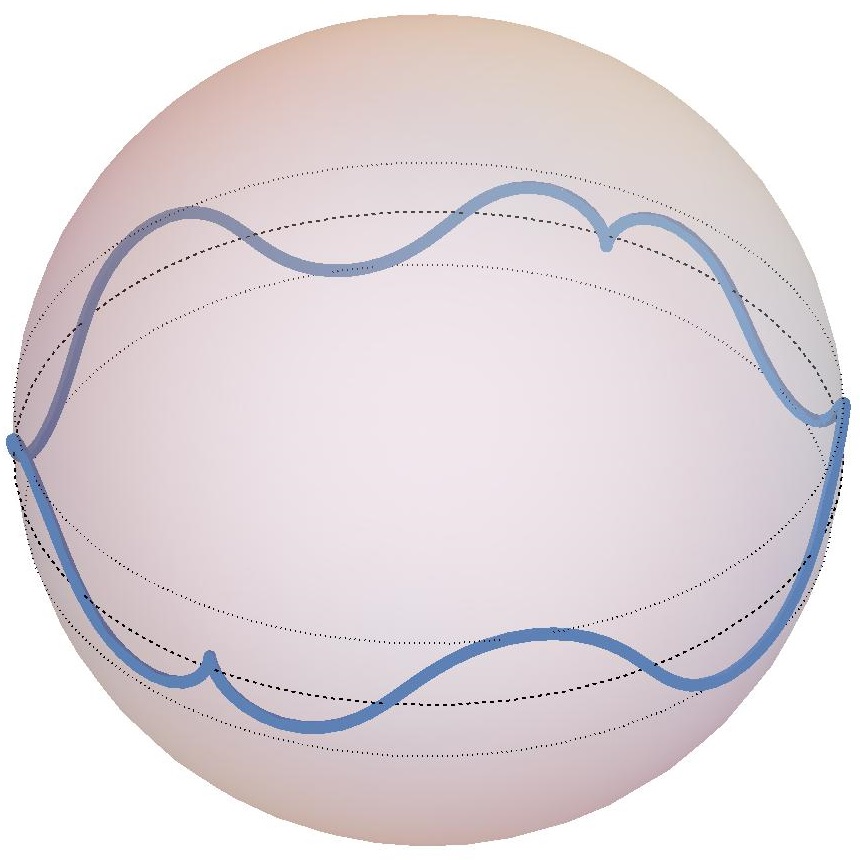}}
\end{picture}
\end{center}
\vspace{-10pt}
\caption{Two closed string solutions with approximate periodicity conditions and $n_2 = 2$}
\vspace{5pt}
\label{fig:dressed_strings_static_n2_2}
\end{figure}

Figure \ref{fig:dressed_strings_approx} depicts six such solutions, one for each class of elliptic strings with an oscillating Pohlmeyer counterpart and two for each class of elliptic strings with a rotating Pohlmeyer counterpart; the latter are one with $\theta_1 < \tilde{\theta}_-$ and one with $\theta_1 > \tilde{\theta}_+$. All solutions of figure \ref{fig:dressed_strings_approx} depict approximate finite closed dressed strings with $n_2=1$. Two indicative examples of dressed solutions with $n_2>1$ are depicted in figure \ref{fig:dressed_strings_static_n2_2}.

The conditions \eqref{eq:asymptotics_condition_ti} and \eqref{eq:asymptotics_condition_st}, which determine the regions where the asymptotic form \eqref{eq:asymptotic_form_theta} and \eqref{eq:asymptotic_form_phi} of the dressed solution is a good approximation, imply that solutions obeying the condition \eqref{eq:finite_approx_condition} are an exponentially good approximation of a finite closed string as long as
\begin{align}
\left| {D\left( {\frac{1}{\beta } + {{\bar v}_0}} \right){\omega _1}} \right|{n_1} = \left| {\left( {\zeta \left( {\tilde a} \right) + \frac{D}{\beta }} \right){\omega _1} - \zeta \left( {{\omega _1}} \right)\tilde a} \right|{n_1} &\gg 1 , \label{eq:approx_finite_condition_ti}\\
\left| {D\left( {\beta  + \frac{1}{{{{\bar v}_1}}}} \right){\omega _1}} \right|{n_1} = \left| {\left( {\zeta \left( {\tilde a} \right) + \beta D} \right){\omega _1} - \zeta \left( {{\omega _1}} \right)\tilde a} \right|{n_1} &\gg 1 , \label{eq:approx_finite_condition_st}
\end{align}
in the case of seed solutions with translationally invariant and static Pohlmeyer counterparts, respectively.

\begin{figure}[ht]
\vspace{10pt}
\begin{center}
\begin{picture}(95,58)
\put(0,30){\includegraphics[width = 0.45\textwidth]{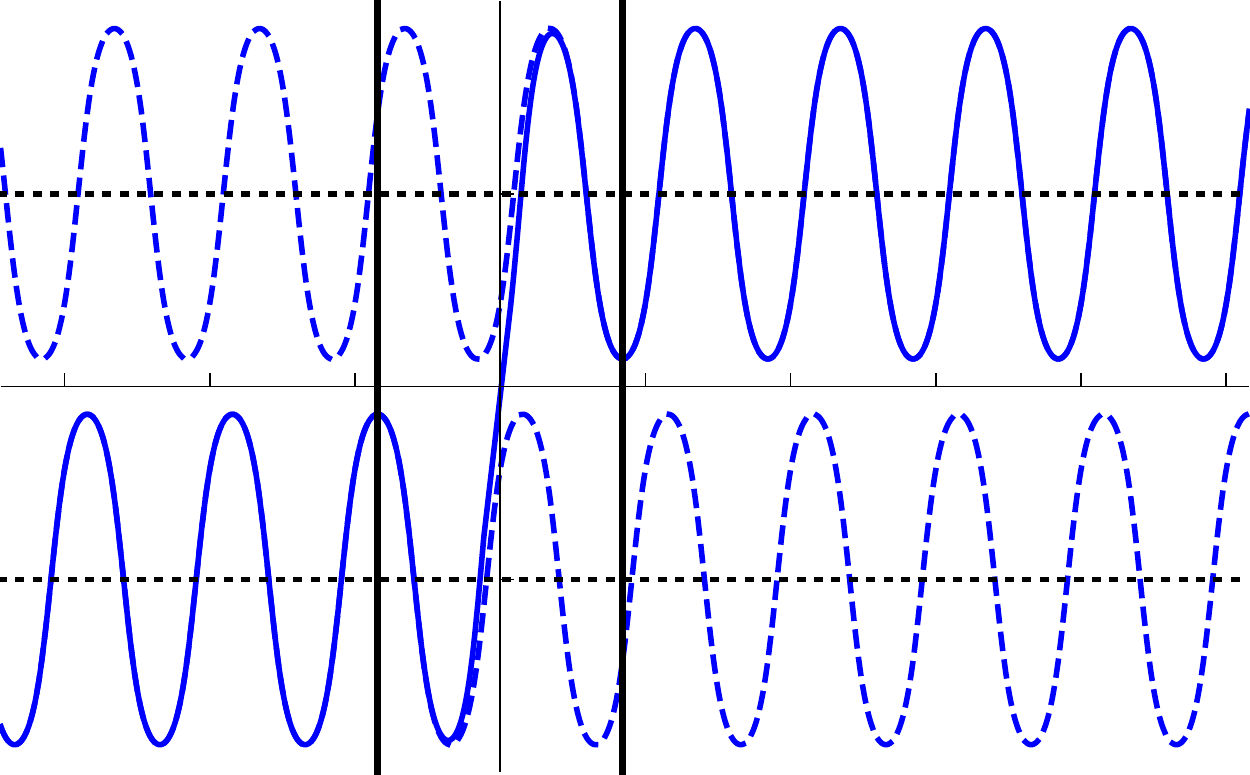}}
\put(50,30){\includegraphics[width = 0.45\textwidth]{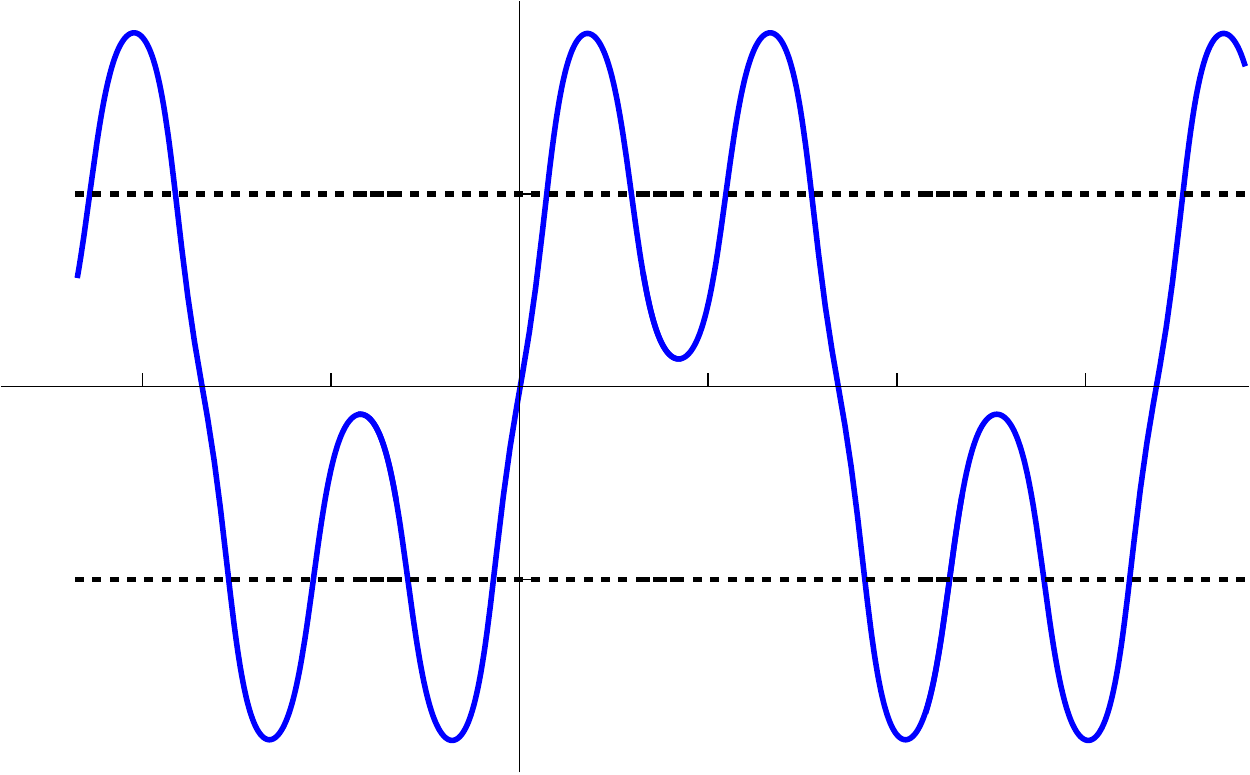}}
\put(0,0){\includegraphics[width = 0.45\textwidth]{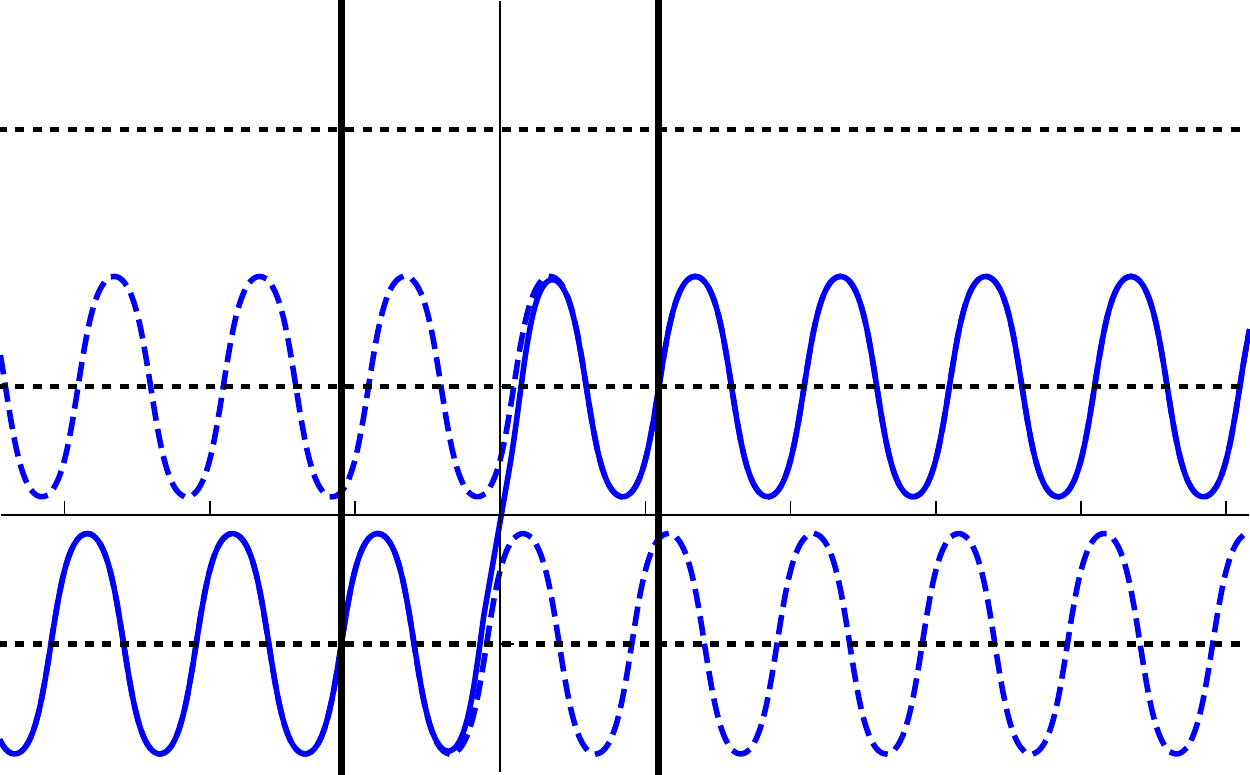}}
\put(50,0){\includegraphics[width = 0.45\textwidth]{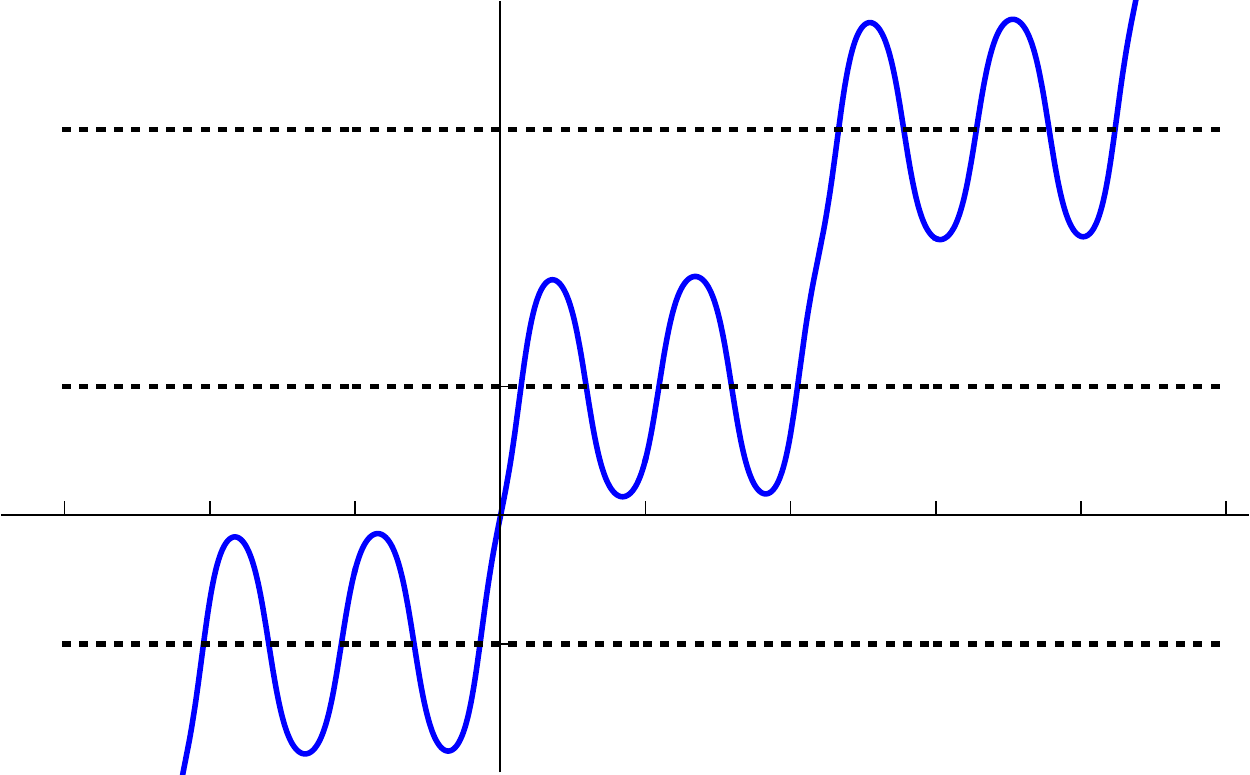}}
\end{picture}
\end{center}
\vspace{-10pt}
\caption{On the left, the kink solution propagating on top of an elliptic background, a degenerate genus two solution of the sine-Gordon equation. The part of this solution between the thick vertical black lines is used to approximate the non-degenerate genus two solution depicted on the right.}
\vspace{5pt}
\label{fig:approximation_SG}
\end{figure}

Such solutions approximate non-degenerate genus two solutions with appropriate periodicity conditions. Figure \ref{fig:approximation_SG} clarifies the performed approximation at the level of the sine-Gordon equation.
The performed approximation is analogous to the fact that solutions of the simple pendulum with energies close to that of the unstable vacuum can be well approximated by a series of patches of appropriate segments of the kink solution. This holds for both oscillatory and rotating solutions of the simple pendulum. In our problem, the former case is depicted in the top row of figure \ref{fig:approximation_SG}, whereas the latter case is depicted in the bottom row of the same figure.

In the top row of figure \ref{fig:approximation_SG}, the non-degenerate genus two solutions that we approximate, has a Pohlmeyer counterpart, which is the non-trivial, non-linear superposition of a train of kinks-antikinks with a train of kinks-antikinks, the latter corresponding to the seed solution. In the bottom row case, it is the superposition of a train of kinks with a train of kinks-antikinks when the seed solution has an oscillatory counterpart and a train of kinks when the seed solution has a rotating counterpart. In a similar manner to the construction of elliptic strings \cite{part1}, where the string solutions with oscillatory Pohlmeyer must obey periodicity conditions corresponding to even integers $n$, the dressed solutions Pohlmeyer counterparts of the kind of the top row of figure \ref{fig:approximation_SG} must have an even value for $n_2$. The string solution depicted on the left of figure \ref{fig:dressed_strings_static_n2_2} has a Pohlmeyer counterpart of the kind of the bottom row of figure \ref{fig:approximation_SG}, whereas the one on right has a Pohlmeyer counterpart of the kind of the top row.

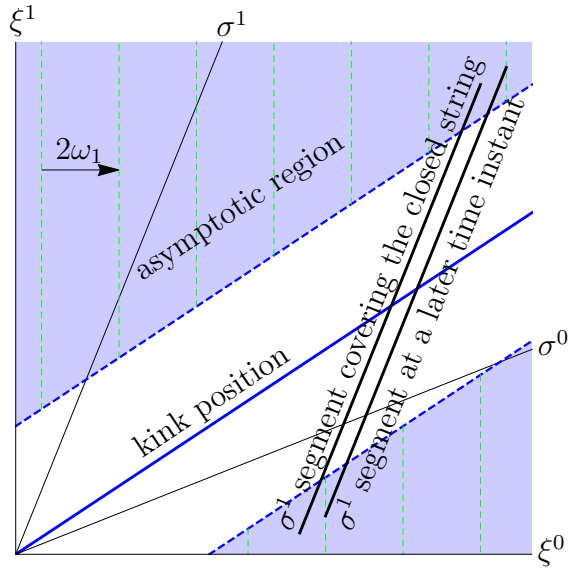
\begin{figure}[ht]
\vspace{10pt}
\begin{center}
\begin{picture}(50,50)
\put(1,1){\includegraphics[width = 0.45\textwidth]{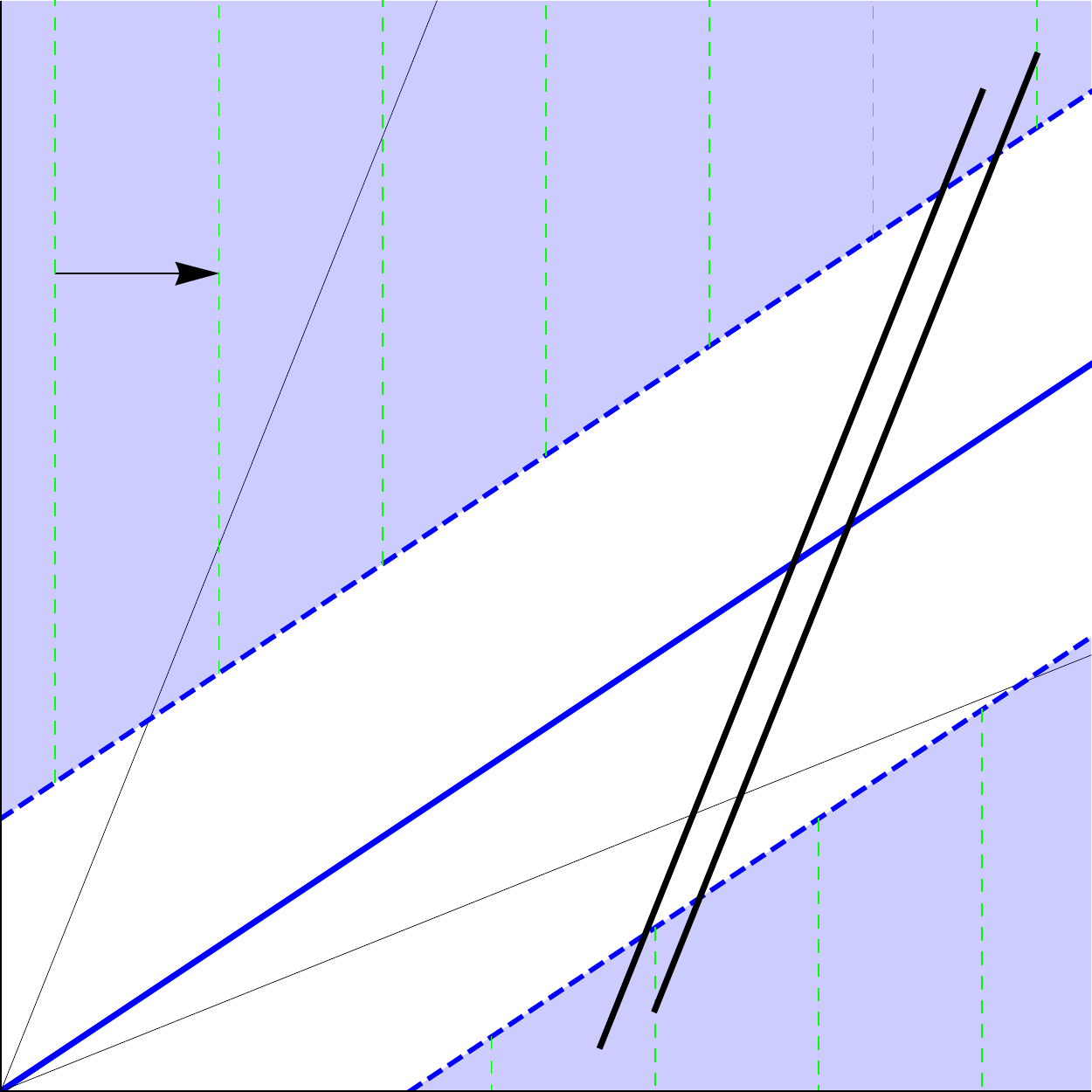}}
\put(46.5,0.5){$\xi^0$}
\put(0.5,47){$\xi^1$}
\put(46.5,18.5){$\sigma^0$}
\put(18.5,46.5){$\sigma^1$}
\put(4.5,35.75){$2\omega_1$}
\put(11,9.5){\rotatebox{33.69}{kink position}}
\put(11,25){\rotatebox{33.69}{asymptotic region}}
\put(23,3){\rotatebox{68.2}{$\sigma^1$ segment covering the closed string}}
\put(28.25,3){\rotatebox{68.2}{$\sigma^1$ segment at a later time instant}}
\end{picture}
\end{center}
\vspace{-10pt}
\caption{Taking advantage of the asymptotic periodicity properties of the sine-Gordon counterpart to form an approximate finite closed string. Notice that the $\sigma^1$ segment that covers the closed string moves with the velocity of the kink and not parallely to the $\sigma^0$ axis.}
\vspace{5pt}
\label{fig:diagram_approximate}
\end{figure}

This picture implies that, as time evolves, the finite segment of the coordinate $\sigma^1$ that parametrizes the finite closed string should move so that the kink in always inside this segment. More specifically the asymptotic formulae \eqref{eq:asymptotic_form_theta_ti}, \eqref{eq:asymptotic_form_phi_ti}, \eqref{eq:asymptotic_form_theta} and \eqref{eq:asymptotic_form_phi} imply that each one of the $n_2$ patches comprising the closed string is parametrized by a $\sigma^1$ segment given by
\begin{align}
{\sigma ^1} &\in \left[ {\frac{{\beta  + {{\bar v}_0}}}{{1 + \beta {{\bar v}_0}}}{\sigma ^0} - \frac{{{n_1}{\omega _1} + s_\Phi \tilde a}}{{\gamma \beta }},\frac{{\beta  + {{\bar v}_0}}}{{1 + \beta {{\bar v}_0}}}{\sigma ^0} + \frac{{{n_1}{\omega _1} + s_\Phi \tilde a}}{{\gamma \beta }}} \right) , \label{eq:span_ti}\\
{\sigma ^1} &\in \left[ {\frac{{\beta  + {{\bar v}_1}}}{{1 + \beta {{\bar v}_1}}}{\sigma ^0} - \frac{{{n_1}{\omega _1} - s_\Phi \tilde a}}{\gamma },\frac{{\beta  + {{\bar v}_1}}}{{1 + \beta {{\bar v}_1}}}{\sigma ^0} + \frac{{{n_1}{\omega _1} - s_\Phi \tilde a}}{\gamma }} \right) , \label{eq:span_st}
\end{align}
in the case of translationally invariant and static seeds, respectively. This is depicted in figure \ref{fig:diagram_approximate} where the $\sigma^1$ segment that is covering the approximate closed finite dressed string solutions is depicted in the original $\xi^{0/1}$ coordinates. The green dashed lines correspond to the periodic properties of the asymptotic limit of the Pohlmeyer counterpart of the solution, i.e. The Pohlmeyer field at all points on the green dashed lines has the same value (or values differing by an integer multiple of $2\pi$).

\subsection{$D^2>0$: Exact Infinite Closed Strings}
\label{subsec:string_infinite_exact}

\begin{figure}[ht]
\vspace{10pt}
\begin{center}
\begin{picture}(35,35)
\put(0,0){\includegraphics[width = 0.35\textwidth]{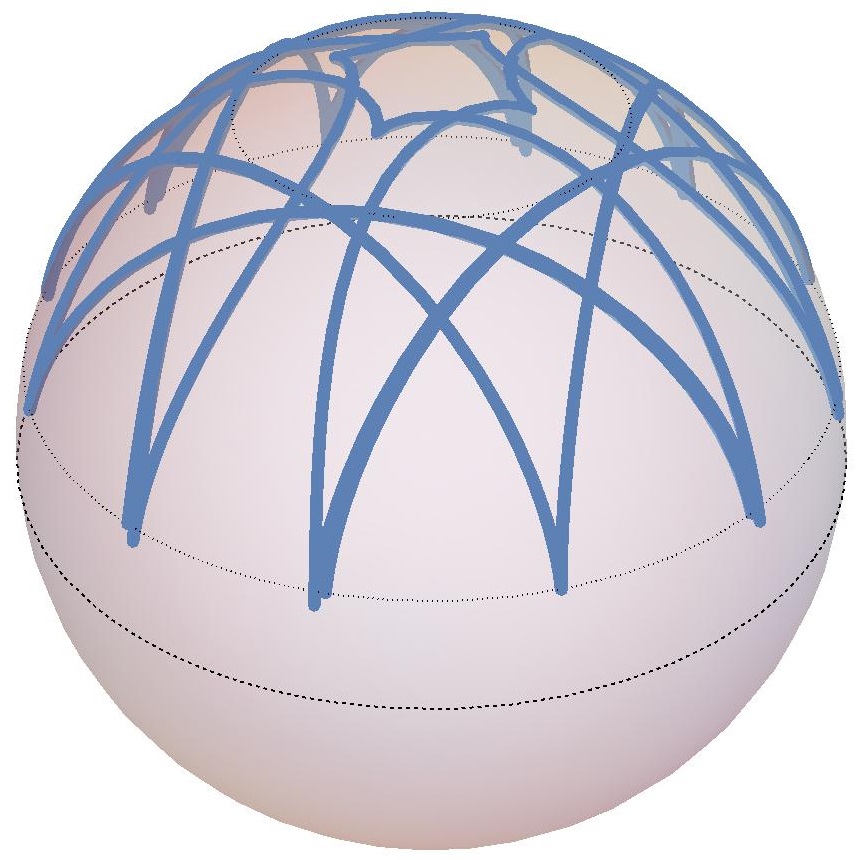}}
\end{picture}
\end{center}
\vspace{-10pt}
\caption{An infinite closed string with exact periodicity conditions and $\delta \varphi = 2 \pi / 3$ and $2 \Delta \varphi = \pi / 2$. The seed solution has a rotating static Pohlmeyer counterpart with $E = 6 \mu^2 / 5$. Four patches of the original seed string solution are required to form the dressed string.}
\vspace{5pt}
\label{fig:exact_periodicity}
\end{figure}

Had we not restricted to finite length strings, we could form infinite strings that obey appropriate and exact periodicity conditions in the same sense as the single spike solution \cite{single_spike_rs2}. Unlike the single spike solution, which far away from the region of the spike tends asymptotically to the equator, thus, providing appropriate boundary conditions at infinity (after infinite self-intersections), this is not the case for dressed elliptic strings. In order to have a well-defined periodic asymptotic behaviour of the dressed string, it is required that $\delta \varphi = 2 \pi m_1 / n_1$, where $m_1$ and $n_1$ are integers. In other words, the seed solution must obey appropriate periodicity conditions (obviously having self-intersections whenever $\gcd \left( m_1 , n_1 \right) = 1$ and $m_1 \neq 1$). A single patch of the dressed string does not form a closed string, even in this case, due to the phase difference of the periodic behaviours of the solution before and after the kink location. However, when $\Delta \varphi = \pi m_2 / n_2$, where $m_2$ and $n_2$ are integers with $\gcd \left( m_2 , n_2 \right) = 1$, it is possible to unite $n_2$ such patches, each one rotated by an angle $2 \pi m_2 / n_2$ in comparison to the previous one. In this way, the asymptotic region of each patch after the location of the kink, coincides with the asymptotic region of the next one before the location of the kink, so that an infinite smooth closed string is formed. An infinite closed dressed elliptic string of this kind is depicted in figure \ref{fig:exact_periodicity}.

These exact infinite closed string solutions can be considered as the $n_1 \to \infty$ limit of the approximate finite closed strings presented in the previous section, with the additional constraint that the seed solution obeys appropriate periodicity conditions so that the asymptotic behaviour of the dressed string is well-defined. In this limit, the conditions \eqref{eq:approx_finite_condition_ti} and \eqref{eq:approx_finite_condition_st} are trivially satisfied and the solutions cease being approximate and becomes exact. It follows that the approximate closed strings of the previous section can also play the role of a regularization scheme for the infinite ones of this section. This will become handy in section \ref{sec:dispersion}, where we will calculate the energy and momentum of the dressed string solutions.

It may appear annoying, that such solutions are parametrized by many infinite patches. However, this is not unexpected. In the literature there are very well-known examples of simpler solutions with similar behaviour, namely the multi-giant magnons. These are degenerate limits of elliptic solutions (the $E \to \mu^2$ limit of the elliptic solutions \eqref{eq:elliptic_solutions_review}). Let us consider an elliptic solution that obeys periodicity conditions with $\delta \varphi = 2 \pi / n$. This solution is parametrized by a segment of $\sigma^1$ which corresponds to $n$ windings around the circle of the torus that corresponds to the real period $2 \omega_1$. At the limit this solution becomes a multi-giant magnon, this period diverges, and, thus the torus is transformed to a cylinder. It follows that appropriate parametrization in this limit, requires the union of $n$ such infinite cylinders, and this is the reason the multi-giant magnon solutions require an infinite range of $\sigma^1$ for the parametrization of each hop. The solutions of this section exhibit the same behaviour. They should be understood as the degeneration of genuine genus two solutions, in the limit when one the two real periods diverges.

\subsection{$D^2<0$: Exact Finite Closed Strings}
\label{subsec:string_finite_exact_D_negative}

When considering dressed string solutions with $D^2 < 0$, the corresponding Pohlmeyer counterpart is not a kink propagating on an elliptic background, but rather a periodic disturbance of the background. This means that the effect of the dressing on the string (as well as in its Pohlmeyer counterpart) is not localized in some region, as it was in the case $D^2>0$. This also implies that there is no limit where the dressed solution tends to become similar to the seed. Thus, in this case, it is not possible to construct an approximate genus two solution, similar to those of section \ref{subsec:string_finite_approximate}.

It is possible to find dressed string solutions with $D^2<0$ that obey exact appropriate periodicity conditions, i.e. it is possible to construct a closed string that corresponds to a finite interval of the space-like parameter $\sigma^1$. The dressing solution contains elliptic functions with argument of the form
\begin{equation}
{\gamma {\sigma ^{0/1}} - \gamma \beta {\sigma ^{1/0}} + {\omega _2}} ,
\label{eq:asymptotics_argument_elliptic}
\end{equation}
inherited by the seed solution. These have the periodicity properties of the seed, i.e. they are periodic in $\sigma^1$, with period $\delta \sigma_0 = 2 \omega_1/(\gamma \beta)$ and $\delta \sigma_1 = 2 \omega_1/\gamma $, respectively. This implies that a closed string that is covered by $\sigma^1 \in \left[ \Sigma_0 , \Sigma_0 + \Delta \Sigma \right)$ should obey
\begin{equation}
\Delta \Sigma = n \delta \sigma_{0/1}, \quad n \in \mathbb{N}.
\end{equation}

Except for this kind of dependence of the dressed solution on the worldsheet variables, there are two angles that appear as arguments in trigonometric functions, one inherited from the seed solution and one from the dressing factor, namely
\begin{align}
{\varphi ^{{\rm{seed}}}}\left( {{\sigma ^0},{\sigma ^1}} \right) &= \ell \left( {\gamma {\sigma ^{1/0}} - \gamma \beta {\sigma ^{0/1}}} \right) - \Phi \left( {\gamma {\sigma ^{0/1}} - \gamma \beta {\sigma ^{1/0}};a} \right) ,\\
{\varphi ^{{\rm{dress}}}}\left( {{\sigma ^0},{\sigma ^1}} \right) &= \sqrt { - {D^2}} \left( {\gamma {\sigma ^{1/0}} - \gamma \beta {\sigma ^{0/1}}} \right) - \Phi \left( {\gamma {\sigma ^{0/1}} - \gamma \beta {\sigma ^{1/0}};\tilde a} \right) .
\end{align}
These functions obey the following quasi-periodicity properties
\begin{align}
\varphi ^{{\rm{seed}}}\left( {{\sigma ^0},{\sigma ^1} + \delta \sigma_{0/1} } \right) &= \varphi ^{{\rm{seed}}}\left( {{\sigma ^0},{\sigma ^1}} \right) + \delta \varphi _{0/1}^{{\rm{seed}}} ,\\
\varphi ^{{\rm{dress}}}\left( {{\sigma ^0},{\sigma ^1} + \delta \sigma_{0/1} } \right) &= \varphi ^{{\rm{dress}}}\left( {{\sigma ^0},{\sigma ^1}} \right) + \delta \varphi _{0/1}^{{\rm{dress}}} ,
\end{align}
where
\begin{align}
\delta \varphi_{0/1}^{{\rm{seed}}} &= \mp 2 \omega_1 \left( {i\zeta \left( {{\omega _1}} \right)\frac{a}{\omega_1} - \left( {i\zeta \left( a \right) + \sqrt {\frac{{\left( {{x_1} - \wp \left( a \right)} \right)\left( {{x_{2/3}} - \wp \left( a \right)} \right)}}{{{x_{3/2}} - \wp \left( a \right)}}} } \right)} \right) , \\
\delta \varphi_{0/1}^{{\rm{dress}}} &= \mp 2 \omega_1 \left( {i\zeta \left( {{\omega _1}} \right)\frac{\tilde{a}}{\omega_1} - \left( {i\zeta \left( \tilde{a} \right) + \sqrt {\frac{{\left( {\wp \left( \tilde{a} \right) - {x_1}} \right)\left( {{x_{2/3}} - \wp \left( a \right)} \right)}}{{{x_{3/2}} - \wp \left( a \right)}}} } \right)} \right) .
\end{align}
Obviously, appropriate periodicity conditions imply 
\begin{align}
\delta \varphi ^{{\rm{seed}}} &= \frac{m^{{\rm{seed}}}}{n^{{\rm{seed}}}} 2\pi , \label{eq:periodicity_Dneg_naked}\\
\delta \varphi ^{{\rm{dress}}} &= \frac{m^{{\rm{dress}}}}{n^{{\rm{dress}}}} 2\pi , \label{eq:periodicity_Dneg_dressed}
\end{align}
where $m^{{\mathrm{seed}}},n^{{\mathrm{seed}}},m^{{\mathrm{dress}}},n^{{\mathrm{dress}}} \in \mathbb{Z}$. If we have selected the above integers, so that $\gcd \left( m^{{\mathrm{seed}}},n^{{\mathrm{seed}}} \right) = \gcd \left( m^{{\mathrm{dress}}},n^{{\mathrm{dress}}} \right) = 1$, then one can obtain a closed string solution with $n = \lcm \left( n^{{\mathrm{seed}}},n^{{\mathrm{dress}}} \right)$.

\begin{figure}[ht]
\vspace{10pt}
\begin{center}
\begin{picture}(50,50)
\put(1,1){\includegraphics[width = 0.45\textwidth]{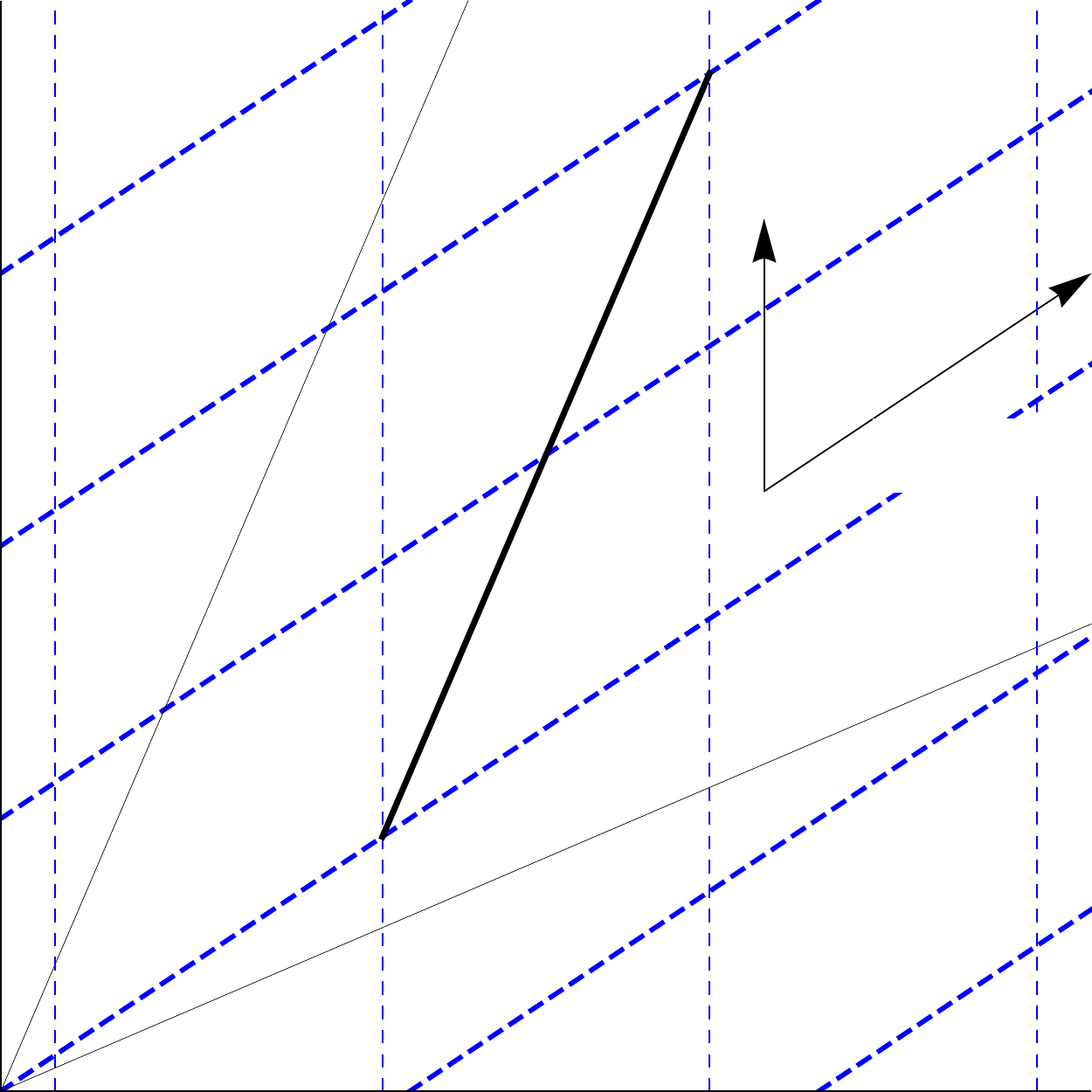}}
\put(46.5,0.5){$\xi^0$}
\put(0.5,47){$\xi^1$}
\put(46.5,20){$\sigma^0$}
\put(20,46.5){$\sigma^1$}
\put(32.75,31.25){$\frac{2\pi}{iD} \hat{\xi}^1$}
\put(36,26){$2\omega_1 ( \hat{\xi}^0+v_{\mathrm{tb}} \hat{\xi}^1 )$}
\put(12,7){\rotatebox{66.8}{$\sigma^1$ segment covering the closed string}}
\end{picture}
\end{center}
\vspace{-10pt}
\caption{The segment of $\sigma^1$ parametrizing a finite closed dressed elliptic string with $D^2<0$, as specified by the periodicity properties of the sine-Gordon counterpart. In the depicted example $n^{\rm{dress}} = 1$ and $m^{\rm{dress}} = 2$.}
\vspace{5pt}
\label{fig:diagram_D_neg}
\end{figure}
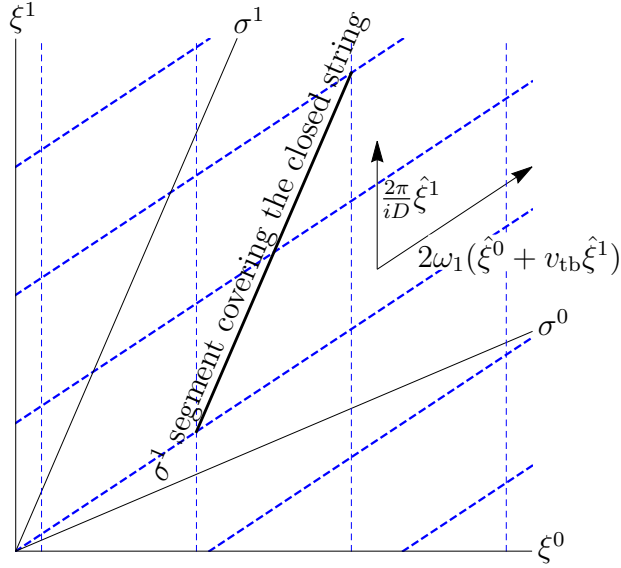

The first of these conditions \eqref{eq:periodicity_Dneg_naked} simply states that the seed solution obeys appropriate periodicity conditions. The second one \eqref{eq:periodicity_Dneg_dressed} is closely related to the periodicity properties of the sine-Gordon counterpart analysed in section \ref{subsec:SG_breathers_periodicity}. More specifically, this condition is equivalent to demanding that the direction of the boosted axis $\sigma^1$ coincides with one of the directions defined by the periodicity lattice of the sine-Gordon counterpart. This can become more transparent expressing the condition \eqref{eq:periodicity_Dneg_dressed} in terms of the velocity of the periodic disturbance $v_{0/1}^{{\rm{tb}}}$, given by equations \eqref{eq:velocity_tb_ti} and \eqref{eq:velocity_tb_st}. It reads
\begin{align}
\delta \varphi _0^{{\rm{dress}}} &= \sqrt { - {D^2}} \left( {\frac{1}{\beta } - v_0^{{\rm{tb}}}} \right)2{\omega _1} = \frac{{{m^{{\rm{dress}}}}}}{{{n^{{\rm{dress}}}}}}2\pi , \\
\delta \varphi _1^{{\rm{dress}}} &=  - \sqrt { - {D^2}} \left( {\beta  - \frac{1}{{v_1^{{\rm{tb}}}}}} \right)2{\omega _1} = \frac{{{m^{{\rm{dress}}}}}}{{{n^{{\rm{dress}}}}}}2\pi ,
\end{align}
which after some algebra results in
\begin{align}
\frac{1}{\beta } &= \frac{{\frac{{2\pi }}{{\sqrt { - {D^2}} }}{m^{{\rm{dress}}}} + 2{\omega _1}v_0^{{\rm{tb}}}{n^{{\rm{dress}}}}}}{{2{\omega _1}{n^{{\rm{dress}}}}}} ,\\
\frac{1}{\beta } &= \frac{{2{\omega _1}{n^{{\rm{dress}}}}}}{{ - \frac{{2\pi }}{{\sqrt { - {D^2}} }}{m^{{\rm{dress}}}} + \frac{{2{\omega _1}}}{{v_1^{{\rm{tb}}}}}{n^{{\rm{dress}}}}}} ,
\end{align}
for solutions whose seeds have a translationally invariant or static Pohlmeyer counterpart, respectively. Bearing in mind, that the sine-Gordon counterpart solution is periodic under the translations \eqref{eq:D2_negative_periodicity_trivial} or \eqref{eq:D2_negative_periodicity_trivial_static} and quasi-periodic under the translations \eqref{eq:D2_negative_periodicity_non_trivial} or \eqref{eq:D2_negative_periodicity_non_trivial_static}, the above equations imply that the $\sigma^1$ axis is lying in the direction of ${m^{{\rm{dress}}}}$ periodic displacements and ${n^{{\rm{dress}}}}$ quasi-periodic displacements on the periodicity lattice of the sine-Gordon counterpart. Figure \ref{fig:diagram_D_neg} visualises the above.

\begin{figure}[ht]
\vspace{10pt}
\begin{center}
\begin{picture}(85,40)
\put(0,0){\includegraphics[width = 0.4\textwidth]{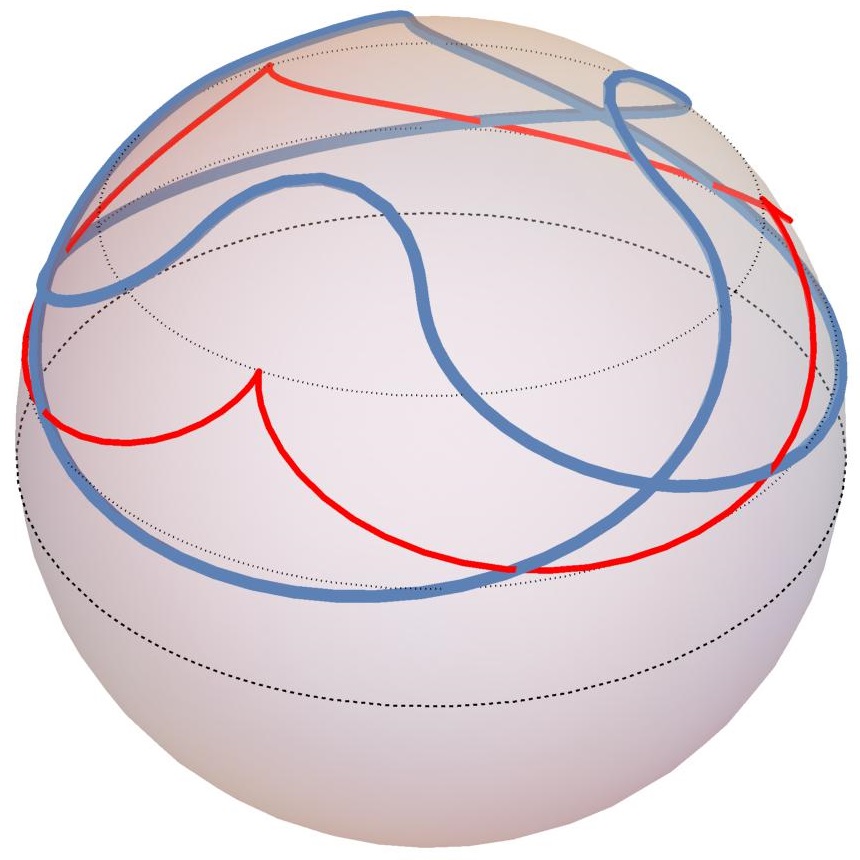}}
\put(45,0){\includegraphics[width = 0.4\textwidth]{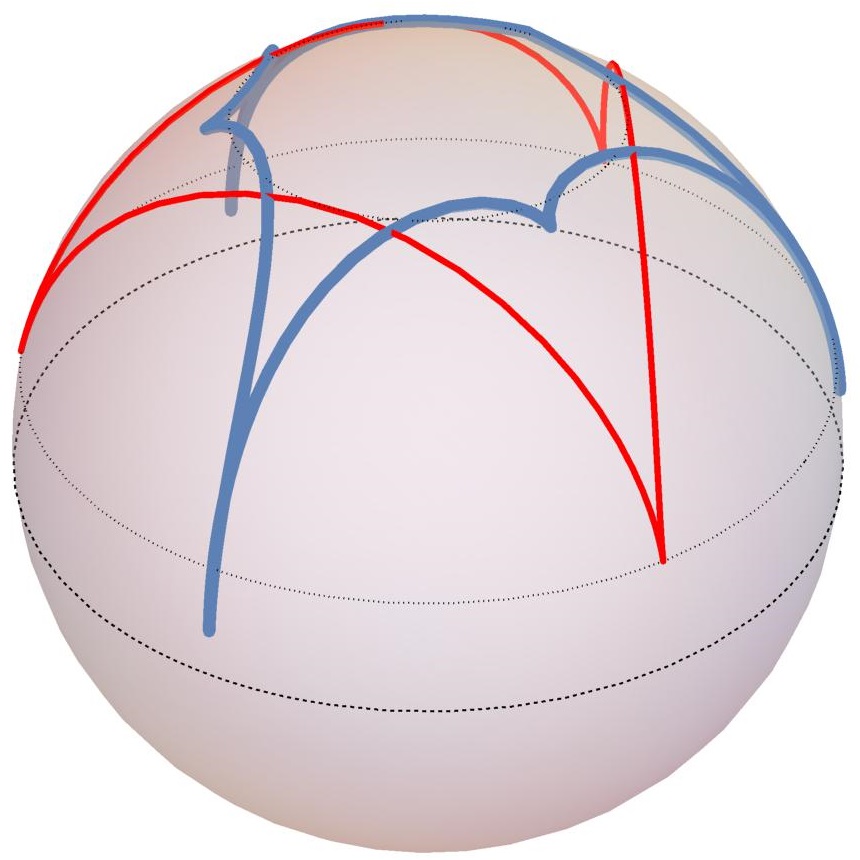}}
\put(32,0){\includegraphics[width = 0.05\textwidth]{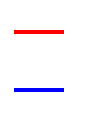}}
\put(36.5,1.5){dressed solution}
\put(36.5,4.5){seed solution}
\end{picture}
\end{center}
\vspace{-10pt}
\caption{Two dressed strings with $D^2<0$ with seed solution having a translationally invariant counterpart (left) and a static counterpart (right)}
\vspace{5pt}
\label{fig:dressed_np_cycle}
\end{figure}

These finite closed string solutions can be considered as the analytic continuation of the exact infinite closed strings that we studied in section \ref{subsec:string_infinite_exact}. However in this case, the resulting strings are of finite size. Similarly to the exact infinite closed strings with $D^2>0$, the seed solution must obey appropriate periodicity conditions, too. However, depending on the integers $n^{{\rm{seed}}}$ and $n^{{\rm{dress}}}$, the dressed string may require several ($\lcm \left( n^{{\rm{seed}}},n^{{\rm{dress}}} \right) / n^{{\rm{seed}}}$) repetitions of the original seed solution in order to complete a closed string. Figure \ref{fig:dressed_np_cycle} depicts an example of such a dressed string solution. 

Similarly to the exact infinite closed string solutions with $D^2>0$, these solutions are also the degenerate limit of genuine genus two solutions. The difference between the two classes of solutions is the fact that the divergent period is the real one in the former case and the imaginary one in the latter. In other words, in this case, the $\sigma^1$ segment parametrizing the string solution corresponds to winding around the compact direction of the cylinder, being the degenerate limit of the torus.

\subsection{$D^2>0$: Special Exact Finite Closed Strings}
\label{subsec:string_finite_exact_special}

In section \ref{subsec:string_finite_approximate}, we showed that under some conditions, it is possible to take advantage of the asymptotic behaviour of the solutions to construct approximate closed dressed elliptic string solutions. The appropriate conditions are given in equations \eqref{eq:approx_finite_condition_ti} and \eqref{eq:approx_finite_condition_st} and it is simple to see that, selecting an adequately large $n$, these conditions can be satisfied, independently of the value of the other parameters. However, there is a special case where this is not possible namely,
\begin{equation}
\beta = - \frac{1}{\bar{v}_{0/1}}.
\label{eq:special_exact_condition}
\end{equation}
In this case, it is not possible to construct such an approximate solution, as the spacelike coordinate $\sigma^1$ follows exactly the motion of the kink, and, thus, no matter how large values $\sigma^1$ takes, a snapshot of the string never reaches the asymptotic region.

In a different approach, in the case $D^2<0$, there is a two-dimensional lattice of symmetries of the sine-Gordon counterpart that allows the construction of periodic and thus, finite string solutions. In the case $D^2>0$, this set of symmetries is one-dimensional and thus, it is not generally possible to use these symmetries for the construction of finite string solutions, unless the $\sigma^1$ axis coincides with the direction of the periodic symmetry of the sine-Gordon counterpart.

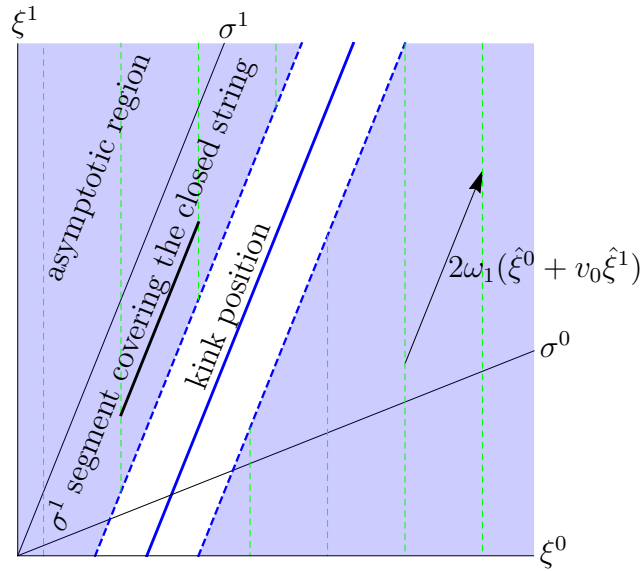
\begin{figure}[ht]
\vspace{10pt}
\begin{center}
\begin{picture}(50,50)
\put(1,1){\includegraphics[width = 0.45\textwidth]{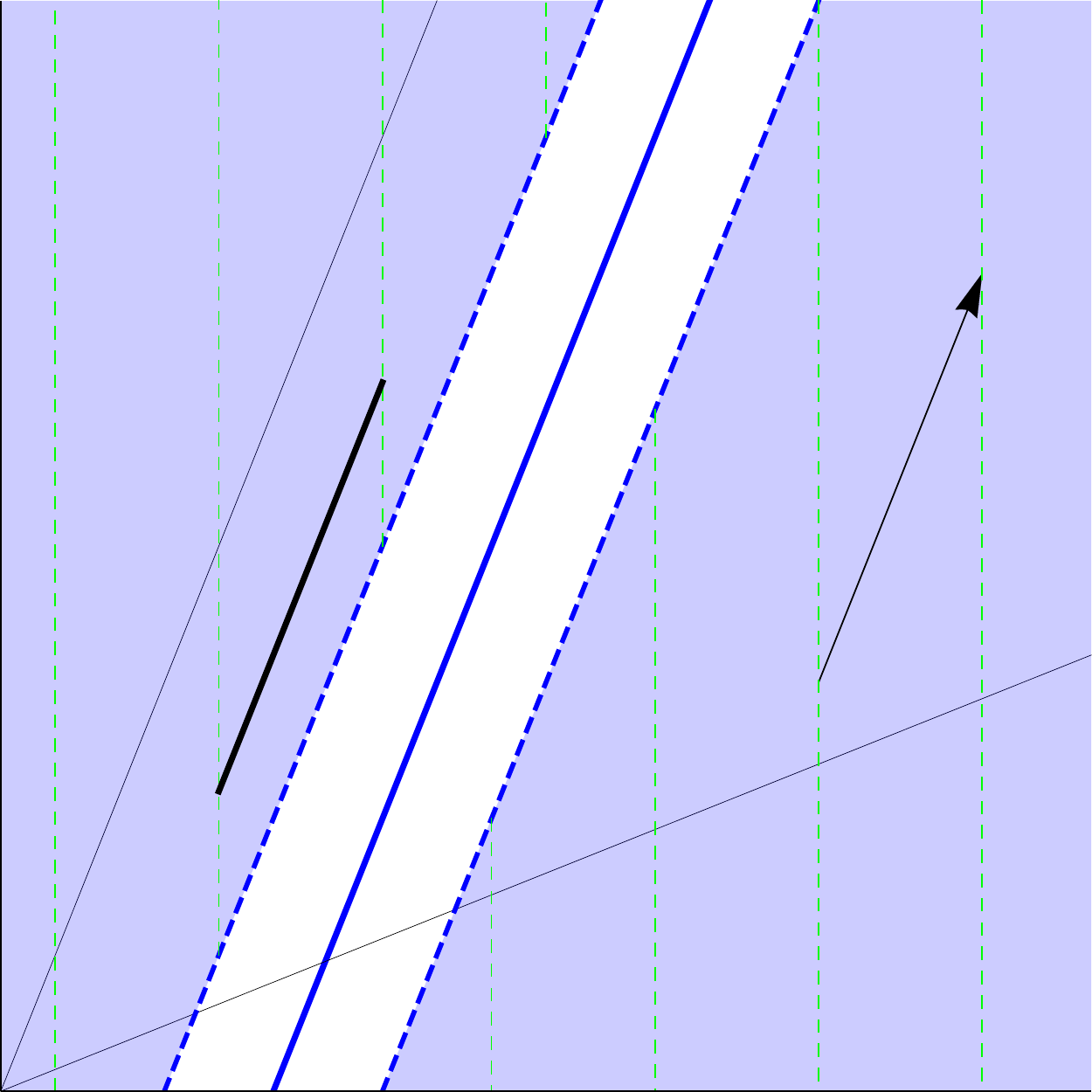}}
\put(46.5,0.5){$\xi^0$}
\put(0.5,47){$\xi^1$}
\put(46.5,18.5){$\sigma^0$}
\put(18.5,46.5){$\sigma^1$}
\put(38.5,25.5){$2\omega_1 ( \hat{\xi}^0+v_{0} \hat{\xi}^1 )$}
\put(15.5,15.5){\rotatebox{68.2}{kink position}}
\put(3,25){\rotatebox{68.2}{asymptotic region}}
\put(3,3){\rotatebox{68.2}{$\sigma^1$ segment covering the closed string}}
\end{picture}
\end{center}
\vspace{-10pt}
\caption{Taking advantage of the periodicity properties of the sine-Gordon counterpart to form a special exact finite closed string}
\vspace{5pt}
\label{fig:diagram_special_exact}
\end{figure}

The condition \eqref{eq:special_exact_condition} corresponds to exactly this case. Therefore, one may use the exact periodic properties of the sine-Gordon counterparts of the dressed elliptic strings \eqref{eq:dressed_elliptic_periodic_Dpos_osc} and \eqref{eq:dressed_elliptic_periodic_Dpos_rot} to construct special exact finite closed string solutions, as shown in picture \ref{fig:diagram_special_exact}.

The condition \eqref{eq:special_exact_condition} is not sufficient to ensure appropriate boundary conditions of the solution. Similarly to the $D^2<0$ case of section \ref{subsec:string_finite_exact_D_negative}, the worldsheet coordinates appear in three combinations in the solution. The first one is trivially $\xi^{0/1}$ or \eqref{eq:asymptotics_argument_elliptic} in terms of $\sigma_{0/1}$, which implies that the possible segment of $\sigma_1$ covering a finite string is given by equations \eqref{eq:periodicity_range_ti} and \eqref{eq:periodicity_range_st} for translationally invariant and static seeds respectively. One should remember that in the case under study it holds $D^2>0$ and thus, the seed may have an oscillating sine-Gordon counterpart. In such a case, in these expressions $2\omega_1$ should be substituted with $4\omega_1$.

Except for this dependence on the worldsheet variable, two more angles appear, namely
\begin{align}
{\varphi_{0/1} ^{{\rm{seed}}}}\left( {{\sigma ^0},{\sigma ^1}} \right) &= \ell \left( {\gamma {\sigma ^{1/0}} - \gamma \beta {\sigma ^{0/1}}} \right) - \Phi \left( {\gamma {\sigma ^{0/1}} - \gamma \beta {\sigma ^{1/0}};a} \right) ,\\
{\varphi_{0/1} ^{{\rm{dress}}}}\left( {{\sigma ^0},{\sigma ^1}} \right) &= D \left( {\gamma {\sigma ^{1/0}} - \gamma \beta {\sigma ^{0/1}}} \right) - \Phi \left( {\gamma {\sigma ^{0/1}} - \gamma \beta {\sigma ^{1/0}};\tilde a} \right) .
\end{align}
The first one appears as argument of trigonometric functions, whereas the second one in hyperbolic functions. Therefore, appropriate periodicity conditions require the condition \eqref{eq:periodicity_Dneg_naked}, whereas the periodicity condition \eqref{eq:periodicity_Dneg_dressed} should be substituted with
\begin{equation}
\delta \varphi ^{{\rm{dress}}} = 0 .
\end{equation}
The first one is equivalent to the seed solution obeying appropriate periodicity conditions, whereas the second one simply implies the condition \eqref{eq:special_exact_condition}. Obviously, such a solution is possible only when the kink propagates with a speed larger than the speed of light.

Both infinite and finite exact periodic string solutions with $D^2>0$ can be considered as the analytic continuation of the exact finite string solution with $D^2<0$. The space or time period of the corresponding sine-Gordon counterparts is equal to $2\pi / \sqrt{-D^2} $. As $D\to 0$ this period diverges. Therefore, naturally the finite strings with $D^2<0$ of section \ref{subsec:string_finite_exact_D_negative} tend to the infinite strings with $D^2>0$ of section \ref{subsec:string_infinite_exact}, unless this vector does not contribute to the $\sigma^1$ direction, i.e. $m^{{\rm{dress}}} = 0$, in which case they tend to the finite exact solutions with $D^2>0$ of this section.

\setcounter{equation}{0}
\section{Time Evolution and Spike Interactions}
\label{sec:spike_interactions}

\subsection{Shape Periodicity}
\subsubsection{$D^2>0$: Approximate Finite and Exact Infinite Strings}

\begin{figure}[p]
\vspace{10pt}
\begin{center}
\begin{picture}(90,113)
\put(2.5,78){\includegraphics[width = 0.35\textwidth]{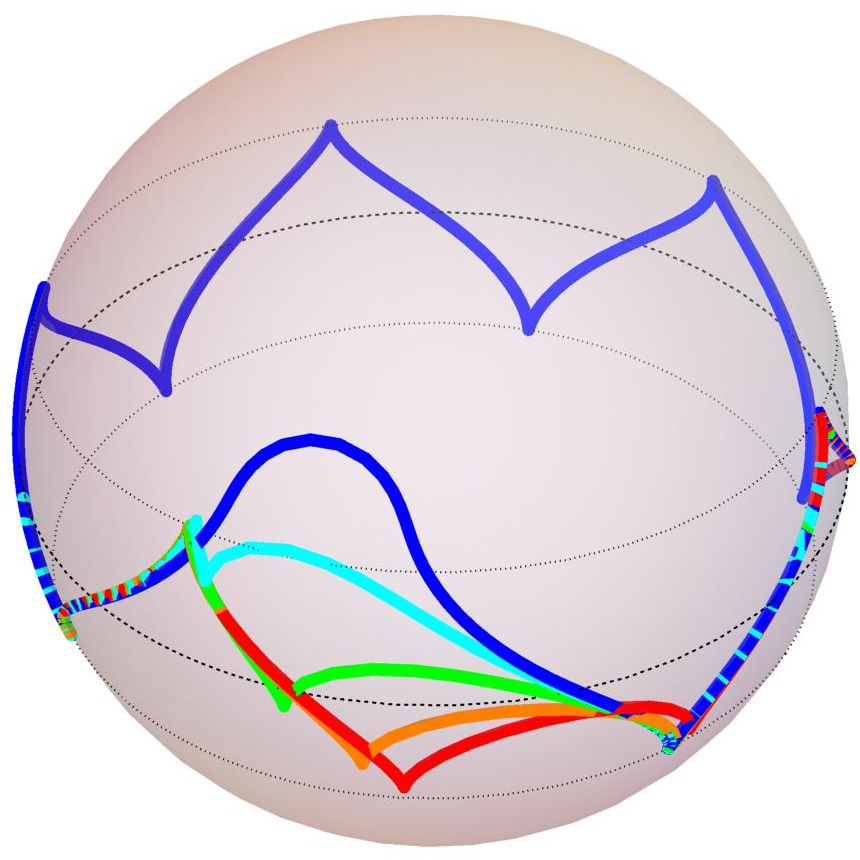}}
\put(47.5,78){\includegraphics[width = 0.35\textwidth]{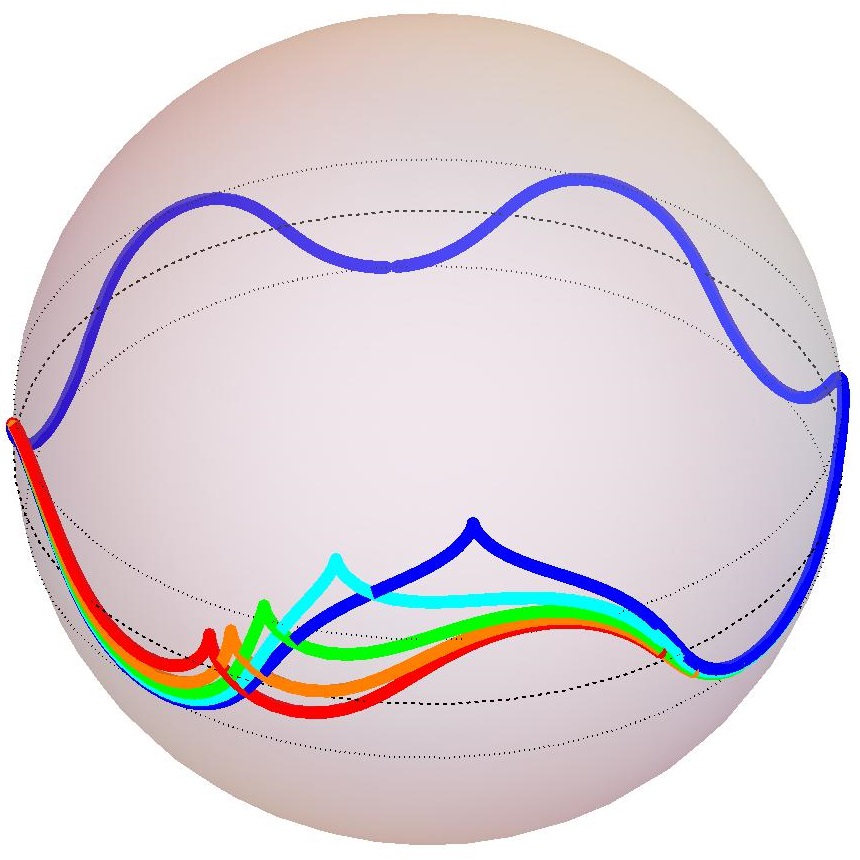}}
\put(2.5,43){\includegraphics[width = 0.35\textwidth]{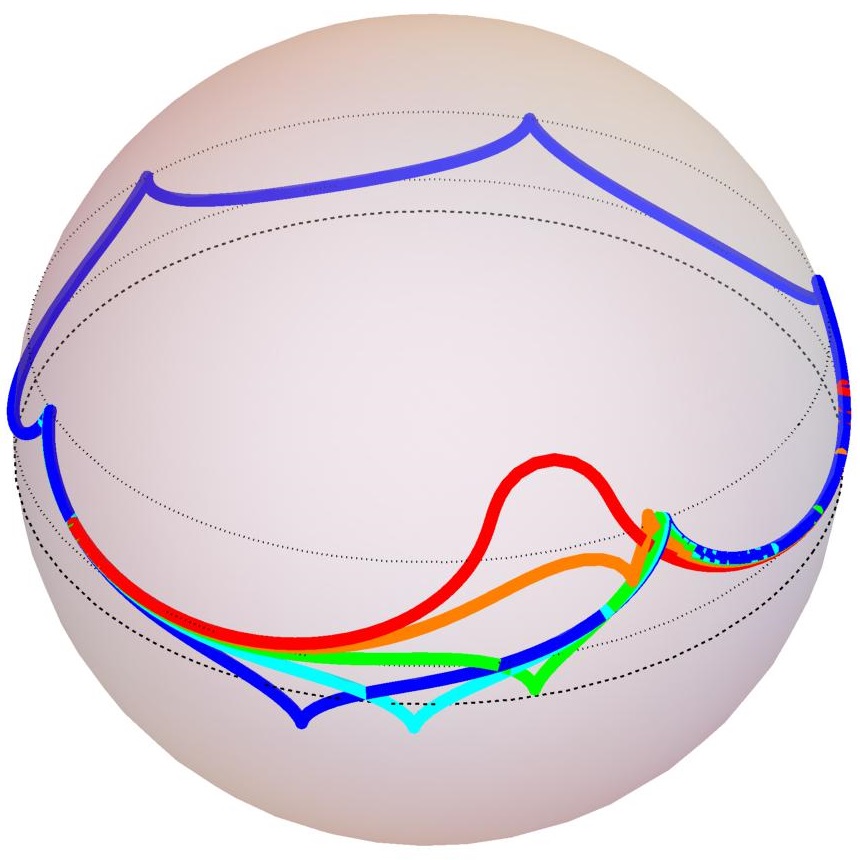}}
\put(47.5,43){\includegraphics[width = 0.35\textwidth]{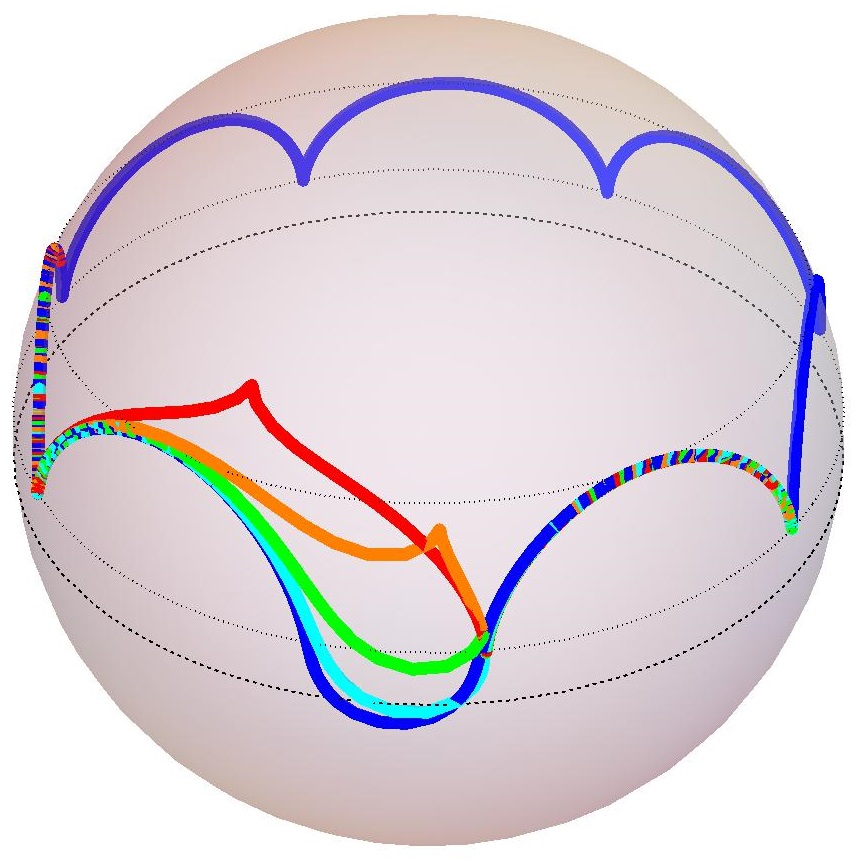}}
\put(2.5,8){\includegraphics[width = 0.35\textwidth]{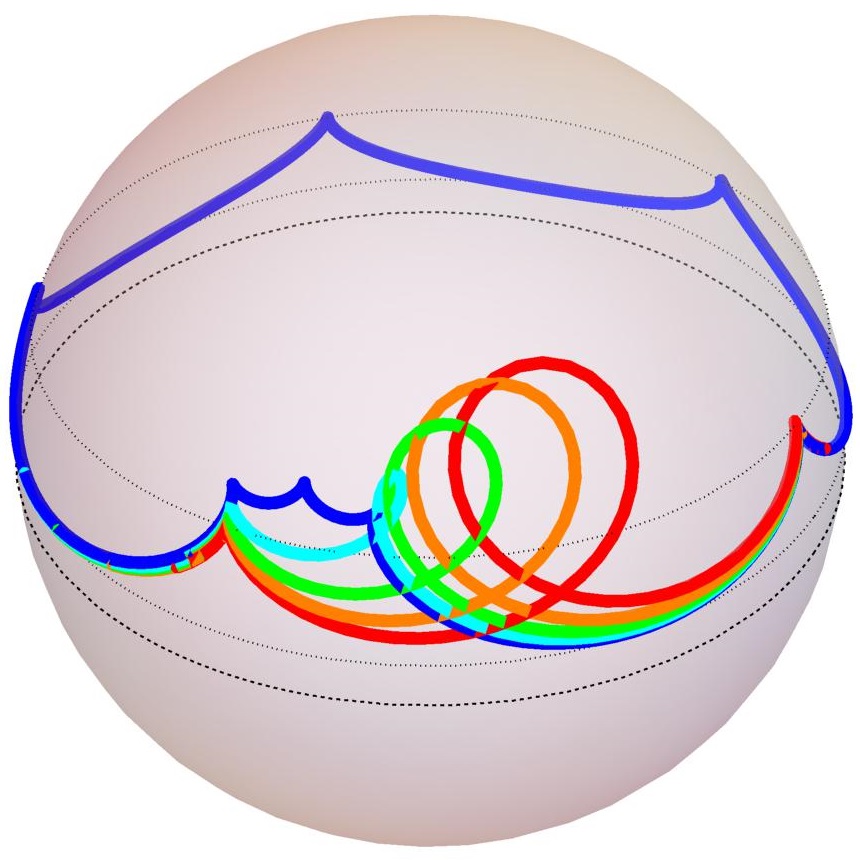}}
\put(47.5,8){\includegraphics[width = 0.35\textwidth]{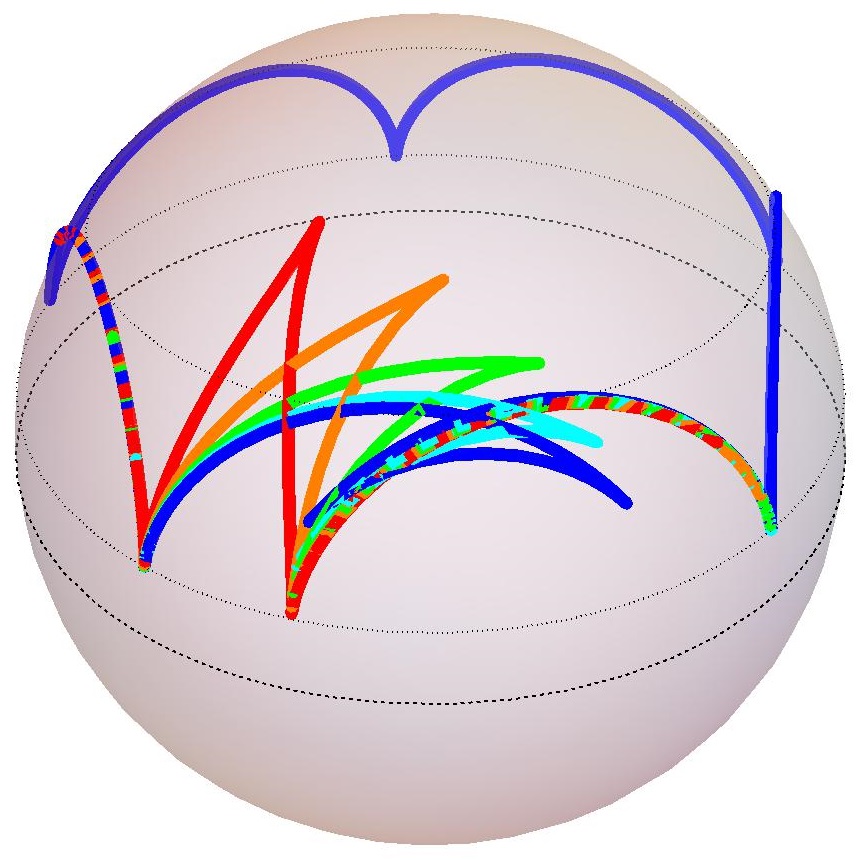}}
\put(34,0){\includegraphics[width = 0.05\textwidth]{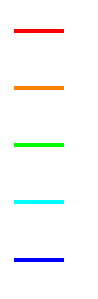}}
\put(39,13.25){$t=0$}
\put(39,10.25){$t=T^{{\rm{shape}}}/8$}
\put(39,7.25){$t=T^{{\rm{shape}}}/4$}
\put(39,4.25){$t=3T^{{\rm{shape}}}/8$}
\put(39,1.25){$t=T^{{\rm{shape}}}/2$}
\end{picture}
\end{center}
\vspace{-10pt}
\caption{The time evolution of the dressed elliptic string solutions depicted in figure \ref{fig:dressed_strings_approx}}
\vspace{5pt}
\label{fig:dressed_strings_static_time}
\end{figure}
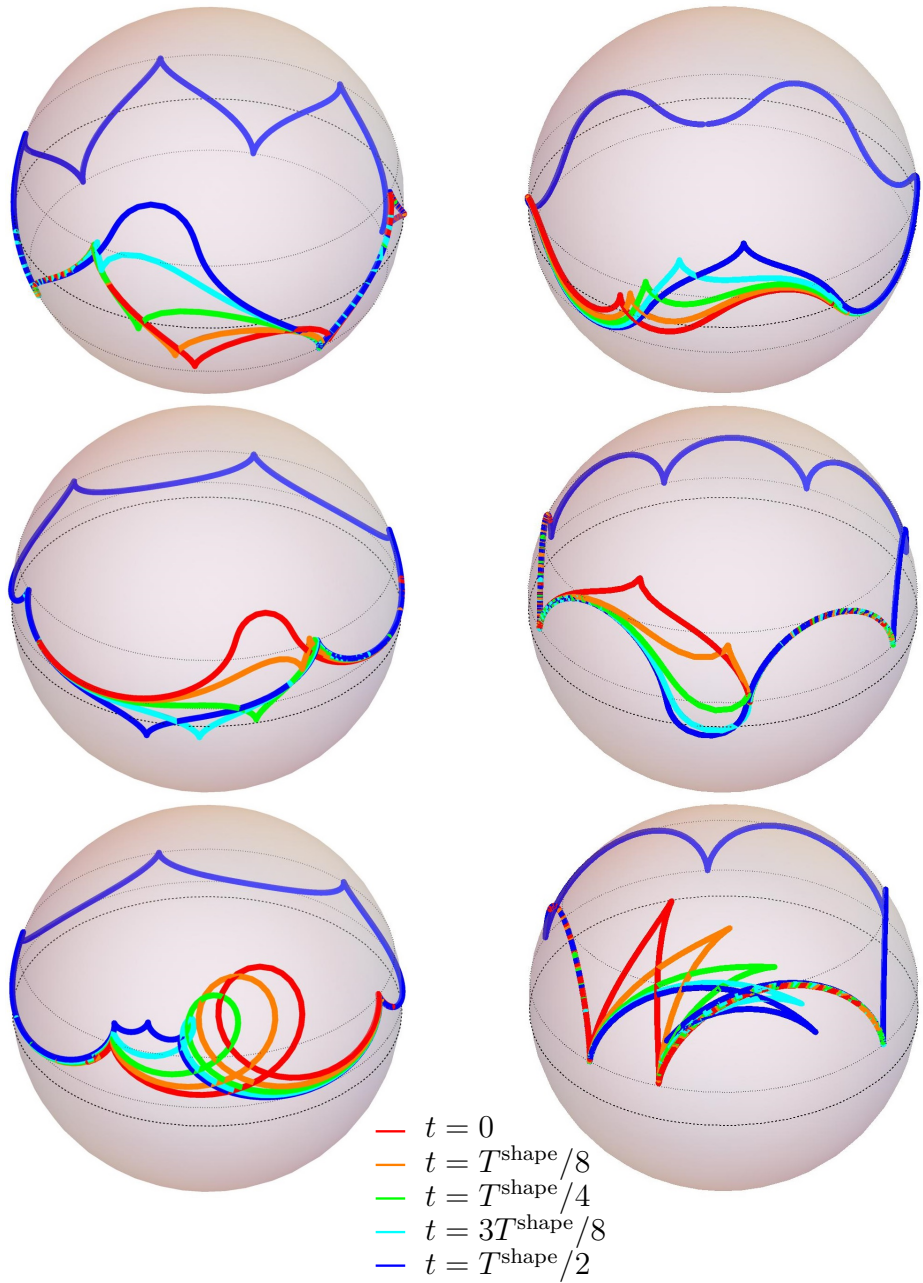

The time evolution of the approximate finite dressed strings with $D^2>0$ is shown in figure \ref{fig:dressed_strings_static_time}. The dressed strings, in the region far away from the extra kink induced by the dressing, are similar to a rotated version of the seed elliptic string solutions. The time evolution of the later is simply a rigid rotation around the $z$-axis with angular velocity equal to \cite{part1}
\begin{equation}
\omega_{0/1}  = \frac{1}{R}\sqrt {\frac{{{x_1} - \wp \left( a \right)}}{{{x_{3/2}} - \wp \left( a \right)}}} .
\label{eq:elliptic_solutions_omega}
\end{equation}
In figure \ref{fig:dressed_strings_static_time}, this rigid rotation has been frozen in order to focus on the change of the shape of the string. The shape of the string alters periodically with period equal to
\begin{equation}
{T_0^{{\rm{shape}}}} = \left( 2 \right) 2 \mu \gamma {\omega _1}\left( {1 - \beta {{{\bar v}_0}}} \right) , \quad
{T_1^{{\rm{shape}}}} = \left( 2 \right) 2 \mu \gamma {\omega _1}\left( {\frac{1}{{{\bar v}_1}} - \beta } \right) ,
\end{equation}
where the extra $2$ applies in the case of oscillatory seed solutions and ${\bar v}_{0/1}$ is given by equations \eqref{eq:kinks_mean_velocity} and \eqref{eq:kinks_mean_velocity_static}, depending on whether the seed solution has a translationally invariant or static counterpart. At the level of the sine-Gordon equation, this formula is simply the time necessary for the kink to travel over a whole period of the elliptic background. This time is directly related to the mean velocity of the kink, as calculated in section \ref{subsec:SG_kink_velocity} in the linear gauge. The above formula is just the appropriate adaptation to the static gauge.

The time evolution of the exact infinite dressed strings with $D^2 > 0$ is similar to the time evolution of the approximate finite strings.

\subsubsection{$D^2<0$}
The question whether the dressed elliptic string solutions with $D^2<0$ are also periodic in time has a similar answer to the same question imposed about the sine-Gordon counterpart. In a similar manner to the periodic in space properties, the dependence of the solution on elliptic functions of $\xi^{0/1}$ implies that a possible period for the motion of the string has to be a multiple of the quantity
\begin{align}
\delta \tau_0  &= \frac{2{\omega _1}}{\gamma }, \\
\delta \tau_1  &= \frac{2{\omega _1}}{\gamma \beta} .
\end{align}
In this case it is not necessary to impose any condition for the angle $\varphi ^{{\rm{seed}}}$. It turns out that the angle $\varphi ^{{\rm{seed}}}$ is altered by an amount that it is independent of $\sigma^1$. Since this angle enter into the solution as an overall rotation via the matrix $U$, such an angle does not correspond to a variation of the shape of the string. On the contrary, periodicity in time requires appropriate condition for the angle $\varphi ^{{\rm{dressed}}}$. This turns out to be 
\begin{align}
\delta \varphi _0^{{\rm{dress}}} &=  - \sqrt { - {D^2}} \left( {\beta  - v_0^{{\rm{tb}}}} \right)2{\omega _1} = \frac{{{m^{{\rm{dress}}}}}}{{{n^{{\rm{dress}}}}}}2\pi , \\
\delta \varphi _1^{{\rm{dress}}} &= \sqrt { - {D^2}} \left( {\frac{1}{\beta } - \frac{1}{{v_1^{{\rm{tb}}}}}} \right)2{\omega _1} = \frac{{{m^{{\rm{dress}}}}}}{{{n^{{\rm{dress}}}}}}2\pi ,
\end{align}
which after some algebra can be written as
\begin{align}
\beta  &= \frac{{ - \frac{{2\pi }}{{\sqrt { - {D^2}} }}{m^{{\rm{dress}}}} + 2{\omega _1}v_0^{{\rm{tb}}}{n^{{\rm{dress}}}}}}{{2{\omega _1}{n^{{\rm{dress}}}}}} ,\\
\beta  &= \frac{{2{\omega _1}{n^{{\rm{dress}}}}}}{{\frac{{2\pi }}{{\sqrt { - {D^2}} }}{m^{{\rm{dress}}}} + \frac{{2{\omega _1}}}{{v_1^{{\rm{tb}}}}}{n^{{\rm{dress}}}}}} .
\end{align}
These equations are implying that the $\sigma^0$ axis coincides to a direction of the periodicity lattice of the sine-Gordon counterpart. Therefore, only in such a case, the string solutions of this class are periodic in time.

\subsection{Spike Dynamics}
\label{subsec:string_properties_interactions}

\begin{figure}[ht]
\vspace{10pt}
\begin{center}
\begin{picture}(90,40)
\put(2.5,0){\includegraphics[width = 0.4\textwidth]{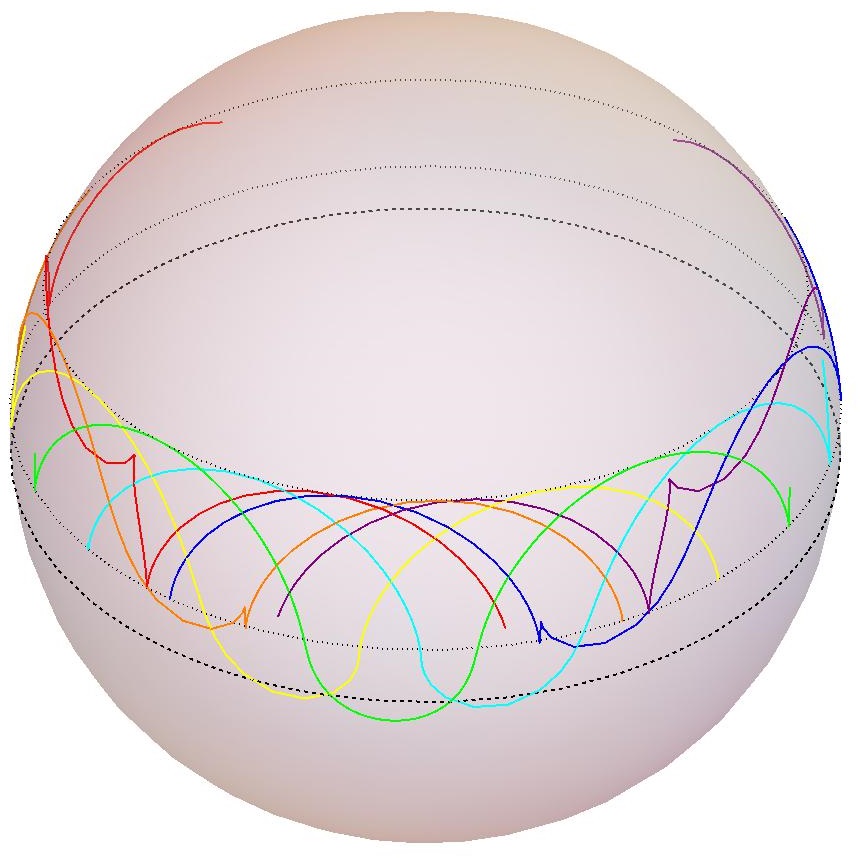}}
\put(47.5,0){\includegraphics[width = 0.4\textwidth]{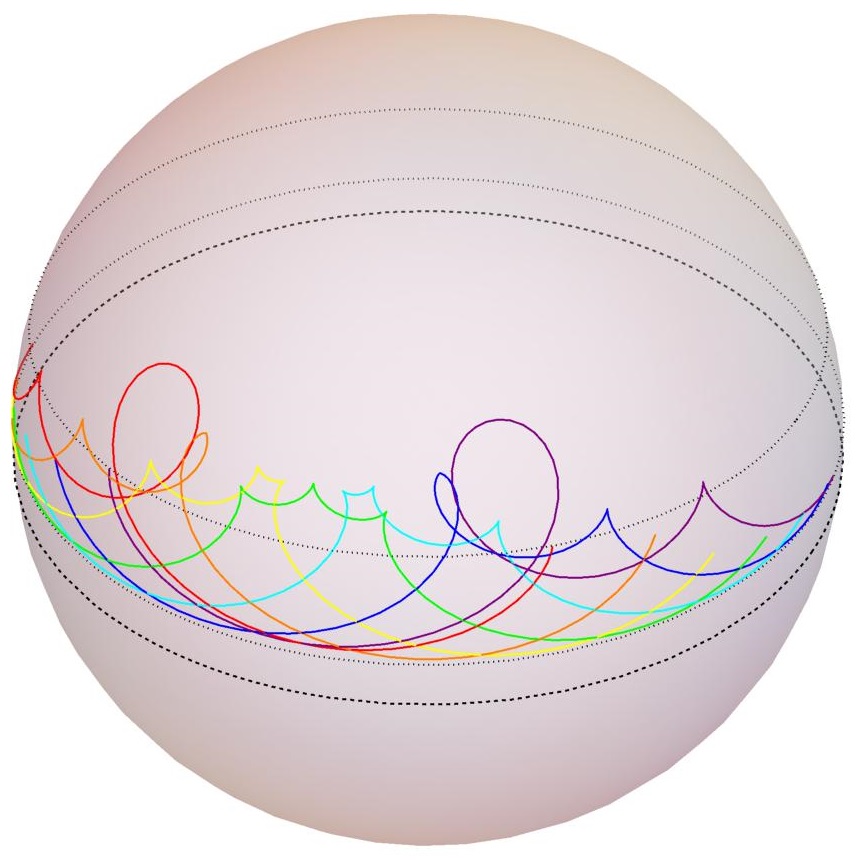}}
\end{picture}
\end{center}
\vspace{-10pt}
\caption{Two kinds of spike interactions. On the left a spike and anti-spike annihilate and regenerate at a different position. On the right a loop dissolves to two spikes. Then, one of those is recombined with another one to form again a loop. The time evolves from the red curve to the purple one.}
\vspace{5pt}
\label{fig:spike_interactions}
\end{figure}

We observe several forms of interaction between the spikes. Two spikes pointing to opposite directions may approach each other until a given time instant when they both disappear. After some time, they reappear at a different position. This is evident in figure \ref{fig:dressed_strings_static_time} top-left, middle-left and middle-right. It is also possible, as shown in the bottom-left part of figure \ref{fig:dressed_strings_static_time}, that a loop shrinks until a time instant when it disappears and two spikes pointing in the same direction appear. Then, the loop reappears in a different position, after the combination of a different pair of spikes. It has to be noted that although the kink induced by the dressing bypasses the kinks of a rotating background, it is possible that the corresponding spikes bypass each other without interacting, as shown in the bottom-right part of figure \ref{fig:dressed_strings_static_time}. A close-up of these kinds of interactions is depicted in figure \ref{fig:spike_interactions}. The time evolution of the string in this figure advances from red to purple. On the left two spikes approach each other and then disappear. It is clear that they cease to exist for a finite time and then, a pair of spikes appears in a symmetric fashion and starts diverging until one of those combines with another spike. On the right the situation is similar, but when the two spikes disappear a loop takes their place.

\begin{figure}[ht]
\vspace{10pt}
\begin{center}
\begin{picture}(90,28)
\put(2.5,0){\includegraphics[width = 0.4\textwidth]{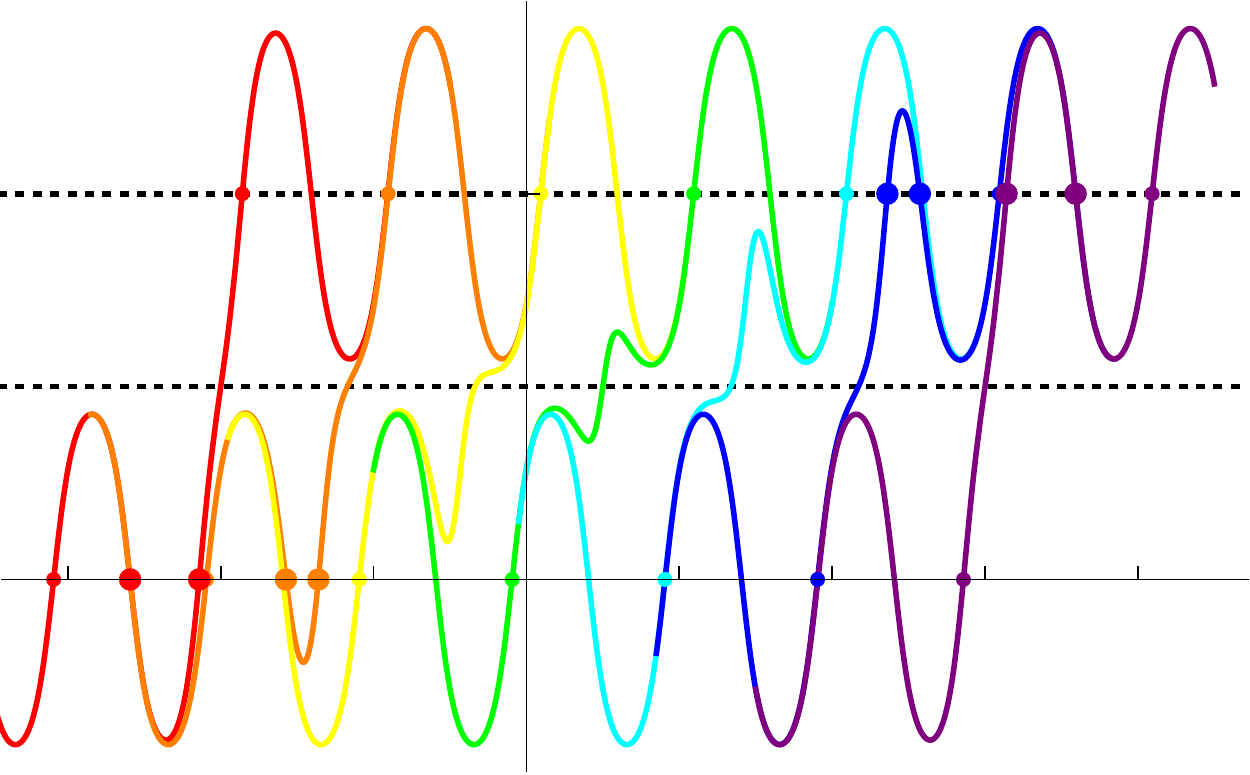}}
\put(47.5,0){\includegraphics[width = 0.4\textwidth]{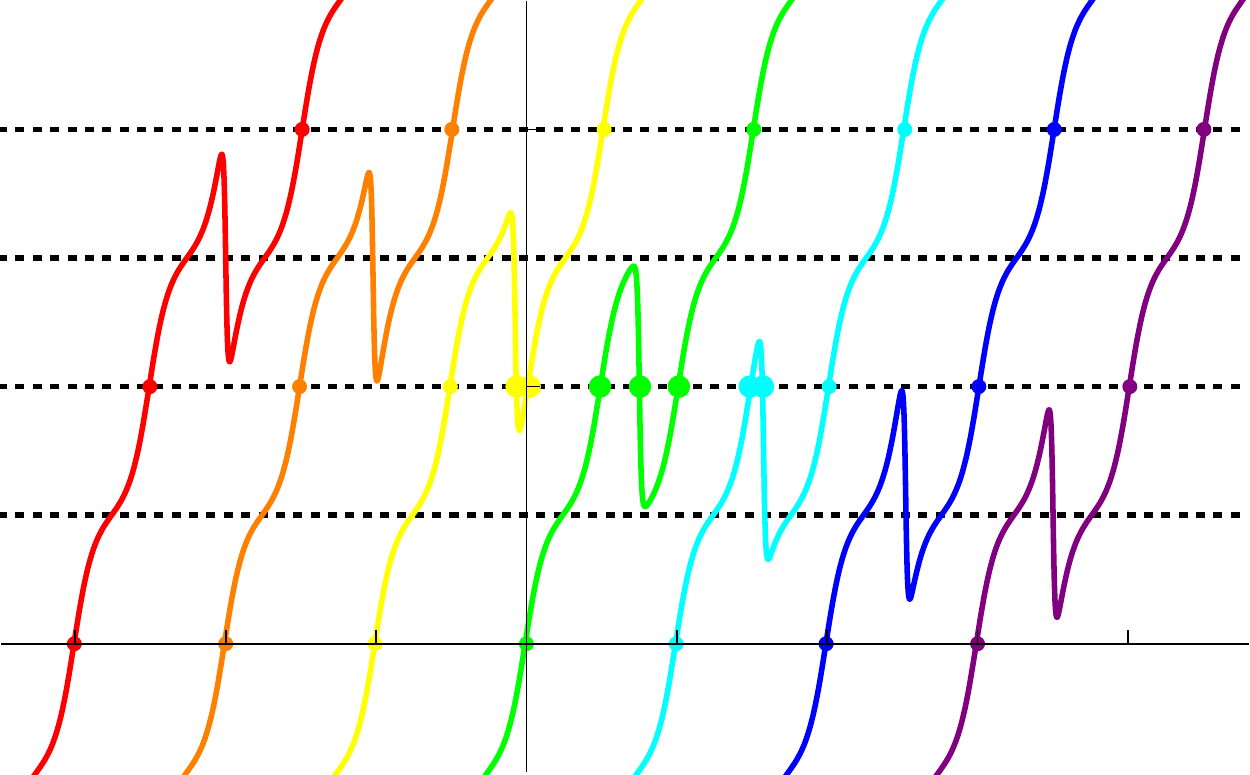}}
\put(1.25,6.25){$0$}
\put(1.25,12.5){$\pi$}
\put(0,19){$2\pi$}
\put(42.75,5.75){$\sigma^1$}
\put(18.5,25.5){$\varphi$}
\put(46.25,4.25){$0$}
\put(45,12.75){$2\pi$}
\put(45,21){$4\pi$}
\put(46.25,8.5){$\pi$}
\put(45,16.75){$3\pi$}
\put(87.75,3.75){$\sigma^1$}
\put(63.5,25.5){$\varphi$}
\end{picture}
\end{center}
\vspace{-10pt}
\caption{The time evolution of a kink propagating on a translationally invariant oscillatory background and an anti-kink propagating on a rotating background being a train of kinks. The dots are positions of spikes. The thick dots are the "interacting" spikes that disappear and reappear in either of the interactions depicted in figure \ref{fig:spike_interactions}.}
\vspace{5pt}
\label{fig:SG_spike_interactions}
\end{figure}
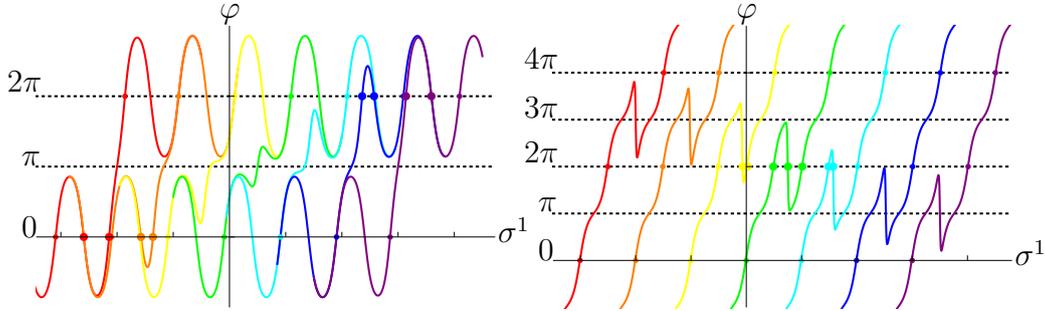

The above processes are quite simple to understand in the language of the sine-Gordon equation. As noted in \cite{part1}, a spike may appear only at positions where the Pohlmeyer field $\varphi$ assumes a value that is an integer multiple of $2 \pi$, as in these positions the derivative $\frac{\partial X}{\partial \sigma^1}$ vanishes. Actually, unless a very special coincidence happens (the second derivative also vanishes), at such points the derivative $\frac{\partial X}{\partial \sigma^1}$ gets inverted, and, thus, these points are positions of spikes. In figure \ref{fig:SG_spike_interactions}, the time evolution of the sine-Gordon counterparts of the dressed elliptic strings is depicted. On the left, the solution is a kink propagating on a translationally invariant oscillating elliptic seed, whereas on the right it is an anti-kink propagating on a train of kinks, i.e. a rotating elliptic seed. As analysed in section \ref{sec:SG_properties}, the shape of the kink alters periodically as it advances in the elliptic background. As the shape changes, it is possible that the solution ceases to cross a $\varphi = 2 n \pi$ horizontal line, or on the opposite may start crossing such a line. Continuity ensures that whenever this happens two points where the solution crosses a $\varphi = 2 n \pi$ line appear or disappear. As these points correspond to spikes, it naturally implies that spikes may interact in pairs that disappear or appear from nothing.
The left part of figure \ref{fig:SG_spike_interactions} depicts the kind of interaction occurring in the top left panel of figure \ref{fig:dressed_strings_static_time}, whereas the right part depicts the kind of interaction happening in the middle row and the bottom right panel of figure \ref{fig:dressed_strings_static_time}. Had one considered the case of a kink propagating on a train of kinks, the situation would be rather different. Such a solution is always monotonous (see figure \ref{fig:phi_jump}), and thus it is not possible that such phenomena occur. Therefore, although the extra spike corresponding to the kink will overpass all other spikes, as the kink advances in the elliptic background, it is not possible that it gets in touch and interacts with any of these spikes. This is the case of the bottom right panel of figure \ref{fig:dressed_strings_static_time}.

The same kinds of spike interactions occur in the time evolution of the other classes of closed strings that we developed in section \ref{sec:string_asymptotics}.

\subsection{A Conservation Law Preserved by Spike Interactions}

In the case of the dressed string solutions with approximate periodicity conditions plotted in figure \ref{fig:dressed_strings_approx}, the space-like worldsheet coordinate $\sigma^1$ runs in a finite interval. Such solutions are characterized by a topological number $N$, being proportional to the difference in the value of the Pohlmeyer field at the endpoints of this interval, obviously being a multiple of $2\pi$,
\begin{equation}
2 \pi N = \int\limits_{{\rm{string}}} {d\sigma {\partial _\sigma }\varphi } , \quad N \in \mathbb{Z} .
\end{equation}
This number is conserved as the string moves due to the continuity of the time evolution of the Pohlmeyer counterpart of the solution.

In the case of the elliptic strings this has been identified to the number of spikes \cite{part1}. However, in this case the spikes never interact with each other, as the time evolution of the elliptic strings is simply a rigid rotation. In the case of dressed elliptic strings, we have seen that spikes may interact in a way that their number is not conserved. Thus, the identification of the topological number in the sine-Gordon equation as the number of spikes cannot be extended beyond the case of the elliptic strings.

The form of these spike interactions guide us to search for a conserved quantity, which receives $\pm 1$ contributions from each spike and $\pm 2$ contributions from each loop. Let us consider the turning number of the closed string. This is a difficult task since the string has singular points (the spikes), where the tangent vector is not well defined. However, it is true that the string contains only this kind of non-smooth points, i.e. points where the tangent gets inverted. Other non-smooth points where the tangent is rotated by an arbitrary angle are not allowed. Therefore, the unoriented tangent to the string is continuous, and, thus, an unoriented turning number can be defined. This is the fundamental group of the mappings from $\mathrm{S}^1$ to the one-dimensional real projective space $\mathbb{R}P^1$. Notice that possible self intersections of the string should not be treated as the same point, where the tangent would not be well-defined, but as separate points. This way the desired turning number is naturally a member of $\pi_1 \left( \mathbb{R}P^1 \right) = \mathbb{Z}$ and must be conserved.

\begin{figure}[ht]
\vspace{10pt}
\begin{center}
\begin{picture}(100,35)
\put(2.5,5){\includegraphics[width = 0.45\textwidth]{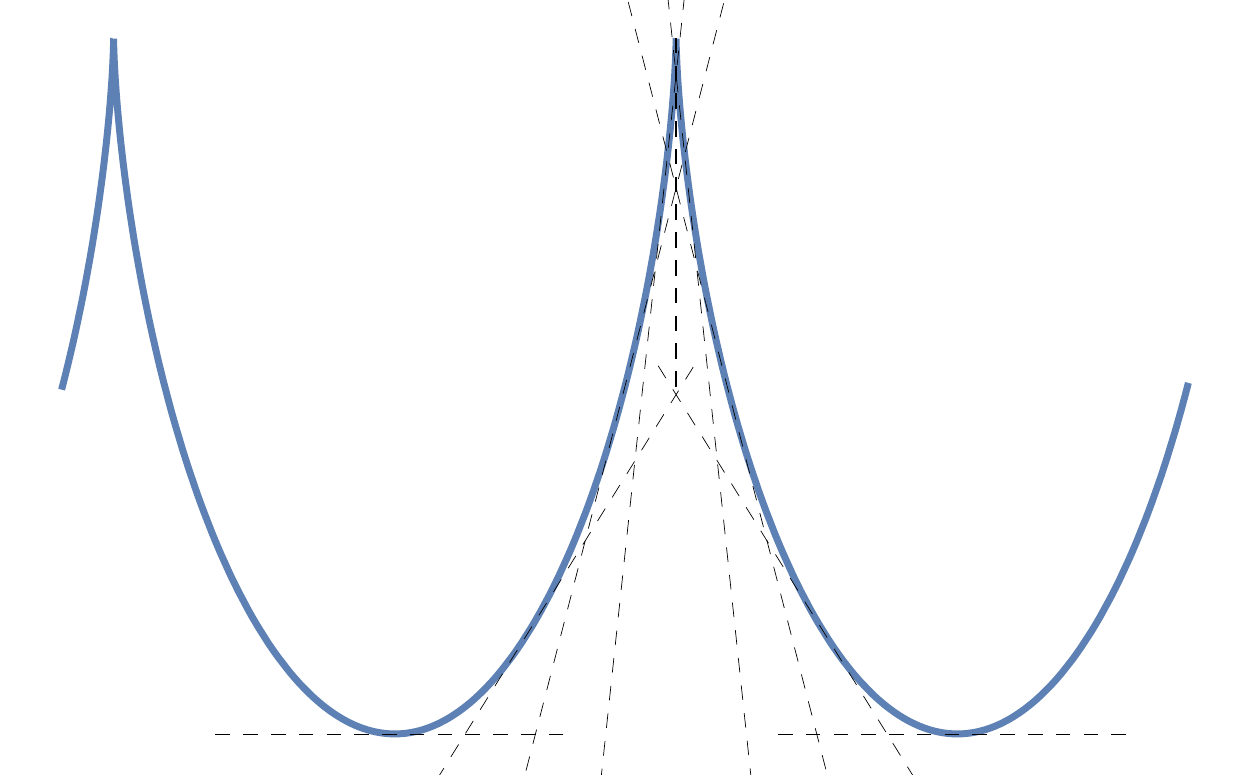}}
\put(52.5,5){\includegraphics[width = 0.45\textwidth]{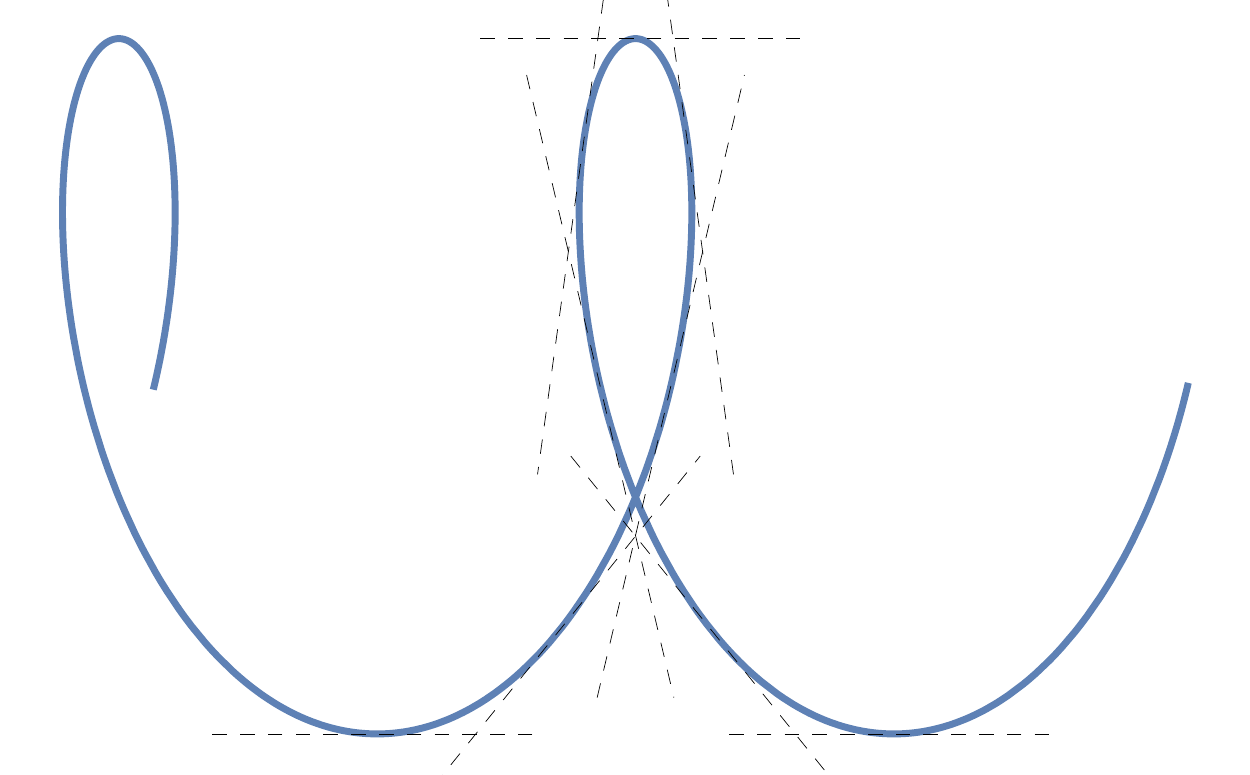}}
\end{picture}
\end{center}
\vspace{-10pt}
\caption{The turning number contributions from a spike (left) and a loop (right)}
\vspace{5pt}
\label{fig:turning_number}
\end{figure}

Figure \ref{fig:turning_number} shows that the existence of a single spike between two points of the string with the same unoriented tangent contributes a $\pm 1$ to this turning number. Similarly, the existence of a loop contributes $\pm 2$. 
This explains the two kinds of interactions we found in section \ref{subsec:string_properties_interactions}. Whenever two spikes with opposite contributions to the unoriented turning number get combined, they just disappear. When two spikes with identical contributions to the turning number get combined they disappear and necessarily the conservation of the turning number implies that a loop must take their place. The above imply that the unoriented turning number and the topological charge of the sine-Gordon equation are in correspondence. They do not have to be equal, but they may differ by an ever integer.

The above are also in line with the effect of the dressing on the shape of the string that we observe in figure \ref{fig:dressed_strings_approx}. In all cases, the action of the dressing procedure on the Pohlmeyer field adds a kink or an antikink to the seed solution, which according to the above should increase or decrease the aforementioned turning number by one. The simplest case is that of a seed solution with a static oscillating counter part (figure \ref{fig:dressed_strings_approx} top-right). In this case the seed solution has no spikes, while the dressed solution has exactly one. In a similar manner, when a seed solution with a translationally invariant oscillating counterpart is considered (figure \ref{fig:dressed_strings_approx} top-left), the seed solution has equal number of spikes that contribute $+1$ and spikes that contribute $-1$ to the turning number, having net turning number $0$, whereas the dressed string has net turning number equal to $1$. In the case of seeds with rotating elliptic counterparts the behaviour is also similar.

\setcounter{equation}{0}
\section{Instabilities of the Elliptic Strings}
\label{sec:instabilities}

When one desires to study the stability of a classical string solution, they usually study the stability of its Pohlmeyer counterpart, as the equations of motion of the reduced system are simpler to study since they contain fewer degrees of freedom and they do not possess any reparametrization symmetry. More specifically, the stability of the elliptic solutions of the sine-Gordon equation has been studied in \cite{SGstability}. It turns out that only the static rotating elliptic solutions of the sine-Gordon equation are stable. Therefore, only one of the four classes of elliptic string solutions on the sphere $\mathrm{S}^2$ is stable.

However, we should be a little concerned about the above result. The stability analysis is performed introducing an arbitrary infinitesimal perturbation to the elliptic solutions of the sine-Gordon equation. However, when a closed elliptic string is considered, appropriate periodicity conditions must be applied, and, thus, only perturbations preserving these conditions should be considered in the stability analysis.

In the following, we will follow a different approach to discover instabilities of the elliptic string solutions. Instead of performing an infinitesimal perturbation to the string solution, we will try to find explicit solutions that tend asymptotically in time to an elliptic string solution, but in general they are not a small perturbation around the latter. Such solutions are the analog, for example, in the case of the simple pendulum, to the trajectories connecting asymptotically two consecutive unstable vacua. The existence of such a solution reveals that the elliptic solution, which is the asymptotic limit of the latter, is unstable.

This class of solutions that reveals instabilities of the elliptic strings may contain solutions with various genuses. However, the simplest case to consider is a degenerate genus two solution, where only one of the two genuses is degenerate. The solution should have a non-degenarate genus, associated with the initial elliptic solution, and furthermore it should have a degenerate one describing the infinite motion that tends asymptotically to the elliptic solution at plus and/or minus infinite time. This is exactly the class of dressed elliptic string solutions.

It turns out that the relevant dressed elliptic solutions are the special finite exact solutions with $D^2>0$ presented in section \ref{subsec:string_finite_exact_special}. These solutions have counterparts with $D^2>0$ being a kink propagating on an elliptic background. Therefore, the sine-Gordon counterparts of these solutions have a specific asymptotic behaviour, namely, far away from the region of the kink they tend to a shifted version of the seed, and similarly the string tends to a rotated version of the seed string solution. In this specific class of solutions, the $\sigma^1$ direction is parallel to the direction that the kink moves in spacetime, thus the asymptotic behaviour of the string is never reached at a snapshot of the string, but it is rather reached asymptotically in time. It follows that these specific string solutions evolve from a rotated version of the seed elliptic spiky string solution to another one, rotated by the opposite angle. Notice that these asymptotic string solutions obey appropriate periodicity conditions and thus they are finite. 

The existence of these solutions indicates that their seed elliptic solutions are unstable. They describe a finite disturbance of a spiky string emerging after an infinitesimal perturbation at minus infinity time.

\begin{figure}[ht]
\vspace{10pt}
\begin{center}
\begin{picture}(100,72.5)
\put(13.75,37.25){\includegraphics[width = 0.35\textwidth]{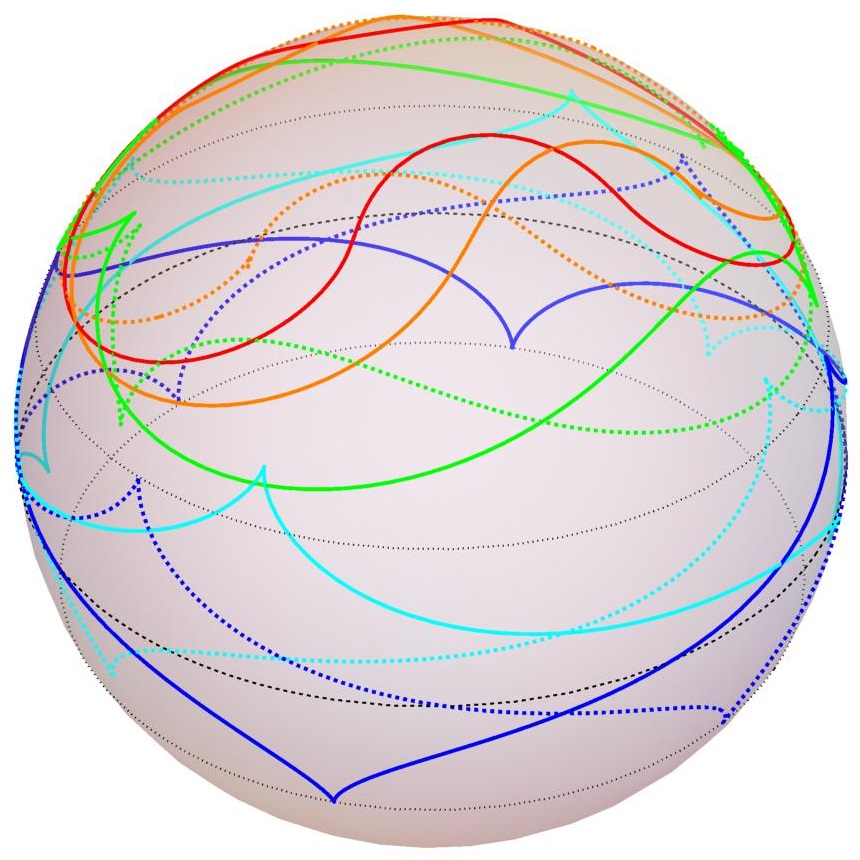}}
\put(51.25,37.25){\includegraphics[width = 0.35\textwidth]{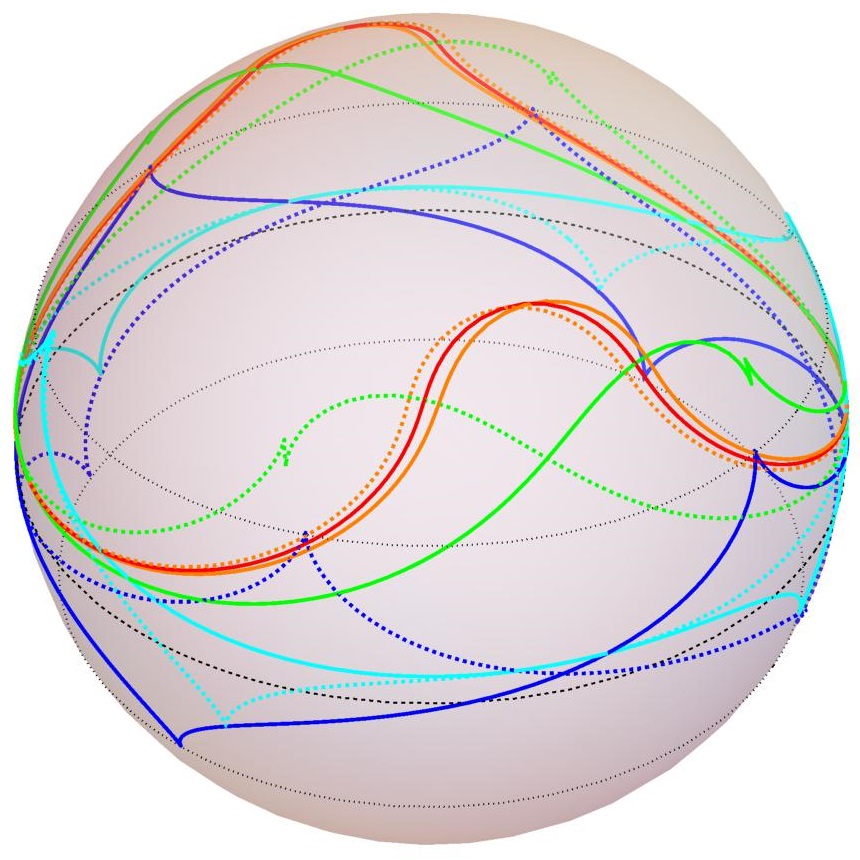}}
\put(13.75,0){\includegraphics[width = 0.35\textwidth]{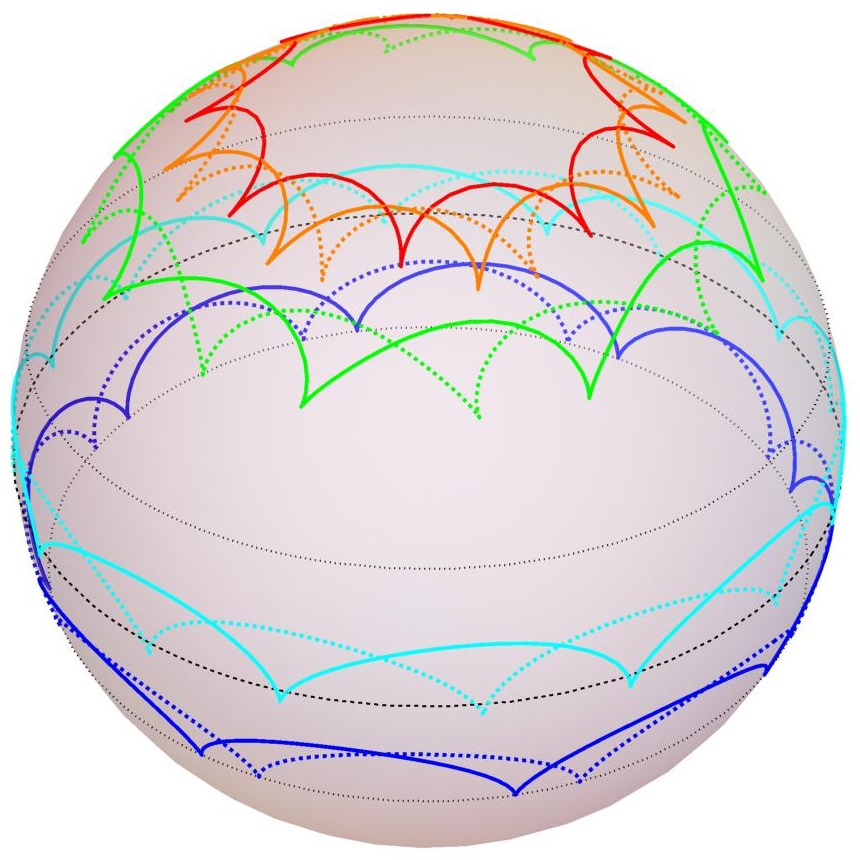}}
\put(51.25,0){\includegraphics[width = 0.35\textwidth]{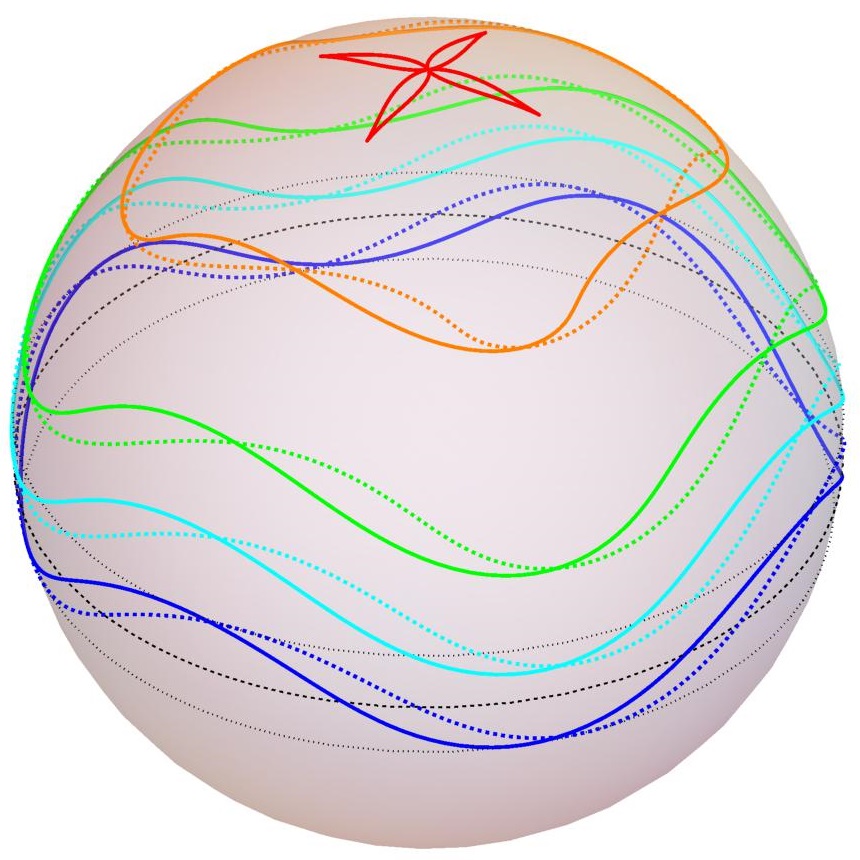}}
\put(0,0){\includegraphics[height = 0.19\textwidth]{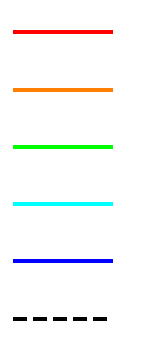}}
\put(2,20){$\sigma^0 / (\frac{\gamma}{\beta D})$}
\put(6.75,16.5){0}
\put(6.75,13.5){1}
\put(6.75,10.5){3}
\put(6.75,7.5){10}
\put(6.75,4.5){50}
\put(6.75,1.5){$<0$}
\put(89,0){\includegraphics[height = 0.19\textwidth]{instability_legend.pdf}}
\put(91,20){$\sigma^0 / (\frac{\gamma}{D})$}
\put(95.75,16.5){0}
\put(95.75,13.5){1}
\put(95.75,10.5){2}
\put(95.75,7.5){3}
\put(95.75,4.5){10}
\put(95.75,1.5){$<0$}
\put(0,50){\includegraphics[height = 0.19\textwidth]{instability_legend.pdf}}
\put(2,70){$\sigma^0 / (\frac{\gamma}{\beta D})$}
\put(6.75,66.5){0}
\put(6.75,63.5){1}
\put(6.75,60.5){3}
\put(6.75,57.5){10}
\put(6.75,54.5){50}
\put(6.75,51.5){$<0$}
\put(89,50){\includegraphics[height = 0.19\textwidth]{instability_legend.pdf}}
\put(91,70){$\sigma^0 / (\frac{\gamma}{\beta D})$}
\put(95.75,66.5){0}
\put(95.75,63.5){1}
\put(95.75,60.5){10}
\put(95.75,57.5){50}
\put(95.75,54.5){200}
\put(95.75,51.5){$<0$}
\end{picture}
\end{center}
\vspace{-10pt}
\caption{The dressed elliptic string solutions that reveal instabilities of their seed elliptic strings. The dashed lines correspond to times opposite to the continuous ones with the same color. On the top row, the two solutions related to an elliptic string, with a translationally invariant oscillatory counterpart with $E=9\mu^2 /10$ and $n=4$, are depicted. The bottom left panel shows the solution related to an elliptic string with a translationally invariant rotating counterpart with $E=3\mu^2 /2$ and $n=8$. Finally, the bottom right panel depicts the solution related to an elliptic string with a static oscillatory counterpart with $E=-\mu^2 /2$ and $n=8$.}
\vspace{5pt}
\label{fig:strings_instabilities}
\end{figure}

The special solutions of this kind emerge only when the kink propagating on top of an elliptic background in the sine-Gordon counterpart of the solution is superluminal, as shown in section \ref{subsec:string_finite_exact_special}. Therefore, following section \ref{subsec:SG_kink_velocity}, only elliptic strings with a translationally invariant sine-Gordon counterpart that is rotating, or oscillating with $E>E_c$, and elliptic strings with a static oscillating sine-Gordon counterpart may expose this kind of instability. Interestingly, as shown in figure \ref{fig:mean_velocity}, the strings with an oscillating translationally invariant counterpart with $E>E_c$ give rise to two distinct dressed string solutions exposing their instability, whereas all other classes give rise to only one. Figure \ref{fig:strings_instabilities} shows the time evolution of the special finite dressed elliptic strings with $D^2>0$ that expose the instability of the seed solution.
The rigid body rotation of the asymptotic elliptic string has been frozen in the figure so that the time evolution is clearly depicted. In all cases the string finally resettles to the same unstable elliptic string configuration but with a delay proportional to $2\left| \tilde{a} \right|$ in comparison to the state it would lie had it followed the simple rigid rotation evolution of the elliptic string.

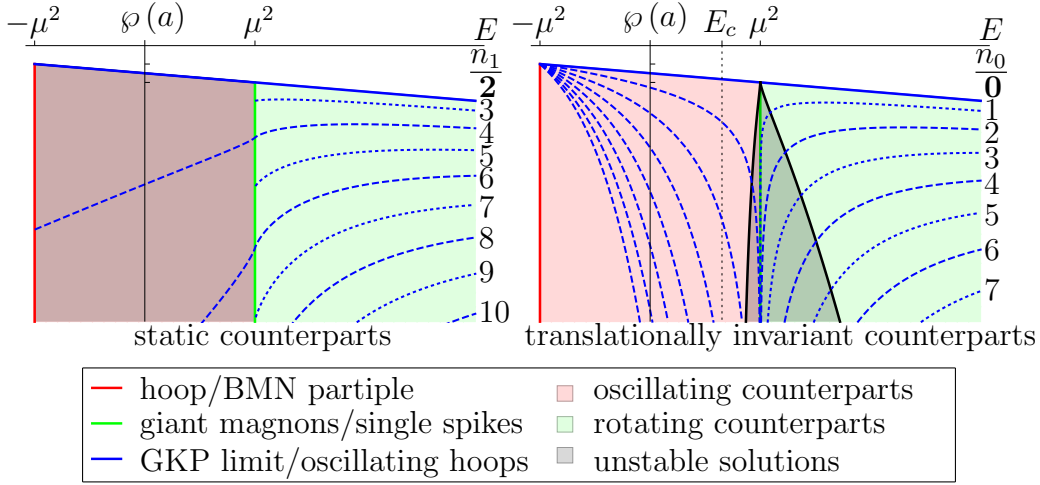
\begin{figure}[ht]
\vspace{10pt}
\begin{center}
\begin{picture}(88,42.5)
\put(1,14){\includegraphics[width = 0.4\textwidth]{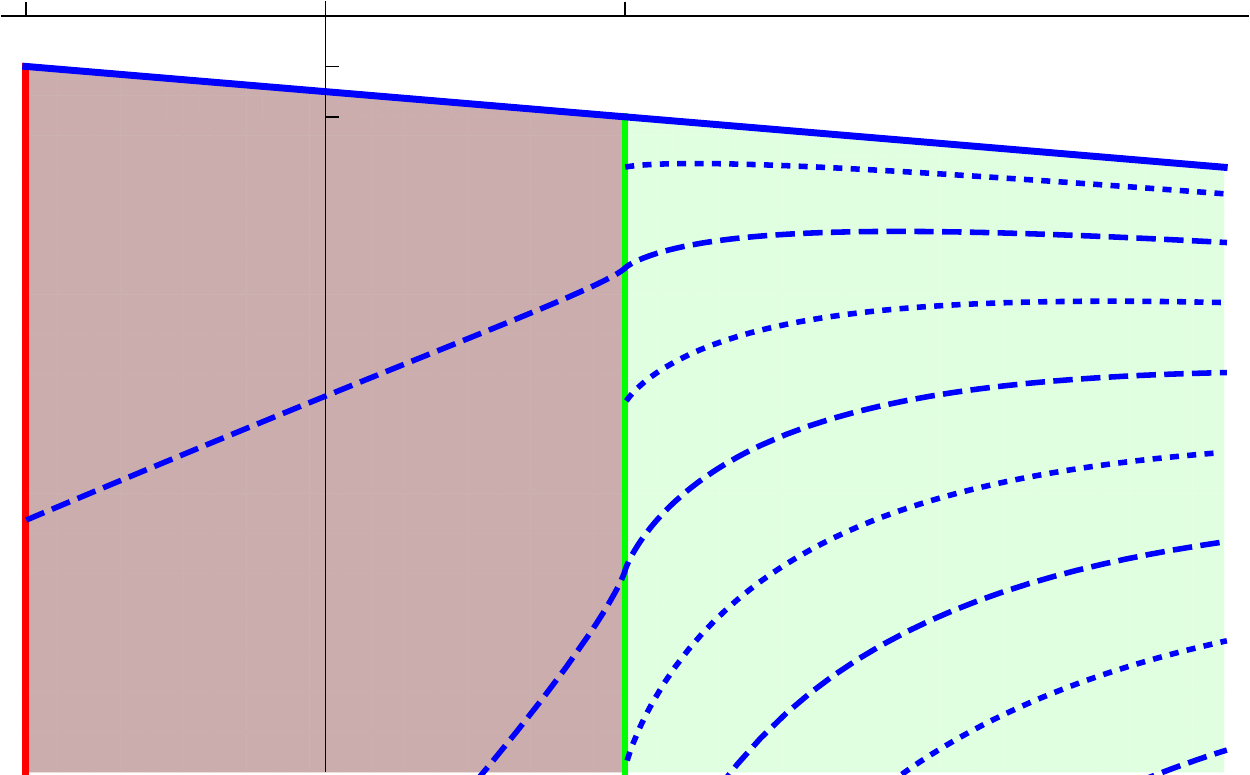}}
\put(45,14){\includegraphics[width = 0.4\textwidth]{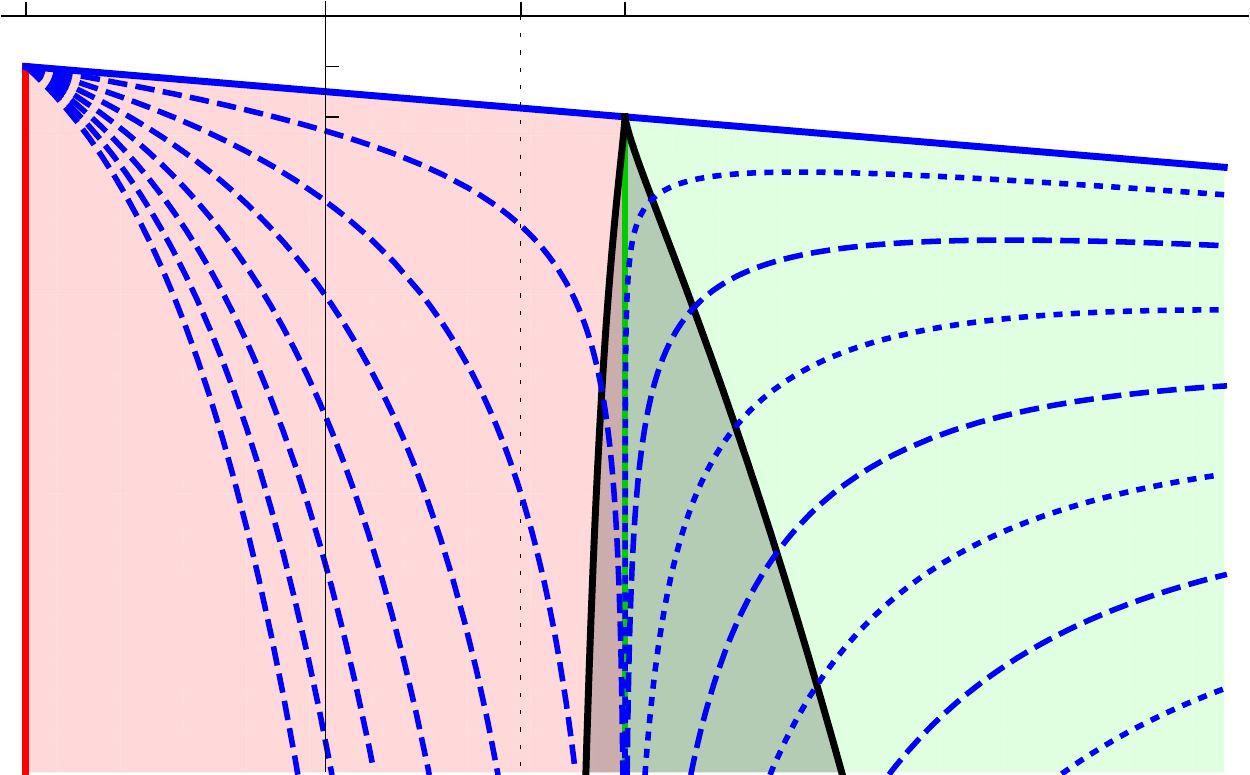}}
\put(10.5,12){static counterparts}
\put(44.5,12){translationally invariant counterparts}
\put(40,38.75){$E$}
\put(-0.75,39.5){$-\mu^2$}
\put(20,39.5){$\mu^2$}
\put(9.25,40){$\wp \left( a \right)$}
\put(39.75,36.75){$\underline{n_1}$}
\put(40.5,33.75){\bf{2}}
\put(40.5,31.75){$3$}
\put(40.5,29.75){$4$}
\put(40.5,27.75){$5$}
\put(40.5,25.75){$6$}
\put(40.5,23.25){$7$}
\put(40.5,20.5){$8$}
\put(40.5,17.5){$9$}
\put(40.5,14){$10$}
\put(84,38.75){$E$}
\put(43.25,39.5){$-\mu^2$}
\put(60,39.5){$E_c$}
\put(64,39.5){$\mu^2$}
\put(53,40){$\wp \left( a \right)$}
\put(83.75,36.75){$\underline{n_0}$}
\put(84.5,33.75){\bf{0}}
\put(84.5,31.75){$1$}
\put(84.5,29.75){$2$}
\put(84.5,27.5){$3$}
\put(84.5,25.25){$4$}
\put(84.5,22.5){$5$}
\put(84.5,19.5){$6$}
\put(84.5,16){$7$}
\put(6,0){\includegraphics[height = 0.1\textwidth]{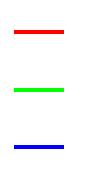}}
\put(11,1){GKP limit/oscillating hoops}
\put(11,4.25){giant magnons/single spikes}
\put(11,7.25){hoop/BMN partiple}
\put(46.5,0){\includegraphics[height = 0.095\textwidth]{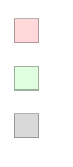}}
\put(50.5,1){unstable solutions}
\put(50.5,4.25){rotating counterparts}
\put(50.5,7.25){oscillating counterparts}
\put(6,0){\line(0,1){10}}
\put(6,0){\line(1,0){76}}
\put(82,0){\line(0,1){10}}
\put(6,10){\line(1,0){76}}
\end{picture}
\end{center}
\vspace{-10pt}
\caption{The set of unstable elliptic string solutions in the moduli space}
\vspace{5pt}
\label{fig:elliptic_string_moduli_instability}
\end{figure}

This observation is in line with the findings of \cite{SGstability}, which support that in general string solutions with sine-Gordon counterparts that can accommodate superluminal kinks are unstable. However, in our case there is a particular difference. The solutions exposing the string instability emerge only when there is a superluminal kink with velocity equal to the inverse of the velocity of the boost connecting the linear and static gauges. This is due to the fact that only such solutions do not disturb the periodicity conditions of the closed seed string solution. Recalling figure \ref{fig:mean_velocity}, the above implies that the elliptic strings with oscillating static counterparts always expose this kind of instability, since the kink velocity diverges at the limit $\tilde{a} \to \omega_1$, and, thus, any possible superluminal kink velocity can be obtained for some value of $\tilde{a}$. On the contrary, for elliptic strings with translationally invariant counterparts, even in the case they can accommodate superluminal kinks, there is a maximum velocity of the latter. This means that, depending on the elliptic string moduli $E$ and $a$, which determine the velocity of the boost connecting the static and linear gauges, this kind of instability may or may not exist. More specifically, given a value of $E$, there is a minimum value of $\wp \left( a \right)$, or in other words, there is a minimum number of spikes required for the existence of the instability. This in turn implies that the ''speeding strings'' limit of the elliptic strings always exposes this kind of instability (when they have translationally invariant counterparts). Figure \ref{fig:elliptic_string_moduli_instability} shows the subset of elliptic strings that present this kind of instabilities within the moduli space of elliptic string solutions as parametrized by the quantities $E$ and $\wp \left( a \right)$.
In the right panel, the thick black line enclosing the unstable elliptic string solutions with oscillating translationally invariant counterparts tends asymptotically to the $E = E_c$ vertical line, where the constant $E_c$ is defined in equation \eqref{eq:kink_E_criterion}.

Of course the above argument is not a proof of the existence of elliptic string solutions, with sine-Gordon counterparts that accommodate superluminal kinks, that are stable; it is possible that more complicated multi-kink generalizations of the above solutions conserve the periodicity conditions and thus give rise to the instability. These should have only one non-degenerate genus, thus, they could emerge from the dressing of the elliptic strings with more complicated dressing factors. The latter can be constructed from the solution of the auxiliary system presented in \cite{Katsinis:2018ewd} in a straightforward manner. Such solutions should not correspond to multiple kinks travelling on top of an elliptic background, as they would have different velocities and thus, their asymptotic behaviour could not be only temporal. They would rather correspond to a single breather propagating on top of an elliptic background. Nevertheless, the stability issue of the spiky strings requires further investigation concerning the constraints originating from the periodicity conditions.

A simple case to consider in particular is the stability of the GKP strings \cite{GKP_string}. These are the elliptic strings with static Pohlmeyer counterparts and modulus $a = \omega_2$ implying that $\beta = 0$, i.e. the linear gauge coincides with the static gauge. It follows that a dressed elliptic solution exposing an instability of a GKP string should have a Pohlmeyer counterpart being a superluminal kink on top of an elliptic background with infinite velocity, in other words a translationally invariant kink. As we have shown in section \ref{subsec:SG_kink_velocity}, the kink velocity on static backgrounds is diverging only in the case of an oscillating seed at the limit $\tilde{a} = \omega_1$. Therefore, the GKP strings with an oscillating Pohlmeyer counterpart are unstable. This is expected since the latter are great circles rotating around the sphere with subluminal velocities and they tend to shrink due to the string tension.

\setcounter{equation}{0}
\section{Energy and Angular Momentum}
\label{sec:dispersion}

\subsection{Approximate Finite and Exact Infinite Strings with $D^2>0$}
The dressed string solutions have a conserved energy and angular momentum as a direct result of the time translation and the rotation symmetries of the NLSM action. The energy is simple to calculate, since
\begin{equation}
E_{0/1} = \left| \frac{\delta L}{\delta \partial_0t} \right| =  T \int\limits_{\textrm{string}} {\frac{{\partial t_{0/1}}}{{\partial \sigma^0 }}d\sigma^1 } =  T \mu \int\limits_{\textrm{string}} {d\sigma^1 } .
\label{eq:charges_energy_general}
\end{equation}
The only non-trivial quantity to be specified is the range of the space-like parameter $\sigma^1$ that covers the whole closed string. In section \ref{sec:string_asymptotics}, apart from the special solutions related to the instabilities of the elliptic strings, we specified two classes of closed dressed elliptic strings with $D^2>0$: those that have finite length and satisfy approximate periodicity conditions and those that are infinite and satisfy exact periodicity conditions. Obviously, the energy of the latter is infinite. The former are covered by $n_2$ patches, of the form
\begin{align}
\sigma ^1 &\in \left[ \bar{\sigma} - \frac{n_1 \omega_1 + s_\Phi \tilde a}{\gamma \beta} , \bar{\sigma} + \frac{n_1 \omega_1 + s_\Phi \tilde a}{\gamma \beta} \right) , \label{eq:span_ti_disp}\\
\sigma ^1 &\in \left[ \bar{\sigma} - \frac{n_1 \omega_1 - s_\Phi \tilde a}{\gamma} , \bar{\sigma} + \frac{n_1 \omega_1 - s_\Phi \tilde a}{\gamma} \right) , \label{eq:span_st_disp}
\end{align}
where $\bar{\sigma}$ is the position of the kink that is induced by the dressing, at any given time\footnote{In section \ref{subsec:string_finite_approximate} as $\bar{\sigma}$ we used the average position of the kink (see equations \eqref{eq:span_ti} and \eqref{eq:span_st}). Equivalently, one could consider the exact position of the kink, i.e. the $\bar{\sigma}$ that obeys $\tilde{\Phi} \left( \sigma^0 , \bar{\sigma} \right) = 0$. Either selection results in the same values for the energy and the angular momentum of the dressed strings.}. Defining as $E_{0/1}^{\textrm{hop}}$ the energy of one hop of the seed elliptic string, it follows that the energy of these strings is equal to
\begin{equation}
E_{0/1} = \frac{{2 T n_2 R {\mu ^2} \left( {n_1 \omega _1 \pm s_\Phi \tilde{a}} \right) }}{\sqrt {{x_{3/2}} - \wp \left( a \right)}} = E_{0/1}^{\textrm{hop}} \left[ n_2 \left( n_1 \pm s_\Phi \frac{\tilde{a}}{\omega_1} \right) \right] .
\label{eq:charges_energy}
\end{equation}

In a similar manner, the angular momentum can in principle be calculated as
\begin{equation}
J = \frac{\delta L}{\delta \partial_0\varphi} = T R^2 \int\limits_{\textrm{string}} {{{\sin }^2}\theta_{0/1} \frac{{\partial \varphi_{0/1} }}{{\partial \sigma^0 }}d\sigma^1 } .
\end{equation}
This requires much more complicated algebra than the calculation of the energy. However, this algebra may be bypassed, since the angular momentum is directly proportional to the sigma model charge. The sigma model charge of the dressed solution differs to that of the seed by a finite amount as described by formula \eqref{eq:dressing_review_change_of_charge}. Therefore, we can calculate easily the angular momentum of the dressed solution given the angular momentum of a segment of the seed solution that corresponds to the range of $\sigma^1$ that covers the closed dressed solution. This is an easy task in the parametrization in terms of the Weierstrass elliptic function, as explained in \cite{part1}.

We will focus on the calculation of the third component of the angular momentum of the string, which presents a certain interest for holographic applications. Before that, let us argue on the reasons we expect the other two components to vanish, when we consider finite closed dressed strings. In the case of elliptic ``naked'' solutions obeying appropriate periodicity conditions, $J_1$ and $J_2$ vanish as a result of the discrete symmetry that these solutions possess. This is also the case when one considers infinite dressed elliptic strings that obey exact periodicity conditions (see figure \ref{fig:exact_periodicity}). However, naively this is not the case when we consider the approximate closed finite dressed solutions with $n_2 = 1$, as the extra spike induced by the dressing breaks this symmetry. Although this symmetry is not present at a given time instant, one should not forget that the dressed strings change shape periodically, while they are simultaneously rotating. Therefore, after a time equal to the period of the string shape, it is expected that the $J_1$ and $J_2$ components will have rotated by an arbitrary angle. As the angular momentum is conserved, the latter implies that $J_1$ and $J_2$ vanish. In the following $J$ denotes the third component of the angular momentum and the indices $0$ and $1$ refer to whether the seed has a translationally invariant or static Pohlmeyer counterpart.

The third component of the angular momentum of the seed solution is given in \cite{part1}. For simplicity, we consider the case of seed solutions with static counterparts. The case of seeds with translationally invariant counterparts can be treated similarly.
\begin{equation}
J_{\text{seed}}=\frac{n_2 \gamma}{\ell}\int_{\bar{\sigma}-\frac{n_1 \omega_1 - s_\Phi \tilde{a}}{\gamma}}^{\bar{\sigma} + \frac{n_1 \omega_1 - s_\Phi \tilde{a}}{\gamma}}\left( \wp\left(\gamma\left(\sigma^1-\beta\sigma^0\right)+\omega_2\right) - x_3 \right) d \sigma^1 .
\end{equation}
Simple algebra yields
\begin{equation}
J_{\text{seed}}= - \frac{n_2}{\ell}\left\{\zeta\left(\sigma_+\right)-\zeta\left(\sigma_-\right)+ 2x_3\left(n_1\omega_1 - s_\Phi \tilde{a}\right)\right\},
\end{equation}
where
\begin{equation}
\sigma_\pm=\gamma\left(\bar{\sigma}-\beta \sigma^0\right)\pm\left(n_1 \omega_1- s_\Phi \tilde{a}\right)+\omega_2.
\label{eq:dispersion_spm}
\end{equation}

The difference between the NLSM charge of the dressed and seed solutions is given by equation \eqref{eq:dressing_review_change_of_charge}. It follows that the difference of the third component of the angular momentum is given by
\begin{equation}
\Delta J=-\frac{1}{2} \Delta \mathcal{Q}_L^{12}.
\end{equation}
It is a matter of algebra to show that the change of the angular momentum induced by the dressing is given by the following expression
\begin{equation}
\Delta J = \frac{n_2}{\ell}\left[-2n_1\zeta\left(\omega_1\right)+\zeta\left(\sigma_+\right)-\zeta\left(\sigma_-\right)+2 s_\Phi \left( \zeta\left( \tilde{a} \right) - D \cos\theta_1 \right) \right] .
\label{eq:dispersion_J_var}
\end{equation}
Thus, the third component of the angular momentum of the dressed solution $J_{\text{dressed}}$ is equal to
\begin{equation}
J_1 = - 2 \frac{n_2}{\ell} \left[ n_1 \left( \zeta\left( \omega_1 \right) + x_3 \omega_1 \right) - s_\Phi \left( \zeta \left( \tilde{a} \right) + x_3 \tilde{a} - D \cos \theta_1 \right) \right].
\label{eq:dispersion_J1_final}
\end{equation}

In a similar manner in the case of translationally invariant seeds we find
\begin{equation}
J_0 = 2 \frac{n_2}{\ell} \left[ n_1 \left( \zeta\left( \omega_1 \right) + x_2 \omega_1 \right) + s_\Phi \left( \zeta \left( \tilde{a} \right) + x_2 \tilde{a} - D \cos \theta_1 \right) \right].
\end{equation}
Defining as $J_{0/1}^{{\rm{hop}}}$ the angular momentum of one hop of the seed solution, the above expressions can be written as
\begin{equation}
{J_{0/1}} = {n_2} J_{0/1}^{{\rm{hop}}} {\left( {{n_1} \pm {s_\Phi }\frac{{\zeta \left( {\tilde a} \right) + {x_{2/3}}\tilde a - D\cos {\theta _1}}}{{\zeta \left( {{\omega _1}} \right) + {x_{2/3}}{\omega _1}}}} \right)} 
\end{equation}

We observe that the dressing parameter $\tilde{a}$ plays in energy and momentum a role similar to that of $\omega_1$. In a similar manner the angle $2 \Delta \varphi$ plays a similar role to the angular opening $\delta	\varphi$, which in the case of elliptic strings is associated to the quasi-momentum in the holographically dual theory. A natural interpretation of these similarities is that the dressed strings are holographic duals of states of the boundary CFT that are characterized by more than one quasi-momenta, interacting with each other in a non-trivial manner. This is not unexpected, since the finite dressed strings approximate genuine genus two solutions.

We observe that the difference of the energy and angular momentum of the dressed solution to those of the seed solution is well-defined and finite, namely,
\begin{align}
\Delta E_{0/1} &= \pm 2 s_\Phi n_2 \frac{{TR {\mu ^2} \tilde{a}}}{\sqrt {{x_{3/2}} - \wp \left( a \right)}} , \label{eq:dispersion_dE}\\
\Delta J_{0/1} &= 2 s_\Phi n_2 \frac{1}{\ell} \left( \zeta \left( \tilde{a} \right) + x_{2/3} \tilde{a} - D \cos \theta_1 \right) . \label{eq:dispersion_dJ}
\end{align}
The exact infinite dressed strings with $D^2>0$ have obviously infinite energy and angular momentum. However, since they are the $n_1 \to \infty$ limit of the approximate solutions, and the above expressions do not depend on $n_1$, the difference of their energy and momentum to those of their elliptic seeds is well-defined, finite and given by \eqref{eq:dispersion_dE} and \eqref{eq:dispersion_dJ}. In other words the finite approximate closed dressed strings may serve as a regularization scheme for the exact infinite closed dressed strings.

Although the above expressions as expressed in terms of transcendental functions, the properties of the elliptic functions allow the specification of the dispersion relation in a closed form whenever the quantities $a$ and $\tilde{a}$ are a rational fraction of $\omega_2$ and $\omega_1$ respectively. The procedure is applied in \cite{part1} for the simpler case of elliptic strings and we will not post further details here.

\subsection{Exact Finite Strings with $D^2>0$ and Strings with $D^2<0$}

The energy and angular momentum of the dressed strings with $D^2<0$ and appropriate periodicity conditions, as well as those of the exact finite dressed strings with $D^2>0$, can be trivially derived from those of the seed solutions. The fact that the solution is periodic in $\sigma^1$ with a period that is an integer multiple of that of the seed, implies that the variation of the energy and angular momentum induced by the dressing is trivially vanishing, as one can read from expressions \eqref{eq:charges_energy_general} and \eqref{eq:dressing_review_change_of_charge}. The energy is trivially equal to
\begin{align}
E_{0/1} &= \frac{{2 T n R {\mu ^2} {\omega _1 }}}{\sqrt {{x_{3/2}} - \wp \left( a \right)}} , \label{eq:charges_energy_non_prop} \\
J_{0/1} &=  \pm \frac{{2 T n R^2 \left( {\zeta \left( {{\omega _1}} \right) + {x_{2/3}}{\omega _1}} \right)}}{{\sqrt {{x_1} - \wp \left( a \right)} }} ,
\label{eq:charges_angular_momentum_non_prop}
\end{align}
where $n$ is equal to $\lcm \left( n^{{\mathrm{seed}}} , n^{{\mathrm{dress}}} \right)$, in the case of dressed strings with $D^2 < 0$, as described in section \ref{subsec:string_finite_exact_D_negative}, and $n = n^{{\mathrm{seed}}}$ in the case of the exact finite elliptic strings with $D^2 > 0$ (or equivalently $n^{{\mathrm{dress}}} = 1$), as described in section \ref{subsec:string_finite_exact_special}.

\begin{figure}[ht]
\vspace{10pt}
\begin{center}
\begin{picture}(100,80)
\put(10,52){\includegraphics[width = 0.40\textwidth]{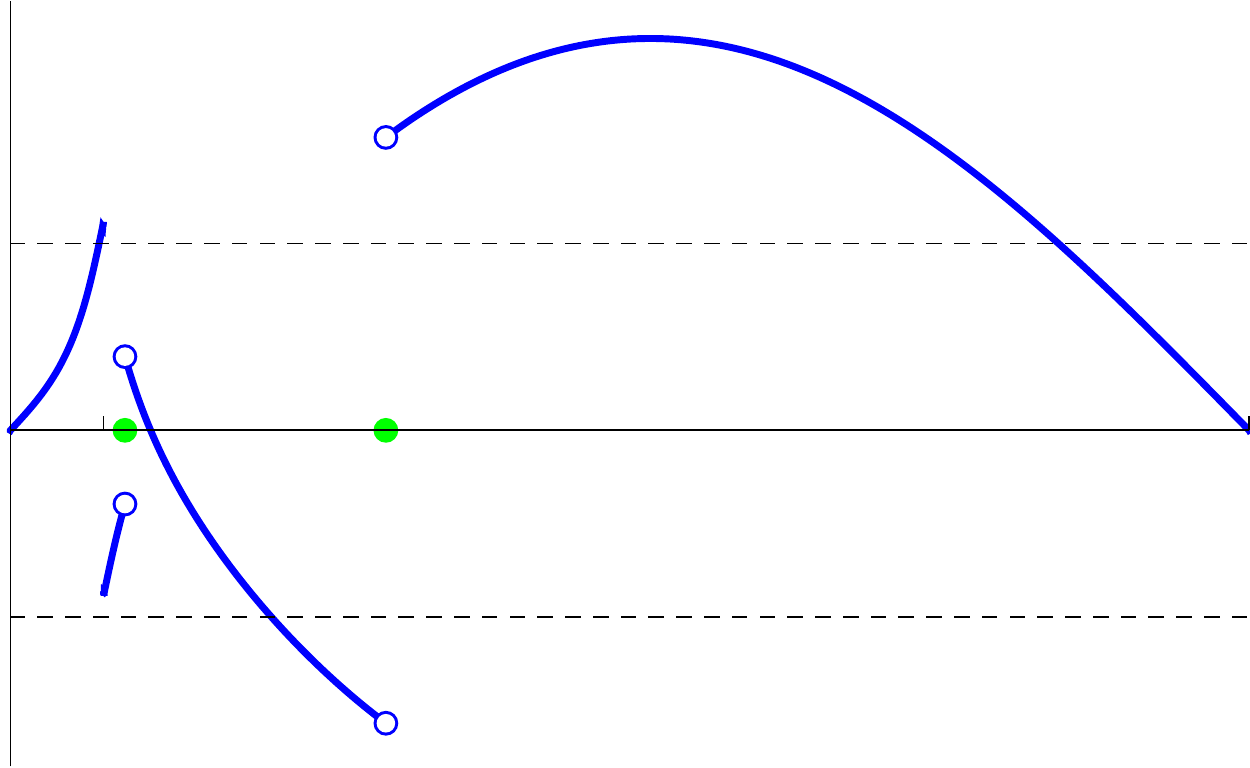}}
\put(57,52){\includegraphics[width = 0.40\textwidth]{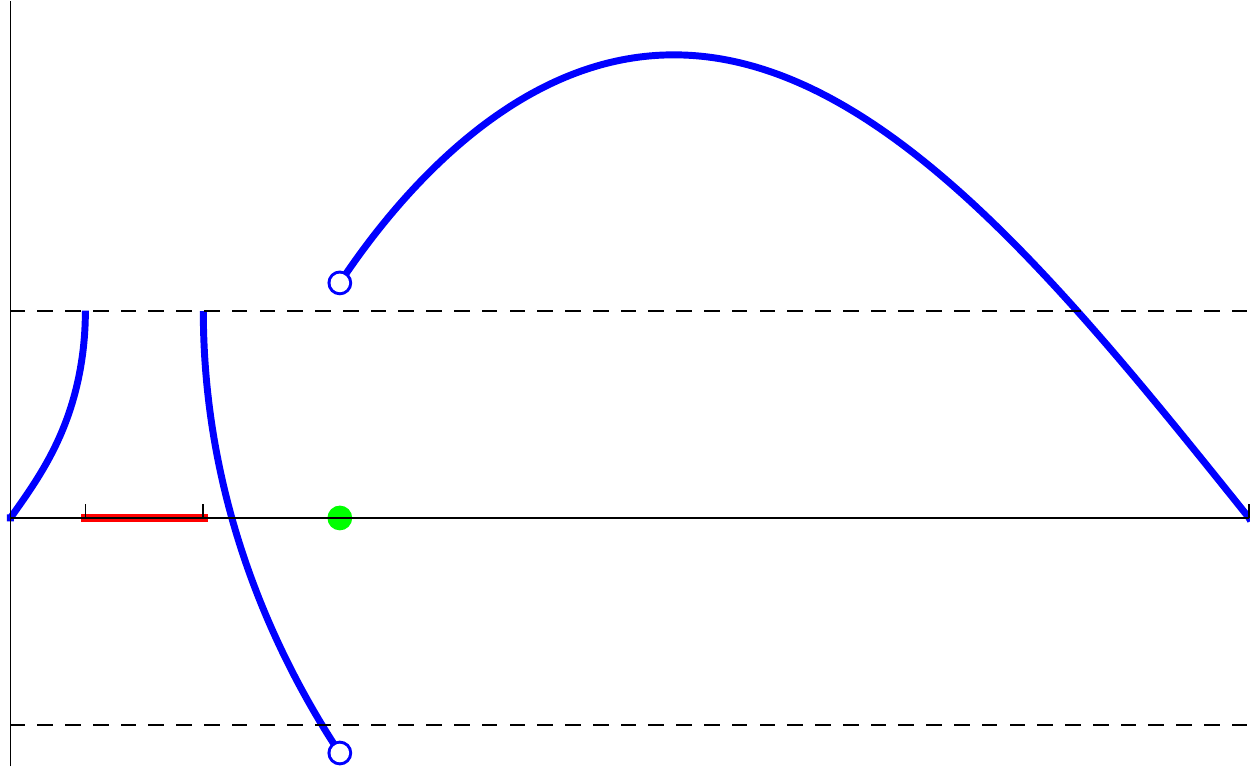}}
\put(10,17){\includegraphics[width = 0.40\textwidth]{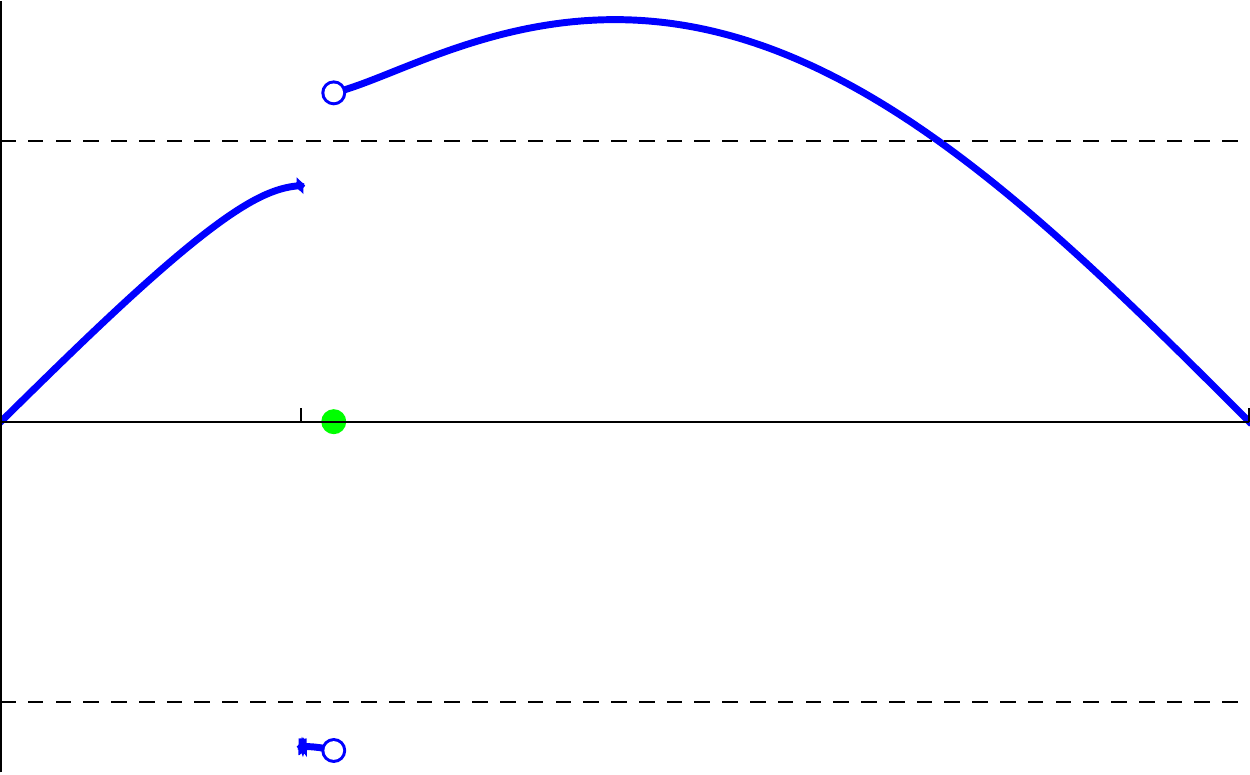}}
\put(57,17){\includegraphics[width = 0.40\textwidth]{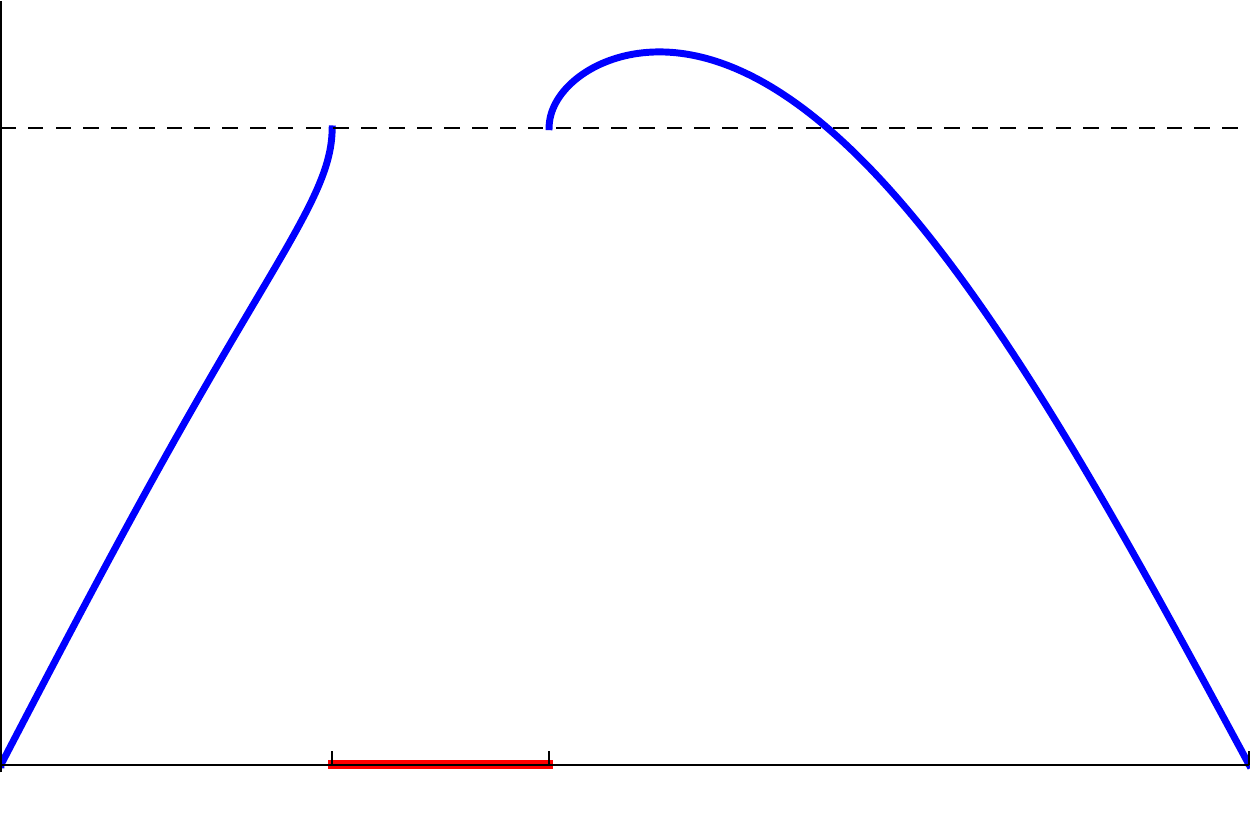}}
\put(10,50){seed with translationally invariant}
\put(57,50){seed with translationally invariant}
\put(17,47.5){oscillating counterpart}
\put(65,47.5){rotating counterpart}
\put(21,14){seed with static}
\put(68,14){seed with static}
\put(17,11.5){oscillating counterpart}
\put(65,11.5){rotating counterpart}
\put(4,77){$\delta \left( E - J \right)$}
\put(51,77){$\delta \left( E - J \right)$}
\put(4,42){$\delta \left( E - J \right)$}
\put(51,43.5){$\delta \left( E - J \right)$}
\put(0.5,68){$\frac{\left( E - J \right)_{\textrm{hop}}}{2}$}
\put(0.5,36.5){$\frac{\left( E - J \right)_{\textrm{hop}}}{2}$}
\put(50.25,62){$\theta_1$}
\put(97.25,59.25){$\theta_1$}
\put(50.25,27.5){$\theta_1$}
\put(97.25,18){$\theta_1$}
\put(11.75,59.75){$\tilde{\theta}$}
\put(18.75,29.25){$\tilde{\theta}$}
\put(58.25,57){$\tilde{\theta}_-$}
\put(62.25,57){$\tilde{\theta}_+$}
\put(66.25,19.75){$\tilde{\theta}_-$}
\put(73.5,19.75){$\tilde{\theta}_+$}
\put(27,0){\includegraphics[height = 0.1\textwidth]{elliptic_str_moduli_legend1.pdf}}
\put(32,1){infinite closed strings with $D^2 > 0$}
\put(32,4.25){finite closed strings with $D^2 > 0$}
\put(32,7.25){finite closed strings with $D^2 < 0$}
\put(27,0){\line(0,1){10}}
\put(27,0){\line(1,0){46.5}}
\put(73.5,0){\line(0,1){10}}
\put(27,10){\line(1,0){46.5}}
\end{picture}
\end{center}
\vspace{-10pt}
\caption{The ${\left( E - J \right)}_{\textrm{dressed}} - {\left( E - J \right)}_{\textrm{seed}}$ as function of the angle $\theta_1$}
\vspace{5pt}
\label{fig:dispersion_EJ}
\end{figure}

The change of the difference of the energy and angular momentum induced by the dressing is plotted versus the dressing parameter $\theta_1$ in figure \ref{fig:dispersion_EJ}.
The plotted graphs for dressed strings whose seeds have translationally invariant counterparts are considered both in the oscillating and rotating case to expose instabilities of the kind presented in section \ref{sec:instabilities}. Had we considered the opposite the graphs would be identical apart from the inversion of the curve between the two instabilities in the case of oscillating counterpart and between $\tilde{\theta}_+$ and the instability in the case of a rotating counterpart, which would be absent.

Furthermore, not all points in the continuous curves of the graphs correspond to closed strings, but only a dense discrete subset of them. The blue lines here correspond to the exact infinite closed strings of section \ref{subsec:string_infinite_exact}. As the expressions for the energy and angular momentum of the approximate finite closed strings of section \ref{subsec:string_finite_approximate} are identical, the relevant plots would be similar apart from two differences:
\begin{itemize}
\item The full continuum of the curves could be set valid, if the parameters of the seed solution were altered appropriately as one moves on the curve so that appropriate periodicity conditions always apply. Otherwise only a dense discrete subset would be valid.
\item A region around each instability point would be invalid since the approximation conditions around the instabilities do not hold.
\end{itemize}

This behaviour implies the existence of an interesting bifurcation in the dispersion relation of the dressed string solution occurring at $E = \mu^2$. The dispersion relation of dressed strings whose seed solutions have oscillating Pohlmeyer counterparts are a non-trivial function of the angle $\theta_1$, which determines the position of the poles of the dressing factor or equivalently specifies the value of the \Backlund parameter $a$. When considering dressed strings whose seeds have rotating counterparts, the dispersion relation is a rather peculiar function of the angle $\theta_1$; there is a range for $\theta_1$ where the dispersion relation does not depend on the latter.

The above is an interesting similarity to the properties of the corresponding solutions of the sine-Gordon equation. As we have seen in section \ref{subsec:SG_breathers_energy}, the mean energy and momentum density of the dressed solution of the sine-Gordon equation with $D^2<0$ is identical to those of the seed solution. It would be interesting to interpret this fact on the side of the holographically dual theory. The difference $E-J$ remains the same after the dressing; however the seed solution is characterised by a single angular opening, i.e. a single quasi-momentum, whereas this is not the case for the dressed solution. A naive interpretation of this solutions could be that they correspond to more complicated excitations, which have formed bound states behaving as a single quasi-momentum state.

There yet another interesting bifurcation of the form of the dispersion relation of the dressed strings in the case of translationally invariant seeds that has to do with the presence of the instabilities. When the seed is unstable, the quantity $\Delta E - \Delta J$ contains further discontinuities related with the inversion of the sign $s_\Phi$. Although the dispersion relations of the dressed strings are too complicated expressions to be directly verifiable in a holographically dual theory, the above discontinuities in the behaviour of the dispersion relation could be in principle detectable.

\setcounter{equation}{0}
\section{Discussion}
\label{sec:discussion}

In the present work, we have carefully studied several physical properties of the dressed elliptic string solutions on $\mathrm{S}^2$, which were derived in \cite{Katsinis:2018ewd}. We have presented these properties in juxtaposition to those of their Pohlmeyer counterparts in an effort to obtain an intuitive understanding of the relativistic string dynamics on the sphere, through the dynamics of the sine-Gordon equation, which is much easier to visualise as a chain of coupled pendulums.

The dressed elliptic solutions have been identified to belong to two large classes depending on the sign of the parameter $D^2$. The ones with $D^2>0$ have Pohlmeyer counterparts which describe localised kinks propagating on top of an elliptic background, whereas those with $D^2<0$ possess Pohlmeyer counterparts which are periodic disturbances on top of an elliptic background. The latter emerge only in the case the seed solution has a rotating Pohlmeyer counterpart.

At first we focused on the necessary conditions that must be obeyed, so that the dressed elliptic strings are closed. We arrived at four specific classes of closed string solutions. One of those is not exact, but these solutions approximate genuine genus two ones, with one of the two genuses being almost singular. The other three classes are exact solutions and can be considered as the analytic continuation of one another as $D^2$ changes sign. One of the latter contains only infinite strings; the approximate class of solutions can serve as a regularization scheme in order to perform the calculations of the conserved charges of the infinite ones.

An interesting feature of the dressed elliptic strings is the existence of interactions between their singular points i.e. their spikes. It has been previously noted in \cite{part1} that in the case of elliptic strings, the number of spikes is identical to the conserved topological charge on the sine-Gordon equation counterpart. However, these solutions have trivial time evolution and the spikes never interact. In the case of the dressed elliptic strings, the form of the allowed interactions between the spikes suggest that the topological charge should not be connected to the number of spikes. It should rather be connected to a more complicated quantity, which receives a $\pm 1$ contribution from each spike and a $\pm 2$ contribution from each loop. This quantity is a carefully defined turning number, which is the homotopy class of the mapping from each point of the string to the \emph{unoriented} direction of the tangent at this point.

The special class of finite exact solutions with $D^2>0$ relates in an interesting way to the stability of the seed elliptic strings. Since these solutions asymptotically interpolate in their dynamical evolution between two versions of the seed elliptic string solution, they reveal that the latter is unstable. It is interesting that such solutions emerge only for the classes of elliptic strings whose sine-Gordon counterparts are unstable \cite{SGstability}. However, the opposite is not true; it is not possible to find such a solution for any elliptic string whose Pohlmeyer counterpart is considered unstable. This may be attributed to the fact that the stability analysis for finite closed string should incorporate only the perturbations that preserve the appropriate periodic conditions. This point deserves further investigation.

The conserved charges of the infinite dressed elliptic strings are divergent, yet one can define a finite difference with respect to the charges of the elliptic seed. This divergence is not surprising, since these string solutions are a long string limit, similar to that of the giant magnons; the latter correspond to genus one solutions with diverging real period, whereas the former are the genus two generalization. As a consequence, they have a dispersion relation that resembles the one of the giant magnons, with an additional free parameter. The two exact finite classes of solutions have identical energy and angular momenta as their seeds.

The dependence of the conserved charges on the moduli of the dressed string solutions exhibits several discontinuities. One of these is related to the qualitative behaviour of the seed solution, whereas the other one is related to the instabilities of the seeds. As the dispersion relation is connected to the anomalous dimensions of operators of the boundary theory, it would be interesting to identify these kinds of bifurcations in the spectrum of the dual theory. The same holds true for the sets of operators, which correspond to the exact finite dressed strings and share the same charges with their seeds.

The techniques that lead to the construction of the dressed elliptic strings on $\mathbb{R} \times \mathrm{S}^2$ have obvious generalizations to other symmetric spaces, such as the AdS, dS, spheres of higher dimensions or tensor products of the latter. Especially the $\mathrm{AdS}_n \times \mathrm{S}^n$ spaces have obvious interest in the framework of the holographic correspondence. The findings of this work suggest that similar phenomena exist in these more interesting cases and deserve further investigation. Similarly, identical techniques can be applied for the study of minimal surfaces in AdS$_4$, which are interesting in the context of the Ryu-Takayanagi conjecture, or Wilson loops.

\subsection*{Acknowledgements}

The authors would like to thank M. Axenides and E. Floratos for useful discussions.

\appendix

\renewcommand{\thesection}{\Alph{section}}
\renewcommand{\thesubsection}{\Alph{section}.\arabic{subsection}}
\renewcommand{\theequation}{\Alph{section}.\arabic{equation}}

\section{On the Asymptotics of the Dressed Elliptic Strings with $D^2>0$}
\label{sec:app_asymptotics}

In the following we present some details of the algebra related to the asymptotic behaviour of the dressed string solutions with a propagating kink Pohlmeyer counterpart ($D^2>0$). For simplicity we consider the case of static seeds. In a similar manner one can study the asymptotic behaviour of dressed strings with translationally invariant seeds.

The equations \eqref{eq:review_E1}, \eqref{eq:review_E2} and \eqref{eq:review_E3} imply that the vectors $E_i$, in the case $\Delta = -D^2 < 0$ can be written as
\begin{align}
{E_1} &= \cosh \left( {D {\xi ^0} + i\Phi \left( {{\xi ^1};\tilde{a}} \right)} \right){e_1} + i\sinh \left( {D {\xi ^0} + i\Phi \left( {{\xi ^1};\tilde{a}} \right)} \right){e_2} ,\\
{E_2} &= i\sinh \left( {D {\xi ^0} + i\Phi \left( {{\xi ^1};\tilde{a}} \right)} \right){e_1} - \cosh \left( {D {\xi ^0} + i\Phi \left( {{\xi ^1};\tilde{a}} \right)} \right){e_2} ,\\
{E_3} &= {e_3} ,
\end{align}
where the vectors $e_1$, $e_2$ and $e_3$ are given by \eqref{eq:review_es}. Far away from the position of the kink, or else when
\begin{equation}
\pm \left( {D {\xi ^0} + i\Phi \left( {{\xi ^1};\tilde{a}} \right)} \right) \equiv \pm \tilde{\Phi} \left( \xi^0 , \xi^1 \right) \gg 1 ,
\end{equation}
these vectors asymptotically assume the form
\begin{align}
{E_1} &\simeq \frac{1}{2} {e^{ \pm \left( {D {\xi ^0} + i\Phi \left( {{\xi ^1};\tilde{a}} \right)} \right)}}\left( {{e_1} \pm i{e_2}} \right) , \\
{E_2} &\simeq \frac{1}{2} {e^{ \pm \left( {D {\xi ^0} + i\Phi \left( {{\xi ^1};\tilde{a}} \right)} \right)}}\left( { - {e_2} \pm i{e_1}} \right) , \\
{E_3} &\simeq {e_3} .
\end{align}
This implies that the solution of the auxiliary system \eqref{eq:dressed_string_auxiliary_solution}, asymptotically equals
\begin{equation}
\Psi  \simeq  - \frac{1}{2} {e^{ \pm \left( {D {\xi ^0} + i\Phi \left( {{\xi ^1};\tilde{a}} \right)} \right)}} \left( {\begin{array}{*{20}{c}}
{{e_1} \pm i{e_2}}\;&{ - {e_2} \pm i{e_1}}\;&0
\end{array}} \right) .
\label{eq:appendix_asymp_Psi}
\end{equation}
It has to be noted that the signs $\pm$ in the above expressions for the asymptotic behaviour of the solution refer to the function $\tilde{\Phi}$ going to $\pm \infty$ and not necessarily the static gauge spacelike coordinate $\sigma^1$. One has to be careful when studying the asymptotic behaviour of the string in identifying the correspondence between the limits of these two parameters. Using the general vector $p$ given by 
\begin{equation}
p = \left( {\begin{array}{*{20}{c}}
{a\cos b}\\
{a\sin b}\\
{ia}
\end{array}} \right) ,
\label{eq:appendix_constant_pvector}
\end{equation}
we may find that the vectors $X_\pm$, defined in equation \eqref{eq:dressed_string_Xpm_def}, can be written as
\begin{align}
{X_ + } &= \Psi \theta p \simeq  - \frac{a}{2} {e^{ \pm \left( {D {\xi ^0} + i\Phi \left( {{\xi ^1};\tilde{a}} \right)} \right) \pm ib}} \left( {{e_1} \pm i{e_2}} \right) , \label{eq:appendix_asymp_Xp}\\
{X_ - } &= \theta \Psi \theta p \simeq  - \frac{a}{2} {e^{ \pm \left( {D {\xi ^0} + i\Phi \left( {{\xi ^1};\tilde{a}} \right)} \right) \pm ib}} \theta \left( {{e_1} \pm i{e_2}} \right) , \label{eq:appendix_asymp_Xm}
\end{align}
which finally implies that the dressed solution, far away from the position of the kink assumes the form
\begin{equation}
X' =  - U \frac{1}{{\wp _A^2}}\left( { \begin{array}{*{20}{c}}
{{\wp _1}\left( {\cos \theta_1 {\wp _a} + \frac{{i\wp '\left( a \right)}}{{2\ell {\wp _a}}}} \right) \mp \frac{{D \wp '\left( {{\xi ^1} + {\omega _2}} \right)}}{{2{\wp _1}{\wp _a}}}}\\
{ \pm D \left( {\cos \theta_1 {\wp _a} + \frac{{i\wp '\left( a \right)}}{{2\ell {\wp _a}}}} \right) + \frac{{\wp '\left( {{\xi ^1} + {\omega _2}} \right)}}{{2{\wp _a}}}}\\
{- \cos \theta_1 \wp _A^2}
\end{array}} \right) ,
\label{eq:appendix_asymp_Xprime}
\end{equation}
where
\begin{align}
\sqrt {\wp \left( {{\xi ^1} + {\omega _2}} \right) - \wp \left( a \right)}  &\equiv {\wp _a} , \\
\sqrt {{x_1} - \wp \left( {{\xi ^1} + {\omega _2}} \right)}  &\equiv {\wp _1} , \\
\sqrt {\wp \left( \tilde{a} \right) - \wp \left( {{\xi ^1} + {\omega _2}} \right)}  &\equiv {\wp _A} 
\end{align}
and the matrix $U$ is given by equation \eqref{eq:review_U}.

The Weierstrass elliptic function obeys the identity
\begin{multline}
{\left[ {\left( {\wp \left( a \right) - \wp \left( c \right)} \right)\,\wp '\left( b \right) \pm \left( {\wp \left( b \right) - \wp \left( c \right)} \right)\,\wp '\left( a \right)} \right]^2}\\
= {\left( {\wp \left( a \right) - \wp \left( b \right)} \right)^2}\left[ {\wp {'^2}\left( c \right) + 4\left( {\wp \left( a \right) - \wp \left( c \right)} \right)\,\left( {\wp \left( b \right) - \wp \left( c \right)} \right)\left( {\wp \left( {a \mp b} \right) - \wp \left( c \right)} \right)} \right] ,
\label{eq:app_Weierstrass_identity_general}
\end{multline}
which is going to be useful in the following. Trivially, if $c$ is any of the half periods, implying that $\wp \left( c \right)$ equals one of the roots $e_i$, the above identity assumes the form
\begin{multline}
{\left[ {\left( {\wp \left( a \right) - {e_i}} \right)\,\wp '\left( b \right) \pm \left( {\wp \left( b \right) - {e_i}} \right)\,\wp '\left( a \right)} \right]^2}\\
 = 4{\left( {\wp \left( a \right) - \wp \left( b \right)} \right)^2}\left( {\wp \left( a \right) - {e_i}} \right)\,\left( {\wp \left( b \right) - {e_i}} \right)\left( {\wp \left( {a \mp b} \right) - {e_i}} \right) .
 \label{eq:app_Weierstrass_identity_specific}
\end{multline}

Writing the dressed solution in spherical coordinates as usual
\begin{equation}
X' = \left( {\begin{array}{*{20}{c}}
{\sin {\theta _{\rm{dressed}}}\cos {\varphi _{\rm{dressed}}}}\\
{\sin {\theta _{\rm{dressed}}}\sin {\varphi _{\rm{dressed}}}}\\
{\cos {\theta _{\rm{dressed}}}}
\end{array}} \right),
\end{equation}
we may read from equation \eqref{eq:appendix_asymp_Xprime} that
\begin{equation}
\ell \cos {\theta _{\rm{dressed}}} =  - \frac{1}{{\wp _A^2}}\left( {\cos \theta_1 {\wp _1}\left( {\wp \left( a \right) - \wp \left( \tilde{a} \right)} \right) - {\wp _1}\frac{{i\wp '\left( a \right)}}{{2\ell }} \pm D \frac{{\wp '\left( {{\xi ^1} + {\omega _2}} \right)}}{{2{\wp _1}}}} \right) .
\end{equation}
The above equation gets significantly simplified using the trigonometric identity $\cos 2x = \left( {1 - {{\tan }^2}x} \right)/\left( {1 + {{\tan }^2}x} \right) = \left( {{{\cot }^2}x - 1} \right)/\left( {{{\cot }^2}x + 1} \right)$
\begin{equation}
\begin{split}
\cos \theta_1 \left( {\wp \left( a \right) - \wp \left( \tilde{a} \right)} \right) &= \cos \theta_1 \left( { - \frac{E}{6} - \frac{{m_ + ^2}}{4} - \frac{{m_ - ^2}}{4} + \frac{E}{6} - \frac{{m_ + ^2}}{4}{{\tan }^2}\frac{\theta_1 }{2} - \frac{{m_ - ^2}}{4}{{\cot }^2}\frac{\theta_1 }{2}} \right)\\
 &=  - \frac{1}{4}\cos \theta_1 \left[ {m_ + ^2\left( {1 + {{\tan }^2}\frac{\theta_1 }{2}} \right) + m_ - ^2\left( {{{\cot }^2}\frac{\theta_1 }{2} + 1} \right)} \right]\\
 &=  - \frac{1}{4}\left[ {m_ + ^2\left( {1 - {{\tan }^2}\frac{\theta_1 }{2}} \right) + m_ - ^2\left( {{{\cot }^2}\frac{\theta_1 }{2} - 1} \right)} \right]\\
 &= \frac{{i\wp '\left( a \right)}}{{2\ell }} + \frac{{\wp '\left( \tilde{a} \right)}}{{2D}} ,
\end{split}
\label{eq:appendix_smart_formula}
\end{equation}
where in the last step we used the equations \eqref{eq:elliptic_solutions_sign_of_a} and \eqref{eq:review_pprime_atilde}. This implies that
\begin{equation}
\begin{split}
\ell \cos {\theta _{\mathrm{dressed}}} &=  - \frac{1}{{\wp _A^2}}\left( {{\wp _1}\frac{{\wp '\left( \tilde{a} \right)}}{{2D }} \pm D \frac{{\wp '\left( {{\xi ^1} + {\omega _2}} \right)}}{{2{\wp _1}}}} \right)\\
&= - \frac{{\left( {\wp \left( {{\xi ^1} + {\omega _2}} \right) - {x_1}} \right)\wp '\left( \tilde{a} \right) \mp \left( {\wp \left( \tilde{a} \right) - {x_1}} \right)\wp '\left( {{\xi ^1} + {\omega _2}} \right)}}{{2\left( {\wp \left( {{\xi ^1} + {\omega _2}} \right) - \wp \left( \tilde{a} \right)} \right)\sqrt {\left( {{x_1} - \wp \left( {{\xi ^1} + {\omega _2}} \right)} \right)\left( {\wp \left( \tilde{a} \right) - {x_1}} \right)} }} .
\end{split}
\label{eq:appendix_costheta}
\end{equation}
Direct application of identity \eqref{eq:app_Weierstrass_identity_specific} results in
\begin{equation}
{\ell ^2}{\cos ^2}{\theta _{{\rm{dressed}}}} = {x_1} - \wp \left( {{\xi ^1} + {\omega _2} \pm \tilde{a}} \right) = {\ell ^2}{\cos ^2}\theta_{{\rm{seed}}} {\left( {{\xi ^1} \pm \tilde{a}} \right)} .
\label{eq:appendix_theta_asymp}
\end{equation}

In the case of seeds with rotating counterparts, where ${\cos}\theta_{\rm{seed}}$ has a given sign for all points of the seed solution, i.e. the whole seed solution lies in a single hemisphere, equation \eqref{eq:appendix_costheta} implies that the asymptotic behaviour of the dressed string has the same property. However, whether this is the same hemisphere is determined by the sign of $- \wp ' \left( \tilde{a} \right)$. In other words when $\tilde{a} > 0$ is positive, the seed and asymptotic behaviour of the dressed solution lie in the same hemisphere, whereas when $\tilde{a} < 0$ they lie in opposite hemispheres. This is exactly the behaviour described in section \ref{subsec:asymptotics}. 

In a similar manner, it is a matter of algebra to show that the azimuthal angle of the dressed solution assumes the form
\begin{equation}
\begin{split}
{\varphi _{{\rm{dressed}}}} &=  \mp \arctan \frac{{- D {P_1} - \ell \left( {{P_2} \pm {P_3}} \right)}}{{\ell {P_1} - D \left( {{P_2} \pm {P_3}} \right)}} + {\varphi _{{\rm{seed}}}}\\
 &=  \mp \arctan \frac{\ell }{{D}} \mp \arctan \frac{{{P_1}}}{{\left( {{P_2} \pm {P_3}} \right)}} + {\varphi _{{\rm{seed}}}} ,
\end{split}
\label{eq:appendix_phi_asymp}
\end{equation}
where
\begin{align}
{P_1} &= \left( {\wp \left( {{\xi ^1} + {\omega _2}} \right) - \wp \left( \tilde{a} \right)} \right)i\wp '\left( a \right) ,\\
{P_2} &= \left( {\wp \left( {{\xi ^1} + {\omega _2}} \right) - \wp \left( a \right)} \right)\wp '\left( \tilde{a} \right) ,\\
{P_3} &= \left( {\wp \left( a \right) - \wp \left( \tilde{a} \right)} \right)\wp '\left( {{\xi ^1} + {\omega _2}} \right) ,
\end{align}
It is trivial that
\begin{equation}
{\partial _0}{\varphi _{{\rm{dressed}}}} = {\partial _0}{\varphi _{{\rm{seed}}}} ,
\end{equation}
while
\begin{equation}
{\partial _1}{\varphi _{{\rm{dressed}}}} = \mp \frac{{\left( {{P_2} \pm {P_3}} \right){\partial _1}{P_1} - {P_1}{\partial _1}\left( {{P_2} \pm {P_3}} \right)}}{{P_1^2 + {{\left( {{P_2} \pm {P_3}} \right)}^2}}} + {\partial _1}{\varphi _{{\rm{seed}}}} .
\end{equation}

The denominator of the fraction in the expression above can be calculated with direct application of the identity \eqref{eq:app_Weierstrass_identity_general},
\begin{equation}
\begin{split}
&P_1^2 + {\left( {{P_2} \pm {P_3}} \right)^2}\\
 &= {\left[ {\left( {\wp \left( {{\xi ^1} + {\omega _2}} \right) - \wp \left( a \right)} \right)\wp '\left( \tilde{a} \right) \mp \left( {\wp \left( \tilde{a} \right) - \wp \left( a \right)} \right)\wp '\left( {{\xi ^1} + {\omega _2}} \right)} \right]^2} \\
 &- {\left( {\wp \left( {{\xi ^1} + {\omega _2}} \right) - \wp \left( \tilde{a} \right)} \right)^2}\wp {'^2}\left( a \right)\\
 &= 4{\left( {\wp \left( {{\xi ^1} + {\omega _2}} \right) - \wp \left( \tilde{a} \right)} \right)^2}\left( {\wp \left( {{\xi ^1} + {\omega _2}} \right) - \wp \left( a \right)} \right)\\
 &\times\left( {\wp \left( \tilde{a} \right) - \wp \left( a \right)} \right)\left( {\wp \left( {{\xi ^1} + {\omega _2} \pm \tilde{a}} \right) - \wp \left( a \right)} \right) .
\end{split}
\end{equation}
The numerator can also be simplified using Weierstrass differential equation and its derivative ${\wp''}\left( x \right) = 6\wp^2 \left( x \right) - {g_2} / 2$
\begin{equation}
\begin{split}
&\left( {{P_2} \pm {P_3}} \right){\partial _1}{P_1} - {P_1}{\partial _1}\left( {{P_2} \pm {P_3}} \right)\\
& =  \pm {{i\wp '\left( a \right)}}\left( {\wp \left( \tilde{a} \right) - \wp \left( a \right)} \right)\\
& \times \left( {\wp ''\left( {{\xi ^1} + {\omega _2}} \right)\left( {\wp \left( {{\xi ^1} + {\omega _2}} \right) - \wp \left( \tilde{a} \right)} \right) - \wp {'^2}\left( {{\xi ^1} + {\omega _2}} \right) \pm \wp '\left( {{\xi ^1} + {\omega _2}} \right)\wp '\left( \tilde{a} \right)} \right)\\
& =  \pm \frac{{i\wp '\left( a \right)}}{2}\left( {\wp \left( \tilde{a} \right) - \wp \left( a \right)} \right)\left[ {2\wp ''\left( {{\xi ^1} + {\omega _2}} \right)\left( {\wp \left( {{\xi ^1} + {\omega _2}} \right) - \wp \left( \tilde{a} \right)} \right)} \right.\\
&\left. { - \wp {'^2}\left( {{\xi ^1} + {\omega _2}} \right) + \wp {'^2}\left( \tilde{a} \right) - {{\left( {\wp '\left( {{\xi ^1} + {\omega _2}} \right) \mp \wp '\left( \tilde{a} \right)} \right)}^2}} \right]\\
 &=  \mp 2 i\wp '\left( a \right)\left( {\wp \left( \tilde{a} \right) - \wp \left( a \right)} \right){\left( {\wp \left( {{\xi ^1} + {\omega _2}} \right) - \wp \left( \tilde{a} \right)} \right)^2}\\
 &\times \left[ {\frac{1}{4}{{\left( {\frac{{\wp '\left( {{\xi ^1} + {\omega _2}} \right) \mp \wp '\left( \tilde{a} \right)}}{{\wp \left( {{\xi ^1} + {\omega _2}} \right) - \wp \left( \tilde{a} \right)}}} \right)}^2} - 2\wp \left( {{\xi ^1} + {\omega _2}} \right) - \wp \left( \tilde{a} \right)} \right]\\
 &=  \mp 2 i\wp '\left( a \right)\left( {\wp \left( \tilde{a} \right) - \wp \left( a \right)} \right){\left( {\wp \left( {{\xi ^1} + {\omega _2}} \right) - \wp \left( \tilde{a} \right)} \right)^2}\left[ {\wp \left( {{\xi ^1} + {\omega _2} \pm \tilde{a}} \right) - \wp \left( {{\xi ^1} + {\omega _2}} \right)} \right] .
\end{split}
\end{equation}

Putting everything together, we yield
\begin{equation}
\begin{split}
&{\partial _1}{\varphi _{{\rm{dressed}}}} = \frac{{i\wp '\left( a \right)\left[ {\wp \left( {{\xi ^1} + {\omega _2} \pm \tilde{a}} \right) - \wp \left( {{\xi ^1} + {\omega _2}} \right)} \right]}}{{2\left( {\wp \left( {{\xi ^1} + {\omega _2}} \right) - \wp \left( a \right)} \right)\,\left( {\wp \left( {{\xi ^1} + {\omega _2} \pm \tilde{a}} \right) - \wp \left( a \right)} \right)}} + {\partial _1}{\varphi _{{\rm{seed}}}}\\
 &= \frac{{i\wp '\left( a \right)}}{{2\left( {\wp \left( {{\xi ^1} + {\omega _2}} \right) - \wp \left( a \right)} \right)\,}} - \frac{{i\wp '\left( a \right)}}{{2\,\left( {\wp \left( {{\xi ^1} + {\omega _2} \pm \tilde{a}} \right) - \wp \left( a \right)} \right)}} + {\partial _1}{\varphi _{{\rm{seed}}}}\\
 &=  - \frac{{i\wp '\left( a \right)}}{{2\,\left( {\wp \left( {{\xi ^1} + {\omega _2} \pm \tilde{a}} \right) - \wp \left( a \right)} \right)}} = {\partial _1}{\varphi _{{\rm{seed}}}}\left( {{{\xi ^0},\xi ^1} \pm \tilde{a}} \right) .
\end{split}
\end{equation}
This finally, implies that
\begin{equation}
{\varphi _{{\rm{dressed}}}}\left( {{\xi ^0},{\xi ^1}} \right) = {\varphi _{{\rm{seed}}}}\left( {{\xi ^0},{\xi ^1} \pm \tilde{a}} \right) + {\varphi _ \pm } .
\label{eq:appendix_phi_asymp_shift}
\end{equation}

The above hold in the case of seed with static counterparts. In a trivial manner one could acquire the analogous asymptotic expressions in the case of translationally invariant seeds. They emerge from equations \eqref{eq:appendix_theta_asymp} and \eqref{eq:appendix_phi_asymp_shift} after the trivial operation $\xi^0 \leftrightarrow \xi^1$. Converting equations \eqref{eq:appendix_theta_asymp} and \eqref{eq:appendix_phi_asymp_shift} to the static gauge trivially results in the asymptotic formulae \eqref{eq:asymptotic_form_theta_ti}, \eqref{eq:asymptotic_form_phi_ti}, \eqref{eq:asymptotic_form_theta} and \eqref{eq:asymptotic_form_phi}.

We would like to determine the constants $\varphi_\pm$. However, the above expressions are given in the linear gauge. Determining the asymptotic behaviour of a snapshot of the string in the physical time $X^0$ requires determining them in the static gauge. The values of the constants at the two gauges are obviously not identical. In the following $\varphi_\pm$ denote the constants in the static gauge. Converting to the latter, we get
\begin{equation}
{\varphi _{{\rm{dressed}}}}\left( {{\sigma ^0},{\sigma ^1}} \right) = {\varphi _{{\rm{seed}}}}\left( {{\sigma ^0},{\sigma ^1} \pm \frac{{\tilde a}}{\gamma }} \right) + {\varphi _ \pm } ,
\end{equation}
Comparing the above to equation \eqref{eq:appendix_phi_asymp} we get
\begin{equation}
{\varphi _ \pm } = {\varphi _{{\rm{dressed}}}}\left( {0,0} \right) - {\varphi _{{\rm{seed}}}}\left( {0, \pm \frac{{\tilde a}}{\gamma }} \right) ,
\end{equation}
where
\begin{align}
{\varphi _{{\rm{dressed}}}}\left( {0,0} \right) &=  \pm \arctan \frac{{D{P_1}\left( {0,0} \right) + \ell {P_2}\left( {0,0} \right)}}{{\ell {P_1}\left( {0,0} \right) - D{P_2}\left( {0,0} \right)}} ,\\
{P_1}\left( {0,0} \right) &= \left( {{x_3} - \wp \left( {\tilde a} \right)} \right)i\wp '\left( a \right) ,\\
{P_2}\left( {0,0} \right) &= \left( {{x_3} - \wp \left( a \right)} \right)\wp '\left( {\tilde a} \right) ,\\
{P_3}\left( {0,0} \right) &= 0  ,
\end{align}
since ${\varphi _{{\rm{seed}}}}\left( {0, 0} \right) = 0$.
Finally, the elliptic solution implies
\begin{equation}
{\varphi _{{\rm{seed}}}}\left( {0, \pm \frac{{\tilde a}}{\gamma }} \right) =  \mp \left( {\ell \beta \tilde a + \Phi \left( {\tilde a;a} \right)} \right) ,
\end{equation}
which in turn results in
\begin{equation}
{\varphi _ \pm } = \pm \left[ {\arctan \frac{{D\left( {{x_3} - \wp \left( {\tilde a} \right)} \right)i\wp '\left( a \right) + \ell \left( {{x_3} - \wp \left( a \right)} \right)\wp '\left( {\tilde a} \right)}}{{\ell \left( {{x_3} - \wp \left( {\tilde a} \right)} \right)i\wp '\left( a \right) - D\left( {{x_3} - \wp \left( a \right)} \right)\wp '\left( {\tilde a} \right)}} + \ell \beta \tilde a + \Phi \left( {\tilde a;a} \right)} \right] .
\end{equation}
It is a matter of algebra and careful use of the appropriate properties of Weierstrass functions to show that this formula is equivalent to the formula \eqref{eq:asymptotic_form_DPhi} for the case of seed solutions with static Pohlmeyer counterparts. In a similar manner one can specify this angle in the case of seeds with translationally invariant counterparts.

\section{Calculation of the Angular Momentum of the Dressed Strings}
\label{sec:appendix_dispersion}

In the following we post some details of the proof of equation \eqref{eq:dispersion_dJ}. The variation of the sigma model charge by the dressing is given by equations \eqref{eq:dressing_review_change_of_charge} and \eqref{eq:dress_review_dressing_factor_two_conjugate_poles_projector}. Using the definitions \eqref{eq:dressed_string_Xpm_def} and \eqref{eq:review_hat_defs}, the projector $P$ assumes the form
\begin{equation}
P = \theta U\theta \frac{{{X_ - }X_ + ^T}}{{X_ + ^T{X_ - }}}\theta {U^T}\theta .
\end{equation}
Taking advantage of the asymptotic form of the vectors $X_\pm$ \eqref{eq:appendix_asymp_Xp} and \eqref{eq:appendix_asymp_Xm}, we find
\begin{equation}
\Delta\mathcal{Q}_L^{12}=2i\sin\theta_1\left(b_+F_1|_{\sigma^1 = {\bar{\sigma} + \frac{n_1 \omega_1 - s_\Phi \tilde{a}}{\gamma}}}-b_-F_1|_{\sigma^1 = {\bar{\sigma}-\frac{n_1 \omega_1 - s_\Phi \tilde{a}}{\gamma}}}\right),
\end{equation}
where
\begin{equation}
b_\pm = \frac{\kappa_0^3\kappa_0^2\mp s_\Phi D\kappa_0^1}{\left(\kappa_0^1\right)^2+\left(\kappa_0^2\right)^2}
\end{equation}
and $F_1$ is given by \eqref{eq:dressed_strings_F1_F2}.

Using the definitions \eqref{eq:defs_kappa_1} and \eqref{eq:defs_kappa_2}
\begin{equation}
i\sin {\theta _1}{b_ \pm } = \frac{{ - k_1^2k_0^3 \pm {s_\Phi }D\left( { k_0^1\cos {\theta _1} - k_1^1} \right)}}{{\wp \left( {{\sigma _ \pm }} \right) - \wp \left( {\tilde a} \right)}} ,
\end{equation}
where $\sigma_\pm$ are given by \eqref{eq:dispersion_spm}. The quantities $k_{0/1}^i$ are determined by the seed elliptic solution via the equations ${U^T}\left( {{\partial _i}U} \right) = k_i^j{T_j}$, where the matrix $U$ is given by \eqref{eq:review_U} and $T_j$ are the generators of the $\mathrm{SO}(3)$ group defined as usual. It is a matter of algebra to show that
\begin{equation}
2i\ell \sin {\theta _1}{b_ \pm }{F_1} =  - \frac{{\wp '\left( {{\sigma _ \pm }} \right) \pm {s_\Phi }D\left[ {2\left( {\wp \left( {{\sigma _ \pm }} \right) - \wp \left( a \right)} \right)\cos {\theta _1} + \frac{{i\wp '\left( a \right)}}{\ell }} \right]}}{{\wp \left( {{\sigma _ \pm }} \right) - \wp \left( {\tilde a} \right)}} .
\end{equation}
Using the formula \eqref{eq:appendix_smart_formula}, the above expression assumes the form
\begin{equation}
2i\ell \sin {\theta _1}{b_ \pm }{F_1} =  - \frac{{\wp '\left( {{\sigma _ \pm }} \right) - \wp '\left( { \pm {s_\Phi }\tilde a} \right)}}{{\wp \left( {{\sigma _ \pm }} \right) - \wp \left( \pm {s_\Phi } {\tilde a} \right)}} \mp 2{s_\Phi }D\cos {\theta _1} ,
\end{equation}
which finally implies that
\begin{equation}
\Delta Q_{L}^{12} = -\frac{2n_2}{\ell}\left[-2n_1\zeta\left(\omega_1\right)+\zeta\left(\sigma_+\right)-\zeta\left(\sigma_-\right)+2 s_\Phi \left( \zeta\left( \tilde{a} \right) - D \cos\theta_1 \right) \right] .
\end{equation}
This equation leads to the equation \eqref{eq:dispersion_J_var} and in turn to equation \eqref{eq:dispersion_J1_final}, which provides the angular momentum of the dressed string. This derivation concerns dressed strings with static seeds. In a similar manner one can repeat the proof for strings with translationally invariant seeds.

The range of the worldsheet coordinate $\sigma^1$ that cover the whole closed dressed string depends on the value of $\tilde{a}$ and the sign $s_\Phi$. This in turn has consequences on the variation of the energy and angular momentum that the dressing procedure has induced. In order to understand how these quantities depend on the position of the poles of the dressing factor, which is determined by the angle $\theta_1$ we have to consider the figures \ref{fig:mean_velocity} and \ref{fig:dressed_strings_static_phi}.

Figure \ref{fig:dressed_strings_static_phi} shows that in all cases the dependence of $\tilde{a}$ on $\theta_1$ is monotonous. When $\theta_1 = 0$, $\tilde{a}$ vanishes. Then, as $\theta_1$ increases, $\tilde{a}$ increases until a given angle $\theta_1 = \tilde{\theta}$ (or $\theta_1 = \tilde{\theta}_-$ in the case of rotating seeds), where $\tilde{a}$ equals the real half-period $\omega_1$. Then, $\tilde{a}$ becomes equal to $-\omega_1$ (the dependence is not discontinuous as the two positions are congruent), either immediately in the case of oscillating seeds, or after some range of $\theta_1$ where it is complex in the case of rotating seeds. Then, it continues increasing until $\theta_1 = \pi$ when it again vanishes.

Returning to figure \ref{fig:mean_velocity}, and bearing in mind that the mean kink velocity is an odd function of $\tilde{a}$, we may conclude the following: in the case of translationally invariant seeds, the sign $s_\phi$ is the sign of $1 + \beta \bar{v}_0$. In our analysis, the parameter $\beta$ is positive and smaller than 1. Therefore, when $E < E_c$, $s_\Phi$ is always positive. When $E_c < E < \mu^2$ and the maximum kink velocity is larger that $1 / \beta$, there are two critical values of $\theta_1$, let them be $\theta_{c1}$ and $\theta_{c2}$, being both larger than $\tilde{\theta}$, since they correspond to negative $\tilde{a}$, and $s_\Phi$ is negative when $\theta_{c1} < \theta_1 < \theta_{c2}$. Similarly, when $E>\mu^2$ and the maximum kink velocity is larger than $1 / \beta$, there is one critical value of $\theta_1$ let it be $\theta_c$, which is larger than $\tilde{\theta}_+$ and $s_\Phi$ is negative when $\tilde{\theta}_+ < \theta_1 < \theta_c$.

\begin{table}[h]
\vspace{10pt}
\begin{center}
\begin{tabular}{ | c || c | c | c | c || c || c | c | c | c |}
\hline
$\theta_1$ & $s_\Phi$ & $\sign \tilde{a}$ & $\sign D^2$ & $s_\Phi \sign \tilde{a}$ & $\theta_1$ & $s_\Phi$ & $\sign \tilde{a}$ & $\sign D^2$ & $s_\Phi \sign \tilde{a}$ \\
\hline\hline
\multicolumn{5}{|c||}{unstable trans. invariant oscillating} & \multicolumn{5}{|c|}{stable trans. invariant rotating}\\
\hline
$ ( 0 , \tilde{\theta} )$ & $+$ & $+$ & $+$ & $+$ & $ ( 0 , \tilde{\theta}_- )$ & $+$ & $+$ & $+$ & $+$\\
\hline
$ ( \tilde{\theta} , \theta_{c1} )$ & $+$ & $-$ & $+$ & $-$ & $ ( \tilde{\theta}_- , \tilde{\theta}_+ )$ &  & $\notin \mathbb{R}$ & $-$ &\\
\hline
$ ( \theta_{c1} , \theta_{c2} )$ & $-$ & $-$ & $+$ & $+$ & $ ( \tilde{\theta}_+ , \pi )$ & $+$ & $-$ & $+$ & $-$\\
\hline
$ ( \theta_{c2} , \pi )$ & $+$ & $-$ & $+$ & $-$ & \multicolumn{5}{|c|}{static oscillating}\\
\hline
\multicolumn{5}{|c||}{stable trans. invariant oscillating} & $ ( 0 , \tilde{\theta} )$ & $-$ & $+$ & $+$ & $-$\\
\hline
$ ( 0 , \tilde{\theta} )$ & $+$ & $+$ & $+$ & $+$ & $ ( \tilde{\theta} , \theta_c )$ & $-$ & $-$ & $+$ & $+$\\
\hline
$ ( \tilde{\theta} , \pi )$ & $+$ & $-$ & $+$ & $-$ & $ ( \theta_c , \pi )$ & $+$ & $-$ & $+$ & $-$\\
\hline
\multicolumn{5}{|c||}{unstable trans. invariant rotating} & \multicolumn{5}{|c|}{static rotating}\\
\hline
$ ( 0 , \tilde{\theta}_- )$ & $+$ & $+$ & $+$ & $+$ & $ ( 0 , \tilde{\theta}_- )$ & $-$ & $+$ & $+$ & $-$\\
\hline
$ ( \tilde{\theta}_- , \tilde{\theta}_+ )$ &  & $\notin \mathbb{R}$ & $-$ & & $ ( \tilde{\theta}_- , \tilde{\theta}_+ )$ &  & $\notin \mathbb{R}$ & $-$ & \\
\hline
$ ( \tilde{\theta}_+ , \theta_c )$ & $-$ & $-$ & $+$ & $+$ & $ ( \tilde{\theta}_+ , \pi )$ & $+$ & $-$ & $+$ & $-$\\
\hline
$ ( \theta_c , \pi )$ & $+$ & $-$ & $+$ & $-$ & \multicolumn{5}{|c|}{ }\\
\hline
\end{tabular}
\vspace{3pt}
\caption{The dependence of the signs of $\tilde{a}$, $D^2$ and the sign $s_\Phi$ on the angle $\theta_1$}
\label{tb:signs}
\end{center}
\end{table}
\begin{figure}[p]
\vspace{10pt}
\begin{center}
\begin{picture}(100,135)
\put(6,109){\includegraphics[width = 0.40\textwidth]{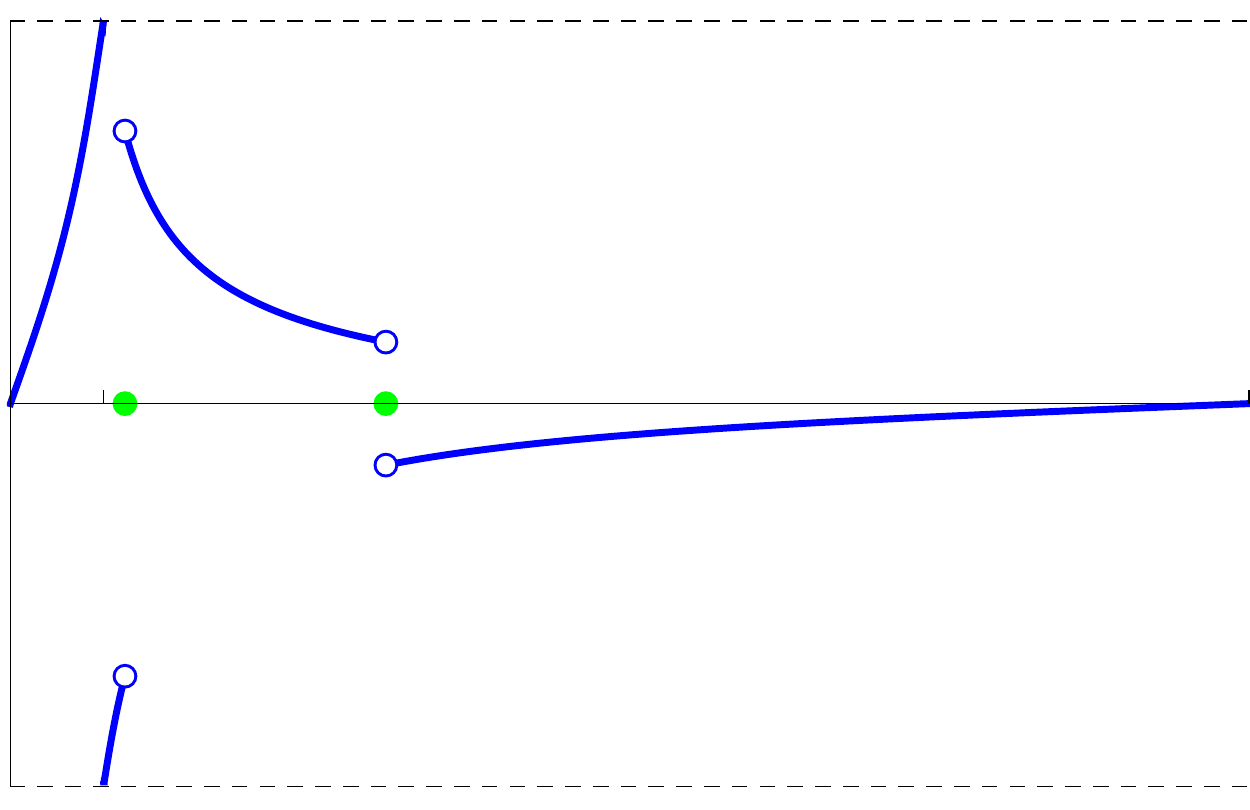}}
\put(57,109){\includegraphics[width = 0.40\textwidth]{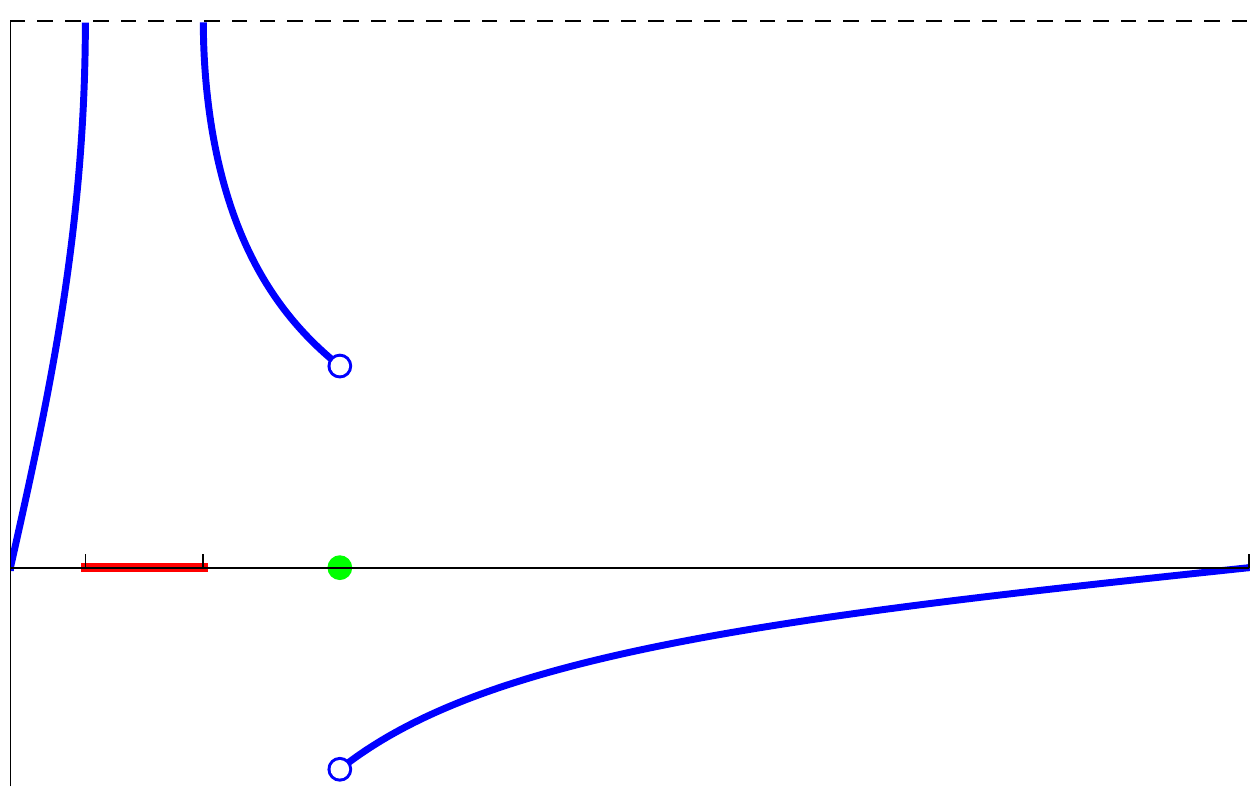}}
\put(6,77){\includegraphics[width = 0.40\textwidth]{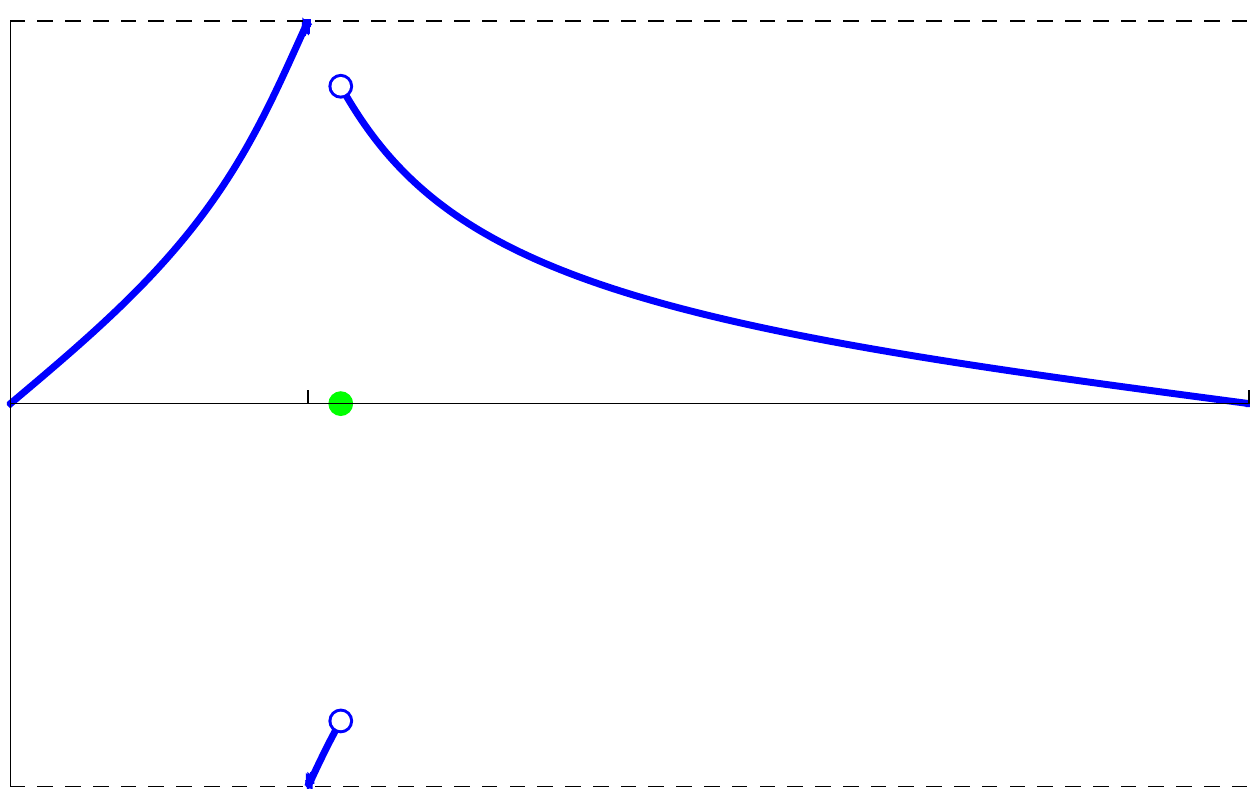}}
\put(57,77){\includegraphics[width = 0.40\textwidth]{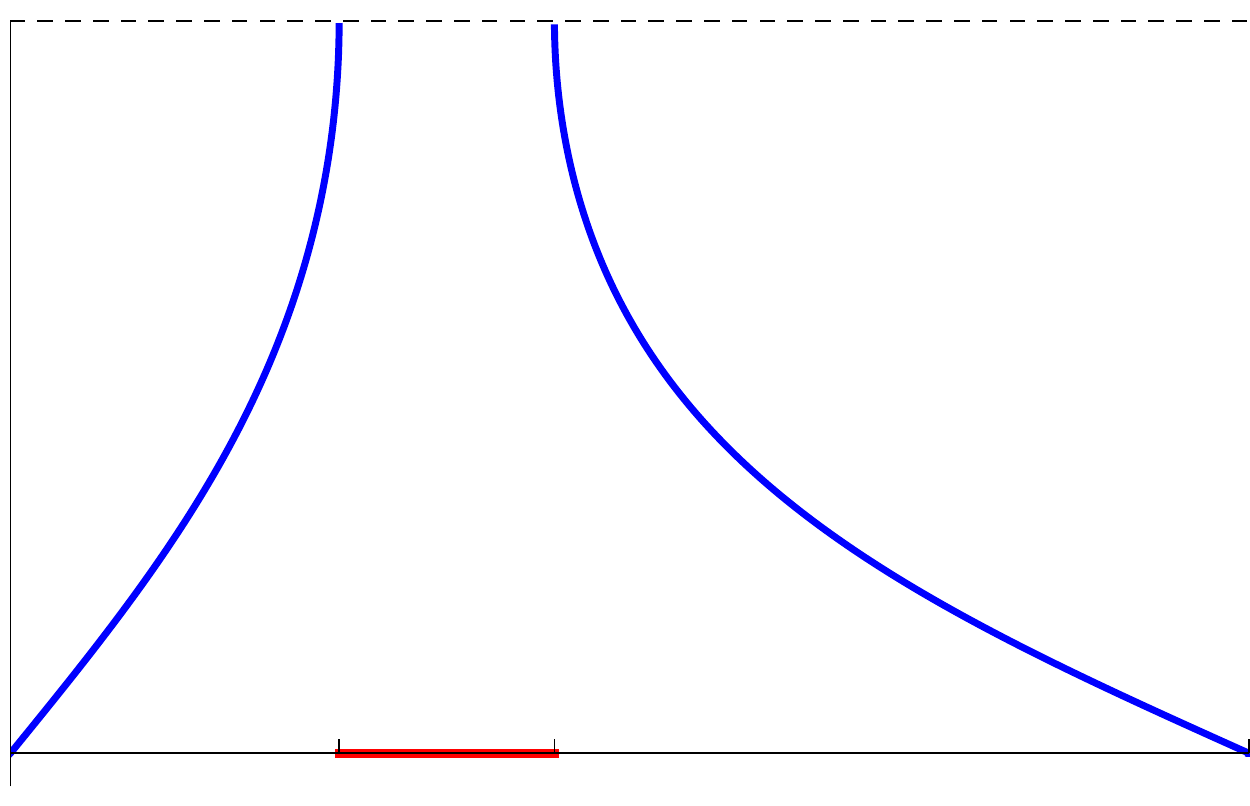}}
\put(2,107){translationally invariant oscillating seed}
\put(55,107){translationally invariant rotating seed}
\put(13,75){static oscillating seed}
\put(66,75){static rotating seed}
\put(6,134.5){$\delta E$}
\put(57,134){$\delta E$}
\put(6,102.5){$\delta E$}
\put(57,102.5){$\delta E$}
\put(0.5,133.5){$\frac{E_{\textrm{hop}}}{2}$}
\put(0.5,101.5){$\frac{E_{\textrm{hop}}}{2}$}
\put(51.5,132.875){$\frac{E_{\textrm{hop}}}{2}$}
\put(51.5,101.375){$\frac{E_{\textrm{hop}}}{2}$}
\put(46.25,121.25){$\theta_1$}
\put(97.25,115.25){$\theta_1$}
\put(46.25,89.25){$\theta_1$}
\put(97.25,77.75){$\theta_1$}
\put(8.5,123.75){$\tilde{\theta}$}
\put(15,90.75){$\tilde{\theta}$}
\put(58.75,117){$\tilde{\theta}_-$}
\put(62.5,117){$\tilde{\theta}_+$}
\put(66.75,79.75){$\tilde{\theta}_-$}
\put(73.75,79.75){$\tilde{\theta}_+$}
\put(27,63){\includegraphics[height = 0.1\textwidth]{elliptic_str_moduli_legend1.pdf}}
\put(32,64){infinite closed strings with $D^2 > 0$}
\put(32,67.25){finite closed strings with $D^2 > 0$}
\put(32,70.25){finite closed strings with $D^2 < 0$}
\put(27,63){\line(0,1){10}}
\put(27,63){\line(1,0){46.5}}
\put(73.5,63){\line(0,1){10}}
\put(27,73){\line(1,0){46.5}}
\put(6,34){\includegraphics[width = 0.40\textwidth]{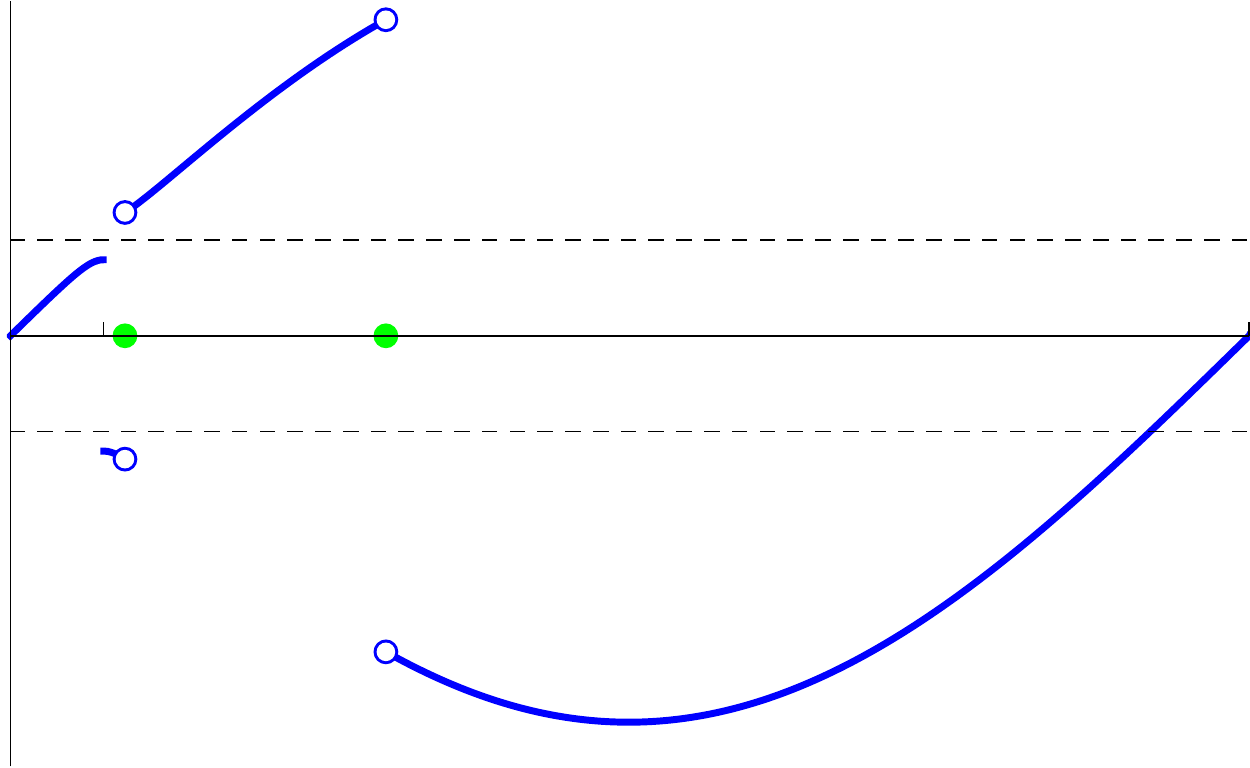}}
\put(57,34){\includegraphics[width = 0.40\textwidth]{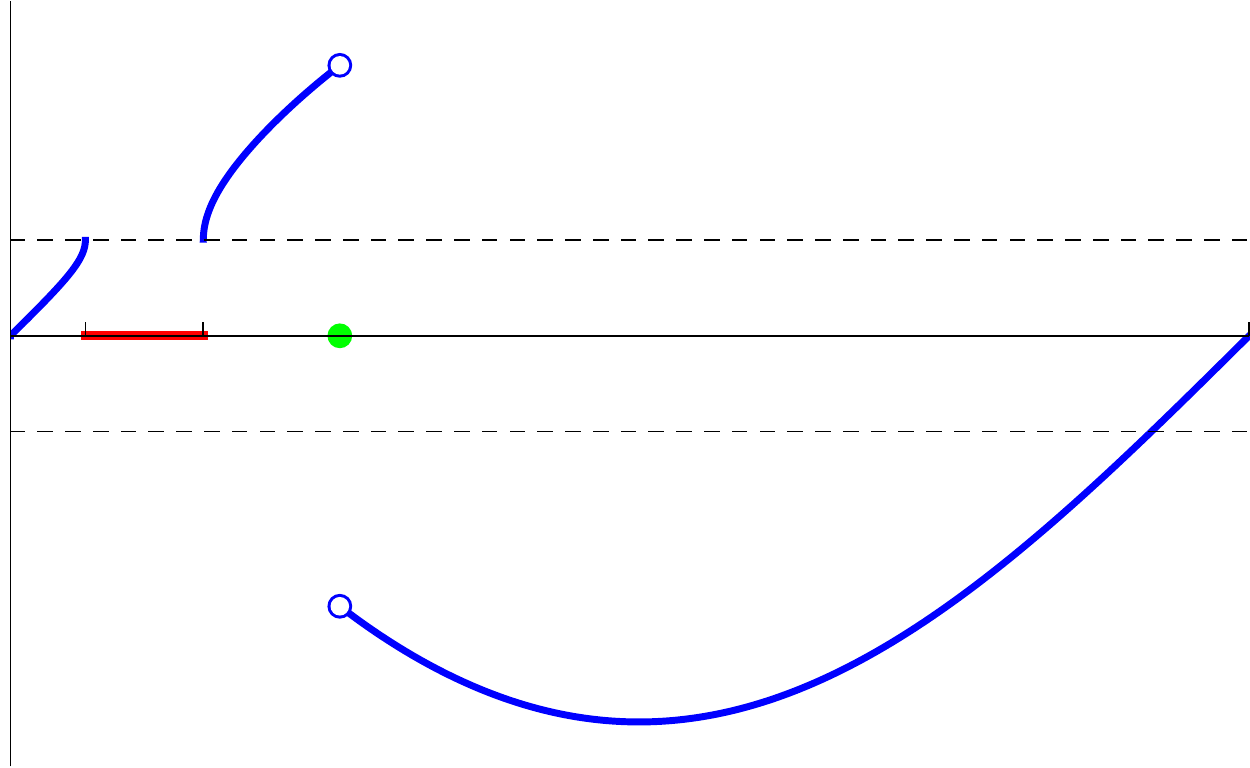}}
\put(6,2){\includegraphics[width = 0.40\textwidth]{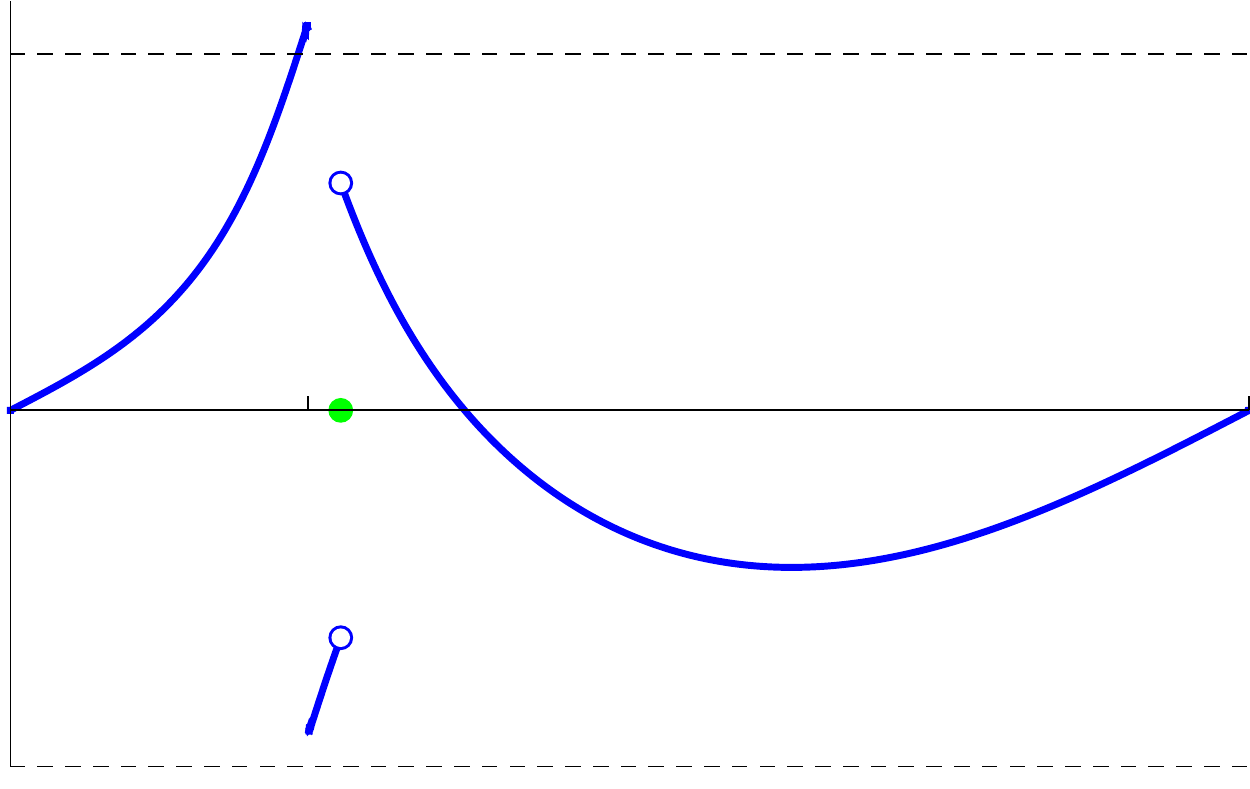}}
\put(57,2){\includegraphics[width = 0.40\textwidth]{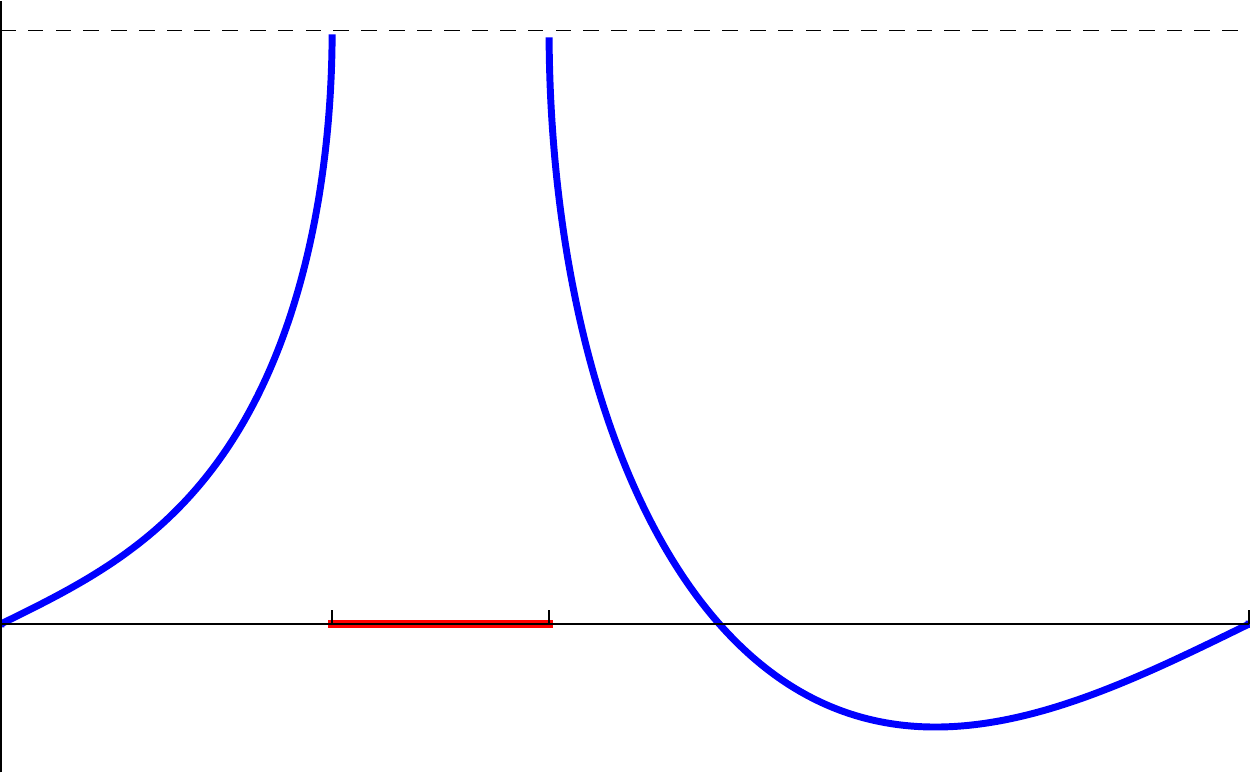}}
\put(2,32){translationally invariant oscillating seed}
\put(55,32){translationally invariant rotating seed}
\put(13,0){static oscillating seed}
\put(66,0){static rotating seed}
\put(5,59){$\delta J$}
\put(56,59){$\delta J$}
\put(5,27.5){$\delta J$}
\put(56,27.25){$\delta J$}
\put(1.5,50.125){$\frac{J_{\textrm{hop}}}{2}$}
\put(1.5,24.75){$\frac{J_{\textrm{hop}}}{2}$}
\put(52.5,50.125){$\frac{J_{\textrm{hop}}}{2}$}
\put(52.5,25.125){$\frac{J_{\textrm{hop}}}{2}$}
\put(46.25,47){$\theta_1$}
\put(97.25,47.25){$\theta_1$}
\put(46.25,13.25){$\theta_1$}
\put(97.25,6){$\theta_1$}
\put(8,44.75){$\tilde{\theta}$}
\put(15,15){$\tilde{\theta}$}
\put(58.75,44.75){$\tilde{\theta}_-$}
\put(62.5,44.75){$\tilde{\theta}_+$}
\put(66.5,7.75){$\tilde{\theta}_-$}
\put(73.5,7.75){$\tilde{\theta}_+$}
\end{picture}
\end{center}
\vspace{-10pt}
\caption{The $E_{\textrm{dressed}} - E_{\textrm{seed}}$ and $J_{\textrm{dressed}} - J_{\textrm{seed}}$ as functions of the angle $\theta_1$}
\vspace{5pt}
\label{fig:dispersion_E}
\end{figure}
In the case of static seeds, the sign $s_\Phi$ is the opposite of the sign of $\beta + 1 / \bar{v}_1$. It follows that $s_\Phi$ is always negative for $\theta_1 < \tilde{\theta}$ in the case of oscillating seeds and $\theta_1 < \tilde{\theta}_-$ in the case of rotating seeds. In the latter case, when $\theta_1 > \tilde{\theta}_+$, $s_\Phi$ is always positive as the mean kink velocity is always subluminal. On the contrary in the former case, there is always a critical $\theta_1$, let it be $\theta_c$ where $\beta + 1 / \bar{v}_1$ vanishes, since the kink velocity diverges as $\tilde{a} \to \pm \omega_1$. The sign $s_\Phi$ is positive when $\theta_1 > \theta_c$ and negative when $\theta_1 < \theta_c$. These are summarized in table \ref{tb:signs}.

The product of the signs of $\tilde{a}$ with $s_\Phi$ directly determines whether the dressed string has larger or smaller energy than its seed, as shown by the equation \eqref{eq:dispersion_dE}. In figure \ref{fig:dispersion_E}, the variation of the energy and angular momentum that got induced by the dressing is plotted versus the angle $\theta_1$.

\end{document}